%% file: paper.tex
\documentclass[11pt]{article}
\usepackage{epsfig,a4}
\newcommand\be{\begin{equation}}
\newcommand\ee{\end{equation}}
\newcommand\bea{\begin{eqnarray}}
\newcommand\eea{\end{eqnarray}}
\newcommand\rarr{\rightarrow}
\newcommand\C{{\rm\kern.24em
    \vrule width.02em height1.4ex depth-.05ex
    \kern-.26em C}}
\newcommand\N{{\rm I\kern-.18em N}}
\newcommand\R{{\rm I\kern-.21em R}}
\newcommand\Z{{\rm\kern.26em
    \vrule width.02em height0.5ex depth 0ex
    \kern.04em
    \vrule width.02em height1.47ex depth-1ex
    \kern-.34em Z}}
\renewcommand\d{{\rm\kern.22em
    \vrule width.02em height1.0ex depth0ex
    \kern-.24em d}}
\newcommand\nn{\nonumber}
\newcommand\fr{\frac}

\newcommand\del{\partial}

\newcommand\tr{\mbox{tr}\,}

\newcommand\real{\mbox{Re}\,}
\newcommand\imag{\mbox{Im}\,}

\newcommand\lbr{\left(}
\newcommand\rbr{\right)}

\newcommand\pbar{\bar{p}}

\newcommand\rhobar{\bar{\rho}}
\newcommand\kf{{\bf k}}
\newcommand\qf{{\bf q}}
\newcommand\lf{{\bf l}}

\newcommand\Af{\mbox{\bf A}}  
\newcommand\Vf{\mbox{\bf V}}  
\newcommand\Ff{\mbox{\bf F}}  

\newcommand\cA{{\cal A}}

\newcommand\cD{{\cal D}}
\newcommand\cS{{\cal S}}
\newcommand\cO{{\cal O}}
\newcommand\cP{{\cal P}}

\renewcommand\O{{\rm\kern.24em
    \vrule width.02em height1.45ex depth-.05ex
    \kern-.26em O}}
\newcommand\Od{{\rm\kern.24em
    \vrule width.02em height1.45ex depth-.05ex
    \kern-.26em O}}
\renewcommand\P{{\rm I\kern-.25em P}}
\newcommand\spommi{{\mbox{\scriptsize\P}}}
\newcommand\soddi{{\mbox{\scriptsize\Od}}}
\newcommand\tpommi{{\mbox{\tiny\P}}}
\newcommand\toddi{{\mbox{\tiny\Od}}}
\newcommand\sbfkl{{\mbox{\scriptsize BFKL}}}

\newdimen\picraise
\newcommand\picbox[1]
{
  \setbox0=\hbox{\input{#1}}
  \picraise=-0.5\ht0 
  \advance\picraise by 0.5\dp0
  \advance\picraise by 3pt     % correct for height of `=' above baseline
  \hbox{\raise\picraise \box0}
}
\begin{document}

\begin{titlepage}
\begin{flushright}
HD-THEP-02-35\\
hep-ph/0306137\\
%\\
\end{flushright}
\hspace*{1cm}
\vfill
\begin{center}
{\LARGE{\bf The Odderon in Quantum Chromodynamics}}\\[.2cm]
\end{center}
\vspace{1.5cm}
\begin{center}
{\bf \Large
Carlo Ewerz
}
\end{center}
\vspace{.2cm}
\begin{center}
{\sl
Institut f\"ur Theoretische Physik, Universit\"at Heidelberg\\
Philosophenweg 16, D-69120 Heidelberg, Germany
\\
\hspace{1cm}\\
email: carlo@thphys.uni-heidelberg.de
}
\end{center}
\vfill
\begin{abstract}
The Odderon is the leading exchange in hadronic scattering processes 
at high energies in which negative charge conjugation and parity 
quantum numbers are transferred in the $t$-channel. 
We review the origin of the Odderon in Regge theory, 
its status in perturbative and nonperturbative Quantum 
Chromodynamics, as well as its phenomenology. 
\vfill
\end{abstract}
%\vspace{5em}
%\hrule width 5.cm
%\vspace*{.5em}
%{\small \noindent 
%$^*$ text
%}
\end{titlepage}

\thispagestyle{empty} 
\hspace{1cm}
\cleardoublepage                                                                

\setcounter{page}{3}

\tableofcontents
\cleardoublepage

\pagenumbering{arabic} 

\section{Introduction}
\label{intro}

Quantum Chromodynamics (QCD) is one of the cornerstones 
of the Standard Model of particle physics. It is the gauge theory 
of color charges describing the interactions of quarks and gluons. 
By now it is beyond any reasonable doubt that QCD is the correct 
microscopic theory of the strong interaction. Quarks and gluons 
interact strongly at large distances which makes it intrinsically 
difficult to calculate observables in QCD to a high precision. 
Therefore the experimental tests of QCD remained on a qualitative 
level for quite some time. The last decade has seen steady progress 
in the area of calculational techniques and many aspects of QCD 
have since been tested experimentally at a really quantitative 
level of a few per cent. We have thus entered an era of precision 
tests of QCD. 
It can be considered a great success of QCD that 
it has passed all these tests. Moreover, the investigation of QCD 
has taught us a lot about quantum field theory in general. 
QCD is thus not only interesting from a phenomenological 
point of view but also because it can serve as a model field 
theory which exhibits an amazingly rich structure. 

Despite the enormous success of QCD there remains a number of  
deep questions to be answered in the field of strong interaction 
physics. The most fundamental problem 
is why quarks and gluons cannot be observed as free particles 
and how exactly they form the observable hadronic bound 
states. A slightly less fundamental but equally challenging 
problem is to understand the high energy behavior of 
hadronic scattering processes. 
The Regge limit of large center--of--mass energy and 
small momentum transfer attracted a lot of interest 
already long before the advent of QCD. 
In pre--QCD times Regge theory was developed and successfully 
applied to high energy scattering. 
The findings of Regge theory in fact belong to the deep truths of 
high energy physics. Regge theory is based on 
a few fundamental assumptions about the scattering amplitude 
like unitarity and analyticity. 
The interaction of the colliding particles is interpreted in 
terms of exchanges between them corresponding to so--called 
Regge poles and Regge cuts. At large center--of--mass energy 
$\sqrt{s}$ the 
leading contribution is given by the exchange of the Pomeron 
(or Pomeranchuk pole) which carries vacuum quantum numbers, 
in particular positive charge parity $C=+1$. 
This exchange was found to result in a slow growth of 
cross sections with energy, 
$\sigma \sim s^{\alpha_{\mbox{\tiny \P}}-1}$, 
with the Pomeron intercept 
$\alpha_{\mbox{\scriptsize\P}} \simeq 1.08$. 
In the 1960's a field theory of interacting Pomerons was 
formulated and widely studied. 
After the discovery of QCD in 1973 
the interest shifted somewhat away 
from Regge theory towards hard scattering processes which 
due to asymptotic freedom can be described in perturbative QCD. 
The interest in the Regge limit was revived during the last decade 
by new experimental data obtained at the HERA collider at DESY. 
Now the main focus is on attempts to derive Regge theory 
from the underlying dynamics of QCD, i.\,e.\ in terms of quarks 
and gluons. Although remarkable progress has been made, 
we are still far from a full understanding of the high energy limit 
of QCD. 

It was for a long time assumed that the cross sections for 
particle and antiparticle scattering become equal at 
very high energies, for example for proton--proton 
and antiproton--proton scattering. This became widely 
known as the Pomeranchuk theorem \cite{Pomeranchuk}. 
Already in 1970 the possibility of processes violating the 
Pomeranchuk theorem was discussed, 
and the corresponding amplitudes were studied 
in detail for the case of asymptotically constant cross sections. 
Motivated by the experimental observation of rising cross 
sections at the highest available energies 
{\L}ukaszuk and Nicolescu \cite{Lukaszuk:1973nt} 
in 1973 considered this possibility also 
in the case of asymptotically rising cross sections 
and established the concept of a Regge exchange with an 
intercept close to one which -- 
in contrast to the Pomeron -- carries odd charge parity $C=-1$. 
Accordingly, the name Odderon was coined for this exchange in 
\cite{Joynson:1975az}. 
The Odderon has experienced a quite varied history. 
It was soon shown that the possible existence of the 
Odderon does not contradict the fundamental theorems of 
Regge theory as was previously often assumed. 
Nevertheless, the Odderon was for a long time widely 
considered a doubtful concept. Even 
if its existence was possible it was thought not to exist or at least 
to be of little importance in reality. 
This scepticism was supported by the experimental results 
which showed only little evidence for the Odderon. 
The search concentrated 
on finding differences between particle cross sections and 
the corresponding antiparticle cross sections at high energy. 
Since the Odderon 
couples differently to particles and antiparticles due to its negative 
charge parity such differences would indicate an Odderon exchange. 
The only trace of the Odderon was 
found in a small difference of the elastic proton--proton 
and proton--antiproton differential cross sections. 
Unfortunately, the statistics of the $p\bar{p}$ data was  
low and the results were thus not unambiguous. 
Other hints were even less convincing and seemed to be 
rather controversial. 

Today the experimental evidence for the Odderon is still 
very unsatisfactory, the best sign still being the difference 
in the differential cross sections for elastic $pp$ and $p\bar{p}$ 
scattering measured at the CERN Intersecting Storage Rings (ISR). 
But the theoretical perspective on the Odderon has 
changed completely. Instead of being thought of as a 
doubtful concept the Odderon is now considered a  
firm prediction of QCD. 
Our QCD--motivated picture of high energy 
scattering is largely based on gluon exchange, 
the Pomeron being a mainly gluonic object 
which in the simplest picture is composed of 
two gluons in a colorless state. 
Although nonperturbative effects necessarily play 
an important r{\^o}le this picture is well established 
also experimentally. From this point of view it would 
be very surprising if the exchange of three gluons in 
a colorless $C=-1$ state, i.\,e.\ an Odderon, would 
not exist. Formally, this picture was supported 
by the investigation of the Regge limit in perturbative QCD: 
In 1980 the existence of a colorless 
three--gluon exchange at high energies with negative 
charge parity was established by Bartels \cite{Bartels:1980pe}, 
Jaroszewicz \cite{Jaroszewicz:1980mq}, and 
Kwieci{\'n}ski and Prasza{\l}owicz \cite{Kwiecinski:1980wb}. 
But even after that it took almost twenty years before the 
corresponding intercept $\alpha_\soddi$ was shown to be close 
to one, thus proving the existence of an Odderon at least in 
perturbative QCD. In addition to this important result 
the investigation of the Odderon in perturbative QCD has 
led to the discovery of beautiful and unexpected 
relations of high energy QCD to the theory of integrable 
models. In this respect the Odderon turned out to be 
a highly interesting object also from a theoretical point of view. 

The rapid progress in the study of the perturbative Odderon 
was in the last few years paralleled by a new development 
in the phenomenology of the Odderon. Earlier 
investigations had mainly concentrated on processes in which 
the Odderon exchange gives one of several contributions 
to the cross section. The corresponding predictions were 
often plagued by the large uncertainties due 
to the unknowns of the other possible exchanges. 
Now the interest has shifted towards more exclusive processes in 
which the Odderon is the only possible exchange, 
typically with the exception of an additional photon 
contribution which is theoretically well under control. 
Hence already the observation of such processes would 
clearly indicate the existence of the Odderon. 
A typical example of such processes is the diffractive 
production of pseudoscalar mesons in electron--proton 
scattering at high energy. The photon which is radiated off 
the electron carries negative charge parity. The positive 
charge parity of the pseudoscalar meson thus requires 
a colorless $C=-1$ exchange producing 
a rapidity gap between the meson and the proton, hence an Odderon. 
The first experimental study \cite{Adloff:2002dw} 
of this process has just been performed by 
the H1 collaboration at the DESY HERA collider, 
and further results are expected for the near future. 
Similar diffractive processes have also been proposed in 
other reactions like proton--(anti)proton collisions or 
photon--photon scattering at high energies. 
The expected cross sections for many of these processes 
are in a range which should make it possible to find the 
Odderon and to study its properties at current and 
future accelerators. 

The high energy limit of QCD has in the recent past attracted 
much interest in a wider context and is presently a 
field in rapid development. At high energies 
the quarks and gluons inside the colliding hadrons 
form a very dense system which is 
largely dominated by nonperturbative effects. 
When probed with sufficiently high momentum, 
however, the partons interact only weakly. 
QCD matter under these conditions has recently 
been termed a color glass condensate: colored 
partons are in a very dense state that resembles 
a glass because its quantum fluctuations are frozen 
over rather long timescales in a frame in which 
the colliding hadrons move very fast. 
It is expected that such a system exhibits new and 
very interesting phenomena. Although the partons interact 
only weakly the usual methods of fixed order perturbation 
theory cannot be applied because of the density of the 
system. Of particular importance is the effect of 
recombinations of different parton cascades in the hadrons. 
Such effects will eventually also tame the growth of 
hadronic cross sections which according 
to unitarity can at most be logarithmic in the high energy 
limit. To understand the details of unitarity restoration 
poses a big challenge. New techniques and different 
approaches to this problem are currently developed. 
Many of them lead to similar results corresponding to 
the perturbative Pomeron when the limit of low densities is taken. 
It will therefore be very important to determine and understand 
the characteristic differences between these approaches by 
going beyond the simplest case of two--gluon exchange. 
It is very likely that the study of the Odderon will be 
very helpful in this context. Although being a more 
theoretical aspect this issue is certainly very important 
for a full understanding of QCD in the high energy limit. 

It is the aim of this article to review the present status of 
the Odderon from a theoretical as well as from a 
phenomenological perspective. We will especially 
emphasize the importance of 
the recent developments regarding the 
perturbative description of the Odderon and of the 
new strategies in its phenomenology. It is our strong 
belief that these aspects offer the most promising 
possibilities for further progress in the near future. 
We start by giving a brief introduction to Regge theory 
in section \ref{pomoddregge} where we discuss in 
particular the asymptotic theorems and 
their relevance to the Odderon. 
We then turn to the theoretical aspects of the Odderon 
in section \ref{theoroddsect}. A large portion of that 
section is devoted to the Odderon in perturbative QCD. 
The discussion of this subject is facilitated by some 
familiarity with the perturbative description of the 
Pomeron of which we present the relevant aspects. 
We then proceed to the Bartels--Kwieci{\'n}ski--Prasza{\l}owicz 
(BKP) equation describing the perturbative Odderon 
and explain its relation to integrable models. 
We present the known solutions to the BKP equation 
and discuss the important question of the Odderon intercept. 
Also in that section we discuss which r{\^o}le the Odderon 
plays in the more general context of high energy QCD. 
We conclude the section with a discussion of some 
results and ideas concerning nonperturbative aspects 
of the Odderon. 
Section \ref{phensect} deals with the phenomenology of 
the Odderon. After presenting general considerations 
we discuss in some detail the experimental 
evidence for the Odderon as it is found in the differential 
cross sections for elastic $pp$ and $p\bar{p}$ scattering. 
We then proceed to review the different observables which 
have been proposed for observing the Odderon and 
describe their advantages and the uncertainties involved 
in the corresponding theoretical predictions. 
Finally, we present our conclusions and some possible 
directions for further study. 

The choice of the material covered in this review is 
influenced by my own interests and prejudices, 
and by the will to keep the review finite. 
I still hope that my attempts at giving fair mention 
to the relevant contributions to the field have not 
failed too badly. 
I apologize to all those whose important work 
is not adequately represented. 

\section{Pomeron and Odderon in Regge Theory}
\label{pomoddregge}
\setcounter{equation}{0} 

The Odderon as well as the Pomeron are objects which originate 
in Regge theory. We will therefore give a brief account of the 
basic concepts of Regge theory, with particular emphasis on those 
issues that are directly relevant to the physics of the Odderon. 
We will then briefly discuss the basic theorems of Regge theory and 
some aspects of the Pomeron. Finally, we describe which place 
the Odderon finds in Regge theory. In particular, we will 
describe the so--called maximal Odderon, that is the maximal 
energy dependence that the Odderon can have without 
violating the Pomeranchuk theorem. 

It should be pointed out that the main focus of the present review 
is on the more recent development related to the Odderon, 
namely its nature in perturbative QCD and its phenomenology 
which will be discussed in the subsequent sections. In favor of 
those aspects we concentrate on the essential issues only in the 
present section on Regge theory 
-- despite the great importance that Regge theory 
has had, and still has, for the understanding of high energy 
scattering and for the development of the concept of the Odderon. 
Therefore no attempt on completeness is made. 
For detailed accounts of Regge theory 
we refer the reader to \cite{Collins:jy}-\cite{Caneschibook}. 

\subsection{Regge Theory and the Pomeron}
\label{reggetheory}

Already before the advent of QCD, hadronic scattering processes 
at high center--of--mass energy were successfully 
described in the framework of Regge theory. 
More specifically, Regge theory is designed 
to describe scattering processes at high center--of--mass 
energy $\sqrt{s}$ much larger than the momentum 
transfer $\sqrt{-t}$ in the reaction. The latter is supposed 
to be of the order of some hadronic mass scale $m_H$, 
\be
 s \gg |t| \simeq m_H^2
\,.
\ee
This limit is known as the Regge limit. 
Regge theory is based on very general properties of the 
scattering matrix, namely its Lorentz invariance, 
its unitarity, and its analyticity. 
The latter postulate determines 
also the properties of the scattering matrix under the 
crossing of initial and final state particles. 
The basic object of interest in Regge theory is the 
scattering amplitude $A$. For a two--to--two scattering 
process $a+b \to c+d$, it is pictorially given by 
\be
\input{reggeabcd.pstex_t}
\vspace*{0.3cm}
\ee
It depends on two of the three Mandelstam variables $s,t,u$ 
defined by the momenta $p_i$ of the four particles as 
\bea
s &=& (p_a + p_b)^2 \\
t &=& (p_a - p_c)^2 \\
u &=& (p_a-p_d)^2
\,,
\eea
which are related via 
\be
\label{studep}
s+t+u=\sum_{i=a}^d m_i^2
\,. 
\ee
One usually chooses as variables the squared center--of--mass 
energy $s$ and the squared momentum transfer $t$, and hence 
considers the scattering amplitude $A(s,t)$. 
The total cross section is related to the elastic forward scattering 
amplitude $A(s,t=0)$ via the optical theorem
\be
\label{opttheorem}
  \sigma_T = 
  \frac{1}{s}\, \mbox{Im}\, A_{\mbox{\scriptsize el}}(s,t=0) 
\,.
\ee
Another observable of interest is the differential cross 
section $d\sigma/dt$ which is obtained from the 
scattering amplitude via 
\be
\label{diffsigmaA}
\frac{d\sigma}{dt} = \frac{1}{16 \pi s^2} \,|A(s,t)|^2
\,.
\ee

The scattering amplitude can be written in a $t$-channel 
partial wave expansion 
\be
\label{partwaves}
A(s,t) = 16 \pi \sum_{l=0}^\infty (2l+1) A_l(t) P_l(z_t)
\ee
with the Legendre polynomials $P_l$ and 
\be
z_t= 1+\frac{2s}{t-4m^2} \simeq 1+\frac{2s}{t} 
\,,
\ee
where we have neglected the masses of the scattering 
particles in the last step. The $A_l(t)$ are called the 
partial waves. 
One then applies the Sommerfeld--Watson transformation 
to write the scattering amplitude as an integral 
over the $t$-channel angular momentum $l$ 
which is understood now as a complex variable\footnote{Here 
we use for simplicity the same symbol $A$ for the scattering amplitude as 
a function of $s$ and as a function of complex angular 
momentum $l$ although these are of course not 
identical functions. Confusion should hardly be possible as 
the variables $s$ for the squared energy and $l$ (or $\omega$ or $J$) 
for angular momentum always indicate which function is meant.}, 
\be
A(s,t) = \frac{1}{2i} \int_C dl (2l+1) \frac{A(l,t)}{\sin \pi l} 
P(l,z_t) 
\,.
\ee
The contour $C$ surrounds the real axis from $0$ 
to $\infty$. 
The Legendre polynomials $P_l$ have a natural 
analytic continuation to complex $l$, $P_l(z_t) \to P(l,z_t)$ 
when they are written in terms of hypergeometric functions. 
In addition, one has to find an analytic continuation $A(l,t)$ 
of the functions $A_l$. In order to find a unique continuation of these 
one has to impose a large $|l|$ behavior in the complex $l$-plane. 
It turns out that this is not possible for a single function $A(l,t)$. 
Instead one has to introduce signatured partial wave amplitudes 
$A^{(+1)}(l,t)$ and  $A^{(-1)}(l,t)$ that are the analytic 
continuations for even and odd partial wave amplitudes, 
respectively. The Sommerfeld--Watson integral then becomes 
\be
\label{sommwat1}
A(s,t) =  \frac{1}{2i} \int_C dl \frac{(2l+1)}{\sin \pi l} 
\sum_{\eta =\pm 1}
\frac{\eta + e^{-i\pi l}}{2} \, A^{(\eta)}(l,t) P(l,z_t)
\,,
\ee
where $\eta=\pm 1$ is called the signature of the partial wave. 
We now deform the contour $C$ to a contour $C'$ that runs 
parallel to the imaginary axis with $\real{l}=-1/2$. 
In general the function $A^{(\eta)}(l,t)$ may have singularities 
in the complex $l$-plane, for example simple poles situated 
at $l=\alpha^{(\eta)}_i (t)$. 
For the moment we will consider only 
simple poles, but will discuss other types of singularities later. 
When deforming the 
contour $C$ to $C'$ we collect $2\pi i$ times the residue 
of a pole when the contour crosses it. 
Accordingly, we get 
\bea
\label{swintcont2}
A(s,t) &=&  \frac{1}{2i} \int_{-\frac{1}{2}-i\infty}^{-\frac{1}{2}+i\infty}
dl \frac{(2l+1)}{\sin \pi l} 
\sum_{\eta =\pm 1}
\frac{\eta + e^{-i\pi l}}{2} \, A^{(\eta)}(l,t) P(l,z_t)
\nn\\
&& 
+ \sum_{\eta =\pm 1}\sum_i
\frac{\eta + e^{-i\pi \alpha^{(\eta)}_i(t)}}{2} 
\frac{\tilde{\beta}^{(\eta)}_i(t)}{\sin \pi \alpha^{(\eta)}_i (t)}
P(\alpha^{(\eta)}_i (t),z_t)
\,.
\eea
The $\tilde{\beta}^{(\eta)}_i (t)$ are 
$\pi (2 \alpha^{(\eta)}_i (t) +1)$ 
times the residues of the poles. The simple poles are called 
Regge poles with signature $\eta$. 
It turns out that the remaining integral in the first 
line of (\ref{swintcont2}) vanishes in the limit $s\gg |t|$ 
due to the behavior of the Legendre polynomials in this limit. 
Using that behavior one can also simplify the second line 
in (\ref{swintcont2}) and obtains for the amplitude 
\be
A(s,t) = \sum_{\eta =\pm 1}\sum_i A_i^{(\eta)}(s,t)
\,,
\ee
and $A_i^{(\eta)}$ denotes the contribution of a pole 
of signature $\eta$ located at $\alpha_i(t)$. 
Its contribution to $A(s,t)$ is (dropping the index $i$) 
\be
\label{polecontr}
A^{(\eta)} (s,t) =
\beta^{(\eta)} (t) \, \Gamma(-\alpha^{(\eta)} (t))\,
\eta\, \xi^{(\eta)}(t) 
\left(\frac{s}{s_0}\right)^{\alpha^{(\eta)} (t)}
\,,
\ee
where $\Gamma$ is the Euler gamma function. 
The factors 
\be
\xi^{(\pm 1)} (t) = 1 \pm e^{-i\pi \alpha(t)} 
\ee
are called signature factors. 
In eq.\ (\ref{polecontr}) we have for simplicity absorbed 
several $t$-dependent factors into the factor $\beta$. 
It can be shown that the $\beta(t)$ are real--valued and 
contain as one factor the residue of the pole. In addition 
we have introduced a scale $s_0$ with respect to which 
the squared energy $s$ is measured. A priori $s_0$ 
cannot be fixed, often one chooses $s_0=1\,\mbox{GeV}^2$. 
Note that instead of $l$ one often uses $\omega$ or $J$ to denote the 
complex angular momentum, and one then 
speaks of the $J$-plane or $\omega$-plane etc. 
We will make use of both notations depending on the context. 

The high energy limit of the 
amplitude (\ref{polecontr}) can be understood as the 
exchange of a so--called reggeon, namely 
an object of angular momentum $\alpha(t)$ 
in the $t$-channel. In general this is not a particle since 
it occurs only as an exchange, and in general it has non--integer 
spin. The general definition of a reggeon is that its exchange 
in the $t$-channel leads to a power--like growth of the 
amplitude with $s$ with exponent $\alpha(t)$. 
It turns out that the amplitude can 
be written in factorized form, that means that $\beta$ 
in (\ref{polecontr}) is a product of two factors which 
depend only on the coupling of the exchanged object to 
the scattering particles
\be
\label{factcouplregge}
\beta(t) = \beta_{ac}(t) \beta_{bd}(t)
\,,
\ee
pictorially, 
\be
\label{reggefactpict}
\input{reggefact.pstex_t}
\,
\ee
These couplings can be taken from one process 
to another when for example the $a\to c$ 
transition is present also in that process and 
the exchanged object is the same. 
The remaining part of the amplitude (\ref{polecontr}) 
is universal and only associated with the exchanged object. 

If one now considers the $t$-channel of the above 
scattering process one would expect to find poles in 
the amplitude reflecting the resonant production of 
real particles with integer spin. It is in fact found that 
the known light mesons ($\rho$, $\omega$, $f_2$, $a_2$, etc.) 
can be associated with 
Regge trajectories (and the same actually holds for baryons as well). 
Their masses $m_i$  and spins $J_i$ 
are related via $\alpha_R(m_i^2) = J_i$, 
and it is found empirically that the function 
$\alpha_R$ is to a remarkable accuracy linear, 
\be
\label{rhotraj}
\alpha_R(t) = \alpha_R(0) + \alpha'_R t
\,.
\ee
Experimentally one finds the parameters 
\bea
\alpha_R(0) &=& 0.5 \\
\alpha'_R &=& 0.9\,\mbox{GeV}^{-2}
\,.
\eea
$\alpha_R(0)$ is called the intercept of the Regge 
trajectory, and $\alpha'_R$ is its slope. 
The exchange in the $s$-channel process associated 
with a reggeon comprises the whole Regge 
trajectory $\alpha_R(t)$, and thus contains 
all possible meson exchanges with the quantum numbers 
of that Regge trajectory. 
The $\rho$-trajectory for example contains mesons 
of isospin $I=1$ and even parity related via the trajectory 
$\alpha_R$. 

From the optical theorem one can find the 
energy dependence that a Regge pole at $\alpha(t)$ 
implies for the total cross section, 
\be
\label{sigtotalpha}
\sigma_T \sim s^{\alpha(0)-1}
\,.
\ee
The exchange of mesonic Regge trajectories with intercept 
$\alpha_R(0) =0.5$ hence decreases with increasing 
energy. Among the known mesons there is no family 
of mesons that would belong to a Regge trajectory of higher 
intercept. 
It is empirically found, however, that at high energy total hadronic 
cross sections rise slowly with energy. This rise cannot be 
associated with a mesonic reggeon exchange. Instead, it is 
attributed to the Pomeron ($\P$) which is by definition the 
leading $t$-channel exchange with vacuum quantum 
numbers. If the Pomeron is assumed to be a simple Regge 
pole it corresponds to the rightmost singularity with vacuum 
quantum numbers in the complex angular momentum plane. 
One finds that the data for total hadronic cross sections are excellently 
described by a reggeon pole (see above) plus a 
Pomeron pole with trajectory 
\be
 \alpha_\spommi (t) = \alpha_\spommi (0) + 
 \alpha'_\spommi t
\ee
with $\alpha_{\mbox{\tiny \P}}=1.09$ \cite{Groom:in}, 
and this is identified with the soft Pomeron \cite{Donnachie:1992ny}. 
The value of the Pomeron slope was found 
\cite{Jaroszkiewicz:ep} to be 
$\alpha'_\spommi = 0.25\,\mbox{GeV}^{-2}$. 
This Pomeron slope and the linearity were 
confirmed at least at low values 
of $t$ in the measured shrinkage of the forward peak 
in elastic $pp$ and $p\bar{p}$ scattering at the ISR and 
at the Tevatron.
We will come back to questions related to the Pomeron 
trajectory in section \ref{nonpertoddsect} when we discuss 
the possible behavior of the Odderon trajectory. 
More recently it was found that actually two Pomeron poles 
are required to fit all presently available data, where in addition to 
the soft Pomeron one introduces a hard Pomeron with intercept 
close to $1.4$ \cite{Donnachie:1998gm,Donnachie:2001xx}. 
Note however that a wide range of models for the 
behavior of the scattering amplitude were investigated and 
fitted to the data in \cite{Cudell:2001pn} with the result that 
the best fit to all available forward ($t=0$) data corresponds 
to a logarithmic behavior of the scattering amplitude. This 
fit is slightly better than the one corresponding to 
the power--like behavior of the 
Pomeron pole. In our opinion the 
current data cannot really distinguish both possibilities, and 
the question of the nature of the Pomeron singularity remains 
open. 

The Pomeron pole is situated near $1$ and 
has positive signature. Starting from (\ref{polecontr}) 
one can show that its contribution to the scattering 
amplitude is predominantly imaginary at small $t$, 
in accordance 
with the fact that it dominates the total hadronic 
cross section via the optical theorem (\ref{opttheorem}) 
into which the imaginary part of the scattering amplitude 
enters. As a Feynman rule for the exchange of a Pomeron 
pole in (\ref{reggefactpict}) one finds the universal behavior 
\be
\label{pomprop}  
  (-i) i 
  \left( \frac{-i s}{s_0}\right)^{\alpha_\tpommi (t)-1} 
\,, 
\ee
which has to be multiplied by suitable (real--valued) 
couplings to the external particles. 
In (\ref{pomprop}) we have left the factors of $i$ for later 
comparison with the Odderon propagator. 
The overall sign of the Pomeron amplitude is fixed by the 
requirement that cross sections are positive. 

Let us briefly turn to the question of other types of Regge 
singularities. So far we have considered simple poles in the 
complex angular momentum plane that give rise to a 
power--like behavior of the scattering amplitude. 
Besides the simple pole the most important possibility 
is that of a Regge cut. Also a cut gives a contribution to 
the scattering amplitude when we go from 
(\ref{sommwat1}) to (\ref{swintcont2}) by deforming the 
integration contour. If the cut starts at $\alpha_c(t)$ 
in the high energy limit 
its contribution is obtained from the discontinuity 
across the cut, 
\be
A_c(s,t) \sim \int^{\alpha_c(t)} dl (2l+1) 
\frac{\mbox{disc} A(l,t)}{\sin \pi l} s^l
\,,
\ee
which leads to a large-$s$ behavior 
\be
A_c(s,t) \sim s^{\alpha_c(t)} \log^{-\gamma(t)} s
\,.
\ee
The power $\gamma$ can be related to the 
behavior of $A(l,t)$ in the vicinity of $\alpha_c(t)$. 
The power--like behavior thus receives logarithmic 
corrections in the case of a Regge cut. 
The case of a double or triple pole will be discussed in section 
\ref{maxoddsect} where we will be concerned with the 
so--called maximal Odderon. 

\subsection{Crossing and the Odderon}
\label{crossoddsect}

A very important consequence of analyticity is crossing symmetry. 
Let us write the amplitude for the scattering process 
\be
\label{abcd}
a + b \to c + d
\ee
as $A^{a+b\to c + d}(s,t,u)$, where we have reinstated the 
dependence on all three Mandelstam variables with the 
understanding that they are not independent, see (\ref{studep}). 
In the physical region of the process (\ref{abcd}) we have 
$s>0$ and $t,u<0$. The amplitude is an analytic function 
of the three variables and we can analytically continue it to the 
physical region of the $t$-channel process 
\be
a +\bar{c} \to \bar{b} + d
\,,
\ee
or to the physical region for the $u$-channel process
\be
\label{uchannrect}
a + \bar{d} \to \bar{b} +c
\,.
\ee
We will be especially interested in the latter case, where the physical 
region is $u>0$ and $s,t<0$. Analyticity then implies that 
\be
\label{ucrossana} 
A^{a + \bar{d} \to \bar{b} +c}(s,t,u) = A^{a+b\to c + d}(u,t,s)
\,.
\ee

The amplitude $A(s,t)$ for the process (\ref{abcd}) 
has branch cuts corresponding to particle thresholds. If the 
masses of the scattering particles are equal, $m_i=m$ for example, 
there will be thresholds at $s=4m^2, 9m^2, \dots$ with branch 
cuts starting at these points. At high $s$ one therefore has to 
specify how the physical amplitude is obtained in the $s$-plane. 
The prescription is to approach the cut along the positive axis 
from above, and the physical amplitude is obtained as 
\be
\label{asepsi}
 \lim_{\epsilon \to 0} A(s+i \epsilon,t)
\,.
\ee
In addition to the thresholds mentioned above the amplitude 
will have cuts on the negative real axis in the $s$-plane related 
to $u$-channel effects starting at $u=4m^2,\dots$, which translates 
into their position in $s$ via (\ref{studep}). 
A prescription analogous to (\ref{asepsi}) holds for the $t$- and 
$u$-channel processes. 

Let us now consider an elastic scattering process, 
\be
\label{abab}
a + b \to a + b
\,,
\ee
with the corresponding amplitude $A^{ab}(s,t)$. 
The elastic process 
\be
\label{abbabb}
a +\bar{b} \to a +\bar{b}
\ee
with amplitude $A^{a\bar{b}}(s,t)$ can be obtained 
from the former by crossing to the $u$-channel. 
We now define two amplitudes $A_\pm$ by
\be
\label{Apmdef}
A_\pm (s,t) = \frac{1}{2}\,(A^{ab}(s,t) \pm A^{a\bar{b}}(s,t))
\,. 
\ee
Under the crossing from the $s$-channel process (\ref{abab}) 
to the $u$-channel process (\ref{abbabb}) the amplitude 
$A_+$ evidently remains unchanged whereas 
the amplitude $A_-$ changes sign. 
Accordingly they are called even--under--crossing and 
odd--under--crossing amplitudes. 

We observe that the amplitude $A_+$ is the same 
for particle--particle and particle--antiparticle 
scattering and thus corresponds to an exchange of 
even (or positive) 
$C$ parity, $C=+1$. The amplitude $A_-$ changes 
sign when going from the particle--particle to the 
particle--antiparticle scattering process and can hence 
be understood to have odd (or negative) $C$-parity, 
$C=-1$. 
The amplitude $A_+$ hence has vacuum quantum numbers, 
and we have already seen that it is dominated at high energy 
by the Pomeron. It can be shown that the exchange of the 
$\rho$ or $\omega$ reggeon 
trajectory (\ref{rhotraj}) is odd under crossing and 
hence contributes to the amplitude $A_-$. The mesonic 
Regge trajectory has an intercept of $\alpha_R(0) = 0.5 $ 
and if it were the leading contribution to $A_-$ the 
odd--under--crossing amplitude would become negligible 
in the Regge limit. Analogously to $A_+$ it is possible that 
there is in addition 
another Regge singularity in $A_-$ whose contribution 
does not vanish rapidly at high energy, 
i.\,e.\ which has an intercept close to $1$. 
This contribution is called the Odderon ($\Od$), 
and is defined as a contribution to the odd--under--crossing 
($C=-1$) amplitude $A_-$ that does not vanish relative 
to the Pomeron contribution or does so only slowly with a 
small power of $s$ or logarithmically in $s$. 
Besides negative charge parity $C$ the Odderon also 
carries negative parity $P=-1$. 
The Odderon was for the first time discussed for the case of a 
theory with asymptotically increasing cross sections in 
\cite{Lukaszuk:1973nt}. 
Originally the name Odderon was attributed only to a 
simple Regge pole at $J=1$ \cite{Joynson:1975az}, 
but it is now used for any type of singularity with those 
properties. 

Let us now consider the phase of the Odderon amplitude. 
When we perform the analytic continuation $s \to s e^{i \pi}$ 
in the amplitude $A^{ab}(s,t)$ we arrive at the complex 
conjugate of the amplitude $A^{a\bar{b}}(s,t)$ for the $u$-channel 
process, 
\be
\label{stouel}
A^{ab}(s  e^{i \pi},t) = (A^{a\bar{b}}(s,t) )^*
\,.
\ee
This can be easily seen from (\ref{ucrossana}) when we 
take into account the relation (\ref{studep}) 
between $s,t$ and $u$. In the Regge limit of both reactions 
$t$ is small as are the masses $m_i$. That 
the analytic continuation $s \to s e^{i \pi}$ brings us to 
the complex conjugate of the $u$-channel amplitude 
is then readily seen from the way in which the 
physical amplitude is defined for the two processes, 
see (\ref{asepsi}), and from the fact that for the 
physical amplitude we have 
$A(s+i\epsilon,t) = (A(s-i\epsilon,t))^*$. 
Let us assume that the amplitude $A_-(s,t)$ is in fact 
dominated by a simple Regge pole situated close to $1$. 
One can now apply (\ref{stouel}) to the amplitude $A^{ab}$ and 
use its decomposition into $A_+$ and $A_-$. If one further 
takes into account that the amplitude $A_+$ is dominated 
by the Pomeron and hence predominantly imaginary one 
can derive that the amplitude $A_-$ associated to the Odderon 
is predominantly real, which is also in agreement with the negative 
signature of the Odderon according to (\ref{polecontr}). 
Stated more generally, any pole 
with negative $C$ parity leads to an additional phase 
$e^{i\pi/2}$ in the amplitude compared to a pole of 
positive $C$ parity at the same position. 
Accordingly, the Feynman rule for Odderon exchange is 
\be
\label{oddprop}
  (-i) \eta_\soddi 
  \left( \frac{-i s}{s_0}\right)^{\alpha_\toddi (t)-1} 
\,,
\ee 
to be compared with the rule (\ref{pomprop}) for 
Pomeron exchange. 
Here we have an additional phase factor $ \eta_\soddi = \pm 1$ 
which a priori cannot be fixed. This is because for the Odderon 
there is no positivity constraint as there was for the Pomeron. 

The first discussion of an Odderon in \cite{Lukaszuk:1973nt} 
did not refer to a simple Regge pole but rather to a more complicated 
singularity of $A_-$ at $J=1$ in the complex angular momentum plane 
which we will describe in section \ref{maxoddsect}. 
In \cite{Kang:1974gt} the case of a general singularity exactly at 
$J=1$ was considered and an interesting derivative relation 
for $A_-$ of the form 
\be
\frac{\pi}{2} \frac{\del}{\del \ln s} \real 
\left[\frac{A_-(s,t)}{s}\right] = 
- \left[ 1 
- \frac{1}{3} \left( \frac{\pi}{2}\frac{\del}{\del \ln s} \right)^2 
- \frac{1}{45} \left( \frac{\pi}{2}\frac{\del}{\del \ln s} \right)^4 - 
\cdots \right] 
\imag \left[\frac{A_-(s,t)}{s}\right] 
\ee
was found from the Sommerfeld--Watson transformation 
together with a relation for the amplitude $A_+$, 
\be
\real \left[\frac{A_+(s,t)}{s}\right] = 
\left[ \frac{\pi}{2}\frac{\del}{\del \ln s} 
+ \frac{1}{3} \left( \frac{\pi}{2}\frac{\del}{\del \ln s} \right)^3 
+ \frac{2}{15}  \left( \frac{\pi}{2}\frac{\del}{\del \ln s} \right)^5 
+ \cdots \right] 
\imag \left[\frac{A_+(s,t)}{s}\right] 
\,.
\ee
From these relations one can derive a number of constraints, for 
example on the high energy behavior of the cross section differences 
of particle--particle and particle--antiparticle scattering. 
We will encounter some of these relations in the next section 
as special cases of asymptotic theorems.  

\subsection{Asymptotic Theorems and the Odderon}
\label{basicoddsect}

Let us now come to the general theorems that have been 
derived in the framework of Regge theory and axiomatic field theory, 
and to their relation to the Odderon in particular. 
Many of these theorems have been derived at a time when the 
Odderon had not yet been invented or was still an almost unknown 
concept. At that time it was therefore often naturally assumed 
that the mesonic reggeon trajectory is the leading contribution 
to the odd--under--crossing amplitude $A_-$, and it was 
hence assumed that necessarily $A_- \to 0$ in the Regge limit. 
Obviously, this assumption excludes the existence of an Odderon. 
Unfortunately, many theorems are still widespread in the literature 
in their simplified form which uses that assumption. 
Especially when the Odderon is discussed a very precise 
formulation of the asymptotic theorems is needed. 

In this section we will state and discuss the theorems for the 
case of $pp$ and $p\pbar$ scattering only. In most cases 
the theorems also hold for other particle--particle and the 
corresponding antiparticle--particle processes. 

A very important theorem is the Froissart--Martin theorem 
\cite{Froissart:ux,Martin:1962rt,Martin:1965jj} which is 
a consequence of unitarity. 
It states that the growth of total hadronic cross sections is 
at most logarithmic with the energy. Specifically, 
\be
\label{fmtheorem}
\sigma_T (s) \le \frac{\pi}{m_\pi^2} \log^2 \left( \frac{s}{s_0}\right)
\,,
\ee
where $s_0$ is an a priori unknown scale. If we assume a
reasonable hadronic scale, $s_0 \simeq 1\,\mbox{GeV}^2$ we find 
that (\ref{fmtheorem}) implies an extremely high 
upper bound of $\sim 10\,\mbox{barns}$ at Tevatron energies, 
for example, which is far away from the actual cross sections. 
It should be pointed out that the Froissart--Martin bound applies 
only to the scattering of stable hadrons. Strictly speaking, it does not 
apply to real or virtual photons, for example. It is widely assumed though 
that a similar bound holds also in this case. 
The Froissart--Martin theorem applies to particle as well as to 
antiparticle scattering. 

The idea of Regge poles at $\alpha(t)$ larger than 
one seems to be incompatible with the Froissart--Martin theorem. 
As we have seen in (\ref{sigtotalpha}) such 
a Regge pole induces an energy dependence 
of the total cross section $\sigma_T \sim s^{\alpha(0)-1}$, and 
hence grows with energy like a power of $s$ if $\alpha(0)-1 >0$. 
Eventually this will violate the asymptotic bound (\ref{fmtheorem}), 
although in practice only at extremely high energies. 
However, at higher energies one has to take into account 
not only the single exchange of that Regge pole in the $t$-channel 
but also the contribution of its iterated (double, triple, etc.) 
exchange to the amplitude. It is the widely accepted, though 
not strictly proven, understanding that the multiple exchange of Regge 
poles (and cuts) will eventually unitarize the scattering 
amplitude, and will lead to a structure in which all singularities 
in the complex angular momentum plane lie below $1$. 
A model of how this might happen was described for example 
in \cite{ChengWu}. 

Next we turn to the Pomeranchuk theorem. 
The original formulation of the theorem 
by Pomeranchuk \cite{Pomeranchuk} was made under 
the assumption that $A_- \to 0$ at high energies. 
With this assumption he showed that 
\be
\label{pomtheorem}
\Delta \sigma = \sigma_T^{\bar{p}p} - \sigma_T^{pp} 
\mathop{\longrightarrow}_{s \to \infty}
0
\,.
\ee
Note that 
\be
\Delta \sigma \sim \frac{1}{s} \,\imag A_-
\,, 
\ee
so that the assumption $A_- \to 0$ is in fact crucial for the 
original form of the Pomeranchuk theorem. 
Already in 1970 possible violations of the Pomeranchuk theorem 
were discussed \cite{Anselm:ag}--\cite{Vishnevsky:jb}, 
and the corresponding amplitudes were investigated in detail 
for the case of asymptotically constant cross sections in 
\cite{Gribov:1970,Gribov:1970ip,Gribov:1970iq}. 
Later the Pomeranchuk theorem was investigated in more detail and 
also formulated for the case that $A_-$ does not 
vanish at high energies. It was shown 
\cite{Eden,Grunberg:mc,Grunberg:xg} that 
\be
\label{genpomtheorem}
\frac{\sigma_T^{\bar{p}p}}{\sigma_T^{pp} } 
\, \mathop{\longrightarrow}_{s \to \infty} \,
1
\,,
\ee
which can be called the general Pomeranchuk theorem. 
That the two forms (\ref{pomtheorem}) and (\ref{genpomtheorem}) 
are not equivalent can be easily seen. Consider the 
following example which assumes a simple 
behavior of the cross sections: 
\bea
\sigma_T^{pp} &=& A \log^2 s + B \log s + C \\
\sigma_T^{\bar{p}p} &=& A \log^2 s + B' \log s + C'
\,.
\eea
This example actually gives rise to the idea of the 
maximal Odderon which we will discuss in the next section. 
If $B \neq B'$ the general Pomeranchuk theorem 
(\ref{genpomtheorem}) is satisfied, but the original 
Pomeranchuk theorem (\ref{pomtheorem}) 
is violated, and instead in this particular example one even has 
$|\Delta \sigma | \to \infty$ for 
$s \to \infty$. 
In general one can use the Froissart--Martin 
theorem to show \cite{Roy:pe,Eden2} that 
the growth of $|\Delta \sigma | $ can be at 
most single--logarithmic at $s \to \infty$ 
\be
\label{deltasigmalog}
 \left| \Delta \sigma \right| \le \mbox{const} \cdot \log s
\,.
\ee
It is further clear that the difference $\Delta \sigma$ of the 
$pp$ and $p\pbar$ total cross sections has to stay smaller than 
the cross sections themselves, 
\be
\Delta \sigma < \sigma^{pp}_T, \,\sigma^{p\pbar}_T
\,.
\ee
Assuming simple Regge poles for the Pomeron and for the 
Odderon with intercepts $\alpha_\spommi$ and 
$\alpha_\soddi$, respectively, we can then conclude that 
$\alpha_\soddi \le \alpha_\spommi$. 

The Cornille--Martin theorem is the analogue 
of the Pomeranchuk theorem for the differential cross sections. 
Assuming again $A_- \to 0$ at large energies one finds 
\be
\Delta \left( \frac{d\sigma}{dt} \right) = 
\frac{d\sigma^{\bar{p}p}}{dt} - \frac{d\sigma^{pp}}{dt} 
\, \mathop{\longrightarrow}_{s \to \infty} \,
0
\,,
\ee
whereas the actual Cornille--Martin theorem \cite{Cornille:wh} 
again holds for the ratio of the two differential cross sections and 
reads 
\be
\label{cmtheoremreal}
\frac{d\sigma^{\bar{p}p}/dt}{d\sigma^{pp}/dt } 
\, \mathop{\longrightarrow}_{s \to \infty} \,
1
\,.
\ee
It holds for $t$-values inside the diffraction peak 
(which in turn shrinks with increasing energy). 

The Khuri--Kinoshita theorem \cite{khuri} 
deals with the so--called $\rho$-parameter 
which is defined as the ratio of the real and imaginary parts of the 
forward scattering amplitude, 
\be
\label{rhodef}
\rho (s) = \frac{\real A(s,t=0)}{\imag A(s,t=0)}
\,.
\ee
One defines the difference of the $\rho$ parameters 
for $\pbar p$ and $pp$ scattering as $\Delta \rho$, 
\be
\label{deltarhodef}
\Delta \rho (s)= \rho\,^{\bar{p}p}(s) - \rho\,^{pp}(s)
\,. 
\ee
The Khuri--Kinoshita theorem makes statements about the 
possible asymptotic values of $\Delta \rho$ for 
different possible behaviors of the scattering amplitudes. 
Under the assumption $A_-\to 0$ at $s\to \infty$ one finds 
that $\Delta \rho \to 0$. If, however, $A_- \not \to 0$ 
it is possible that
\be
 \Delta \rho \mathop{\not \rightarrow}_{s \to \infty} 0
\,,
\ee
although it is not necessary. 

Finally, there is an interesting theorem on the correlation of the 
signs of $\Delta \sigma$ and $\Delta \rho$. 
It was proposed in \cite{Giffon:1995ah} 
for a restricted class of asymptotic behaviors and generalized 
in \cite{Gauron:1996yt}. 
The authors make the rather general but still non--trivial assumption 
that the scattering amplitudes do not oscillate indefinitely, 
i.\,e.\ that they approach their limiting behavior monotonically 
above some energy $s_1$. With this assumption 
the following result holds. If 
\be
|\real A_-(s) | > \,\mbox{const.} \cdot s 
\,,
\ee
as is for instance the case for an Odderon pole, one finds that 
\be
\delta(s) = \Delta \sigma (s) \cdot \Delta \rho(s) < 0
\,.
\ee
If on the other hand the leading singularity in $A_-$ 
is a Regge pole with intercept $\alpha$, and if 
$0<\alpha<1$ (as is the case for the mesonic Regge 
trajectory), then $\delta(s) >0$. 
In this way a measurement of the sign of $\delta(s)$ can 
help identify the Odderon in forward scattering data 
and total cross sections. 

\subsection{The Maximal Odderon}
\label{maxoddsect}

The concept of the maximal Odderon is based on a the idea of a 
maximality principle for the strong interaction. This principle 
postulates that the strong interactions are at high energies 
as strong as the asymptotic theorems allow them to be. 
According to this idea the cross sections should at high 
energies saturate the asymptotic bounds in their functional 
form. This means for example that according to the 
maximality principle the behavior of the total cross section 
should be 
\be
\sigma_T \longrightarrow C \log^2 s 
\ee
for $s \to \infty$ with a positive constant $C$. 
Similarly, one would expect that 
\be
\Delta \sigma  \longrightarrow C_\Delta \log s 
\ee
as  $s \to \infty$. 
The maximality principle was first formulated in 
\cite{Lukaszuk:1973nt} where it was motivated 
by the discovery of rising total $pp$ cross section at the CERN ISR. 
The authors applied the idea of maximality then 
also to the odd--under--crossing amplitude $A_-$ 
and suggested that it could grow at high energies as 
rapidly as $A_-(s) \sim s \log^2 s$ which clearly corresponds 
to an Odderon. Historically the concept of the Odderon hence 
goes back to the assumption of the maximality of the strong 
interaction. This idea was studied further in \cite{Kang:1974gt}, 
and in \cite{Joynson:1975az} also other types of asymptotic 
behaviors were studied like for example an Odderon of 
Regge pole type. The type of Odderon that follows from the 
assumption of the maximality principle has been further 
elaborated in \cite{Gauron:1985gj,Gauron:1986nk,Gauron:1990cs} 
and is now known as the maximal Odderon. In \cite{Gauron:1992zc} 
it was shown that the maximal Odderon hypothesis is consistent 
with all general principles of axiomatic field theory. 

Here we will describe the amplitudes associated with the 
maximal Odderon in the form suggested in 
\cite{Gauron:1986nk,Gauron:1990cs}. 
The idea of the approach is to have two types of contributions 
to the scattering amplitude. One type of contribution is given 
by simple Regge poles and Regge cuts 
with an intercept not larger than one. 
These terms are supposed to be the most important ones at 
energies up to ISR energies of about $\sqrt{s} \simeq 50 \,\mbox{GeV}$. 
In addition there is another contribution that functionally 
saturates the asymptotic bounds and is supposed to become
the dominant one at energies in or above the TeV range. 
The principle of maximality in this sense is applied not 
only to the amplitude $A_-$ but also to the amplitude 
$A_+$ which is conventionally thought to be dominated 
by a Pomeron pole with intercept above one. Here, however, 
one assumes the Pomeron pole to be situated exactly at one, 
and the second type of contribution saturates the asymptotic 
bound which follows from the Froissart--Martin theorem. 
Accordingly, the latter contribution has been termed the 
Froissaron. 

Specifically, the amplitudes $A_+$ and $A_-$ are split 
into their normal part $A_\pm^N$ containing the conventional 
Regge poles and cuts (though here only with intercept at or below one), 
and their asymptotic parts $A_\pm^{AS}$ which contain 
the Froissaron and maximal Odderon terms, 
\be
A_\pm = A_\pm^N + A_\pm^{AS}
\,.
\ee
The amplitudes $A_\pm^{AS}$ are then constructed in such a way that 
at $t=0$ they have the form 
\bea
\label{froissaronterm}
A_+^{AS} &=& 
i s \left( F_1 \log^2 \bar{s} + F_2 \log \bar{s} + F_3 \right)
\\
\label{maxodderonterm}
A_-^{AS}&=& 
 s \left( O_1 \log^2 \bar{s} + O_2 \log \bar{s} + O_3 \right)
\,,
\eea
where the $F_i$ and $O_i$ are constants and 
\be
 \bar{s} = \frac{s}{s_0} \exp\left(-\frac{1}{2} i \pi\right) 
\,,
\ee
with the choice $s_0=1\,\mbox{GeV}^2$. 
The term (\ref{froissaronterm}) refers to the Froissaron, whereas 
(\ref{maxodderonterm}) refers to the maximal Odderon term. 
At $t=0$ the Froissaron corresponds to a triple pole at $J=1$ 
in the complex angular momentum plane, 
and the maximal Odderon term corresponds to a double pole 
at $J=1$. In \cite{Gauron:1985gj} the amplitudes were extended 
to $t\neq 0$ by choosing appropriate $t$-dependent $J$-plane 
singularities that collapse at $t=0$ to the triple and double pole 
of the Froissaron and maximal Odderon, respectively. 
This cannot be done arbitrarily. Instead, the $t$-dependence 
away from the forward direction is controlled by the 
Auberson--Kinoshita--Martin (AKM) theorem \cite{Auberson:ru}. 
Originally, it is formulated for general amplitudes having 
Froissart growth, but it can also be applied to amplitudes with 
definite crossing symmetry. It then 
states that in the limit $s\to \infty$ 
\be
\label{akmformula}
A_\pm (s,t) \longrightarrow A_\pm (s,0) g_\pm (\tau)
\,,
\ee
where $g_\pm(\tau)$ are entire functions of order $1/2$ of 
$\tau^2$ with the scaling variable 
\be
\tau = \mbox{const}\cdot \sqrt{-t} \, \log s
\,.
\ee
In \cite{Gauron:1985gj,Gauron:1986nk} it was shown that 
the simplest possible choice in agreement with the AKM theorem 
is obtained by choosing the following singularities in the 
complex angular momentum plane:
\bea
\label{froissaroncont}
A^{AS}_+(J,t) &=& 
\frac{\beta_+(J,t)}{\left[ (J-1)^2 - tR_+^2\right]^{\frac{3}{2}}}
\\
A^{AS}_-(J,t) &=& 
\frac{\beta_-(J,t)}{(J-1)^2 - tR_-^2}
\,,
\eea
with real and positive constants $R_\pm$. The residue functions 
$\beta_\pm$ are assumed to be slowly varying functions of $J$ 
and to have a simple exponential $t$-dependence. With a 
suitable choice for the branch cuts in (\ref{froissaroncont}) 
one can then perform the Mellin transformation leading 
to the amplitudes as functions of $s$. 
They become 
\bea
\label{froiterm}
\frac{1}{is} A_+^{AS} (s,t) &=& 
F_1 \log^2 \bar{s} \, \frac{2 J_1(K_+\bar{\tau})}{K_+\bar{\tau}} 
\exp(b_1^+ t) 
+ F_2 \log \bar{s} \, J_0(K_+\bar{\tau}) \exp(b_2^+ t) 
\nn \\
&&+ F_3 [ J_0(K_+\bar{\tau}) - K_+\bar{\tau}J_1(K_+\bar{\tau})]
\exp(b_3^+ t) 
\\
\label{moterm}
\frac{1}{s} A_-^{AS} (s,t) &=& 
O_1 \log^2 \bar{s} \, \frac{\sin (K_-\bar{\tau})}{K_-\bar{\tau}}
\exp(b_1^- t) 
+O_2 \log \bar{s} \, \cos(K_-\bar{\tau}) \exp(b_2^- t) 
\nn\\
&& + O_3 \exp(b_3^- t) 
\,,
\eea
where $J_0$, $J_1$ are Bessel functions, 
\be
 \bar{\tau} = \sqrt{-\frac{t}{t_0}} \log \bar{s} 
\ee
with $t_0 = 1\,\mbox{GeV}^2$, and we have 
constants $F_i$, $O_i$ and $b^\pm_i$. 

In addition to these asymptotic terms one has also the normal 
terms $A_\pm^N$ which contain different Regge pole and 
Regge cut contributions to the amplitude. 
In the version of the maximal Odderon \cite{Gauron:1990cs}
which is mostly used these were chosen to be a Pomeron pole 
and a Pomeron--Pomeron cut for $A_+^N$, and 
an Odderon pole, a Pomeron--Odderon cut as well as the usual 
reggeon contributions and reggeon--Pomeron cuts for $A_-^N$. 
Both the Pomeron and the Odderon pole have an intercept 
of exactly one, and we will discuss the reason for this choice 
momentarily. Also the terms in the normal amplitudes come 
with a number of parameters 
so that the total number of parameters of the maximal Odderon 
model (when used together with the maximality principle for 
$A_+$) exceeds 20 already without the reggeon terms. With 
the reggeon terms the total number of parameters of the model 
is almost 40. These parameters can then be fitted using a variety 
of data for different observables. 

In adding different asymptotic and Regge pole contributions 
one might wonder whether there is some double counting 
inherent in this procedure. This question is closely related to the 
question of how the fulfillment of the Froissart--Martin 
theorem works here as compared to the usual approach that 
uses only Regge poles and Regge cuts. There we have seen 
that Regge poles are in general allowed to have intercepts 
larger than one, and it is their iterated exchange which 
is supposed to lead to a unitarization of the amplitude, i.\,e.\ leads 
to a unitarized amplitude with singularities only at or below one. 
The iterated exchange of Regge poles appears very natural 
in conjunction with their interpretation as Regge trajectories 
relating their exchange to resonances in the crossed channel. 
In the case of the maximal Odderon approach the situation is 
quite different. Here the asymptotic terms (the Froissaron 
and the maximal Odderon term) are not simple poles and 
thus cannot be associated with any particles in the crossed 
channel at $t>0$. The asymptotic terms hence cannot simply be 
interpreted in terms of a $t$-channel exchange, and they 
must not be iterated in the maximal Odderon approach. 
Instead the amplitudes fulfill the asymptotic theorems 
by construction. The normal Regge contributions in the 
amplitudes $A_\pm^N$ on the other hand are usual 
Regge poles, and should eventually be iterated. In order that 
this iteration does not interfere with the asymptotic terms 
these normal Regge singularities have to be situated at exactly 
one (for the Pomeron and Odderon poles) or below one 
(for the reggeon pole trajectory) in the complex angular 
momentum plane. It is then argued that with these requirements 
any double counting of terms in the two contributions 
$A_\pm^N$ and $A_\pm^{AS}$ to the amplitudes is avoided. 
Further general discussion of unitarity constraints on the 
Odderon in the framework of Regge theory can be 
found for instance in \cite{Finkelstein:1989mf}--\cite{Martynov:1991ib}. 

The energy dependence of the maximal Odderon is quite 
different from that of an Odderon of Regge pole type. 
These two possible types of Odderon can in some sense 
be viewed as two extreme cases of models for the Odderon. 
Among different models for the Odderon the maximal 
Odderon predicts the most dramatic phenomenological 
effects due to its strong energy dependence. Some 
phenomenological aspects of the maximal Odderon will 
be discussed in sections \ref{elasticppsect} and \ref{rhosect}. 

\section{Theoretical Aspects of the Odderon in QCD}
\label{theoroddsect}
\setcounter{equation}{0}

With the knowledge that QCD is the
correct microscopic theory of the strong interactions it is 
natural to ask whether the Regge limit can 
be understood in terms of the elementary degrees of freedom 
of QCD. The ultimate goal would of course be to derive Regge 
theory from QCD and especially to determine the positions of 
Regge singularities from first principles. 
The first step towards an understanding of the 
high energy limit in terms of quarks and gluons 
was made by Low \cite{Low:1975sv} 
and Nussinov \cite{Nussinov:mw} who proposed a simple 
model of the Pomeron in which it consists of a colorless state 
of two gluons. Since their proposal the theoretical 
picture of the Pomeron has been refined in many ways, 
but we are still far from a full understanding. 
The basic problem is that most hadronic scattering processes 
at high energy are dominated by low momentum scales 
and thus by soft interactions. Our understanding of QCD 
on the other hand is best in situations in which we can 
apply perturbation theory, namely in hard scattering 
processes involving large momentum scales, or equivalently 
small distances. At low momenta the strong coupling 
constant $\alpha_s$ becomes large and one cannot apply 
perturbation theory. 

Fortunately, there are a few scattering processes which 
can be approached in perturbation theory also at 
high energies. These processes can be characterized 
as scattering processes of two small color dipoles. Examples are 
heavy onium scattering or collisions of highly virtual photons.
These processes involve a large momentum scale (the heavy quark
mass or the photon virtuality, respectively) and can be treated
perturbatively even at high energy. One therefore hopes that by
studying these processes some essential features of the dynamics
of high energy QCD can be discovered using perturbative methods. 
Of course one has to be careful in using the results in situations 
in which the applicability of the perturbative approach is less certain. 
But even apart from the phenomenological applicability the 
study of the Regge limit in perturbative QCD has proven to 
be so rich that it is already an interesting subject on its own. 

The perturbative approach to the high energy limit is based 
on the concept of resummation. In a perturbative situation 
the coupling constant $\alpha_s$ is small. However, at 
large energies there are configurations in which 
the smallness of the coupling constant can be compensated 
by large logarithms of the energy, $\log s$. The aim 
of resummation, when applied to high energy scattering, 
is to include all contributions of the 
order $(\alpha_s \log s)^n$. The corresponding 
approximation scheme is called the leading logarithmic 
approximation (LLA). 
For QCD this resummation has been performed 
by Balitzkii, Fadin, Kuraev and Lipatov 
\cite{Kuraev:fs,Balitsky:ic}, and the result is known 
as the BFKL Pomeron. It describes the exchange of 
colorless state of two interacting reggeized gluons 
in the $t$-channel. In fact it can be shown that the 
exchange of quarks is suppressed by powers of the 
energy and does not play any r\^ole in the high 
energy limit of hadronic scattering in this approximation. 

The Odderon can in QCD be described in a similar 
way. It can be obtained as a colorless exchange of 
three interacting reggeized gluons in the $t$-channel. 
The corresponding resummation collects terms which 
are suppressed by an additional factor of $\alpha_s$ as 
compared to the LLA, and the resummation is cast 
into the form of the Bartels--Kwieci\'nski--Prasza{\l}owicz 
(BKP) equation. The approximation scheme is known 
as the generalized leading logarithmic approximation (GLLA) 
and has also been extended to exchanges with more than three 
gluons. These compound states of $N$ reggeized gluons 
have an amazingly rich structure. Quite unexpected 
relations have been found with the theory of integrable 
models and conformal field theory, for example. 
The Odderon is a special case of these results. Using them 
it was now possible to solve a longstanding problem, namely 
to find exact solutions of the BKP equation and to 
determine the Odderon intercept in perturbative QCD. 
The result is that the intercept is close to one (or exactly one, 
depending on the scattering process). This can be interpreted 
as strong indication that the Odderon should also be 
present at least in semiperturbative situations at high energy. 

In this section we will present the BKP equation for the Odderon 
and its known solutions. The BKP equation can be understood 
as a generalization of the BFKL equation and shares with it 
many features, and we explain most of them first for the simpler 
case of the BFKL equation in section \ref{bfklsect}. 
One of the central properties of 
perturbative QCD in the high energy limit is the reggeization 
of the gluon. A recently found exact solution of the BKP equation 
indicates that reggeization is of vital importance for a full 
understanding of the perturbative Odderon. We therefore 
try to give in section \ref{reggeizationsect} 
an unconventional view of reggeization which 
in our opinion is particularly helpful for understanding the 
structure of certain solutions of the BKP equation. 

One of the main findings concerning the perturbative Odderon 
is its integrability. In fact it was shown, that the Odderon 
in the GLLA is equivalent to a completely integrable system, namely the 
XXX Heisenberg model of $\mbox{SL}(2,\C)$ spin zero. 
This result has opened the possibility of applying the 
powerful tools that have been developed in mathematical 
physics for the investigation of integrable models. 
In fact it was the application of these methods which has 
led to a much better understanding of the perturbative Odderon 
and its spectrum. 
A full account of these interesting results 
would go beyond the scope of the present review, and we therefore 
restrict ourselves to the presentation of the main results. 
We will however make one exception, namely we will explain 
in some detail the equivalence of the Odderon with the 
XXX Heisenberg model. 
Since this equivalence holds in general 
in the large-$N_c$ limit of the GLLA we will discuss it for 
the general case of compound states of $N$ reggeized gluons 
in section \ref{gllasect}. 
It should be emphasized that in the case of the Odderon, that is for 
$N=3$, the equivalence is exact and holds also for finite $N_c$. 

In the GLLA one considers only exchanges in the $t$-channel 
in which the number of gluons stays constant. More difficult 
to describe but also more interesting is the case in which the number 
of gluons is allowed to fluctuate during the $t$-channel 
evolution. The corresponding amplitudes are obtained in the 
extended GLLA. Also here the Odderon is an important 
object. An interesting result that is directly relevant 
to the Odderon and also to its phenomenology is the existence 
of a perturbative Pomeron--Odderon--Odderon vertex. 
This vertex has been calculated in the extended GLLA and 
we will explain how it emerges there. Both the GLLA and 
the extended GLLA are discussed in section 
\ref{oddunitaritysect} where we also explain their relation 
to the unitarity of high energy scattering. 

It should be emphasized that the use of perturbation theory 
in high energy scattering has its limitations even if the scattering 
particles provide hard momentum scales in the process. 
It can be shown that eventually there will be a large 
contribution from the nonperturbative small--momentum 
region in the limit $s \to \infty$ 
--- no matter how hard the external momentum scales are. 
This effect is related to the diffusion of transverse gluon momenta. 
It has been studied in detail in the context of the BFKL Pomeron, 
but is of more general nature and applies also to the Odderon. 
This effect is very important conceptually, and we therefore 
describe it in some detail for the simplest and best--studied case 
of the BFKL Pomeron in section \ref{bfklapplicabsection}. 
It is not only of theoretical interest but also relevant to the 
phenomenological applicability of the BFKL Pomeron and 
of the BKP Odderon. 

In general the understanding of nonperturbative QCD effects in 
high energy scattering is still rather poor. This is particularly 
unfortunate since the wealth of hadronic scattering processes 
is in fact dominated by soft momenta. This is also true for many 
processes involving the Odderon. A number of models has been 
devised to approach this difficult problem of soft high energy 
processes, and we describe in section \ref{nonpertoddsect} 
especially those methods and models which can also be applied 
to the Odderon. 

\subsection{High Energy Scattering in Perturbative QCD and the BFKL Pomeron}
\label{bfklsect}

The aim of this section is to present the basic notions and results 
of the leading logarithmic approximation. We will put some 
emphasis on those aspects that are immediately relevant to the 
Odderon. For more detailed accounts of the BFKL approach 
we refer the reader to \cite{Forshaw:dc,Lipatov:1996ts,Lipatovinbook}. 
 
\subsubsection{The BFKL Equation}
\label{bfkleqsect}

Let us briefly recall the concept of resummation in the 
more familiar case of the 
Dokshitzer--Gribov--Lipatov--Altarelli--Parisi (DGLAP) 
evolution equation \cite{Gribov:ri,Altarelli:1977zs,Dokshitzer:sg}. 
In a high energy scattering process the colliding particles 
have light--like momenta $p_1$ and $p_2$, $p_1^2=p_2^2 =0$, 
where we neglect here the particle masses. 
Any momentum $k$ can be decomposed into 
longitudinal momenta in the $p_1$-$p_2$ plane 
and a transverse momentum $\kf$ 
according to a Sudakov decomposition
\be
  k = x p_1 + w p_2 + \kf
\,,
\ee
where $x$ and $w$ are the longitudinal 
momentum fractions. 

\begin{figure}[ht]
\begin{center}
\input{halbleiter.pstex_t}
\end{center}
\caption{Real gluon emissions from the $t$-channel gluon 
\label{emissions}}
\end{figure}
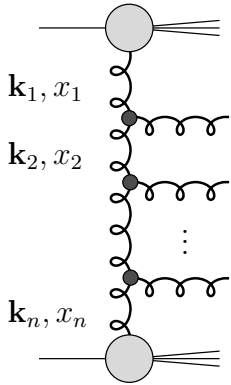

In a hard process in which a gluon is exchanged 
other gluons can be radiated off this gluon, see figure \ref{emissions}. 
In order to calculate the cross section for such a process 
one has to square the amplitude. The DGLAP resummation 
collects all diagrams which give a contribution of the 
form $(\alpha_s \log Q^2)^n$, where $Q^2$ is the external 
momentum scale at which the exchanged gluon is probed. 
In these terms the smallness of the coupling constant 
is compensated by large logarithms of the momentum scale. 
It can be shown that these terms are exactly given by 
those diagrams in which the transverse gluon momenta along 
the ladder are strongly ordered, 
\be
\kf_1^2 \gg \kf_2^2 \gg \dots \gg \kf_n^2
\,.
\ee
Precisely these configurations of gluon momenta lead to 
a logarithmic enhancement of the phase space integration. 
The corresponding diagrams for the 
cross section have the famous ladder structure. 

Let us now turn to the BFKL resummation of leading logarithms 
of the energy. Here one also assumes that the coupling constant 
is small, but logarithms of the energy can compensate this smallness, 
\be
  \alpha_s \ll 1 \,;\:\:\: \alpha_s \log(s) \sim 1 
\,.
\ee
Now one collects all contributions of the order $(\alpha_s \log s)^n$. 
It turns out that these contributions correspond to 
diagrams in which the longitudinal momenta of the 
gluons along the ladder are strongly ordered, 
\be
x_1 \gg x_2 \gg \dots \gg x_n
\,,
\ee
where the $x_i$ are the longitudinal momentum fractions 
of the gluons in the ladder, see figure \ref{emissions}. Here it is 
the integration over the longitudinal phase space that leads 
to the logarithmic enhancement. 

\begin{figure}[ht]
\begin{center}
\input{regge2gluon.pstex_t}
\caption{Factorization of the perturbative Pomeron amplitude 
in the high energy limit \label{figpomfact}
}
\end{center}
\end{figure}
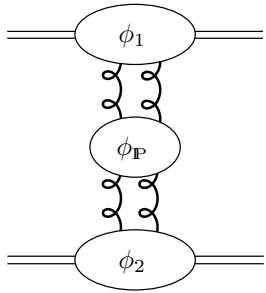

In the high energy limit the longitudinal and transverse 
degrees of freedom 
decouple and the dynamics takes place in transverse space only. 
Perturbative high energy factorization ensures that the 
amplitude can be written in factorized form, 
\be
\label{highenergyfactpom}
A(\omega,t) = \int \frac{d^2\kf}{(2\pi)^3} 
\frac{d^2\kf'}{(2\pi)^3} 
\phi_\spommi(\kf,\kf';\qf) \phi_1(\kf,\qf-\kf)
\phi_2(\kf',\qf-\kf')
\,.
\ee
This is illustrated in figure \ref{figpomfact}. 
Here $\qf$ is the transverse momentum transferred in the 
$t$-channel, and we have $t=-\qf^2$. 
The functions $\phi_1$, $\phi_2$ are the impact factors 
of the scattering colorless states and describe their coupling 
to the two gluons. The color neutrality 
implies 
\be
\phi_{1,2}(\kf=0,\qf) = \phi_{1,2}(\kf=\qf,\qf) = 0
\,,
\ee
which means that gluons of zero momentum (and hence of infinitely long 
wavelength) cannot resolve the color neutral scattering states. 
This property is important for the infrared finiteness of the amplitude. 
The function $\phi_\spommi$ can be interpreted as the 
partial wave amplitude for the scattering of two virtual 
gluons with virtualities $-\kf^2$, $-(\qf-\kf)^2$ to two virtual 
gluons with virtualities $-\kf'^2$, and $-(\qf-\kf')^2$. It depends on 
the complex angular momentum $\omega$, but for brevity 
we do not write out this dependence explicitly. The 
partial wave amplitude $\phi_\spommi$ 
is described by the BFKL equation. In momentum space it will 
be more convenient to write the BFKL equation for the 
amputated amplitude $f_\spommi$, 
\be
\label{amputationofphip}
 f_\spommi (\kf,\kf';\qf) 
= \kf^2 (\qf - \kf)^2 \phi_\spommi(\kf,\kf';\qf) 
\,.
\ee
The BFKL equation is an integral equation in 
the two--dimensional space of transverse momenta 
and of Bethe--Salpeter type. In detail, the BFKL equation has the form 
\be
\label{BFKLeq}
  \omega f_\spommi (\kf,\kf';\qf) =
 f^0(\kf,\kf';\qf) 
+ \int \frac{d^2\lf}{(2\pi)^3} \,
 \frac{1}{\lf^2 (\qf-\lf)^2} 
 K_{\mbox{\scriptsize BFKL}}(\lf,\qf-\lf;\kf,\qf-\kf) 
\,f_\spommi (\lf,\kf';\qf) 
\,.
\ee
For the convolution in this equation we will later use a 
shorthand notation, 
\be
\label{convolutionbfklkernel}
K_{\mbox{\scriptsize BFKL}} \otimes f_\spommi (\kf,\kf';\qf) 
= \int \frac{d^2\lf}{(2\pi)^3} \,
 \frac{1}{\lf^2 (\qf-\lf)^2} 
 K_{\mbox{\scriptsize BFKL}}(\lf,\qf-\lf;\kf,\qf-\kf) 
\,f_\spommi (\lf,\kf';\qf)
\,.
\ee
In the BFKL equation $f^0$ is an inhomogeneous term. It is 
the amputated version of 
\be
 \phi^0(\kf,\kf';\qf) = 
\frac{1}{\kf^2(\qf-\kf)^2} \,f^0 (\kf,\kf';\qf)
= \frac{\delta(\kf-\kf')}{\kf^2(\qf-\kf)^2}
\,,
\ee
which describes the free propagation of two elementary gluons. 
The integral kernel, 
the so--called BFKL kernel or Lipatov kernel, is given by 
\bea
\label{Lipatovkernel}
K_{\mbox{\scriptsize BFKL}}(\lf,\qf-\lf;\kf,\qf-\kf) &=& 
 -  N_c g^2 \left[ \qf^2 - \frac{\kf^2(\qf-\lf)^2}{(\kf-\lf)^2} 
 - \frac{(\qf-\kf)^2 \lf^2}{(\kf-\lf)^2} \right] \\
 & & +(2\pi)^3 \kf^2 (\qf-\kf)^2 
 \left[ \,\beta(\kf) + \beta(\qf - \kf) \right] 
 \delta^{(2)}(\kf-\lf)
  \,.
\nn
\eea
The first of the two terms in this sum corresponds to real gluon emission, 
whereas the second term corresponds to virtual corrections. The latter 
are closely related to the phenomenon of gluon reggeization 
which we will discuss further below. 
The strong coupling constant is normalized to 
$\alpha_s= \frac{g^2}{4\pi}$. The function $\beta$ in 
the kernel is defined as 
\be
\label{gluontraj} 
 \beta(\kf^2) = - \frac{N_c}{2} g^2  \int \frac{d^2\lf}{(2 \pi)^3} 
          \frac{\kf^2}{\lf^2 (\lf -\kf)^2} 
\,.
\label{traject}
\ee
The function 
\be
 \alpha(\kf^2) = 1 + \beta(\kf^2) 
\label{alphatraject}
\ee
is known as the gluon trajectory function. It passes through 
the physical 
spin $1$ of the gluon at vanishing argument $\kf^2=0$ because 
$\beta(\kf^2=0)=0$. 

The factor $(-N_c)$ which comes with the term describing 
real gluon emission in the BFKL kernel, 
i.\,e.\ with the first term in square brackets 
in (\ref{Lipatovkernel}), is a color factor. If the 
two gluons entering the amplitude $\phi_\spommi$ are not in a 
color singlet state the color factor $C_I$ will be different. 
(If the two gluons are not in a color singlet state 
the amplitude is not infrared finite. It is then necessary to 
introduce a regularization. 
When dimensional regularization is used, for example,  
the calculation 
has to be performed in $2+\epsilon$ dimensions.) 
In general $C_I$ depends on the 
irreducible representation $I$ of the two gluons. If 
$N_c=3$ the factor $C_I$ equals $-3$, $-\frac{3}{2}$, 
$-\frac{3}{2}$, $0$, $1$ 
for the irreducible representations ${\bf 1}$, ${\bf 8_A}$, ${\bf 8_S}$, 
${\bf 10}+{\bf \overline{10}}$, ${\bf 27}$, respectively.

The general form of the solution of the BFKL equation 
can be derived from the integral equation by iteration. 
Accordingly, we find exactly the ladder structure that we have 
already mentioned before, 
\be
 \lim_{s \to \infty} \quad \input{limit2to2.pstex_t} 
= \sum_{\mbox{\tiny number} \atop \mbox{\tiny of rungs}} 
 \,\,\input{ladder.pstex_t}
\,,
\label{ladder}
\ee
and the ladder rungs represent BFKL kernels. 
The vertical gluons in the ladder are not elementary but 
reggeized gluons. 
Further, the part of the BFKL kernel corresponding to 
real gluon emissions contains not only simple triple--gluon 
vertices but represents an infinite number of elementary 
diagrams as a result of resummation. 

For vanishing momentum transfer $t$ one obtains 
the BFKL equation in the forward direction. Using 
polar coordinates $\kf=(|\kf|,\phi)$ in the transverse 
momentum plane the BFKL equation can be 
diagonalized by power functions 
\be
\label{kleinebfkl}
  e^{(\nu,n)} (\kf) = \frac{1}{\pi \sqrt{2}} \,
( \kf ^2)^{-\frac{1}{2} + i\nu}
 e^{i n\phi}
\ee
with $\nu \in \R$ and $n\in \Z$. The corresponding eigenvalues are 
\be
\label{eigenvalues}
 \frac{N_c\alpha_s}{\pi}\chi(\nu,n) = \frac{N_c\alpha_s}{\pi}
\left[ 2 \psi(1) - \psi\left(\frac{1 + |n|}{2} + i \nu \right)
 - \psi\left(\frac{1 + |n|}{2} - i \nu \right) \right]
\,,
\ee
where $\psi$ is the logarithmic derivative of the Euler 
$\Gamma$-function. The function $\chi(\nu,n)$ is 
often called the Lipatov characteristic function. 
The full solution of the BFKL equation in the case of 
forward scattering reads 
\be
f_\spommi(\kf,\kf',\qf=0) = \sum_{n=-\infty}^{+\infty}
\int_{-\infty}^{+\infty} \frac{d\nu}{2\pi} 
\frac{1}{\omega - \frac{N_c \alpha_s}{\pi}\chi(\nu,n)} 
e^{(\nu,n)} (\kf)e^{(\nu,n)*} (\kf')
\label{fullsolution}
\,.
\ee
The eigenfunctions for non--vanishing momentum 
transfer $t$ have a very complicated structure in momentum 
space. In order to study that case it is far more convenient 
to treat the BFKL equation in impact parameter space. 

The high energy asymptotics of the BFKL amplitude 
is determined by the rightmost singularity in the 
plane of complex angular momentum $\omega$. 
Since the function $\chi(\nu,n)$ (see (\ref{eigenvalues})) 
decreases with increasing $|n|$ one can neglect the contributions 
with $n \neq 0$ in the high energy limit. Further it is possible to 
make an expansion of $\chi(\nu,0)$ in $\nu$ around zero, 
\be
\chi(\nu,0) = 4 \ln 2 - 14 \, \zeta(3) \nu^2 + {\cal O}(\nu^4)
\,,
\ee
to find the leading singularity 
in (\ref{fullsolution}). Due to the continuous parameter $\nu$ 
the spectrum of the BFKL kernel is gapless and the 
BFKL Pomeron corresponds to a fixed cut 
singularity in the complex angular momentum plane. 
It leads to a power--like growth of the amplitude 
\be
  A \sim s^{(1+\omega_{\mbox{\tiny BFKL}})}
\,,
\label{amplitudewithexponent}
\ee
and the exponent as obtained from that calculation is 
\be 
 \omega_{{\mbox{\scriptsize BFKL}}} 
= \frac{\alpha_s N_c}{\pi} \,4 \ln 2 
   \simeq 0.5  
\,,
\label{BFKLexponent}
\ee
where the latter value is obtained when assuming that the 
strong coupling constant has a value of $\alpha_s \simeq 0.2$, 
typical for the momentum scales involved in high energy 
reactions. 
Consequently, the total cross section in the leading logarithmic 
approximation grows like 
\be
  \sigma_{\mbox{\scriptsize tot}}
  \sim s^{\omega_{\mbox {\tiny BFKL}}}
\,.
\ee
It should be noted that the cross section resulting from the 
BFKL resummation violates the Froissart theorem 
(\ref{fmtheorem}). We will come back to that problem 
in section \ref{oddunitaritysect} below. 

In an effort that lasted for almost a decade the BFKL 
equation has been extended to next--to--leading 
logarithmic approximation (NLLA), now including terms of 
the order $\alpha_s (\alpha_s \log s)^n$, 
see \cite{Fadin:1998py,Ciafaloni:1998gs} 
and references therein. The corrections were found to 
be rather large, giving for the characteristic 
exponent $\omega_\sbfkl=\alpha_\tpommi -1$ 
of the energy dependence 
\be
 \omega_{\sbfkl} \simeq 2.65 \, \alpha_s (1 - 6.18 \,\alpha_s)
\,,
\ee
where the first term corresponds to the intercept in LLA. 
The large correction indicates a poor convergence 
of the perturbative series in this case. Initially this 
result led to serious doubts about the BFKL approach 
in NLLA. These doubts have been considerably weakened 
after the problem was subsequently studied in more detail. 
The large corrections were found to originate from 
collinear divergences due to the emission of real gluons 
that are close to each other in rapidity. 
Several methods have been proposed to circumvent 
this problem, among them a renormalization 
group improvement of the BFKL equation resumming 
additional large logarithms of the transverse 
momentum \cite{Ciafaloni:1999yw}, 
the application of a Brodsky--Lepage--Mackenzie 
(BLM) scale setting procedure \cite{Brodsky:1998kn}, 
and a method to veto the emission of gluon pairs 
close in rapidity in a Monte Carlo implementation 
of the BFKL equation \cite{Schmidt:1999mz}. 
Those methods lead to stable results for the intercept, 
although the precise values differ slightly for 
the different methods. 
With those improvements the NLL BFKL 
equation is now widely considered a reasonable 
approximation scheme. For typical values of $\alpha_s$ 
around $0.2$ one now obtains an intercept of $1.2$ to $1.3$, 
to be compared with the LLA intercept of $1.5$. 

\subsubsection{Conformal Invariance of the BFKL Equation}
\label{bfklconfsect}

A remarkable property of the BFKL Pomeron is its conformal 
invariance in two--dimensional impact parameter space 
\cite{Lipatov:1985uk}. A practical consequence of this 
symmetry is the possibility to find the solutions of the BFKL 
equation also in the non--forward direction $t \neq 0$. 
The Fourier transformation of the partial wave amplitude 
$\phi_\spommi$ is defined as 
\bea
\label{fouriervonampli}
\delta(\qf-\qf') \phi_\spommi(\kf,\kf';\qf) &=& 
\int d^2\rho_1 d^2\rho_2 d^2\rho_{1'} d^2\rho_{2'}
\phi_\spommi(\rho_1,\rho_2;\rho_{1'},\rho_{2'})\times
\\
&& \quad \times \exp\left( i\kf \rho_1+i(\qf-\kf)\rho_2
-i\kf'\rho_{1'} - i(\qf'-\kf')\rho_{2'}\right)
\,.
\nn
\eea
Fourier transformation then also defines the BFKL equation in 
two--dimensional impact parameter space. 
It is convenient to write the vectors in this space 
in complex notation, 
\be
\rho = \rho_x + i \rho_y
\,.
\ee
Accordingly, the notation in (\ref{fouriervonampli}) should be 
understood as $\kf \rho=k_x \rho_x + k_y \rho_y$. 
The complex numbers $\rho$ are the holomorphic coordinates, 
and we define antiholomorphic coordinates by 
\be
\rhobar = \rho_x  - i \rho_y
\,.
\ee
We also define the derivatives $\del= \del/\del \rho$ and 
$\bar{\del}=\del/\del \bar{\rho}$. 

The BFKL equation in impact parameter space reads 
\be
\label{BFKLfullimpact}
\omega \phi_\spommi = 
\phi_\spommi^{(0)} + {\cal H}_\spommi \phi_\spommi 
\,
\ee
and the Hamiltonian ${\cal H}_\spommi$ obtained from the BFKL kernel 
can be split into two parts, 
\be
\label{twopartsbfklh}
 {\cal H}_\spommi = \frac{\alpha_s N_c}{2 \pi} 
(H_\spommi +\bar{H}_\spommi)  
\,,
\ee
with the operator 
\be
\label{kernelk}
H_\spommi (\rho_1,\rho_2) 
= \log[(\rho_1-\rho_2)^2 \del_1] + \log[(\rho_1-\rho_2)^2 \del_2]
   - 2 \log (\rho_1-\rho_2) - 2 \psi(1)
\,.
\ee
Here $\psi$ denotes the logarithmic derivative of the Euler gamma 
function. The operator $\bar{H}_\spommi$ is defined in analogy 
to $H_\spommi$ but 
with the antiholomorphic coordinates $\rhobar_i$. 
The BFKL Hamiltonian can hence 
be decomposed into one part which acts only on the holomorphic 
coordinates $\rho_i$ and one that acts only on the antiholomorphic 
coordinates $\rhobar_i$. This holomorphic separability implies that 
the eigenfunctions of the Hamiltonian are products of factors which 
depend only on holomorphic and antiholomorphic coordinates, 
as will in fact be the case, see (\ref{bfkleigenfu}) below. 

The BFKL Hamiltonian can be shown to be invariant 
under M\"obius transformations 
\be
\label{Moebius}
   \rho \rightarrow \rho' = 
  \frac{a \rho +b}{c \rho +d}\,;\:\:\:\:
   ad -bc = 1  
\,,
\ee
and similar transformations for $\rhobar$. 
These transformations are characterized by the group 
\be
  \left(
  \begin{array}{cc}
  {a}&{b}\\
  {c}&{d}
  \end{array}
  \right)
  \in \mbox{SL}(2,\C) / Z_2 \,,
\ee
i.\,e.\ the group of projective conformal transformations. 
An arbitrary conformal transformation can always 
be obtained as the 
superposition of the following basic transformations: 
\begin{tabbing}
\hspace*{3cm}\=translations: \hspace{.7cm} \=$\rho \rarr \rho + b$ \\
\>rotations: \>$\rho \rarr a \rho; \quad |a|=1$\\
\>dilatations: \>$\rho \rarr \lambda \rho; \quad \lambda \in \R_+ $\\
\>inversions: \> $\rho \rarr \rho^{-1}$
\,.
\end{tabbing}

We should point out that 
the conformal invariance of the BFKL equation is broken 
when one includes NLL corrections. But it has been found 
that this breaking is mild in the sense that it occurs 
only due to the running of the gauge coupling $\alpha_s$ 
which in NLLA is no longer constant but receives corrections 
reflecting its running according to the renormalization group 
equation in one loop approximation. 
The conformal invariance of the BFKL equation in LLA 
therefore remains very valuable for the understanding of 
the high energy limit in perturbative QCD even when 
NLL corrections are included. 

\subsubsection{Solutions of the BFKL Equation in Impact Parameter Space}
\label{bfklsolutionsect}

A formal solution of the BFKL equation (\ref{BFKLfullimpact}) is 
\be
\phi_\spommi = \frac{1}{\omega - {\cal H}_\spommi} \, \phi_\spommi^{(0)}
\,.
\ee
The problem of finding the solution can therefore be 
reduced to finding the eigenstates $\varphi_\spommi$ of the Hamiltionian 
${\cal H}_\spommi$, 
\be
  {\cal H}_\spommi \varphi_\spommi (\rho, \rhobar) = 
E  \varphi_\spommi (\rho, \rhobar)
\,,
\ee
where by $\rho$ we denote the set of holomorphic coordinates, 
$\rho_1$ and $\rho_2$, and analogously for $\rhobar$. 
With a complete set of two--gluon states $\varphi_{\spommi \alpha}$, 
with $\alpha$ denoting the quantum numbers of the states, 
the full solution of the BFKL equation can be found as 
a superposition of solutions obtained from the eigenstates of 
${\cal H}_\spommi$. With a suitably defined scalar product we 
can write 
\be
\label{bfklsolutionwitheigen}
\phi_\spommi = 
\sum_\alpha
\frac{1}{\omega - E_\alpha} 
\left| \varphi_{\spommi \alpha} \right\rangle
\left\langle \varphi_{\spommi \alpha} \right| \, 
\phi_\spommi^{(0)}
\,.
\ee
The summation symbol indicates the sum over all discrete and the 
integration over continuous quantum numbers. 

Due to the conformal invariance of the BFKL Hamiltonian 
${\cal H}_\spommi$ the eigenstates correspond to the principal series 
representation of the group $\mbox{SL}(2,\C)$. The eigenfunctions 
$\varphi_\spommi$ can therefore be identified with the representation 
functions $E^{(\nu,n)}$ given by 
\be
\label{bfkleigenfu}
E^{(\nu,n)}(\rho_{10},\rho_{20})
= 
\left( \frac{\rho_{12}}{\rho_{10}
\rho_{20}}\right)^h
\left( \frac{\rhobar_{12}}{\rhobar_{10}
\rhobar_{20}}\right)^{\bar{h}}
\,.
\ee
Here we use 
\be
\rho_{ij}= \rho_i - \rho_j\,,
\ee
and an analogous definition for $\rhobar_{ij}$. 
The coordinate $\rho_0$ represents an additional parameter 
of the functions $E^{(\nu,n)}$. Note that, as is implied by the 
holomorphic separability (\ref{twopartsbfklh}) of the Hamiltonian, 
the eigenfunctions are products of factors depending only on 
holomorphic and antiholomorphic coordintes, respectively. 
The parameters $h$ and $\bar{h}$ 
are the conformal weights of the representation, and we have 
\be
 \bar{h} = 1 - h^*
\,.
\ee
The representation theory of the group $\mbox{SL}(2,\C)$ 
implies that the conformal weight $h$ is quantized as 
\be
\label{principalquantofh}
h = \frac{1 + n}{2} + i \nu,  \hspace*{1cm} 
n \in \Z,\,\,\,\, \nu \in \R
\,.
\ee
The combination 
\be
h- \bar{h} = n \in \Z
\label{hminush}
\ee
is the integer conformal spin of the state and its scaling dimension 
is given by 
\be
 h +\bar{h} = 1 + 2 \nu \in \R
\,.
\label{hplush}
\ee                                                                    
These two relations also explain the notation $E^{(\nu,n)}$ 
for the representation functions. The eigenvalues corresponding 
to the representation functions $E^{(\nu,n)}$ are exactly the 
ones given in eq.\ (\ref{eigenvalues}). 

The full solution of the BFKL equation expanded in conformal 
partial waves then reads (with a suitable 
normalization of the eigenfunctions) 
\bea
\label{bfklsolutionpartial}
\phi_\spommi (\rho_1,\rho_2;\rho_{1'},\rho_{2'})
&=& \sum_{n=-\infty}^{+\infty} \int_{-\infty}^{+\infty}
\frac{d\nu}{2\pi}
\frac{16 \nu^2 + 4 n^2}{[4\nu^2+ (n-1)^2][4\nu^2+ (n+1)^2]}
 \times 
\\
&& 
\times
\frac{1}{\omega - \frac{N_c \alpha_s}{\pi} \chi(\nu,n)}
\int d^2\rho_0
E^{(\nu,n)}(\rho_{10},\rho_{20}) 
E^{(\nu,n)*}(\rho_{1'0},\rho_{2'0})
\,.
\nn
\eea
The leading singularity of the BFKL amplitude is obtained 
for $h= \bar{h}=\frac{1}{2}$, as discussed in section 
\ref{bfkleqsect}. 

The representation functions can be understood as 
three--point correlation functions of a conformal field theory 
\cite{Belavin:1984vu} (for a review see \cite{Ginsparg:1988ui}), 
\be
E^{(\nu,n)} (\rho_{10},\rho_{20})
= \langle \phi_{0,0}(\rho_1,\bar{\rho}_1)
\phi_{0,0}(\rho_2,\bar{\rho}_2)
O_{h,\bar{h}}(\rho_0,\bar{\rho}_0) \rangle
\,,
\ee
with the identities (\ref{hminush}),(\ref{hplush}) relating 
$\nu$, $n$, and $h$. Conformal invariance in fact fixes 
the form of the three--point function in a conformal field 
theory up to an overall normalization factor. 
Pictorially, the three--point function can be represented as 
\be
E^{(\nu,n)}
\:\:\:= \:\:\:
\input{poperator.pstex_t} 
\:\:\:= \:\:\:
\langle \,\phi_1 \,\phi_2 \,\,{\cal O}^\spommi \,\rangle
\,.
\ee
The operators $\phi_{0,0}$ can be interpreteted as elementary 
fields representing reggeized gluons. They have conformal 
weight zero. The operator $O_{h,\bar{h}}$ represents a composite 
state of two reggeized gluons that emerges from the dynamics 
of the theory. 
Similarly, one can interpret the solution (\ref{bfklsolutionpartial}) 
of the BFKL equation as a four--point function, 
\be
\label{pgreen}
\phi_\spommi 
\:\:\:= \:\:\:
\input{pgreen.pstex_t} 
\:\:\:= \:\:\:
\langle \,\phi_1 \,\phi_2 \,\phi_{1'}\,\phi_{2'} \,\rangle
\,.
\ee

\subsubsection{Reggeization of the Gluon}
\label{reggeizationsect}

We have seen the phenomenon of reggeization already in 
section \ref{reggetheory}. Basically, the reggeization of 
a particle of mass $M$ and spin $J$ means that the 
amplitude $A$ for a scattering process corresponding 
to the $t$-channel exchange of the quantum numbers 
of that particle behaves at large energy $\sqrt{s}$ as 
\be
 A \sim s^{\alpha(t)}\,,
\ee
where the trajectory $\alpha(t)$ satisfies 
$\alpha(M^2)=J$. The latter condition just means that 
the particle itself lies on the trajectory. 
In a perturbative approach the particles exchanged 
in hadronic scattering processes 
are quarks and gluons, and one can expect that also 
these particles reggeize at high energy. 
This expectation has been confirmed 
for both quarks and gluons. Of particular interest 
in the context of the Odderon is the reggeization of 
the gluon. Gluon reggeization was first shown to 
two-- and three--loop order 
\cite{Tyburski:1975mr}--\cite{Lo:py} 
and then also to all orders \cite{Kuraev:ge} 
in the leading logarithmic approximation 
(and using somewhat different methods also in 
\cite{Mason:1976fr,Cheng:gt}). 

To see how the reggeization of the gluon emerges let us 
consider the BFKL equation in the color octet\footnote{We 
speak of 'octet' to mean the adjoint representation 
also for general $N_c$.} channel, i.\,e.\ for two gluons 
in an antisymmetric color octet state. 
It should be noted that in this color representation the 
amplitude is not infrared finite and a regularization has 
to be applied. 
For antisymmetric color octet exchange the color factor 
in the BFKL kernel (\ref{Lipatovkernel}) 
is $N_c/2$ instead of $N_c$. 
Let us further assume that the inhomogeneous term $\phi^{(0)}$ 
is a function of $(\kf_1+\kf_2)$. 
In this situation the BFKL equation exhibits a special solution, 
\be
\label{Pole}
\phi_{\bf 8_A} (\kf_1+\kf_2) = 
\frac{\phi_{\bf 8_A}^{(0)} (\kf_1+\kf_2)}{\omega -\beta(\kf_1+\kf_2)} 
\,, 
\ee
as is easily verified. 
This solution has a pole and can be interpreted as describing 
the propagation of a single particle with momentum $(\kf_1+\kf_2)$ 
and the quantum numbers of a gluon. 
The gluon can hence be associated with the trajectory 
$\alpha_g(\kf^2) = 1 + \beta(\kf^2)$ with $t=-\kf^2$ and 
$\beta(\kf^2)$ is given in (\ref{gluontraj}). 
In particular, the gluon reggeizes. 
As expected the gluon trajectory $\alpha_g$ passes through 
the physical spin $1$ of gluon at $t=0$. 

For the understanding of the Odderon and in particular 
of the Bartels--Lipatov--Vacca solution (see section 
\ref{blvsolutionsect}) it will be helpful to look at this 
result from another perspective. We started from the 
BFKL equation describing two gluons in a color octet 
state which are exchanged in the $t$-channel. 
The solution (\ref{Pole}), however, describes the propagation 
of a single particle in a color octet state, hence with gluon quantum 
numbers. The above result (\ref{Pole}) can therefore be 
interpreted in the sense that the gluon turns out to be a bound 
state of two gluons here. Actually already the two gluons 
in the BFKL equation are reggeized gluons representing 
an infinite set of Feynman diagrams involving elementary gluons. 
The fact that the gluon is a composite state of gluons is 
often termed 'bootstrap'. It indicates that the correct degrees 
of freedom in high energy QCD are not elementary gluons but 
rather reggeized gluons. From this perspective the reggeized gluon 
can be understood as a collective excitation of the gauge field. 
The reggeized gluon contains an infinite number of what can 
be interpreted as higher Fock states. In eq.\ (\ref{Pole}) we see 
the first nontrivial Fock state of the reggeized gluon, namely the 
two--gluon state. In \cite{Ewerz:2001fb} it was shown that 
this picture can be consistently extended to higher Fock states 
of the reggeized gluon which appear in the study of amplitudes 
with more than two gluons in the $t$-channel in the 
(extended) generalized logarithmic approximation, see section 
\ref{oddunitaritysect}. There also other color channels have 
been studied, including in particular the symmetric color 
octet channel. 

An important result related to the reggeization of the gluon 
concerns the exchange of three gluons in a $C=+1$ state. 
When such a state is coupled to a virtual photon impact 
factor consisting of a quark loop, the three--gluon state 
has the same analytic properties as a Pomeron consisting 
of two gluons \cite{Bartels:1992ym}. We will address this 
point in a bit more detail in section \ref{egllasect} 
in the context of unitarity corrections to the BFKL 
Pomeron and its phenomenological implications 
also in section \ref{phensect}. 

As an aside let us point out another interesting property 
of the reggeized gluon that can be extracted from the 
discussion above. 
When we interchange the two gluons in the amplitude 
above we find that due to the antisymmetric color state 
the amplitude changes sign. This fact gives rise to the 
notion of signature which has turned out to be 
very useful in the investigation of scattering amplitudes 
in the high energy limit in general. 
It characterizes the behavior under the exchange 
of two gluons, that is the simultaneous interchange of color and 
momentum labels. The reggeized gluons obviously carries 
negative signature. 

\subsubsection{Applicability of the BFKL Pomeron}
\label{bfklapplicabsection}

The resummation of leading logarithms of the energy 
which defines the BFKL Pomeron is a particular 
approximation scheme. As such it has a certain 
range of applicability which needs to be determined. 
This is primarily a phenomenological question, 
but in the case of the BFKL Pomeron the limitations 
have an origin which is also very interesting from 
a theoretical point of view. It is related to the contribution 
that the BFKL evolution receives from the infrared (IR) 
region of low transverse gluon momenta. Similar effects will 
be relevant for the perturbative Odderon as well. 
In view of later applications of the Odderon we will 
concentrate on rather direct applications 
of the BFKL Pomeron. By this we mean processes 
in which the Pomeron is exchanged on the level of the amplitude 
or the cross section (the squared amplitude). 
It should be pointed out that notwithstanding the following 
discussion of IR effects the BFKL equation is a valuable tool for 
computing the small-$x$ anomalous dimension of the gluon. 
Let us first collect the obvious requirements for the application 
of the BFKL Pomeron which have already been stated 
more or less explicitly. 
First of all, the scattering process in which we want to 
apply the BFKL Pomeron 
should be determined by hard momentum scales in order 
to make the use of perturbation theory possible. 
Next it is required that the total 
energy $\sqrt{s}$ of the scattering process is much larger 
than the momentum transfer $\sqrt{-t}$. 
This point is crucial for the validity of the high energy 
factorization of the amplitude (see eq.\ (\ref{highenergyfactpom})). 
Which energy is required for this factorization to hold 
depends in general on the process and is not always easy to determine. 
The cleanest but often rather cumbersome 
way to answer this question is to compare with 
a fixed order calculation that naturally contains also terms 
which break factorization. 
The BFKL equation is an evolution equation in the energy, 
or equivalently in the rapidity or the longitudinal momentum 
of the exchanged gluon in the ladder. In contrast to the DGLAP equation 
there is no evolution in the transverse momentum. From this 
we have to conclude that the BFKL equation should only be applied 
to processes that are governed by only one hard momentum scale. 
In short this can be phrased in the condition to have two small 
color dipoles of the same size colliding. 
That ensures that there is no evolution in transverse momentum 
that is not described properly by the BFKL Pomeron. 
The above requirements in fact reduce the number of ideal 
processes for the observation of the BFKL Pomeron drastically. 
The following are examples of processes\footnote{Here we give 
only a few representative references.} 
that meet all requirements and are in fact 
the ones that are most actively studied in the context of the 
perturbative Pomeron: 
the total hadronic cross section in the collision of two virtual 
photons \cite{Bartels:1996ke}--\cite{Brodsky:1997sd}, 
Mueller--Navelet jets in hadron--hadron collisions \cite{Mueller:ey}, 
forward jets in deep inelastic scattering \cite{Bartels:1996gr}, 
and the quasi--diffractive process $\gamma^*\gamma^* \to J/\psi \,J/\psi$ 
\cite{Kwiecinski:1998sa} in which the photons can be virtual or real 
(in the latter case the hard scale is given by the mass of the $J/\psi$). 
The total hadronic $\gamma^*\gamma^*$ cross section for example 
can be measured in $e^+e^-$ collisions in which both the 
scattered electron and positron are tagged. In this way the virtualities 
of the two virtual photons can be measured, and suitable cuts can 
be chosen to fix them at (roughly) the same value. 
Then the process is indeed a one--scale process as required. 
The requirement of large energy would clearly favor to study 
this process at a future linear collider, but it has already been studied 
at LEP \cite{Abbiendi:2001tv,Achard:2001kr}. 
In this process the Pomeron is exchanged 
on the level of the squared amplitude whereas the other three processes 
involve Pomeron exchange on the amplitude level. 
Also those processes are one--scale processes after 
suitable cuts are applied. 
One can of course try to use the BFKL Pomeron also in cases in 
which the above requirements are only partially fulfilled. But then 
one obviously has to expect corrections to the result which might 
be difficult to control in the BFKL framework. 

\begin{figure}[ht]
\vspace*{0.4cm}
\begin{center}
\input{diffusion.pstex_t}
\caption{Diffusion of transverse momenta in the BFKL Pomeron
\label{figdiffusion}}
\end{center}
\end{figure}
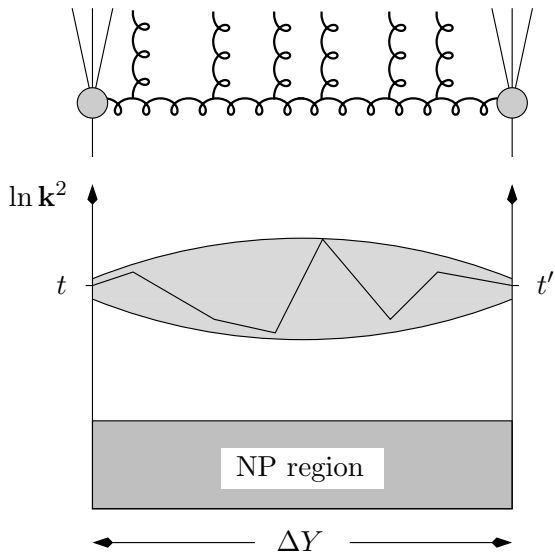

\begin{figure}[ht]
\vspace*{0.4cm}
\centering
\includegraphics[width=0.4\textwidth,clip]{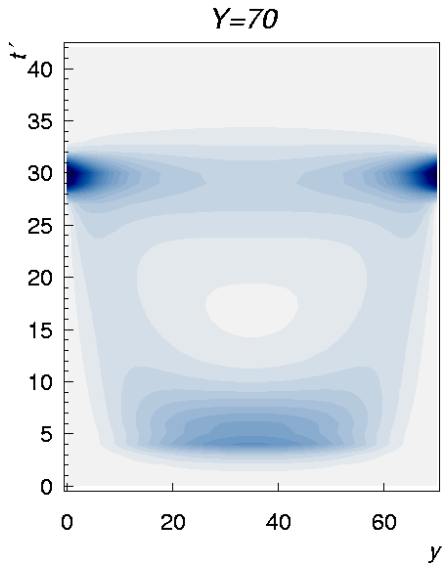}
\caption{Transition from cigar--type evolution to 
tunneling--type evolution, figure from \protect\cite{Ciafaloni:2002xk}
\label{figalien}}
\end{figure}

The above conditions for the applicability of the BFKL Pomeron 
can at least in principle be fufilled by carefully choosing the 
observable and the corresponding experimental cuts. 
Now we turn to a problem that can only be suppressed but 
not completely avoided. 
The leading logarithms of the energy to be resummed in the 
BFKL equation are obtained 
from configurations in which the emitted gluons along 
the ladder are strongly ordered in longitudinal momentum. 
At the same time there is no ordering in the transverse 
momenta of those gluons, and there is in principle 
no restriction preventing them from having arbitrarily small 
transverse momenta. Let us consider a process in which the 
external particles provide hard scales, for example in the 
scattering of two virtual photons of equal virtualities $Q^2$. 
One can then perform a numerical simulation of the BFKL 
evolution and observe the behavior of the emitted gluons along 
the ladder. This is illustrated in figure \ref{figdiffusion}. 
The horizontal axis represents the rapidity interval between the 
ends of the ladder, $\Delta Y \sim \ln(s)$. The vertical axis shows 
the logarithm of the squared transverse gluon momenta, $\ln\kf^2$. 
At the ends of the ladder the gluon momenta are fixed by the 
external particles, $t=t'=\ln Q^2$. Along the ladder the real gluon 
emissions lead to a random walk in $\ln\kf^2$. 
In fact one can show analytically that the BFKL becomes equivalent 
to a diffusion equation in the limit of large energy. Sampling a large 
number of such random walks one can find the probability distribution of 
the typical transverse momenta along the BFKL ladder. The contour 
corresponding to the mean width of the distribution is given by the 
shaded area in the figure. The exact shape of the probability distribution 
has been studied for a number of processes, including cases in which 
$t\neq t'$, in 
\cite{Bartels:1993du,Bartels:1995yk}. For obvious reasons the 
probability distribution is known as the Bartels cigar. As the energy 
$\sqrt{s}$ and thus $\Delta Y$ becomes larger the cigar becomes 
wider in the middle. Eventually, the cigar will touch the nonperturbative 
(NP) region. In other words, the probability of finding gluons 
with very low transverse momenta in the ladder will become large. 
But in this situation the perturbative approach on which the BFKL 
equation is based is no longer valid. In realistic situations it turns out 
that the contribution of small transverse momenta is considerable. 
That IR contribution can only be suppressed by fixing $t$ and $t'$ at 
large values experimentally (which however implies a much smaller 
cross section). But even for large $t$ and $t'$ the momenta will 
eventually move into the NP region in the limit of large $\sqrt{s}$. 
The problem could even be much more severe as was shown in 
\cite{Ciafaloni:2002xk}. When running coupling effects are taken 
into account the diffusion process is no longer symmetric. Since 
$\alpha_s$ is larger at low momenta the probability of small 
gluon momenta becomes larger, and the cigar is deformed into 
a banana--like momentum distribution along the ladder. 
At very large rapidity separations there can even be a tunneling 
transition: the first emission brings the $t$-channel gluon into 
the NP region where it stays for the whole evolution up to the 
last emission. The transition from the cigar--type evolution 
to the tunneling--type evolution is shown in figure \ref{figalien}. 
In such a situation the whole perturbative approach breaks down 
and the process is completely determined by the soft Pomeron. 

The problem of diffusion in the BFKL Pomeron should always be 
kept in mind in phenomenological applications, in particular when 
the momentum scales of the external particles are only moderately 
large. So far the diffusion problem in the BFKL equation has only been 
studied in the LLA. It would be very interesting to see how the 
phenomenon is possibly modified in NLLA, and whether and at which 
energies the transition to a tunneling--like evolution occurs. 

\subsection{The Odderon in Perturbative QCD}
\label{oddpertsect}

In perturbative QCD the Odderon is a state of three 
interacting reggeized gluons exchanged in the $t$-channel. 
Let us first see how and why it is possible that an Odderon can be 
constructed as an object consisting of three gluons. 
Consider for simplicity a purely perturbative situation with 
small gluon fields $\Af_\mu(x) = A^a_\mu(x) t^a$ 
where $t^a$ are the generators of the gauge group, 
and we consider the general case that the gauge group is 
$\mbox{SU}(N_c)$. 
Under a charge conjugation transformation 
the behavior of the gluon fields is 
\be
\label{aunderC}
\Af_\mu (x) \longrightarrow - \Af_\mu^T(x)
\,.
\ee
It is quite obvious that there is only one possibility to form 
a color singlet state of two gluon operators, namely 
\bea
\cP_{\mu \nu} (x,y) &=& \tr ( \Af_\mu(x) \Af_\nu(y)) 
\nn\\
&=&\frac{1}{2} \, \delta_{ab} A_\mu(x) A_\nu(y)
\,.
\eea
From (\ref{aunderC}) we see that this state remains 
unchanged under charge conjugation and thus has 
$C=+1$. The exchange of two gluons in a color singlet 
state is then described by the free--field correlation 
function of the product of two such operators, 
\be
  \langle \mbox{T} \,\cP_{\mu' \nu'} (x',y') \cP_{\mu \nu} (x,y) 
\rangle
\,,
\ee
This is what we have called the Green function of the Pomeron 
in eq.\ (\ref{pgreen})\footnote{We use a different notation here since in 
that equation we were considering the BFKL Pomeron 
in two--dimensional transverse space only.}. 
The exchange carries $C=+1$ and can be associated with the Pomeron. 
In the case of three gluons there are two ways of constructing 
a color singlet state. The first possibility is 
\bea
\label{3gluonopp}
\cP_{\mu \nu \rho} (x,y,z) &=& 
-i\,\tr ( [ \Af_\mu(x), \Af_\nu(y)] \Af_\rho(z) )
\nn \\
&=& \frac{1}{2} \,f_{abc} A^a_\mu(x) A^b_\nu(y) A^c_\rho(z)
\eea
with the structure constants $f_{abc}$ 
defined via the Lie algebra of $\mbox{SU}(N_c)$ 
\be
[ t^a, t^b ] = i f_{abc} t^c
\,.
\ee
The $f_{abc}$ are totally antisymmetric, and one finds that also the 
operator $\cP_{\mu \nu \rho} (x,y,z)$ is even under charge 
conjugation. Hence the corresponding exchange of three 
gluons has $C=+1$. 
(For a discussion of this exchange see the last paragraphs of 
sections \ref{bkpsect} and \ref{genconsidsect}.) 
The other possibility to form a color singlet state out of three 
gluon operators is 
\bea
\label{3gluonopo}
\cO_{\mu \nu \rho} (x,y,z) &=& 
\tr ( \{ \Af_\mu(x), \Af_\nu(y)\} \Af_\rho(z) )
\nn \\
&=&\frac{1}{2} \, d_{abc} A^a_\mu(x) A^b_\nu(y) A^c_\rho(z)
\,,
\eea
with the totally symmetric tensors 
\be
\label{dabcsun}
d_{abc}= 2\, [ \mbox{tr} ( t^a t^b t^c )+\mbox{tr} ( t^c t^b t^a ) ] 
\,.
\ee
Now one finds from (\ref{aunderC}) that this state is odd under 
charge conjugation, and the corresponding exchange carries 
$C=-1$. It can hence be associated with an Odderon. 
The crucial point is that the existence of the symmetric tensor 
$d_{abc}$ requires that the gauge group has a rank higher than 
one. This is the case for $\mbox{SU}(N_c)$ as soon as $N_c\ge 3$, 
and hence the operator corresponding to the Odderon exists in QCD. 
In an $\mbox{SU}(2)$ gauge theory, however, it would not be possible 
to construct an operator corresponding to (\ref{3gluonopo})
since there the analogue of the tensor (\ref{dabcsun}) vanishes, 
such that the antisymmetric $\epsilon_{ijk}$ is the only tensor 
that can be constructed of three generators of $\mbox{SU}(2)$. 

Note that in the argument above we only made use of very general 
properties of the gluon fields, namely their behavior under a charge 
conjugation (\ref{aunderC}). This behavior will also hold for 
large fields $A$, and the argument should therefore also apply 
to nonperturbative situations. However, here the interpretation 
of the operator constructed from nonperturbative gluon fields 
might be ambiguous without specifying a nonperturbative 
framework for their exact definition and use. 

Obviously, the above argument works for three elementary 
and non--interacting gluons. Accordingly, the use of a simple 
three--gluon exchange model satisfies the requirements 
for an Odderon. But the exchange of three non--interacting 
gluons leads to an Odderon intercept of exactly one, and it 
corresponds to a pole in the complex angular momentum plane. 
In the case of the BFKL Pomeron we have seen that resummation 
can lead to a quite different energy behavior and can hence 
have very significant effects. It is therefore obviously desirable 
to study resummation also in the case of the Odderon. 

\subsubsection{The Bartels-Kwieci{\'n}ski-Prasza{\l}owicz Equation}
\label{bkpsect}

The resummation of large logarithms of the energy has for 
the Odderon been performed by Bartels \cite{Bartels:1980pe}, 
Jaroszewicz \cite{Jaroszewicz:1980mq}, and 
Kwieci{\'n}ski and Prasza{\l}owicz \cite{Kwiecinski:1980wb}. 
The corresponding equation is known as the 
Bartels--Kwieci{\'n}ski--Prasza{\l}owicz (BKP) equation. 

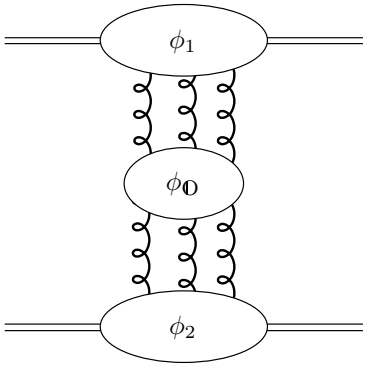
\begin{figure}[ht]
\begin{center}
\input{regge3gluon.pstex_t}
\caption{Factorization of the perturbative Odderon amplitude 
in the high energy limit \label{figoddfact}
}
\end{center}
\end{figure}

The conceptual basis is again perturbative high energy 
factorization which separates the dynamics in transverse 
space from the longitudinal degrees of freedom. 
The amplitude is given as a convolution of the Odderon 
as a three--gluon compound state with the impact factors 
coupling it to the external particles. Diagrammatically 
this is shown in figure \ref{figoddfact}. 
The corresponding equation will be given in eq.\ (\ref{Aoddifact}) 
further below when we turn to phenomenological applications. 
Here we will first concentrate on the amplitude $\phi_\soddi$ 
describing the Odderon exchanged between the scattering particles. 

The BKP equation resums terms of the order $\alpha_s (\alpha_s \log s)^n$ 
with arbitrary $n$ in which three gluons in a $C=-1$ state are exchanged 
in the $t$-channel. It is in many respects very similar to the BFKL 
equation. It is an integral equation for the Green function of 
three reggeized gluons in the $t$-channel. It acts in two--dimensional 
transverse space and describes the pairwise interaction of the three gluons.  
As in the case of the BFKL equation we again consider partial wave 
amplitudes which are functions of the complex angular momentum $\omega$. 
(This dependence will always be understood but not written out 
explicitly.) The energy dependence of the amplitude at high energies 
can then be inferred from the singularities in the $\omega$-plane. 

Let $\phi_\soddi(\kf_1,\kf_2,\kf_3;\kf'_1,\kf'_2,\kf'_3)$ be 
the amplitude for the scattering of three reggeized gluons with 
momenta $\kf'_1$, $\kf'_2$, $\kf'_3$ to three reggeized gluons 
with momenta $\kf_1$, $\kf_2$, $\kf_3$. 
We again denote the transverse momentum transferred in the $t$-channel 
by $\qf$, such that we have $\sum_{i=1}^3 \kf'_i = \sum_{i=1}^3 \kf_i = \qf$. 
For brevity we will sometimes suppress the arguments $\kf'_1,\kf'_2,\kf'_3$ 
of the amplitude $\phi_\soddi$ in our notation. 
The amplitude $\phi_\soddi$ carries color labels $a_1$, $a_2$, $a_3$ for 
the three outgoing gluons. As we will see momentarily we do not need a 
second set of color labels for the incoming gluons. 

We now give the BKP equation in momentum space in a form which 
can be easily generalized to the case of more than three gluons in 
the $t$-channel. That case will be discussed in more detail in section 
\ref{oddunitaritysect}. Further below we will see that the BKP equation 
can be written in simpler notation in impact parameter space. 
In momentum space it is convenient to use the amputated amplitude 
$f_\soddi$ defined as 
\be
\label{amputationofphio}
f_\soddi (\kf_1,\kf_2,\kf_3) = \kf_1^2 \kf_2^2 \kf_3^2 \,
\phi_\soddi (\kf_1,\kf_2,\kf_3)
\,.
\ee
The BKP equation for $f_\soddi$ then reads 
\bea
\label{bkpoddigl}
\left(\omega -\sum_{i=1}^3 \beta(\kf_i^2) \right) 
f_\soddi^{a_1 a_2 a_3} (\kf_1,\kf_2,\kf_3)
&=& f_\soddi^{(0)\,a_1 a_2 a_3}(\kf_1,\kf_2,\kf_3) \\
&&+ \sum K^{ \{b\} \rarr \{a\} }_{2\rarr 2} \otimes 
f_\soddi^{b_1 b_2 b_3} (\kf_1,\kf_2,\kf_3)
\,. \nn
\eea
The function $\beta$ corresponds to the Regge trajectory of the gluon 
and is given explicitly in (\ref{gluontraj}). 
We choose the inhomogeneous term $f_\soddi^{(0)}$ as 
\be
 f_\soddi^{(0)\,a_1 a_2 a_3}(\kf_1,\kf_2,\kf_3) = 
d_{a_1 a_2 a_3}  
\delta(\kf'_1 - \kf_1) \delta(\kf'_2 - \kf_2) \delta(\kf'_3 - \kf_3)
\ee
such that the lowest order term $\phi_\soddi^{(0)}$ of 
the Odderon amplitude $\phi_\soddi$ describes the 
free propagation of three gluons, 
\be
\phi_\soddi^{(0)\,a_1 a_2 a_3} (\kf_1,\kf_2,\kf_3;\kf'_1, \kf'_2, \kf'_3)= 
d_{a_1 a_2 a_3} 
\prod_{i=1}^3 \frac{1}{\kf_i^2} \,\delta(\kf'_i - \kf_i)
\,.
\ee
Here we have already specified the color tensor for the Odderon state. 
As discussed in the previous section the three gluons in the Odderon 
are in a symmetric color singlet state, hence the 
color tensor for the inhomogeneous term is $d_{a_1 a_2 a_3}$, 
and the same is true for the full amplitude, 
\be
\phi_\soddi^{a_1 a_2 a_3} (\kf_1,\kf_2,\kf_3)= 
d_{a_1 a_2 a_3} \phi_\soddi(\kf_1,\kf_2,\kf_3)
\,,
\ee
and we use an analogous notation for the amputated amplitude $f_\soddi$. 
The color tensor remains (up to a constant factor) 
unchanged under the action of the kernel in 
the last term of the BKP equation to which we now turn. 
The interaction of the three gluons in the Odderon is pairwise, and 
the sum in the last term of the BKP equation (\ref{bkpoddigl}) 
extends over all three pairs of gluons. We will give the kernel 
$K^{ \{b\} \rarr \{a\} }_{2\rarr 2}$ explicitly for the case that 
the first two gluons interact, the other two combinations are 
readily obtained by a permutation of the labels. 
The superscript $\{b\} \rarr \{a\} $ of the kernel refers to 
the color part of the interaction of two reggeized gluons.
The color structure of the kernel is 
\be
\label{colorpartofk2to2}
K^{ \{b\} \rarr \{a\} }_{2\rarr 2} = 
f_{b_1a_1c} f_{ca_2b_2} \delta_{b_3a_3}
K_{2\rarr 2} (\lf_1,\lf_2;\kf_1,\kf_2)
\,.
\ee
When contracted with the color tensor of the Odderon amplitude 
this yields a factor of $-N_c/2$, 
\be
f_{b_1a_1c} f_{ca_2b_2} \delta_{b_3a_3} d_{b_1b_2b_3} = 
-\frac{N_c}{2} \, d_{a_1a_2a_3}
\,,
\ee
which is one half of the color factor occurring in the BFKL 
equation. As a side remark we note that the same factor $-N_c/2$ 
would be obtained if the three gluons were in an antisymmetric color 
state. 

The momentum part $K_{2\rarr 2} (\lf_1,\lf_2;\kf_1,\kf_2)$ of 
the kernel is given by 
\be
K_{2\rarr 2} (\lf_1,\lf_2;\kf_1,\kf_2) = g^2 \left[
(\kf_1 + \kf_2)^2 
- \frac{\kf_1^2 \lf_2^2}{(\kf_2 -\lf_2)^2}
- \frac{\kf_2^2 \lf_1^2}{(\kf_1-\lf_1)^2}
\right]
\,,
\ee
and is up to the color factor identical to the part of the 
BFKL kernel (\ref{Lipatovkernel}) corresponding to real 
gluon emission. The convolution $\otimes$ is the same as in 
the case of the BFKL kernel in (\ref{convolutionbfklkernel}), 
i.\,e.\ stands for a two--dimensional integration over the 
momentum $\lf_2-\lf_1$ with measure $1/(2 \pi)^3$ and 
the propagators $1/(\lf_1^2 \lf_2^2)$ of the two gluons 
involved in the interaction. 

In the BKP equation (\ref{bkpoddigl}) the terms containing the function 
$\beta$ correspond to virtual corrections, whereas the terms 
containing the kernel $K_{2\rarr 2}$ correspond to real 
gluon emission. They are not infrared finite separately, but 
the divergenes cancel between them. It turns out that the 
BKP equation can be written in terms of full BFKL kernels 
in which the cancellation of infrared divergences is already 
known. To see this we start from eq.\ (\ref{bkpoddigl}) and 
bring the terms involving the function $\beta$ to the RHS. 
There is one of these terms for each reggeized gluon. 
There are three terms with kernels $K_{2\rarr 2}$ on the RHS, 
and the latter are up to the color factor identical to the real gluon 
emission part of the BFKL kernel. 
Each gluon is affected by the interaction described by two of 
these terms. 
We can hence distribute the functions $\beta$ in such 
a way that each kernel is associated with a term $\beta/2$ 
for each of the two gluons entering the kernel. The factor 
$1/2$ matches exactly the additional factor $1/2$ in the color 
factor of the kernel in the case of the Odderon. 
The interaction between two gluons is 
therefore given by $1/2$ times a full BFKL kernel (including 
virtual corrections) as given in (\ref{Lipatovkernel}). 

\begin{figure}[ht]
\begin{center}
\input{oddleiter.pstex_t}
\caption{Ladder structure of the BKP Odderon 
\label{oddileiter}
}
\end{center}
\end{figure}
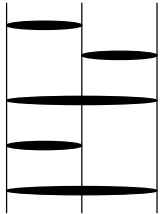

The diagrammatic structure of the perturbative Odderon can be inferred by 
iteration of the BKP equation. That iteration yields ladder type diagrams 
in which the vertical lines represent reggeized gluons, and 
the ladder rungs represent BFKL kernels (now multiplied by $1/2$), 
see figure \ref{oddileiter}. 

It is again very convenient to write the BKP equation in two--dimensional 
impact parameter space instead of transverse momentum space. 
We use the same notation as introduced in the previous section, 
in particular we have a set of holomorphic coordinates, denoted by $\rho$, 
and a set of antiholomorphic coordinates $\rhobar$ for the three incoming 
and outgoing gluons. The partial wave amplitude 
$\phi_\soddi(\rho_i,\rhobar_i;\rho_{j'},\rhobar_{j'})$ 
in impact parameter is obtained via Fourier transformation in 
complete analogy to eq.\ (\ref{fouriervonampli}). The coordinates 
$\rho_i$ ($i=1,2,3$) refer to the gluons with momenta $\kf_i$, 
and analogously for the primed coordinates and momenta. 

The BKP equation in impact parameter space becomes 
\be
\label{bkpinpositionspace}
\omega \phi_\soddi = \phi_\soddi^{(0)} + {\cal H}_\soddi \phi_\soddi 
\,.
\ee
The Hamiltonian ${\cal H}_\soddi$ can be split into two parts 
which act only on the holomorphic or antiholomorphic 
coordinates, respectively,
\be
\label{holosepfuerodd}
{\cal H}_\soddi= 
\frac{\alpha_s N_c}{4 \pi} \left( H_\soddi + \bar{H}_\soddi \right)
\,,
\ee
and the two Hamiltonians $H_\soddi$ and $\bar{H}_\soddi$ 
commute with each other, $[H_\soddi, \bar{H}_\soddi]=0$. 
The Hamiltonian hence exhibits holomorphic separability. 
The part acting on the holomorphic coordinates $\rho$ 
consists of three terms, 
\be
H_\soddi = H_{12} + H_{23} + H_{31}
\,,
\ee
describing the pairwise interaction of the three reggeized gluons. 
The pairwise interaction is given by BFKL kernels up to a factor 
of $1/2$ which we have already taken into account in the 
prefactor in (\ref{holosepfuerodd}). Hence each of the 
three terms in $H_\soddi$ is given by the BFKL kernel 
 (\ref{kernelk}) which can also be written as 
\be
H_{ij} = H_\spommi(\rho_i,\rho_j) = 
-\psi(-J_{ij}) - \psi(1 +J_{ij}) + 2 \psi(1)
\,.
\ee
The operator $J_{ij}$ is defined as an operator solution 
of the equation 
\be
\label{lijdef}
J_{ij} (1 + J_{ij}) = L_{ij} = - (\rho_i-\rho_j)^2 \del_i \del_j
\,.
\ee
Analogous relations hold for the part $\bar{H}_\soddi$ of the Hamiltionian 
that acts on the antiholomorphic coordinates. 

In analogy to the case of the BFKL equation a formal solution of the 
BKP equation (\ref{bkpinpositionspace}) is obtained via 
\be
\label{formaloddisol}
\phi_\soddi = \frac{1}{\omega - {\cal H}_\soddi} \, \phi_\soddi^{(0)}
\,.
\ee
The problem of finding the full solution $\phi_\soddi$ can therefore 
be formulated as the simpler problem of finding the eigenstates of 
the BKP Hamiltonian ${\cal H}_\soddi$, 
\be
\label{schroedingerbkp}
{\cal H}_\soddi \varphi_\soddi (\rho,\rhobar) 
= E \varphi_\soddi (\rho,\rhobar)
\,,
\ee
and according to (\ref{formaloddisol}) the Odderon intercept is given 
by the largest eigenvalue $E$. 
Let us denote the set of quantum numbers of the three--gluon states 
satisfying this eigenvalue equation by $\{\alpha \}$ and the corresponding 
eigenvalues by $E_{\{ \alpha \} }$. With a complete set of 
such eigenstates $\varphi_{\soddi \{\alpha \}}$ and a suitable 
scalar product on the corresponding Hilbert space the solution 
of the BKP equation can then be written as 
\be
\label{bkpsolutionwitheigen}
\phi_\soddi = 
\sum_{\{ \alpha\}}
\frac{1}{\omega - E_{\{ \alpha \} }} 
\left| \varphi_{\soddi \{ \alpha \} } \right\rangle
\left\langle \varphi_{\soddi \{ \alpha\} } \right| \, 
\phi_\soddi^{(0)}
\,,
\ee
with the summation symbol indicating the sum over discrete 
and the integration over continuous quantum numbers and parameters. 
In the case of the BFKL equation the quantum numbers of 
the eigenstates are given by the conformal weight $h$, 
see (\ref{bfkleigenfu}). In the case of the BKP equation the 
situation is more complicated. As we will discuss in detail in section 
\ref{XXXsect} below the eigenstates of the BKP Hamiltonian carry an 
additional quantum number corresponding to the operator $q_3$. 
It turns out to be a non--trivial problem to identify those eigenstates 
of $q_3$ which are physically meaningful and should hence be 
included in the full solution (\ref{bkpsolutionwitheigen}). 
Although considerable progress has been made recently, a generally 
accepted answer to this question has yet to be found. 
We will discuss this problem in 
the following sections, see in particular section \ref{pertinterceptsect}. 

Due to the holomorphic separability of the BKP Hamiltonian 
${\cal H}_\soddi$ the three--gluon eigenstates $\varphi_\soddi$ 
can be factorized into a product of functions depending only 
on holomorphic or antiholomorphic coordinates, 
\be
\varphi_\soddi (\rho,\rhobar) = \varphi (\rho) \bar{\varphi} (\rhobar)
\,.
\ee
The Schr\"odinger equation (\ref{schroedingerbkp}) can hence be 
replaced by two separate equations for $\varphi$ and $\bar{\varphi}$, 
\bea
 H_\soddi \varphi(\rho) &=& \epsilon \varphi (\rho)\\
 \bar{H}_\soddi \bar{\varphi}(\rhobar)
&=& \bar{\epsilon} \bar{\varphi} (\rhobar)
\,,
\eea
and we have 
\be
E= \frac{\alpha_s N_c}{4 \pi} ( \epsilon + \bar{\epsilon} )
\,.
\ee
Thereby the description of the perturbative Odderon has been 
reduced to the problem of solving two (identical) eigenvalue problems 
for the holomorphic and antiholomorphic Hamiltonians $H_\soddi$ 
and $\bar{H}_\soddi$. 

Concluding this section a remark is in order concerning the exchange 
of three gluons in the $t$-channel carrying 
positive charge parity $C=+1$. 
We have already seen the corresponding three--gluon 
operator in (\ref{3gluonopp}). Because of its 
quantum numbers this state can be considered 
a Pomeron state. It has not yet been studied in 
much detail so far, but we expect it to be quite 
interesting and important. 
In contrast to the Odderon, the three gluons are 
in an antisymmetric color state given by the 
color tensor $f_{abc}$ of $\mbox{su(3)}$. 
In perturbation theory this state is described by 
a BKP equation in the $C=+1$ sector. 
The color factors in the kernels of the BKP equation are 
the same for symmetric and antisymmetric octet 
states, and therefore the resulting BKP equation 
for the momentum part of the $C=+1$ three--gluon 
state is identical to the one for the Odderon, 
but the solutions have different symmetry properties. 
The energy levels of the corresponding Hamiltonian 
are therefore not identical to those of the Odderon, 
although some degeneracies are not excluded. 
The contribution of the three--gluon Pomeron or 
Odderon state to scattering cross sections strongly depends 
on the particular scattering process even in cases 
where both exchanges are present. The latter is the 
case when the initial and final states are not eigenstates 
of $C$-parity. 
In general the impact factors describing the coupling to the 
external particles project out different contributions 
in the $C=+1$ and $C=-1$ channels. 
For further discussion of the $C=+1$ three--gluon exchange 
we refer the reader to section \ref{egllasect} and to the last 
paragraph of section \ref{genconsidsect}. 

\subsubsection{Conformal Invariance of the BKP Equation}
\label{bkpsymmsect}

As we have just seen the BKP Hamiltonian can 
be decomposed into BFKL Hamiltonians. The latter are invariant 
under conformal transformations in two--dimensional impact 
parameter space, and so is hence the BKP Hamiltonian. 
Technically speaking, the Hamiltonian ${\cal H}_\soddi$ 
commutes with the operator 
\be
\label{q2def}
q_2 = L^2 = L_{12}^2 + L_{23}^2 + L_{31}^2 
\ee
with $L_{ij}$ defined in (\ref{lijdef}) and an operator $\bar{q}_2$ 
defined similarly in the antiholomorphic sector. This 
holds separately for the holomorphic part of ${\cal H}_\soddi$, 
\be
[q_2,H_\soddi] = [\bar{q}_2,H_\soddi] =0
\,,
\ee
and its antiholomorphic counterpart $\bar{H}_\soddi$. 
The operator $q_2$ is the quadratic Casimir operator of 
$\mbox{SL}(2,\C)$ on the space of three--gluon states, and 
according to the group theory of $\mbox{SL}(2,\C)$ its eigenvalues 
can take the values\footnote{Here and in the following sections 
we denote the operator and its eigenvalue by the same symbol, 
assuming that the meaning is clear from the context.} 
\be
q_2 = - h (h-1)
\,,
\ee
where $h$ is the conformal weight of the corresponding eigenstate. 
The conformal weight is again quantized as $h= (1+n)/2 + i \nu$ 
with $n \in \Z$ and $\nu \in \R$. 

Conformal invariance places very strong constraints on the 
eigenfunctions of the Hamiltonian ${\cal H}_\soddi$. We recall 
that in the case of the BFKL equation the eigenfunctions with 
given conformal weight are fixed up to a normalization, see 
(\ref{bfkleigenfu}). In the case of the three--gluon states 
of the Odderon there is slightly more freedom in the dependence of the 
eigenstates on the coordinates. Here the eigenfunctions of 
${\cal H}_\soddi$ parametrized by conformal weights $h$ 
and $\bar{h}=1-h^*$ have the general form 
\cite{Lipatov:1990zb,Gauron:1991cg,Lipatov:1993qn} 
\be
\label{odderonwavefct}
 \varphi^{(h,\bar{h})}_\soddi(\rho,\rhobar;\rho_0) = 
\left( \frac{\rho_{12}\rho_{13}\rho_{23}}{\rho_{10}^2
\rho_{20}^2\rho_{30}^2} \right)^{h/3}
\left( \frac{\rhobar_{12}\rhobar_{13}\rhobar_{23}}{\rhobar_{10}^2
\rhobar_{20}^2\rhobar_{30}^2} \right)^{\bar{h}/3}
\phi^{(h,\bar{h})}(z,\bar{z})
\,,
\ee
in which the dependence on the coordinates is fixed up 
to a function of the anharmonic ratio 
\be
z = \frac{\rho_{12} \rho_{30}}{\rho_{13}\rho_{20}}
\ee
and the similarly defined ratio $\bar{z}$. 
As in the case of the BFKL equation the coordinate $\rho_0$ 
is an additional parameter of the eigenstates. 
Due to the holomorphic separability of the Odderon Hamiltonian 
${\cal H}_\soddi$ the function $\phi^{(h,\bar{h})}(z,\bar{z})$ in 
(\ref{odderonwavefct}) can be further factorized as 
\be
\label{furtherfactphi}
\phi^{(h,\bar{h})}(z,\bar{z}) = 
\phi^{(h)}(z) \, \bar{\phi}^{(\bar{h})} (\bar{z})
\,.
\ee

The function $\phi^{(h,\bar{h})}(z,\bar{z})$ is not arbitrary 
but has to satisfy constraints related to the operator $q_3$ 
to which we turn in the next section. Assuming a non--singular 
behavior of $\phi^{(h,\bar{h})}(z,\bar{z})$ one can deduce from 
(\ref{odderonwavefct}) that the Odderon state 
$\varphi^{(h,\bar{h})}_\soddi$ vanishes if two of the gluon 
coordinates $\rho_i$ coincide. This is in fact true for the 
case that the eigenvalue of the operator $q_3$ does not vanish. 
In the case that $q_3=0$, however, the function 
$\varphi^{(h,\bar{h})}_\soddi$ can be highly singular and 
can even be a distribution rather than a function. We will discuss 
a solution of that type in section \ref{blvsolutionsect}. 

The eigenfunctions $\varphi^{(h,\bar{h})}_\soddi$ can be interpreted 
as four--point functions of a conformal field theory. This can be 
illustrated pictorially as 
\be
\varphi^{(h,\bar{h})}_\soddi
\:\:\:= \:\:\:
\input{oddioperator.pstex_t} 
\:\:\:= \:\:\:
\langle \phi_1 \,\phi_2 \,\phi_3 \,{\cal O}^\soddi \rangle
\ee
where $\rho_0$ can be identified with the coordinate of the 
three--gluon compound state corresponding to the operator 
${\cal O}^\soddi$. The full Green function $\phi_\soddi$ of the 
Odderon can then be understood as a six--point function, 
pictorially 
\be
\label{oddigreen}
\phi_\soddi 
\:\:\:= \:\:\:\input{oddigreen.pstex_t} 
\:\:\:= \:\:\:
\langle \,\phi_1 \,\phi_2 \,\phi_3 \,\phi_{1'}\,\phi_{2'} \,\phi_{3'} 
\,\rangle
\,.
\ee
This can also be interpreted as a diagrammatic form of 
eq.\ (\ref{bkpsolutionwitheigen}). 

\subsubsection{The Odderon as an Integrable Model}
\label{XXXsect}

Remarkably, the Hamiltonian ${\cal H}_\soddi$ associated with the 
perturbative Odderon possesses an additional hidden symmetry 
and turns out to be equivalent to a completely integrable 
model. The discovery of this equivalence was an important 
breakthrough in the understanding of the perturbative Odderon. 
In fact most of the recent results on the Odderon rest on its 
integrability. Most importantly, the development triggered by that 
discovery has lead to the precise determination of the perturbative 
Odderon intercept. 

We have already seen in the previous section that the Odderon 
Hamiltonian ${\cal H}_\soddi$ possesses an integral 
of motion $q_2$ related to the conformal invariance in 
impact parameter space. In addition, the total transverse 
momentum of the three reggeized gluons, 
\be
P= -i \del_1 -i \del_2 - i \del_3
\,,
\ee
is a conserved quantity in the $t$-channel evolution, 
i.\,e.\ under the action of the Odderon Hamiltonian, 
$[P,{\cal H}_\soddi]= 0$. With $q_2$ and $P$ we have two 
conserved charges in the holomorphic sector, and also the 
corresponding operators $\bar{q}_2$ and $\bar{P}$ in the 
antiholomorphic sector are conserved. For the complete 
integrability of the problem one more integral of motion and its 
antiholomorphic counterpart are required. 
It was discovered in \cite{Lipatov:1993yb} that the Odderon 
indeed has an additional conserved charge associated with the 
operator $q_3$ defined as 
\be
q_3 = - i \rho_{12} \rho_{23} \rho_{31} \del_1 \del_2 \del_3
\,,
\ee
and a similarly defined operator $\bar{q}_3$ in the antiholomorphic 
sector. These operators commute with the Hamiltonian $H_\soddi$, 
\be
[q_3,H_\soddi] = [\bar{q}_3, H_\soddi] = 0
\,,
\ee
as well as with $\bar{H}_\soddi$ 
and hence with the full Odderon Hamiltonian ${\cal H}_\soddi$. 

In \cite{Faddeev:1994zg} (see also \cite{Korchemsky:1994um}) 
it was subsequently shown that the origin of the conserved 
charge is the equivalence of the perturbative Odderon with the 
XXX Heisenberg model of spin $s=0$, consisting of a closed 
spin chain containing three sites. It was found that this 
equivalence actually goes far beyond the Odderon itself. 
It extends to the large-$N_c$ limit of the general case of 
an arbitrary but fixed number of reggeized gluons in the 
generalized leading logarithmic approximation. Such a 
system is described by a straightforward generalization of the 
BKP equation for the Odderon to a larger number of 
reggeized gluons. That system will be the subject of section \ref{gllasect}, 
where we will also give a detailed account of the equivalence 
with the XXX Heisenberg model. It is important to note that 
for an arbitrary number of reggeized gluons the equivalence 
holds only in the large-$N_c$ limit. But in the case of the 
Odderon, i.\,e.\ for three reggeized gluons, the equivalence is exact 
already for finite $N_c$ and does not require any further assumptions. 

The complete integrability implies that the Odderon Hamiltonian 
is a function of the conserved charges $P$, $q_2$, 
$q_3$ and their antiholomorphic counterparts. As a consequence 
of the M\"obius invariance the Hamiltonian does not depend 
on the total momentum $P$ of the three reggeized gluons, 
and we have for example for the holomorphic sector 
\be
\label{hdeponq}
H_\soddi = H_\soddi(q_2,q_3) 
\,.
\ee
Hence the determination of the eigenvalues of the Hamiltonian 
is equivalent to the simultaneous diagonalization of the 
operators $P$, $q_2$, and $q_3$. 
The possible eigenvalues of the operator $q_2$ are fixed by 
the representation theory of the symmetry group $\mbox{SL}(2,\C)$ 
as discussed above. Therefore the spectrum of the Odderon Hamiltonian 
can be found by calculating the spectrum of the operator $q_3$. 
We should add though that 
the dependence of the Hamiltonian $H_\soddi$ on the conserved charges 
is very complicated, and there is no simple explicit form of 
(\ref{hdeponq}). 

The quantization condition for the possible eigenvalues of the 
operator $q_3$ comes from the requirement that the 
Odderon wave function has to be single--valued. So far we have 
considered the holomorphic and antiholomorphic parts of the 
Odderon Hamiltonian separately. But the physical 
wave function of the Odderon must clearly be single--valued 
in the complete transverse impact parameter plane, i.\,e.\ for 
the physical region in which $\rhobar = \rho^*$, and by this condition 
the holomorphic and the antiholomorphic sector are tied together. 
Two further requirements are the normalizability of the wave function 
and its Bose symmetry. It turns out that the latter two 
requirements are easily fulfilled, and it is the single--valuedness 
condition which is instrumental in finding the quantization 
condition for the conserved charge $q_3$. 

Motivated by the complete integrability of the Odderon the use of a 
generalized Bethe ansatz was suggested for this problem 
in \cite{Faddeev:1994zg,Korchemsky:1994um}. That method is 
based on the Baxter $Q$-operator and the separation of variables 
in the Sklyanin representation of the wave function. 
It can be successfully applied also to the more general case of 
a state of a fixed number of reggeized gluons in the large-$N_c$ limit 
of the generalized leading logarithmic approximation, 
see section \ref{gllasect}. Again we emphasize that in the case of the 
Odderon the method works also for finite $N_c$. 
The method relies on the existence of the Baxter $Q$-operator 
${\bf Q}(\lambda)$. It depends on a complex spectral parameter $\lambda$, 
and commutes with itself for different values of the spectral parameter, 
$[{\bf Q}(\mu),{\bf Q}(\lambda)]=0$, as well as with the conserved 
charges, $[{\bf Q}(\lambda),q_2] = [{\bf Q}(\lambda),q_3] = 0$. 
The eigenvalues $Q(\lambda)$ of the Baxter $Q$-operator satisfy the 
Baxter equation which in the case of the Odderon reads 
\be
\label{BaxforQ}
(2 \lambda^3 + q_2 \lambda + q_3) Q(\lambda) = 
(\lambda + i)^3 Q(\lambda + i) + (\lambda - i)^3 Q(\lambda - i) 
\,,
\ee
where $q_2$ and $q_3$ here denote the eigenvalues of the corresponding 
operators. In addition to the Baxter equation there are two more 
conditions which the Baxter $Q$-operator has to fulfill: it is a 
meromorphic function with known pole structure, and it has to exhibit 
a known asymptotic behavior at infinity. 
Analogous conditions hold for the Baxter operator in the antiholomorphic 
sector. It can be shown that the Baxter equation (\ref{BaxforQ}) and 
the other two conditions fix the Baxter $Q$-operator uniquely, and 
consequently also determine the quantization of the conserved charges 
$q_2$ and $q_3$. For given eigenvalues $q_2$ and $q_3$ one can 
then obtain the eigenvalue $\epsilon$ of the Hamiltonian $H_\soddi$ via 
\be
\label{energyausbaxterQ}
\epsilon (q_2,q_3) = 
-6 + i \frac{d}{d\lambda} \log 
\left. \frac{Q(\lambda-i)}{Q(\lambda+i)} \right|_{\lambda=0}
\,,
\ee
and one can hence compute the intercept as well as the full 
spectrum of the Odderon from the eigenvalues of the Baxter operator. 
A useful way to deal with the Baxter equation is to write $Q(\lambda)$ 
in terms of a contour integral \cite{Faddeev:1994zg} 
\be
\label{contourQ}
Q(\lambda) = \int_C \frac{dx}{2 \pi i} \,x^{-i \lambda -1} 
(x-1)^{i \lambda -1} \tilde{Q}(x)
\,,
\ee
with a closed path $C$ in the complex $x$-plane, after 
which the Baxter equation (\ref{BaxforQ}) is transformed into 
a third--order differential equation for $\tilde{Q}$. The condition 
on the asymptotic behavior of $Q$ mentioned above implies that 
$\tilde{Q}(x) \to x^{h-2}$ as $x \to \infty$. 

The problem of the quantization condition for the conserved 
charge $q_3$ has been considered in a number of papers 
\cite{Korchemsky:1994um}--\cite{Derkachov:2002pb}. Of these 
the papers \cite{Korchemsky:1994um}--\cite{Korchemsky:1996kh} deal with 
different aspects of the Baxter equation in the Bethe ansatz approach. 
In \cite{Lipatov:1998as} the quantization condition was studied 
in a more direct approach. Here the single--valuedness condition 
for the wave function was discussed in the context of a further 
duality symmetry of the Odderon Hamiltonian to which we will turn 
in more detail in the next section. 
Also the investigation of the perturbative Odderon problem 
in the quasiclassical (WKB) approximation in 
\cite{Korchemsky:1995be,Korchemsky:1996kh,Korchemsky:1997yy,Korchemsky:1997ve} 
was fruitful in several respects. Firstly, it has turned out that the exact 
spectra of the conserved charge $q_3$ and of the Odderon Hamiltonian 
${\cal H}_\soddi$ are reproduced to surprising accuracy in 
that approximation \cite{Derkachov:2002wz,Derkachov:2002pb}. 
This renders the WKB approximation a very valuable tool  in the study 
of the perturbative Odderon problem. Secondly, the WKB approximation 
has made it possible to establish that the perturbative Odderon bears 
interesting relations to the integrable structures found in 
Seiberg--Witten dynamics of ${\cal N}=2$ supersymmetric 
QCD \cite{Korchemsky:1996kh} as well as to the dynamics of 
solitonic waves \cite{Korchemsky:1997yy}. 

The intercept of the solutions with $q_3 \neq 0$ was first found 
in \cite{Wosiek:1996bf} and \cite{Janik:1998xj}. In \cite{Wosiek:1996bf} 
an exact method was devised to solve the Baxter equation for 
arbitrary values of the conserved charges $q_2$ and $q_3$. This method 
uses a double contour integral representation of the eigenvalue of the 
Baxter operator in which the latter is written as the sum of two terms 
of the form (\ref{contourQ}). With suitably chosen contours this allows 
one to implement the cancellation of boundary terms arising from 
integration by parts. That cancellation is not possible in a 
single--contour integral representation due to the 
particular structure of the Riemann surface of the integrand. 
In this way it is possible to find a unique solution of the Baxter 
equation within the ansatz of the double--contour integration. 
A more direct approach was then used in \cite{Janik:1998xj} to 
construct the quantization condition for the conserved charge $q_3$ 
and to determine the corresponding eigenfunction of the Odderon 
Hamiltonian. That construction will be described in section 
\ref{jwsolutionsect} below. From the values for $q_3$ found by this 
construction it was then possible to compute the Odderon intercept via 
the method of \cite{Wosiek:1996bf} outlined above. In these studies 
it was assumed that the maximal eigenvalue of the Odderon 
Hamiltonian is obtained for the conformal weight $h=1/2$. 
Subsequent studies of the Odderon intercept for arbitrary 
conformal weights $h$ in \cite{Praszalowicz:1998pz,Kotanski:2001iq} 
have confirmed that assumption. 

In \cite{DeVega:2001pu,deVega:2002im} and in 
\cite{Derkachov:2001yn,Korchemsky:2001nx,Derkachov:2002wz} 
finally the full spectrum of the operator $q_3$ and of the Odderon 
Hamiltonian ${\cal H}_\soddi$ was computed using the method 
of the Baxter ${\bf Q}$-operator. The results for the spectrum 
were compared to the corresponding results in the WKB 
approximation in \cite{Derkachov:2002pb}, and again a good 
agreement was found. There is, however, a fundamental difference 
between \cite{deVega:2002im} and 
\cite{Korchemsky:2001nx,Derkachov:2002wz} 
concerning the physical interpretation of the results which we will 
discuss momentarily. 

A detailed account of all studies of the integrability properties 
of the Odderon mentioned above would clearly go beyond 
the scope of the present review. We therefore restrict ourselves to the 
description of the main results and of the presently open problems. 
In sections \ref{jwsolutionsect} and \ref{blvsolutionsect} 
we will discuss the explicitly known solutions of the BKP equation. 
The significance of these two solutions and the Odderon intercept 
will then be the subject of section \ref{pertinterceptsect}. 
Concluding the present section we now discuss the possible 
eigenvalues of the operator $q_3$. 

There exist eigenfunctions of the Hamiltonian $H_\soddi$ 
both for the quantum number $q_3=0$ as well as for $q_3 \neq 0$. As 
we will see in detail below their properties are quite different. 
The solutions with $q_3=0$ (see section \ref{blvsolutionsect}) 
require to extend the Hilbert space of the Odderon Hamiltonian 
such that also distributions and not only smooth functions are 
included. Such an extension appears very natural due to the 
phenomenon of reggeization. Although this extension needs 
further study the solutions with $q_3=0$ are generally accepted as 
physical solutions. The situation is more complicated in the case 
of solutions with $q_3 \neq 0$. For this case the full spectrum 
of $q_3$ has been computed in \cite{DeVega:2001pu,deVega:2002im} 
and in \cite{Derkachov:2001yn,Korchemsky:2001nx,Derkachov:2002wz}. 
The results of these two groups agree as far as purely imaginary 
values of $q_3$ are concerned. However, there is a disagreement 
concerning the case of general complex values of $q_3$ with nonvanishing 
real part. The authors of \cite{deVega:2002im} find that the corresponding 
eigenfunctions of the Hamiltonian are unphysical, whereas in 
\cite{Derkachov:2002wz} it is argued that completeness of the 
eigenfunctions requires to include them. Clearly, further study 
is needed to clarify this important question. 
Note however that this disagreement does not affect the determination 
of the Odderon intercept because the state with the highest 
eigenvalue $E$ corresponds to a purely imaginary value of $q_3$. 

\subsubsection{Modular Invariance of the BKP Equation}
\label{modularsect}

A very interesting symmetry of the Odderon wave function 
was observed in \cite{Gauron:1991cg} 
and further discussed in \cite{Janik:1995kb,Janik:1996mm}. 
Both the Hamiltonian $H_\soddi$ of the Odderon as well as the 
conserved charge $q_3$ are invariant under cyclic permutations 
of the gluon coordinates $\rho_i$. This invariance is a direct 
consequence of Bose symmetry for the three--gluon state. 
We should note here that the invariance under cyclic permutations 
of the gluon coordinates also applies to the more general case 
of $N$-gluon states in the large-$N_c$ limit of the GLLA. 
The situation is, however, slightly different for more than $3$ 
gluons since in that case the cyclic permutations together with the 
inversion of the order do not exhaust the complete permutation 
group as in the case of the Odderon. 

Under the repeated cyclic permutation $\rho_1 \to \rho_2 \to \rho_3$ 
the anharmonic ratio $z$ transforms as 
\be
 z \,\,\,\longrightarrow \,\,\,
1-\frac{1}{z} \,\,\,
\longrightarrow \,\,\,
\frac{1}{1-z}
\,,
\ee
and the analogous transformation property holds for $\bar{z}$ in the 
antiholomorphic sector. The invariance under this transformation 
is called modular invariance. 
As shown in \cite{Janik:1995kb,Janik:1996mm} the invariance 
under cyclic permutations of the gluon coordinates in fact allows one 
to relate the Odderon problem to the modular invariance in the 
theory of elliptic curves. 
It can be shown that as a result of 
modular invariance the function $\phi^{(h)}(z)$ 
in (\ref{furtherfactphi}) depends on $z$ in a special way, 
namely it is a function of the ratio 
\be
\label{modvarxi}
\xi = \frac{i}{3 \sqrt{3}} \,\frac{(z-2)(z+1)(2z-1)}{z(z-1)}
\ee
only, and a similar result holds for $\bar{\phi}^{(\bar{h})}(\bar{z})$. 
This property was later used in \cite{Praszalowicz:1998pz,Kotanski:2001iq} 
for constructing the quantization conditions for the conserved charge 
$q_3$. The invariance of the Odderon under cyclic permutations 
was also used in \cite{Korchemsky:1999is} for finding 
a new representation for the Odderon wave function. 
It classifies it according to a new quantum number 
triality $\lambda$ for which $\lambda^3=1$. 
In this way simultaneous eigenfunctions of $q_2$ 
and $q_3$ were constructed which possess definite 
triality. Also here the quantization condition 
for $q_3$ can be studied by making use of this 
decomposition. The new representation is also useful for 
calculating the eigenvalues of a wide class of operators. 

There is an interesting consequence of the modular invariance of the 
Odderon for the spectrum of the operator $q_3$. The eigenvalues of 
$q_3$ found in \cite{Korchemsky:2001nx,Derkachov:2002wz} have 
a lattice--like structure in the complex plane. 
(Note however that the physical significance of those 
eigenstates of $q_3$ with eigenvalues that are not purely 
imaginary is still unclear, see the discussion in 
sections \ref{XXXsect} and \ref{pertinterceptsect}.) 
The origin of that lattice--like structure 
becomes particularly transparent in the quasiclassical (WKB) 
approximation \cite{Derkachov:2002pb}. There it is shown that 
the modular invariance can be related to the properties 
of the Riemann surface defined by the spectral curve of the 
resulting integrable model. 
The same Riemann surface is also 
of central interest in the attempt to interpret the Odderon 
and higher $N$-gluon states in the GLLA in terms of string 
theory amplitudes \cite{Gorsky:2002ju}. 

Another interesting symmetry of the Odderon was discovered in 
\cite{Lipatov:1998as}. It was found there that the conserved charges 
$q_2$ and $q_3$ and hence also the Odderon Hamiltonian $H_\soddi$ 
are invariant under a duality transformation which relates the 
coordinates to the corresponding derivative operators, 
i.\,e.\ to the momenta of the reggeized gluons. 
Specifically, the duality transformation is given by 
\be
\label{lipdua}
\rho_{i-1,i} \longrightarrow i \del_i \longrightarrow 
\rho_{i,i+1}
\ee
(with the cyclic identification $\rho_4=\rho_1$) and the simultaneous 
reversing of the order of the operator multiplication. 
If the duality transformation is performed twice one is 
led to the cyclic permutation of the gluon coordinates discussed above. 
The invariance under cyclic permutations and the resulting 
modular invariance is hence closely related also to the 
invariance under the duality transformation. 

This duality symmetry has an immediate consequence for the Odderon 
wave function. In the next section we will discuss the form 
of the wave function for the class of solutions for which $q_3 \neq 0$. 
In the representation that we introduce there the wave function is 
determined by three parameters. It turns out that in this representation 
the condition resulting from the invariance under the duality 
transformation has a particularly simple form. It says that the ratio 
of two of the wave function parameters must equal the modulus 
of the quantized eigenvalue of the operator $q_3$ for the corresponding 
solution. The invariance under the duality transformation is hence 
closely related also to the quantization condition for the operator $q_3$. 

Let us finally note that the invariance under the duality transformation 
holds also for the more general case of a fixed number $N$ of reggeized 
gluons in the large-$N_c$ limit of the GLLA. Interestingly, the above 
duality transformation coincides with a supersymmetric translation 
in the limit $N \to \infty$. It has been suggested in \cite{Lipatov:1998as} 
that this indicates a close relationship of the integrability 
of the $N$-gluon states in the large-$N_c$ limit of the GLLA with 
supersymmetric field theories and superstring theories. 

\subsubsection{The Janik-Wosiek Solution}
\label{jwsolutionsect}

Exact eigenfunctions of the operator $q_3$ and of the BKP 
Odderon Hamiltonian have been found by Janik and Wosiek 
in \cite{Janik:1998xj}, and we will call them JW solutions. 
These eigenfunctions have quantum 
numbers $q_3 \neq 0$. The advantage of the method of 
\cite{Janik:1998xj} is that it not only allows one to find the 
quantization of the eigenvalues of $q_3$ and the Odderon 
intercept but also to obtain the Odderon 
wave function in explicit form. 

Since the operators $q_2$ and $q_3$ commute one can find a set 
of simultaneous eigenfunctions for these operators. 
We have seen in section \ref{bkpsymmsect} that the eigenfunctions 
of the Odderon Hamiltonian can be written in conformally invariant 
form diagonalizing $q_2$, see (\ref{odderonwavefct}) together 
with (\ref{furtherfactphi}). One can use that form of the eigenfunctions 
and find the quantization condition of the operator $q_3$. 
For fixed $q_2$ the eigenvalue equation for the operator 
$q_3$ can be translated into a third--order linear 
differential equation \cite{Lipatov:1993qn} which then reads 
\be
\left[
a(z) \frac{d^3}{dz^3} + b(z) \frac{d^2}{dz^2} 
+c(z) \frac{d}{dz} +d(z)
\right]
\phi^{(h)} (z) =0
\,,
\ee
where the coefficient functions are defined as 
\bea
a(z)&=& z^3(1-z)^3
\\
b(z)&=&2 z^2 (1-z)^2 (1-2z)
\\
c(z)&=&z(z-1) 
\left(z(z-1)(3\mu+2)(\mu -1) + 3\mu^2 -\mu \right)
\\
d(z)&=&\mu^2(1-\mu)(z+1)(z-2)(2z-1) - iq_3 z(1-z)
\,,
\eea
and we have used $\mu=h/3$. This differential equation has regular 
singular points at $z=0,1$ and $\infty$ and can be solved by 
standard methods in the case $q_3\neq 0$. 

For the wave function $\phi^{(h,\bar{h})}(z,\bar{z})$ in the conformally 
invariant ansatz (\ref{odderonwavefct}) one can make the ansatz 
\be
\label{jwansatz0}
\psi(z,\bar{z}) = \sum_{i,j} \bar{u}_i(\bar{z}) A_{ij}^{(0)} u_j(z)
\,,
\ee
where $u_i(z)$ and $\bar{u}_i(\bar{z}) $ are eigenfunctions of the 
operators $q_3$ and $\bar{q}_3$. The quantization of $q_3$ follows from 
the conditions that the wave function is single--valued, must 
be normalizable and must satisfy Bose symmetry. As already emphasized, 
the first assumption is the crucial point in the quantization of $q_3$. 
It turns out to be possible to find a solution to these conditions by 
studying the differential equation in the vicinity of the singular points. 

One can construct a set of linearly independent solutions $u_i^{(0)}$ 
around one of the singular points, say $z=0$. The largest eigenvalue 
of $E$ of the Odderon Hamiltonian is expected for $h=1/2$, and for 
that choice one obtains the three independent solutions 
$u^{(0)}_1 \sim z^{1/3}$, $u^{(0)}_2 \sim z^{5/6}$ and 
$u^{(0)}_3 \sim z^{5/6} \log z + z^{-1/3}$. 
From this and from the uniqueness of the wave function one can conclude that 
the coefficient matrix $A^{(0)}$ in (\ref{jwansatz0}) has to have the form 
\be
A^{(0)}=
 \left(
  \begin{array}{ccc}
  {\alpha}&{0}&{0}\\
  {0}&{\beta }&\gamma\\
 0 & \gamma & 0
  \end{array}
  \right)
\,.
\ee
This form also automatically ensures the normalizability of the 
wave function. 
Next, one imposes the same requirements around the singular point $z=1$. 
This can be done by considering the solutions 
$u^{(1)}_i(z) = u^{(0)}_i(1-1/z)$ which have a similar behavior around 
$z=1$. Further, one can numerically calculate the analytic transition 
matrices $\Gamma$ in $u^{(0)}_i(z) = \Gamma_{ij} u^{(1)}_j(z)$. 
Then the wave function $\psi(z,\bar{z})$ is expressed in terms of 
the solutions $u^{(1)}_i$ in a form analogous to (\ref{jwansatz0}). 
It follows that the transformed coefficient matrix 
$A^{(1)} = \bar{\Gamma} A^{(0)} \Gamma$ has to have the same 
form as $A^{(0)}$. This conditions leads to a system 
of equations for the three parameters $\alpha$, $\beta$, 
and $\gamma$. That system turns out to be overconstrained and 
therefore fixes not only those three parameters but also the allowed 
values of $q_3$. 
If the coefficient matrices $A^{(0)}$ and $A^{(1)}$ coincide the 
requirement of Bose symmetry can finally be implemented by 
adding the corresponding wave function with the eigenvalues 
$(-q_3,-\bar{q}_3)$, 
\be
\psi(z,\bar{z}) =
\psi_{q_3, q_3^*}(z,\bar{z}) + \psi_{-q_3, -q_3^*}(z,\bar{z}) 
\,.
\ee
Having determined the possible eigenvalues of the operator $q_3$ 
in this way one can finally use the method developed in 
\cite{Wosiek:1996bf} to compute the corresponding intercept 
as described in section \ref{XXXsect}. 

A number of possible solutions were found in \cite{Janik:1998xj}. 
The state with the highest intercept has $h=1/2$ and 
$q_3= \pm 0.20526\, i$. The corresponding intercept is 
\be
\alpha_\soddi = 1 - 0.24717 \, \frac{\alpha_s N_c}{\pi}
\,,
\ee
which for a realistic $\alpha_s \simeq 0.2$ yields 
$\alpha_\soddi = 0.96$. 
The corresponding wave function parameters are 
$\alpha = 0.7096$, $\beta= - 0.6894$ and 
$\gamma= 0.1457$. 
In \cite{Praszalowicz:1998pz,Kotanski:2001iq} also arbitrary 
conformal weights $h$ have been investigated, and the assumption 
has been confirmed that the maximal intercept is obtained 
for $h=1/2$. 
In this review we will usually refer to this particular solution 
with the maximal intercept as the JW solution, having in mind 
that the JW construction applies to a whole set of solutions 
with $q_3 \neq 0$. 

The intercept of the JW solution is only slightly below one, 
and it should therefore stay relevant up to rather high 
energies. It was found, however, that the JW solution does 
not couple to all phenomenologically relevant impact factors. 
The coupling to the $\gamma \Od \eta_c$ impact factor 
for example vanishes in leading order. 
This is actually true for any solution with $q_3\neq 0$ 
of the form (\ref{odderonwavefct}). This problem will be 
discussed in more detail in see section \ref{epdiffsect}. 

In section \ref{modularsect} we have discussed the 
invariance of the Odderon Hamiltonian under the duality 
transformation discovered in \cite{Lipatov:1998as}. 
This invariance has a very interesting consequence for the 
wave function parameters of the JW solution. Namely, 
one can show that the parameters $\alpha$ and $\gamma$ 
introduced above are related by $\gamma = |q_3| \alpha$. 
This relation has in fact been confirmed for all solutions 
of the JW type that have been found, see for example 
\cite{Janik:1998xj}. This also applies to the solutions found in 
\cite{Derkachov:2002wz} (see the discussion at the end of section 
\ref{XXXsect}) which have eigenvalues $q_3$ that 
are not purely imaginary. 

\subsubsection{The Bartels-Lipatov-Vacca Solution}
\label{blvsolutionsect}

Another exact solution of the BKP equation was constructed in 
\cite{Bartels:1999yt}. This Bartels--Lipatov--Vacca (BLV) solution 
has the quantum number $q_3=0$ and its properties differ 
quite significantly from those of the solutions with $q_3 \neq 0$ 
discussed in the previous section. Initially the case $q_3=0$ had 
been considered unphysical and became generally accepted as 
a valid solution of the BKP equation only after the BLV solution was 
discovered and its phenomenological significance was shown. 

The BLV solution is most conveniently written in momentum space. 
Let us denote by $E^{(\nu,n)} (\kf_1,\kf_2)$ the momentum space 
representation of the eigenfunctions of the BFKL Hamiltonian 
${\cal H}_\spommi$, i.\,e.\ the Fourier transforms of the functions 
given in (\ref{bfkleigenfu}). For the special case of the forward 
direction $\kf_1 = -\kf_2 = \kf$ they are (up to a normalization 
factor) simply given by (\ref{kleinebfkl}). In the nonforward 
direction the $E^{(\nu,n)} (\kf_1,\kf_2)$ are more complicated, 
their explicit form can be found for example in \cite{Bartels:2001hw}. 
Exact eigenfunctions of the BKP Hamiltonian ${\cal H}_\soddi$ 
are then given by 
\be
\label{blvsolution}
E_3^{(\nu,n)} (\kf_1,\kf_2,\kf_3) = 
d(\nu,n) \sum_{(ijk)} \frac{(\kf_i+ \kf_j)^2}{\kf_i^2 \kf_j^2} 
E^{(\nu,n)} (\kf_i+\kf_j,\kf_k)
\ee
where the sum goes over cyclic permutations of the three 
momenta. It can be shown that these solutions exist only for 
odd values of the conformal spin $n$. The normalization factor 
$d(\nu,n)$ is chosen in \cite{Bartels:1999yt}  as 
\be
d(\nu,n) = g_s \frac{N_c}{\sqrt{N_c^2-4}} \,
\frac{1}{\sqrt{-3\frac{N_c \alpha_s}{\pi}\chi(\nu,n)}}
\,.
\ee

The solution of the BKP equation corresponding to these eigenfunctions 
of the Hamiltonian has quite interesting properties. We first observe 
that the BLV solution exhibits the phenomenon of reggeization, which we 
have discussed in detail in section \ref{reggeizationsect}. As is explicit 
in (\ref{blvsolution}) this three--gluon state is the superposition of 
two--gluon (BFKL) states. In each of them two of the three 
gluon momenta enter only as a sum. 
If one takes the Fourier transform of a function that depends 
only on the sum of two momenta one readily gets a function 
in impact parameter space that is proportional to a delta function 
of the difference of the corresponding coordinates. The two gluons 
corresponding to the two momenta in the sum are hence at the 
same position in configuration space. 
Consequently, one can view them as forming a more composite 
gluon in the sense of reggeization. Note that the same momentum 
structure, and hence reggeization, also occurs in the three--gluon 
amplitude carrying positive $C$-parity in the extended GLLA 
which we discuss in section \ref{egllasect} below, see equation 
(\ref{d3reggeizes}) there. 

The special momentum structure of the BLV solution is also the origin of 
its second important property. As we have just seen two of the three 
gluon coordinates coincide in each of the terms in the sum in 
(\ref{blvsolution}), and in impact parameter space the solutions 
thus contain delta function terms of the difference of two coordinates. 
Accordingly one easily finds that the 
BLV solution does not vanish in the case that two gluon 
coordinates are identical as it was the case for the JW solution. 
A consequence of that observation is that the BLV solution cannot 
be written in the form (\ref{odderonwavefct}) if $\phi^{(h,\bar{h})}$ 
in that equation is chosen as a nonsingular function. The Hilbert space 
of admissible solutions of the BKP equation hence has to be extended 
in order to accommodate the BLV solution. The important problem 
of defining the proper Hilbert space for the solutions of the BKP 
equation is discussed in \cite{Bartels:1999yt} but certainly 
deserves further study. 

The intercept $\alpha_\soddi$ of the BLV Odderon solution is exactly 
equal to one, $\alpha_\soddi=1$. 
In particular, it is larger than the intercept of the 
JW solution discussed above. Also the BLV solution contains the 
conformal weight $h$ and hence depends on the continuous 
parameter $\nu$. As a consequence also this solution corresponds 
to a cut in the complex angular momentum plane. 
We will compare the phenomenological relevance of the BLV 
and JW solutions in the next section and also in section 
\ref{epdiffsect}. Here we already point out that the special momentum 
structure of the BLV solution has important consequences for the 
coupling of the Odderon to external particles. 
In particular, it couples to the $\gamma \Od \eta_c$ impact factor, 
in contrast to the JW solution. 

Concluding this section we note that the BLV solution is a special 
case of a whole family of solutions of the BKP equation for the 
systems of $N$ reggeized gluons, considered in the large-$N_c$ limit 
when $N \ge 4$. This set of solutions was constructed in 
\cite{Vacca:2000bk}. There it was shown that a an exact solution for 
the $N$-reggeon BKP equation can be obtained from the solution 
of the BKP equation for $(N-1)$ reggeized gluons. The corresponding 
construction is analogous to the one in (\ref{blvsolution}) and hence 
rests on the reggeization property of the gluon.  
Also these solutions have, like the BLV Odderon solution, an intercept 
of exactly one. These solutions and especially their quantum 
numbers (see section \ref{gllasect}) have also been discussed in 
\cite{Derkachov:2002wz}. There these states are called descendent states 
because of their construction in which the $N$-gluon solution 
is obtained from the $(N-1)$-gluon solution. 

\subsubsection{The Odderon Intercept and the Odderon Spectrum}
\label{pertinterceptsect}

We have seen that there are two classes of solutions of the 
BKP equation which have quite different properties. These 
two classes are characterized by their quantum number 
$q_3$. In the case of the JW solution we have $q_3 \neq 0$ 
whereas the BLV solution has  $q_3 =0$. The maximal intercept 
for the solutions of JW type is found to be $\alpha_\soddi= 0.96$, 
whereas the BLV solution has an intercept of exactly one, 
$\alpha_\soddi= 1$. Accordingly one should expect that 
the BLV solution dominates at very high energies. However, besides 
the intercept one also has to consider the coupling of the two solutions
to external particles in order to determine their relative 
phenomenological relevance. Here it turns out that the couplings 
strongly depend on the scattering process under consideration. 
The JW solution for example does not couple to the 
$\gamma \eta_c \Od$ impact factor in leading order, 
whereas the BLV solution does so, see section \ref{epdiffsect}. 
There are also cases in which both solutions couple, for example 
the perturbative Pomeron--Odderon--Odderon vertex, see section 
\ref{poosect}, and a similar situation is also to be expected in general for 
the coupling of the Odderon to complicated hadronic bound states 
like the proton. It is feasible that there exist also cases in which 
only the JW solution couples to the external particles, but the 
BLV solution does not. 
In all cases where the BLV solution is relevant it will be the 
dominant solution at sufficiently high energies, and energy behavior 
will then be given by the BLV intercept of exactly one. 
It should be kept in mind though that the 
difference between the two intercepts for the JW and BLV solutions 
is rather small such that the relative magnitude of the 
impact factors of the two solutions is quite important. 

Another important issue is the type of the Odderon singularity. 
In the generalized leading logarithmic approximation one 
usually finds solutions which correspond to cuts in the 
complex angular momentum plane. This is true for the 
BFKL Pomeron, and it also holds for both known solutions 
for the BKP Odderon. The JW solution as well as the BLV 
solution correspond to cuts in the complex angular momentum 
plane, and the intercept refers to the point where the cut 
starts. The reason for these cuts is the fact that the set of 
solutions for the Pomeron as well as for the Odderon depend 
on the conformal weight $h$. According to the 
representation theory of the underlying symmetry group 
$\mbox{SL}(2,\C)$ the parameter $h$ is continuous. 
Hence the spectrum of the BFKL and BKP Hamiltonians is 
gapless, which induces cuts in the complex angular momentum plane. 
In the case of the 
BFKL Pomeron it is believed that the cut is transformed 
into a pole singularity when running coupling effects are 
taken into account. However, this result depends on the way 
in which the running coupling constant is cut off at low 
momenta. A similar mechanism might also apply to the 
Odderon, but this has not yet been investigated.  

Also of phenomenological interest is the twist of the Odderon 
solutions which determines for example the dependence on the 
photon virtuality in processes involving the $\gamma^* \eta_c \Od$ 
impact factor. This question has not been considered in any detail 
in the literature so far. There are indications 
\cite{Grishaprivcomm} that the BLV and JW solutions have 
different twist, namely that the BLV solution corresponds to twist three 
whereas the JW solution corresponds to twist four. 

We should point out again that there is presently 
a disagreement in the literature concerning the question 
which condition the eigenvalues of the operator $q_3$ have 
to satisfy in order to correspond to physical Odderon states. 
In \cite{deVega:2002im} states with $q_3 \neq 0$ and 
a nonvanishing real part of $q_3$ are considered 
unphysical, but in \cite{Derkachov:2002wz} arguments are given 
that these states should be included in the physical Odderon spectrum. 
Though being conceptually quite important this point does not affect 
the issue of the Odderon intercept since in any case the leading 
singularity in the JW type solutions corresponds to a purely 
imaginary value of $q_3$. 

Before the exact solutions to the Odderon Hamiltonian 
were found the Odderon intercept was subject to a number 
of studies using variational methods 
\cite{Gauron:ic}--\cite{Braun:1998uu}. 
After the exact solutions were found these studies have lost most 
of their importance although they were very 
valuable at the time and have led to interesting 
results about the Odderon in general. We will 
not discuss the variational approaches in detail here. 
Interestingly, most attempts to determine 
the perturbative Odderon intercept in variational approaches had 
found that the Odderon intercept is larger than one, 
$\alpha_\soddi >1$, and have thus failed to find 
the actual value. However, after the exact Janik--Wosiek 
solution was found another attempt was made 
\cite{Braun:1998tg,Braun:1998mg} 
in the variational approach with an improved set of 
functions the choice of which was based on the 
Janik--Wosiek solution. Indeed their result 
for the intercept was confirmed. 
The BLV solution has not been found in any variational 
approach. This is not surprising because of its singular 
dependence on the gluon coordinates which just has to 
be missed by any variational approach that starts from 
smooth functions. 

\subsection{The Odderon and Unitarity of High Energy Scattering}
\label{oddunitaritysect}

The perturbative approach to high energy scattering leads in 
the leading logarithmic approximation to a rather rapid growth 
of hadronic cross sections at high energy. In LLA the cross section 
is dominated by ladder type diagrams with two interacting reggeized 
gluons in the $t$-channel. The corresponding cross sections 
exhibit a powerlike growth $\sim s^{\omega_{\mbox{\tiny BFKL}}}$, 
with the BFKL intercept of around $1+\omega_{\mbox{\scriptsize BFKL}}
\simeq 1.5$, and even in NLLA the BFKL intercept 
is still around $1.2$. If this growth would continue to asymptotically 
large energies $\sqrt{s}$  it would clearly contradict the 
Froissart--Martin bound (\ref{fmtheorem}) which allows at most 
a logarithmic growth, and hence would violate unitarity. 
In practice this violation of the Froissart--Martin bound happens 
only at asymptotically large energies far beyond the string scale. 
But a full understanding of the high energy limit of QCD clearly 
requires to establish a description satisfying unitarity bounds. 
The solution to the problem is --- in principle --- well known. 
It is easy to show that at higher and higher energies exchanges 
with more than two reggeized gluons become more important 
and eventually dominant. Moreover, such diagrams can give 
a sizable contribution to scattering processes already at 
intermediate energies and can hence be phenomenologically important. 
In the limit $s \to \infty$ an infinite number of such exchanges 
with arbitrary numbers of gluons in the $t$-channel has 
to be included. It is expected that with these exchanges included 
the cross section will unitarize, i.\,e.\ will no longer violate the 
Froissart--Martin bound. 
There are two ways of doing this in practice. Both go 
under the name of the generalized logarithmic approximation 
(GLLA). 
Since the difference between the two will be important for us 
we will here call them GLLA and extended GLLA. In the GLLA 
\cite{Bartels:1980pe,Jaroszewicz:1980mq,Kwiecinski:1980wb}, 
\cite{Bartels:1978fc} 
one takes into account exchanges with arbitrary numbers 
of gluons in the $t$-channel, but keeps the number fixed for each 
exchange. In the extended GLLA one includes also the possibility 
that the number of reggeized gluons fluctuates during the evolution 
in the $t$-channel. The GLLA is expected to unitarize the total 
cross section whereas the extended GLLA also satisfies unitarity bounds 
in all possible subchannels. 

The perturbative BFKL approach is only 
applicable to scattering processes of two small color dipoles, 
the size of which provides a hard momentum scale. In this sense 
the scattering of two highly virtual photons is the ideal process for 
the perturbative study of the Regge limit. Strictly speaking, however, 
the Froissart--Martin theorem cannot be proven for this process, 
although it is widely believed to hold also here. From this point of view 
a unitarization of the BFKL Pomeron would not be required in this process. 
But independently of the validity of the Froissart--Martin theorem 
in this particular process 
the exchanges of more than two reggeized gluons have turned out 
to be very interesting by themselves. Moreover, they can well give 
considerable contributions to high energy scattering processes in 
general and thus be of great phenomenological interest. 

The subject of unitarity in the perturbative approach to 
the Regge limit would offer enough material for a separate review. 
Here we will concentrate only on those aspects that are directly relevant 
to the Odderon. We will explain the equivalence of the 
large-$N_c$ limit of the GLLA to an integrable model, namely 
to the XXX Heisenberg model of spin $s=0$. The Odderon is a 
special case of this equivalence. We will then proceed to the 
extended GLLA. The emphasis will here be on the number 
changing vertices that couple different $N$-gluon states to 
each other. In particular we will describe how these can be 
used to derive the perturbative Pomeron--Odderon--Odderon 
vertex. 
Finally, we briefly mention other (mainly perturbative) 
approaches to the problem of the Regge limit and discuss the 
r{\^o}le of the Odderon in these approaches. 

\subsubsection{$N$-Reggeon States in the GLLA}
\label{gllasect}

A system of $N$ reggeized gluons (or reggeons) 
in the $t$-channel is in the 
GLLA described by the $N$-particle BKP equation. 
This is a straightforward generalization of the BKP equation 
for the Odderon (\ref{bkpoddigl}) to more than three gluons. 
It includes all pairwise interactions of the $N$ reggeons. 
However, for more than three gluons the color structure 
is nontrivial and actually makes the problem intractable, 
at least with currently available methods. 

An enormous simplification occurs when one considers 
the large-$N_c$ limit of the problem. This limit 
corresponds to a situation in which only the gluons 
next to each other interact via BFKL kernels. Moreover, 
neighboring gluons are in this limit in a pairwise color octet state 
such that the complicated color structure reduces to a simple 
factor of $-N_c/2$ in the BFKL kernel. 
It has been shown in \cite{Lipatov:1993yb} that this system 
has a sufficient number of hidden conserved charges to be 
completely integrable. In \cite{Faddeev:1994zg} it was 
then identified as the XXX Heisenberg model of spin $s=0$. 
Here we will give a brief account of this amazing result. 
The case of the Odderon is then easily seen to arise for $N=3$. 

The BKP equation for $N$ reggeons in the large-$N_c$ limit 
can be treated in analogy to the case $N=3$, see section \ref{bkpsect}. 
The eigenvalue equation for the Hamiltonian in impact parameter 
space is 
\be
{\cal H}_N \varphi_N = E_N \varphi_N 
\,,
\ee
where the wave function depends on the holomorphic 
and antiholomorphic coordinates $\rho_i$ and $\rhobar_i$ of 
the $N$ reggeons. The Hamiltonian exhibits holomorphic 
separability, 
\be
\label{holoN}
{\cal H}_N = \frac{\alpha_s N_c}{4 \pi} \left( H_N + \bar{H}_N \right)
\,,
\ee
and $H_N$ describes the interaction of nearest neighbors 
in the holomorphic sector, 
\be
H_N = \sum_{k=1}^N H(\rho_k,\rho_{k+1}) 
\,.
\ee
The pairwise interaction is given by 
\be
\label{refertothis}
H_{12} = H(\rho_1,\rho_2) = 
-\psi(-J_{12}) - \psi(1 +J_{12}) + 2 \psi(1)
\,. 
\ee
Here $J_{12}$ is an operator solution of 
\be
J_{12} (1 + J_{12}) = L_{12} = - (\rho_1-\rho_2)^2 \del_1 \del_2
\,.
\ee
Due to the holomorphic separability (\ref{holoN}) of the Hamiltonian 
${\cal H}_N$ the eigenstates can be factorized as 
\be
\varphi_N (\rho,\rhobar) = \varphi (\rho) \bar{\varphi}(\rhobar)
\,.
\ee
Accordingly, the eigenvalues $E_N$ are obtained as 
\be
E_N= \frac{\alpha_s N_c}{4 \pi} ( \epsilon + \bar{\epsilon} ) 
\ee
with 
\bea
H_N \varphi(\rho) &=& \epsilon \varphi (\rho)\\
\bar{H}_N \bar{\varphi}(\rhobar)
&=& \bar{\epsilon} \bar{\varphi} (\rhobar)
\,. 
\eea

We recall the well--known Heisenberg model with $N$ sites 
of spin $1/2$, 
\be
\label{heisenberhalb}
H_N^{s=1/2} = -\sum_{m=1}^N 
\left( \vec{S}_m \cdot \vec{S}_{m+1} -\frac{1}{4} \right)
\,,
\ee
where $\vec{S}_m$ are the $\mbox{SU}(2)$ generators for spin $s=1/2$ 
acting on the spin at the $m$-th site. We assume periodic boundary 
conditions, $\vec{S}_{N+1}= \vec{S}_1$. 
The generalization of the model to arbitrary complex spins is 
nontrivial. The case of arbitrary spin in general corresponds to 
noncompact, unitary representations of the group $\mbox{SL}(2,\C)$. 
In particular the quantum Hilbert space of the model becomes infinite 
dimensional and there is no longer a highest weight state. 

Let us consider a chain of $N$ spins with periodic boundary 
conditions, and the number of sites coincides with the number 
of reggeons which we are considering. Each site is parametrized 
by two--dimensional holomorphic and 
antiholomorphic coordinates, $\rho_m$ and $\rhobar_m$ with 
$m=1,\dots , N$. In the following we will mainly concentrate 
on the holomorphic coordinates, and there are analogous relations 
for the antiholomorphic coordinates. 
One can assign to each site $k$ the spin operators $S_k^{\pm,3}$ and 
$\bar{S}_k^{\pm,3}$ which are the generators of the 
principal series of the group $\mbox{SL}(2,\C)$ for spin $s$, 
\be
\label{heisenspins}
S_k^+= \rho_k^2 \del_k- 2s \rho_k\,,\hspace*{1cm}
S_k^-= -\del_k \,,\hspace*{1cm}
S_k^3= \rho_k \del_k -s
\,,
\ee
and we have $S^{\pm} = S^1 \pm i S^2$. The operators $\bar{S}_k^{\pm,3}$ 
are obtained by replacing $\rho_k \to \rhobar_k$ and $ s\to \bar{s}=1-s^*$. 
The pair of complex parameters $(s,\bar{s})$ specifies the 
$\mbox{SL}(2,\C)$ representation. 

The definition of the XXX Heisenberg spin chain is now based 
on the existence of the operator $R_{km}(\lambda)$, 
the so--called $R$-matrix for the group $\mbox{SL}(2,\C)$. 
It depends on an arbitrary complex parameter $\lambda$ and 
satisfies the Yang--Baxter equation
\be
R_{km}(\mu) R_{kl}(\rho) R_{ml}(\mu-\rho)=
R_{ml}(\mu-\rho) R_{kl}(\rho) R_{km}(\mu)
\,,
\ee
in which $\mu$ and $\rho$ are again arbitrary complex spectral 
parameters, and $k$, $m$, $l$ denote different sites in the spin chain. 
The solution of the Yang--Baxter equation for an arbitrary 
complex $s$ is 
\be
\label{ybsolution}
R_{km}(\lambda) = 
\frac{\Gamma(i\lambda - 2s) \Gamma(i \lambda + 2s +1)
}{
\Gamma(i \lambda - J_{km})\Gamma(i \lambda + J_{km} +1)
}
\,,
\ee
with the operator $J_{km}$ acting on the holomorphic coordinates 
at the sites $k$ and $m$. It satisfies 
\be
J_{km} (1+ J_{km}) = (\vec{S}_k + \vec{S}_m)^2 
= 2 \vec{S}_k \vec{S}_m + 2s(s+1)
\,.
\ee
Then the Hamiltonian for the XXX Heisenberg magnet for 
spin $s$ is defined as 
\be
H_N^s = \sum_{m=1}^N H_{m,m+1}
\,,
\ee
where 
\be
\label{hmmvonr}
H_{m,m+1} = 
\left. 
- i \frac{d}{d\lambda} \ln R_{m,m+1} (\lambda)
\right|_{\lambda=0}
\,.
\ee

Let us first briefly check that this definition yields the usual 
Hamiltonian for the well--known Heisenberg chain 
in the case of $s=1/2$. We observe that in this case the 
operator $J_{km}$ has only two eigenvalues $0$ and $1$, 
and we can decompose it into the projector operators onto 
the corresponding subspaces, 
\bea
 \Pi_0 &=& - \vec{S}_k\cdot \vec{S}_m +\frac{1}{4}
\\
\Pi_1 &=& \vec{S}_k \cdot \vec{S}_m + \frac{3}{4} 
\,.
\eea
Applying (\ref{ybsolution}) one finds 
\be
R_{km} (\lambda) = \Pi_0 \,\frac{i \lambda +1}{i \lambda -1}
\,.
\ee
From this one indeed obtains with (\ref{hmmvonr}) the original 
form (\ref{heisenberhalb}) of the XXX Heisenberg spin chain. 

Let us now turn to the case $s=0$. With the explicit expressions 
(\ref{heisenspins}) for the spins one obtains from (\ref{hmmvonr}) 
in fact exactly the Hamiltonian (\ref{refertothis}). 
This proves the equivalence of the two Hamiltonians for the 
$N$ reggeon system in the large-$N_c$ limit of the GLLA 
and the XXX Heisenberg chain with $s=0$. 
Here it should be noted that so far one has only shown that the 
two Hamiltonians for the $N$ reggeon system in the large-$N_c$ 
limit and for the XXX Heisenberg chain of spin $s=0$ 
are equivalent. 
The physical Hilbert spaces of allowed functions for the two systems 
are not necessarily identical. This is in fact a crucial point which 
is currently discussed and still needs further clarification. 

The most important property of the XXX Heisenberg chain 
is that it is an integrable system. This means that it possesses 
a family of hidden mutually commuting conserved charges $q_k$, 
\be
\label{heisenvertausch}
 [q_k,q_j] = [q_k,H_N] = 0
\,.
\ee
In particular it is important that their number is large enough 
for the system to be completely integrable. 
In order to find the conserved charges for the system of 
$N$ interacting reggeons in the large-$N_c$ limit 
one can apply the quantum inverse scattering method. 
To do this one assigns to each site a Lax operator 
\be
\label{lax}
L_k (\lambda) = 
  \left(
  \begin{array}{cc}
  {\lambda+i S_k^3}&{iS_k^-}\\
  {iS_k^+}&{\lambda -iS_k^3}
  \end{array}
  \right)
=
\lambda \cdot {\bf 1} + i {1 \choose \rho_k} 
\otimes (\rho_k,-1) \del_k
\,,
\ee
where again $\lambda$ is an arbitrary complex parameter, 
and the $S^{\alpha}_k$ are the spin $s=0$ generators of 
the group $\mbox{SL}(2,\C)$ acting on the holomorphic 
coordinates. They are obtained from (\ref{heisenspins}) 
with $s=0$. 
With the help of the Lax operators one can construct the 
monodromy matrix
\be
T(\lambda) = L_N(\lambda) L_{N-1}(\lambda) \cdots L_1(\lambda) 
= 
  \left(
  \begin{array}{cc}
  {A(\lambda)}&{B(\lambda)}\\
  {C(\lambda)}&{D(\lambda)}
  \end{array}
  \right)
\,.
\ee
Here the operators $A$, $B$, $C$, and $D$ act on the 
holomorphic coordinates of the $N$ reggeons and 
satisfy the Yang--Baxter equation. Further one 
obtains the transfer matrix 
\be
\Lambda(\lambda) = \tr T(\lambda) = A(\lambda) + D(\lambda)
\,,
\ee
and verifies that it is a polynomial of degree $N$ in $\lambda$ 
of the form 
\be
\label{heisenpoly}
\Lambda(\lambda) = 
2 \lambda^N + q_2 \lambda^{N-2} + \dots + q_N
\,.
\ee
The $q_k$ are operators acting on the holomorphic coordinates 
of the $N$ reggeons. In fact it follows from the Yang--Baxter 
equation for the monodromy matrix $T(\lambda)$ that 
the $q_k$ satisfy (\ref{heisenvertausch}). Hence the operators 
$q_2, \dots, q_N$ form a set of $N-1$ conserved charges 
for the system of $N$ interacting reggeons in the large-$N_c$ limit. 
For the number of conserved charges to match the total number 
of reggeons we need one more conserved charge. 
This last conserved charge is associated with the center--of--mass 
motion of the $N$-reggeon compound state and is equal to the 
total reggeon momentum 
\be
P = \pi_1 + \pi_2  \dots + \pi_N= i (S_1^- + S_2^- + \dots + S_N^-) 
\,,
\ee
with $\pi_j= -i \del_j$ being the holomorphic component of the 
transverse momentum of the $j$th reggeon. The explicit form 
of the other $N-1$ conserved charges can be found 
starting from the Lax operator (\ref{lax}) using the explicit 
expressions for the spin operators (\ref{heisenspins}) with $s=0$. 
One finds 
\be
q_k =\sum_{N\ge j_1 > j_2>\dots>j_k \ge 1} 
 i^k  \rho_{j_1 j_2} \rho_{j_2 j_3} \dots \rho_{j_k j_1} 
\del_{j_1} \del_{j_2} \dots \del_{j_k}
\,,
\ee
where again $\rho_{jk}= \rho_j -\rho_k$. 
These are exactly the conserved charges that were identified already 
in \cite{Lipatov:1993yb}. 
Note that we have
\be
q_2 = \sum_{N\ge j > k \ge 1} \rho_{jk}^2 \del_j \del_k
= -h (h-1)
\,,
\ee
which is the quadratic Casimir operator of $\mbox{SL}(2,\C)$, 
and its eigenvalue $h$ defines the conformal weight of the 
holomorphic wave function $\varphi(\rho_1,\dots , \rho_N)$ 
of the $N$-reggeon compound state. The reggeon wave function 
belongs to the principal series representation of the group 
$\mbox{SL}(2,\C)$. As we have already seen in the cases of 
the BFKL Pomeron and the BKP Odderon 
this implies that its conformal weight is quantized as 
\be
h = \frac{1 + n}{2} + i \nu,  \hspace*{1cm} 
n \in \Z,\,\,\,\, \nu \in \R
\,.
\ee
The integer $n$ defines the Lorentz spin of the $N$-reggeon state 
corresponding to rotations in two--dimensional impact parameter 
space, $\varphi_N \to e^{in\alpha}\varphi_N$ as 
$\rho_j \to e^{i \alpha}\rho_j$ and 
$\rhobar_j \to e^{-i \alpha}\rhobar_j$. 
In principle it is now possible, though complicated, 
to express the reggeon Hamiltonian as a function of the 
conserved charges, 
\be
H_N = H_N(q_2, \dots, q_N) 
\,.
\ee
Note that due to the M\"obius invariance the Hamiltonian does 
not depend on the total momentum $P$. 
Now the problem of solving the full Schr\"odinger 
equation has been translated into the simpler problem of 
the simultaneous diagonalization of the operators 
$P,q_2,\dots , q_N$. The eigenvalues of these operators form 
a complete set of quantum numbers parametrizing the 
$N$ reggeon compound state $\varphi_{N, \{ q \} }(\rho_1, \dots, \rho_N)$. 

The $N$-reggeon compound states with $N \ge 4$ in the large-$N_c$ 
limit of the GLLA have been studied further in a number of papers, see for 
example \cite{Korchemsky:1994um}--\cite{Korchemsky:1996kh}, 
\cite{Korchemsky:1997yy,Korchemsky:1997ve}, 
\cite{Lipatov:1998as,DeVega:2001pu,Derkachov:2001yn}, 
\cite{Korchemsky:2001nx}--\cite{Derkachov:2002pb}, 
\cite{Gorsky:2002ju}. 
Following \cite{Faddeev:1994zg,Korchemsky:1994um} 
these papers use the method of the Baxter $Q$-operator 
and the corresponding separation of variables 
in the Sklyanin representation of the wave function $\varphi_N$. 
The Baxter $Q$-operator ${\bf Q}(\lambda)$ 
depends on a complex spectral parameter $\lambda$. 
It commutes with itself for different values of the spectral parameter, 
$[{\bf Q}(\mu),{\bf Q}(\lambda)]=0$, as well as with the conserved 
charges $q_k$, $[{\bf Q}(\lambda),q_k] = 0$. 
The eigenvalues $Q(\lambda)$ of the Baxter $Q$-operator satisfy the 
Baxter equation 
\be
\label{baxterforN}
\Lambda(\lambda) Q(\lambda) = 
(\lambda + i)^N Q(\lambda + i) + (\lambda - i)^N Q(\lambda - i) 
\,,
\ee
where $\Lambda(\lambda)$ is the transfer matrix given in 
terms of the conserved charges in (\ref{heisenpoly}). 
One can perform a change of variables from the coordinates 
$\rho$ to the separated variables $x$ 
(and similarly for the antiholomorphic sector) 
via a unitary transformation which can be constructed explicitly. 
In terms of the separated variables the wave function $\varphi(x)$ 
is then basically given by a product of the eigenvalues of the Baxter 
$Q$-operator, $\varphi(x) \sim Q(x_1) \cdot \dots \cdot Q(x_{N-1})$, and 
an additional factor $\exp(i P x_N)$ 
describing the center--of--mass motion. 
With the help of the integral transformation (\ref{contourQ}) 
the Baxter equation (\ref{baxterforN}) is transformed into an 
$N$-th order differential equation. The eigenvalues of the Baxter 
$Q$-operator have a known asymptotic behavior and known pole 
structure. These conditions fix $Q(\lambda)$ uniquely and provide 
the quantization condition for the conserved charges $q_k$. 
The holomorphic energy of the $N$-reggeon state can finally be obtained 
from $Q(\lambda)$ via 
\be
\label{energyNausbaxterQ}
\epsilon (q_2, \dots, q_N) = 
i \frac{d}{d\lambda} \log 
\left. \frac{(\lambda-i)^N  Q(\lambda-i)}{(\lambda +i)^N Q(\lambda+i)} 
\right|_{\lambda=0}
\,. 
\ee
With the help of this method the spectrum of the $N$-reggeon states 
has been computed for up to $N=4$ in \cite{deVega:2002im} and 
even for up to $N=8$ in \cite{Korchemsky:2001nx,Derkachov:2002wz}. 
As in the case of the Odderon (see section \ref{XXXsect}) there is a 
disagreement between these papers concerning the interpretation 
of the results. In \cite{deVega:2002im} it is argued that for 
given $N$ only those eigenstates of the Hamiltonian are physically 
meaningful for which $q_N$ is purely real (imaginary) for even (odd) $N$. 
In \cite{Korchemsky:2001nx,Derkachov:2002wz}, on the other hand, 
it is argued that all states with general complex eigenvalues of $q_N$ 
are physical. In addition, there is a disagreement between 
\cite{Korchemsky:2001nx,Derkachov:2002wz} and \cite{deVega:2002im} 
concerning the intercept of the four--gluon state, for which different 
values are found. The origin of this discrepancy is a technical difference 
in the definition of the Baxter operator. Further study is needed in 
order to settle these open questions. We should point out again that 
the above problems do not affect the Odderon intercept but only the 
higher excited states of the Odderon. 

The results of \cite{Korchemsky:2001nx,Derkachov:2002wz} show 
that in the large-$N_c$ limit of the GLLA the intercept of the 
$N$-reggeon states increases (decreases) with $N$ for odd (even) $N$, 
approaching one in the limit $N\to \infty$. As can be easily seen, however, 
for finite $N_c$ and for even $N$ the intercept has to be larger 
than the intercept of the $N/2$-Pomeron cut, 
$1+N (\alpha_\spommi -1)/2$. 
This implies that the large-$N_c$ limit is not a good approximation for 
larger values of $N$, at least in the case that $N$ is even. 
Nevertheless, due to its integrability the large-$N_c$ limit can possibly 
still be a good starting point for studying the case of finite $N_c$. 

In \cite{Vacca:2000bk} another family of $N$-reggeon 
compound states in the large-$N_c$ limit of the GLLA has been 
found. It is shown there that for any given $N$ one can construct a 
solution of the $N$-reggeon BKP equation from solutions of the 
$(N-1)$-reggeon BKP equation. A characteristic property of these 
descendant states \cite{Derkachov:2002wz} 
is that the corresponding eigenvalue of the conserved 
charge $q_N$ vanishes. We have already seen such a solution in the 
case of the Odderon, namely the BLV solution discussed in section 
\ref{blvsolutionsect}. The explicit construction for general $N$ is 
again based on the reggeization of the gluon and is 
analogous to (\ref{blvsolution}). 

Let us finally note that very interesting relations of the problem 
of the $N$-reggeon states in the GLLA with topological field theory 
\cite{Ellis:1998us} and with string theory \cite{Gorsky:2002ju} 
have been discovered. Furthermore, there are indications 
\cite{Gorsky:2002ju} that recently discovered gauge theory / string 
theory dualities might offer a new way of approaching the problem 
of the GLLA, and possibly also the problem of the Regge limit of QCD 
in general. In our opinion these directions offer very promising 
possibilities for further research. 

\subsubsection{Extended GLLA}
\label{egllasect}

In the GLLA one considers only exchanges in the $t$-channel 
in which the number of reggeized gluons is constant during 
the $t$-channel evolution. This corresponds to the quantum 
mechanical problem of $N$-reggeon states. In the extended 
GLLA\footnote{We should stress again that in the literature 
also this approximation scheme is often called GLLA. For 
clarity we distinguish between GLLA and extended GLLA.} 
\cite{Bartels:unp} 
one now takes into account also the possibility that 
the number of gluons fluctuates during the $t$-channel 
evolution. The step from the GLLA to the extended GLLA 
hence corresponds to the transition to a quantum 
field theory of interacting reggeons. 

The objects of interest in the extended GLLA are amplitudes 
describing the production of $n$ reggeized gluons in the 
$t$-channel. For any given $n$ one then resums all diagrams 
of the perturbative series which contain the maximal number 
of logarithms of the energy for that $n$. For technical reasons 
the extended GLLA has been studied explicitly only for the case 
that the system of reggeized gluons is coupled to a virtual photon 
impact factor, and we will adopt the corresponding notation 
as it has been used for example in \cite{Bartels:1993ih,Bartels:1994jj}. 
It is expected that due to high energy factorization the results obtained 
in that special case are universal. In particular, the interaction 
vertices between states with different numbers of reggeized gluons 
have a universal meaning independent of the impact factor. 

The amplitudes $D_n^{a_1\dots a_n}(\kf_1, \dots, \kf_n)$ describe 
the production of $n$ reggeized gluons in the $t$-channel carrying 
momenta $\kf_i$ and color labels $a_i$ in the adjoint representation 
of $\mbox{SU}(N_c)$. The lowest order term of these amplitudes 
is given by the virtual photon impact factor consisting of a quark loop 
to which the $n$ gluons are coupled. But there are also terms in 
which less than $n$ gluons are coupled to the quark loop and 
there are transitions to more gluons during the $t$-channel 
evolution. Technically this means that the amplitudes $D_n$ obey 
a tower of coupled integral equations in which the equation for a given 
$n$ involves all amplitudes $D_m$ with $m<n$. The first of these 
integral equations (i.\,e.\ the one for $D_2$) is identical to 
the BFKL equation. In the higher equations new transition kernels 
occur which are generalizations of the BFKL kernel which have 
been derived in \cite{Bartels:unp}. 
The system of coupled integral equations is rather complicated. 
Nevertheless, it is possible to extract very valuable 
information about the structure of the amplitudes \cite{Ewerz:2001fb}, 
\cite{Bartels:1992ym}, \cite{Bartels:1993ih}--\cite{Ewerz:2001uq}.  
In short, the structure of the solutions is such that they involve only 
states with fixed even numbers of gluons which are coupled to each 
other by effective transition vertices that can be computed explicitly. 
As we will see momentarily it is the reggeization of the gluon that 
leads to the fact that in the solutions of the integral equations for the 
amplitudes $D_n$ the states with fixed odd numbers of gluons do 
not occur. 
The only pieces missing for a full analytic solution of the 
integral equations are the analytic properties (their intercepts etc.) 
for the states with fixed even numbers of reggeized gluons in the GLLA. 
As we have seen in the preceding section the latter can be found 
in the large-$N_c$ limit, but only little is known for finite $N_c$. 
Unfortunately, the large-$N_c$ limit is not very useful in the 
extended GLLA since the transition vertices between states 
with different numbers of gluons are subleading in the expansion 
in $1/N_c$. Therefore at least the naive application of the large-$N_c$ 
limit would immediately reduce the extended GLLA to the simple GLLA. 

Let us now briefly summarize the most important results for the $n$-gluon 
amplitudes $D_n$ obtained in \cite{Ewerz:2001fb}, \cite{Bartels:1992ym}, 
\cite{Bartels:1993ih}--\cite{Ewerz:2001uq}. 
The first result is the reggeization of the gluon in the amplitudes $D_n$. 
The simplest example of this is the three--gluon amplitude $D_3$. 
The corresponding integral equation can be solved analytically 
\cite{Bartels:1992ym}, giving 
\be
\label{d3reggeizes}
D^{abc}_3 (\kf_1,\kf_2,\kf_3) = g f_{abc} 
\left[
D_2(\kf_1+\kf_2,\kf_3) - D_2(\kf_1+\kf_3,\kf_2) 
+ D_2(\kf_1,\kf_2+\kf_3) 
\right]
\,.
\ee
The three--gluon amplitude is thus a superposition of 
two--gluon amplitudes $D_2$. As a consequence, an actual 
three--gluon state in the $t$-channel does not occur. 
In each of the three terms 
in this expression the momenta of two gluons enter only 
as a sum. This means that the two gluons are at the same 
point in impact parameter space. They can be regarded as 
forming a `more composite' reggeized gluon that occurs 
in the amplitude $D_2$. This is exactly the process known 
as reggeization, see section \ref{reggeizationsect}. 
Reggeization occurs also in all higher amplitudes $D_n$, 
all of which contain a contribution which can be decomposed 
into two--gluon amplitudes in a way similar to 
(\ref{d3reggeizes}). 
In the higher amplitudes also more than two gluons can form 
a more composite reggeized gluon. Therefore these amplitudes 
are well suited for studying in detail the process of reggeization and 
in particular the corresponding behavior of the color degrees 
of freedom \cite{Ewerz:2001fb}. 

The three--gluon amplitude can be written completely in terms 
of two--gluon amplitudes. In the higher amplitudes with $n\ge 4$ 
gluons additional contributions occur. The four--gluon 
amplitude can be shown to have the following structure 
\cite{Bartels:1993ih,Bartels:1994jj}: 
\be
\label{d4solutionpics}
D_4 = \sum \input{solutiond41.pstex_t} + \input{solu42.pstex_t} 
\hspace*{.5cm}\,.
\ee
Here the first term is again the reggeizing part consisting of a 
superposition of two--gluon amplitudes in a way similar to 
(\ref{d3reggeizes}). In the second term the $t$-channel evolution 
starts with a two--gluon state that is coupled to a full four--gluon 
state via a new effective 2-to-4 gluon transition vertex 
$V_{2 \rightarrow 4}$ the explicit form of which was found in 
\cite{Bartels:1993ih,Bartels:1994jj}. 
The full four--gluon state includes all 
pairwise interactions of the four gluons and is exactly the 
one described by the GLLA. As can be seen from (\ref{d4solutionpics}) 
the four--gluon state does not couple directly to the quark box diagram 
of the photon impact factor. 
The structure emerging here is that of a quantum field theory in which 
states with different numbers of gluons are coupled to each other 
via effective transition vertices like $V_{2 \rightarrow 4}$. 
This structure has been shown to persist also to higher amplitudes in 
\cite{Bartels:1999aw}. There the amplitudes with up to six gluons 
were studied. It turns out that the five--gluon amplitude can 
be written completely in terms of two--gluon amplitudes and 
four--gluon amplitudes containing the vertex $V_{2 \rightarrow 4}$. 
Here it can be observed that reggeization is universal, i.\,e.\ takes 
places in different $m$-gluon states in the same way. 
The mechanism of reggeization can be shown to emerge in all 
amplitudes $D_n$ described by the integral equations of the 
extended GLLA \cite{Bartels:1999aw}, and it is expected 
that all amplitudes $D_n$ with an odd number $n$ of gluons 
are superpositions of lower amplitudes with even numbers of gluons. 
The six--gluon amplitude $D_6$ was shown to consist of two 
reggeizing parts, being superpositions of two-- and four--gluon 
amplitudes, and a part in which a two--gluon state is coupled to a 
six--gluon state via a new effective transition vertex $V_{2 \to 6}$ 
which has been computed explicitly in \cite{Bartels:1999aw}. 
Thus also the six--gluon amplitude exhibits the structure of 
a field theory of interacting $m$-gluon states in the $t$-channel 
evolution. 
As we will discuss in the next section the new effective vertex 
$V_{2 \to 6}$ is directly relevant to the Odderon. 
A still open question concerning the six--gluon amplitude 
as well as higher amplitudes is how exactly the number changing 
2-to-$m$ vertices behave in the case that the two incoming gluons 
are not in a color singlet state, for a detailed discussion see 
\cite{Bartels:1999aw}. 

A most remarkable property of the effective transition vertices is their 
conformal invariance in impact parameter space. This symmetry was 
shown for the vertex $V_{2 \to 4}$ in \cite{Bartels:1995kf} , 
and for $V_{2 \to 6}$ in \cite{Ewerz:2001uq} where also the potential 
form of higher vertices $V_{2 \to 2m}$ is discussed. Also the 
$n$-gluon states of the GLLA which occur in the effective field 
theory are known to be conformally invariant. 
There are in fact strong indications that the whole set of amplitudes 
can be cast into the form of a conformally invariant field theory 
in $2\!+\!1$ dimensions, with rapidity acting as a time--like variable.  
It should be emphasized that the conformal invariance applies 
to the two--dimensional impact parameter space for each value of 
the rapidity\footnote{In the mathematical literature one would 
therefore speak of a two--dimensional conformal field theory 
with one real--valued parameter.}. 
The identification of that conformal field theory would certainly 
be a major step towards a better understanding of the Regge 
limit in perturbative QCD. 

\subsubsection{The Perturbative Pomeron-Odderon-Odderon Vertex}
\label{poosect}

The quantum numbers of the Odderon and of the Pomeron are such 
that a Pomeron--Odderon--Odderon vertex can exist. In the 
present section we will show how the perturbative $\P \Od \Od$ 
vertex can be computed from the effective two--to--six reggeized 
gluon vertex of the extended GLLA discussed in the previous section. 
The $\P \Od \Od$ vertex is not only of theoretical interest but also has 
phenomenological applications which we will discuss in 
sections \ref{epdiffsect} and \ref{gammagammasect} below. 

The perturbative $\P \Od \Od$ vertex was for the first time 
considered in \cite{Bartels:1999aw} as a direct application of the 
two--to--six reggeon vertex $V_{2 \to 6}$ arising in the 
amplitude $D_6$ of the extended GLLA. In that amplitude the two 
gluons entering that vertex are by construction in a BFKL 
Pomeron state which is coupled to a virtual photon impact factor. 
Due to high energy factorization the vertex itself is independent of 
the particular impact factor and we can replace the amputated 
amplitude $D_2$ to which the vertex $V_{2 \to 6}$ couples by a 
simple Pomeron amplitude without the impact factor. We will here 
nevertheless write the vertex for the amplitude $D_2$. This is because 
one can define the $\P \Od \Od$ also in a different way, namely including 
also contributions from the reggeizing parts of the six--gluon amplitude, 
for a discussion of the latter see \cite{Bartels:2003zu}. In order to 
avoid confusion we present here the vertex with the full 
amplitude $D_2$ for the Pomeron. The vertex 
$V_{2 \to 6}$ couples that Pomeron amplitude to a general 
six--gluon state. In order to obtain the $\P \Od \Od$ vertex 
one replaces the six--gluon state by a state consisting of two 
BKP Odderons. Equivalently, one can project the vertex 
$V_{2 \to 6}$ onto a product of two eigenfunctions of the 
BKP Odderon Hamiltonian, 
\bea
\label{poointegral}
V_{\spommi \soddi \soddi}
&=&\int \left( \, \prod_{i=1}^6 d^2\kf_i \right) 
  (V_{2 \to 6}^{a_1a_2a_3a_4a_5a_6}D_2)
(\kf_1,\kf_2,\kf_3,\kf_4,\kf_5,\kf_6) \times 
\nn \\
&&\hspace{.6cm}
\times \, d_{a_1a_2a_3} d_{a_4a_5a_6}
\varphi_{\soddi, \alpha_1} (\kf_1,\kf_2,\kf_3) 
\varphi_{\soddi, \alpha_2} (\kf_4,\kf_5,\kf_6) 
\,,
\eea
where $\varphi_{\soddi, \alpha_i}$ are the momentum space 
eigenfunctions of the BKP Hamiltonian carrying the quantum 
numbers $\alpha=(h,q_3)$. 
This procedure is completely analogous to the construction of the 
perturbative triple--Pomeron vertex from the two--to--four 
reggeon vertex $V_{2 \to 4}$ \cite{Lotter:1996vk}. For that vertex 
the exact value for the leading Pomeron states was 
computed subsequently in \cite{Korchemsky:1997fy}. That has not 
yet fully been done for the $\P \Od \Od$ vertex, but it would of course be 
very interesting to know its coupling strength numerically. 

The detailed form of the vertex $V_{2 \to 6}$ can be found in 
\cite{Bartels:1999aw}. Here we mention only some of its properties 
relevant to the discussion below. 
$V_{2 \to 6}$ depends on the transverse momenta $\kf_i$ of the 
outgoing gluons and carries six color labels $a_i$. 
The vertex is understood as an integral operator acting 
on the two--gluon--amplitude $D_2$. Due to its symmetry 
properties it can be written as a sum over certain permutations, 
\bea
\label{newpiece}
\lefteqn{
(V_{2 \to 6}^{a_1a_2a_3a_4a_5a_6}D_2)
(\kf_1,\kf_2,\kf_3,\kf_4,\kf_5,\kf_6) = 
} \nn \\
&& \hspace*{1cm}   
= \sum   d_{a_1a_2a_3} d_{a_4a_5a_6} 
(W D_2)(\kf_1,\kf_2,\kf_3;\kf_4,\kf_5,\kf_6) 
\,.
\eea
The sum extends over all (ten) partitions of the six gluons 
into two groups containing three gluons each, and in the terms 
in the sum the indices of the gluon momenta and of the 
color labels are permutated as 
\bea
\label{explsumoverw}
\lefteqn{
\sum   d_{a_1a_2a_3} d_{a_4a_5a_6} 
(W D_2)(\kf_1,\kf_2,\kf_3;\kf_4,\kf_5,\kf_6) =
} \nn \\
& & \hspace*{.7cm} 
= d_{a_1a_2a_3} d_{a_4a_5a_6} 
(W D_2)(\kf_1,\kf_2,\kf_3;\kf_4,\kf_5,\kf_6) + 
\nn \\
&&\hspace*{1.1cm} 
+ \, d_{a_1a_2a_4} d_{a_3a_5a_6} 
(W D_2)(\kf_1,\kf_2,\kf_4;\kf_3,\kf_5,\kf_6)
+ \dots +
\nn \\
&&\hspace*{1.1cm} 
+ \,d_{a_1a_5a_6} d_{a_2a_3a_4} 
(W D_2)(\kf_1,\kf_5,\kf_6;\kf_2,\kf_3,\kf_4)
\,.
\eea
The function $W D_2$ is the same in all permutations. Its detailed 
form is rather complex and can be found in \cite{Bartels:1999aw}. 
In \cite{Ewerz:2001uq} it was shown to exhibit an interesting 
regularity which was then used to prove the conformal invariance 
of the two--to--six reggeon vertex in impact parameter space. 

There are two known types of Odderon solutions of the BKP equation, 
namely the JW solution and the BLV solution described in sections 
\ref{jwsolutionsect} and \ref{blvsolutionsect}, respectively. 
It turns out that one obtains a nonvanishing $\P \Od \Od$ vertex 
for either choice of the Odderon solution for $\varphi_\soddi$ in 
(\ref{poointegral}). 

The $\P \Od \Od$ vertex becomes particularly simple if the 
two Odderons states $\varphi_{\soddi,\alpha_i}$ 
are given by eigenstates corresponding to the 
JW solution. We recall that for the JW solution $q_3 \neq 0$, and 
due to conformal invariance the eigenstate has the form 
(\ref{odderonwavefct}) with a regular function $\phi(z,\bar{z})$. 
As can be seen from that form the states $\varphi_{\soddi,\alpha_i}$ 
vanish if two gluon coordinates coincide. In the function $W D_2$, 
on the other hand, there are many terms which depend on the 
sum of two gluon momenta only, which means that the positions 
of the two gluons in impact parameter space coincide. 
Accordingly, many terms in $W D_2$ 
vanish when convoluted with the JW Odderon states. 
In fact it was found in \cite{Bartels:1999aw} that the vertex 
reduces to 
\bea
\label{POOvertex}
V_{\spommi \soddi \soddi}&=&
g^6 \,\frac{9}{4} \,
\frac{(N_c^2-4)^2  (N_c^2 -1)}{N_c^2} 
\int  \!\left( \,\prod_{i=1}^6 d^2\kf_i \right) \,
\varphi_{\soddi, \alpha_1} (\kf_1,\kf_2,\kf_3) 
\varphi_{\soddi, \alpha_2} (\kf_4,\kf_5,\kf_6) \times 
\nn \\
&& \times [ 2 a(\kf_1 + \kf_4,\kf_2 + \kf_5,\kf_3+\kf_6) 
- s(\kf_1+\kf_4,\kf_2+\kf_5,\kf_3+\kf_6) + 
\nn \\
&& \hspace*{1cm}
- s(\kf_1+\kf_4,\kf_3+\kf_6,\kf_2+\kf_5) ] 
\,. 
\eea
Here we have used the abbreviations 
\bea
a(\kf_1,\kf_2,\kf_3) &=&  \int \fr{d^2\lf}{(2 \pi)^3}  
       \fr{\kf_1^2}{(\lf-\kf_2)^2 [\lf-(\kf_1+\kf_2)]^2}
      \,D_2\!\left(\lf, \sum_{j=1}^{3}\kf_j-\lf \right) \,,
\\
s(\kf_1,\kf_2,\kf_3) &=& 
- \fr{2}{N_c g^2} \beta(\kf_1)  D_2(\kf_1+\kf_2,\kf_3) 
\,,
\eea
where $\beta$ is given in (\ref{gluontraj}). 
The two functions $a$ and $s$ are infrared divergent separately, 
but the divergences cancel in the combination that occurs in 
(\ref{POOvertex}). Their impact parameter space representation 
can be found for example in \cite{Volkerthesis}. 
An important property of the vertex (\ref{POOvertex}) is that 
it contains only one color coefficient. Note that in the sum 
(\ref{explsumoverw}) there is one term in which the color structure 
matches exactly the color structure of the two Odderon states 
in (\ref{poointegral}). Interestingly, exactly this term vanishes 
completely in the convolution and only the other nine terms give 
a contribution to (\ref{POOvertex}). As already mentioned above 
it would be very interesting to compute the numerical value 
of  (\ref{POOvertex}) using the known JW solutions of the BKP 
equation. 

In the case of the BLV solution the situation is more difficult. 
An interesting observation is that the momentum structure of the 
first three and of the last three momenta in the function 
$W D_2$ in the full two--to--six gluon vertex $V_{2 \to 6}$ 
is identical to the momentum structure of the BLV solution 
(\ref{blvsolution}). 
The construction of the BLV solution in \cite{Bartels:1999yt} 
was in fact motivated by the inspection of the function 
$W D_2$. Already this observation suggests that 
a perturbative $\P \Od \Od$ vertex exists also for the 
BLV Odderon solution. This was confirmed explicitly in 
\cite{Bartels:2003zu} where that case was considered. 
Here the simplifications due to the 
special form of the JW solution do not take place and the result 
is hence rather complicated. In particular, there are contributions 
from all color structures in (\ref{explsumoverw}) such that there 
is also a term with the color coefficient 
$(d_{abc} d_{abc})^2= (N_c^2-4)^2 (N_c^2-1)^2/N_c^2$, 
containing two more powers of $N_c$ as compared to the 
color coefficient in (\ref{POOvertex}). In \cite{Bartels:2003zu} 
only this leading term in the $1/N_c$ expansion is considered. It is 
obtained by inserting only the first (but still complicated) 
term in the sum (\ref{explsumoverw}) into (\ref{poointegral}). 
Note that the BLV solution is constructed from eigenfunctions 
of the BFKL kernel. The resulting $\P \Od \Od$ vertex hence 
becomes an integral which only contains BFKL eigenfunctions, 
and can thus be evaluated using a saddle point approximation. 
For a more detailed discussion of that calculation we refer the 
reader to \cite{Bartels:2003zu}. Also in the case of the BLV solution 
it would be very desirable to compute the exact numerical 
value of the $\P \Od \Od$ vertex, including the subleading terms 
in the $N_c$ expansion. 

\subsubsection{Other Approaches}

There are many approaches to the problem of the Regge limit 
in QCD using perturbative or semi--perturbative methods like 
for example the color glass condensate approach initiated in 
\cite{McLerran:1993ka,McLerran:1993ni,McLerran:1994vd} or 
the operator expansion method based on a description of high 
energy scattering using Wilson line operators \cite{Balitsky:1998ya}. 
It turns out that many of these approaches, although on first sight 
very different, lead to identical results for certain quantities. 
The BFKL Pomeron is for example reproduced in all of these approaches 
when the limit of a sufficiently dilute system or of sufficiently weak 
color fields is considered, corresponding to a situation in which 
nonlinearities like contributions containing a triple--Pomeron 
vertex can be neglected. Characteristic differences between 
the different approaches are therefore hard to isolate when only 
the Pomeron approximation is used to compare them. 
Since the different approaches have very different 
starting points and the full solutions are not known for most 
of them it would be helpful to have a further comparatively simple 
object that can be considered in different frameworks. 
In this respect the study of the Odderon can offer new ways of 
comparing different approaches and help to distinguish their 
characteristic features. 

To the best of our knowledge the BKP Odderon has not yet been studied 
in any perturbative approach other than the GLLA and the 
extended GLLA. In many of the other approaches there is at least 
no fundamental obstacle to considering the Odderon. 
An exception might be the color dipole approach of 
\cite{Mueller:1993rr}--\cite{Chen:1995pa}, at least in its simplest 
version. There the problem is approached by studying the evolution of 
a small color dipole in the large-$N_c$ limit. It is investigated 
how the dipole emits gluons, and the emission of these gluons is 
described by the iterated splitting of a dipole into two dipoles. 
The interaction of two such systems is then reduced to the elementary 
dipole--dipole scattering via gluon exchange. Since the dipoles carry 
always the same charge parity it seems unlikely 
that in this approach an Odderon can be observed. 
Probably one needs to include higher multipoles in order to 
accommodate Odderon exchanges. 
As this example indicates, the occurrence or non--occurrence of the 
Odderon can give valuable insight in the structure of the different 
approaches to high energy QCD. 

\subsection{The Odderon in Nonperturbative QCD}
\label{nonpertoddsect}

Very little is known about nonperturbative effects 
on the Odderon and its intercept. 
From a theoretical point of view 
that is also true for the nonperturbative Pomeron, 
but in the case of the Pomeron we are in the 
comfortable situation that we have a rather 
precise picture in terms of successful phenomenological 
fits to a wealth of data obtained in many different scattering 
processes. In this way the intercept of the soft Pomeron, 
for example, is known to rather high accuracy. 
This important piece of information is missing almost 
completely for the nonperturbative or soft Odderon, 
the available information being by far not as precise and 
reliable as for the Pomeron. 
That situation is extremely unfortunate, since most 
phenomenological predictions necessarily 
have to rely on more or less plausible assumptions 
or speculations as far as this point is concerned. 
The importance of studying nonperturbative effects 
on the Odderon and on its intercept in particular 
(and of course on the Pomeron as well) 
cannot be overemphasized. So far, however, only 
few studies of this problem have been performed. 
In the following we summarize some methods to 
approach the problem of nonperturbative effects 
on the Odderon. 

\subsubsection{The Regge Picture}
\label{npreggesect}

Let us recall what is known about the interplay of soft 
and hard physics in the case of the Pomeron intercept. 
Soft reactions show in general a slow rise of the cross 
section at high energies, 
$\sigma \sim s^{\alpha_{\mbox{\tiny \P}}-1}$ 
with $\alpha_{\mbox{\tiny \P}}=1.09$ \cite{Groom:in}, 
which is identified with the soft Pomeron \cite{Donnachie:1992ny}. 
In reactions involving a hard momentum scale, however, 
the energy dependence becomes steeper as the momentum 
scale is increased, see for example \cite{Adloff:2001rw}. 
The simplest way of describing the available data is to assume 
the existence of two Pomerons, a soft one with intercept $1.09$, 
and a hard one with intercept around $1.4$ 
\cite{Donnachie:1998gm,Donnachie:2001xx}. 
It is obviously conceivable that things are much more 
complicated in reality, but this simple picture gives us at 
least a guideline for the minimal complications we should 
realistically expect in the case of the Odderon. 

Although here we are already entering the realm 
of speculations it would not be too surprising if there were 
two Odderons, a soft one and a hard one. As in the case 
of the Pomeron, the hard Odderon would be expected 
to be described at least approximately by a 
perturbative resummation of the leading logarithms in 
the energy, i.\,e.\ by the BKP equation or eventually 
by a NLLA version of it. Even if the intercept of the 
soft Odderon should be so low as to make it 
invisible at high energy, the hard Odderon would not 
necessarily be affected by this. 
In the contrary, in suitable processes which are 
dominated by hard momentum scales only, there 
is no obvious reason why the perturbative Odderon 
should not be applicable -- although of course 
with the same caveats as the perturbative Pomeron, 
see section \ref{bfklapplicabsection}. 
For the Odderon, there are even less of these optimal 
processes than for the Pomeron, and we will 
describe the best probe of the perturbative Odderon 
in section \ref{gammagammasect} below. 
But in most situations a considerable 
contribution of the soft Odderon is unavoidable. 

The simplest picture of a nonperturbative Odderon 
is a Regge parametrization corresponding 
to a simple pole (see also section \ref{pomoddregge}) 
in which the Odderon propagator is given by
\be
  (-i) \eta_\soddi 
  \left( \frac{-i s}{s_0}\right)^{\alpha_\toddi (t)-1} 
\ee 
with the a priori unknown Odderon phase 
$\eta_\soddi = \pm 1$ and a fixed energy scale 
$s_0$ chosen for example as $1\,\mbox{GeV}^2$. 
The Odderon trajectory $\alpha_\soddi (t)$ is 
usually assumed to be linear, 
\be
 \alpha_\soddi (t) = \alpha_\soddi (0) + \alpha'_\soddi t
\,.
\ee
The couplings of the Odderon to external particles have 
to be fixed phenomenologically, 
for examples see section \ref{Oddprotoncouplsect}. 
Obviously, the unknown couplings to the external 
particles induce a considerable uncertainty. 
Also possible are more complicated Regge 
parametrizations of the Odderon corresponding 
to other types of singularities like for example 
the double pole of the maximal Odderon 
(see section \ref{basicoddsect}). 

This uncertainty can to some extent be avoided 
in processes involving a hard momentum scale. 
In such cases one can use a three--gluon model 
of the Odderon, and the coupling of the individual 
gluons to external particles can be calculated from 
first principles in some cases, for example 
for virtual photons or for heavy mesons. The 
coupling of the three gluons to a proton or to other 
complicated hadronic bound states still requires 
some model assumptions, see the discussion in section 
\ref{Oddprotoncouplsect}. The exchange of 
three noninteracting gluons does not induce any 
energy dependence, but one can combine it with a 
Regge--like Odderon. This is achieved by making a 
phenomenological ansatz for a nonperturbative 
Odderon by supplementing 
a simple three--gluon model with 
a powerlike energy dependence, 
\be
  \phi_{\soddi} \:\: \longrightarrow \:\: 
\phi_\soddi \, \left( \frac{s}{s_0}\right)^{\alpha_\toddi (t)-1}
\,,
\ee
where $\phi_\soddi$ here stands for the three--gluon 
exchange model for the Odderon (possibly with some 
sort of nonperturbative gluon propagators, see next paragraph). 
One can then try to fit the data on $pp$ and $p\bar{p}$ 
elastic scattering to find the Odderon intercept $\alpha_\soddi$. 
The resulting energy dependence can then be compared with 
the intercept of the resummed perturbative (BKP) Odderon 
in order to estimate the nonperturbative effects in the 
scattering process under consideration. 
The difference between the differential cross sections for 
elastic $pp$ and $p\bar{p}$ scattering would in principle 
be a good example for a process to which this method 
could be applied. The interesting region in the squared 
momentum transfer $t$ is at the lowest edge of applicability 
of perturbation theory, and nonperturbative effects are thus 
expected to be sizable. Unfortunately, the presently available data 
do not really allow one to extract a precise value for 
$\alpha_\soddi$, for more details on this point 
see section \ref{elasticppsect}. 

\subsubsection{Nonperturbative Gluon Propagators}
\label{nppropagatorsect}

The perturbative picture of the Odderon can also be 
extended in another direction, namely by trying 
to include as much of the soft physics of low transverse 
momenta as possible. 
Let us first consider the simple three--gluon exchange 
model for the Odderon in which the three gluons are 
in a colorless $C=-1$ state and do not interact 
with each other. In a sense this can be called 
an abelian model for the Odderon and is 
analogous to the Low--Nussinov picture 
\cite{Low:1975sv,Nussinov:mw} of the Pomeron. 
One can now modify the perturbative behavior of the 
gluon propagators at small momenta 
where nonperturbative effects are expected to dominate. 
For the Pomeron such a model was constructed by 
Landshoff and Nachtmann in \cite{Landshoff:1986yj} 
(see also \cite{Nachtmann:1991ua}) 
in the context of studying the quark counting in the 
coupling of the Pomeron to hadrons. 
At low momenta the $1/k^2$ behavior was replaced 
by a nonperturbative propagator $D_{\mbox{\tiny np}}(k^2)$. 
The latter was then related to the nonlocal gluon 
condensate\footnote{We understand the nonperturbative 
gluon propagator to contain a factor of $\alpha_s$. 
Accordingly, this equation contains an additional factor 
of $\alpha_s$ compared to the original paper 
\protect\cite{Landshoff:1986yj}.}, 
\be
\label{lnpomtogluon}
\langle \, \alpha_s :\!F_{\mu \nu}(x) F^{\mu \nu}(y)\!:\, \rangle
= -i \int \frac{d^4k}{(2 \pi)^4} \, e^{-ik(x-y)} 
\,6 k^2  D_{\mbox{\tiny np}}(k^2) \,.
\ee
From this expression the local gluon condensate 
introduced in \cite{Shifman:bx}--\cite{Shifman:bw} 
is obtained in the limit $y \to x$. Since this condensate 
must be finite the integral in (\ref{lnpomtogluon}) 
has to be convergent. That requires that the 
nonperturbative propagator $D_{\mbox{\tiny np}}(k^2)$ 
falls off faster than $1/k^6$ at large $k^2$. 
Therefore the perturbative gluon propagator should 
take over at large $k^2$ in the corresponding model 
for the Pomeron or the Odderon. 
It was further shown in \cite{Landshoff:1986yj} and 
\cite{Nachtmann:1991ua} 
that in the special situation of hadronic scattering 
at large energy and small momentum transfer 
the exchange of two of these nonperturbative gluons 
involves only the dependence of the nonperturbative gluon 
propagator $D_{\mbox{\tiny np}}$ on the transverse momentum 
components $\kf^2$, whereas the dependence on the longitudinal 
components of the gluon momenta becomes trivial. 
The same will hold in an analogous model for the Odderon. 
This is in agreement with the picture arising in 
perturbative QCD according to high energy factorization. 
There the dynamics is in the high energy limit 
found to take place in transverse space only. 
Each of the three gluons exchanged in our abelian model 
of the Odderon is hence described by 
a perturbative propagator $\sim 1/\kf^2$ in 
two--dimensional transverse momentum space. 
One can also start from this perturbative picture and 
modify the perturbative behavior of the 
gluon propagators at small transverse momenta. 
Since each gluon exchange comes with a factor 
of the strong coupling constant $\alpha_s$ a change of the 
gluon propagator can also be viewed as a modification 
of the strong coupling in the infrared region. 
The gluon propagator $D(\kf^2)$ is hence modified as 
\be
\label{freezeprop}
D(\kf^2) = \frac{\alpha_s}{\kf^2}\:\:\: \longrightarrow 
\:\:\: \tilde{D}(\kf^2) = \frac{\alpha_{\mbox{\tiny eff}}(\kf^2)}{\kf^2}
\,.
\ee
Here $\alpha_{\mbox{\tiny eff}}$ denotes an effective 
strong coupling constant which at large momenta 
reproduces the perturbative running of $\alpha_s$. 
Such an effective strong coupling constant is widely 
used in the dispersive approach to power corrections 
in QCD \cite{Dokshitzer:1995qm} -- \cite{Korchemsky:1995zm}. 
There it appears that the definition of an effective  
running coupling constant at very low momenta is possible 
in the sense that its integral moments have a universal 
meaning. A number of possible models for the coupling 
have been constructed, a typical example being 
\be
\alpha_{\mbox{\tiny eff}}(\kf^2)  = 
\fr{4 \pi}{\lbr 11-\fr{2}{3} n_f \rbr \, \ln
  ( \kf^2 / \Lambda^2_{\mbox{\tiny QCD}} +a)}
\ee
with for example $a=6$. If $a=0$ instead, the model reproduces 
exactly the running of the strong coupling constant 
in one--loop approximation, with $n_f$ denoting the 
number of active quark flavors and 
$\Lambda_{\mbox{\tiny QCD}} \simeq 250 \, \mbox{MeV}$.  
In the description of the Odderon a modified gluon 
propagator was used for example in \cite{Donnachie:1984hf}. 
There it was however mainly required due to an 
unfortunate choice of the Odderon--proton impact 
factor leading to a singular behavior at small gluon momenta. 
Accordingly, the modified $\tilde{D}(\kf^2)$ was chosen 
to vanish at $\kf^2=0$ in order to regularize that singularity 
(see also the discussion in section \ref{elasticppsect}). 
The phenomenological consequences of 
modifications of the type (\ref{freezeprop}) have not yet 
been studied systematically for the Odderon (for one particular 
application see section \ref{elasticppsect} below), 
and in fact not even for the Pomeron, 
but they might well be of importance. 
As far as the Odderon intercept is concerned, however, 
the modifications considered so far are not relevant. 
The reason for this is the abelian character of the 
model from which we started. A nontrivial energy dependence 
requires that the exchanged gluons interact with each other. 
So far, the modification of gluon propagators in the case 
that the exchanged gluons interact has only been studied 
in the case of the Pomeron. 
Here one basically starts from the ladder diagrams of the 
BFKL equation and inserts nonperturbative models for the 
propagators of the gluons. The corresponding effects on the 
BFKL intercept have been studied in 
\cite{Hancock:xh}--\cite{Nikolaev:1994kw}. 
Another possibility is to cut off the singular behavior of 
the perturbative gluon propagator at some fixed momentum 
scale $\kf_0^2$. The effects of this cutoff on the BFKL intercept 
have been studied in 
\cite{Collins:1991nk,McDermott:1995jq,McDermott:1996nb}. 
Analogous investigations are clearly possible for the Odderon 
as well, but have not been performed so far. Here one would 
start from the BKP ladder diagrams and cut off the 
perturbative gluon propagators or alternatively replace them 
by nonperturbative ones. 
It should be emphasized that there are no solid theorems 
concerning the use of such models for the nonperturbative 
gluon propagator in the context of Pomeron and Odderon 
exchanges. Nevertheless, that approach is probably a 
step in the right direction, and accounts for an important 
part of the nonperturbative effects. 

\subsubsection{The Stochastic Vacuum Model}
\label{npheidelbergsect}

The use of a three--gluon exchange model for the Odderon 
and in particular its perturbative coupling to the external 
hadronic particles are questionable for soft processes, 
and the description of these processes requires a 
fully nonperturbative framework. 
A treatment of high energy scattering which is well 
suited for the implementation of nonperturbative models 
was developed by Nachtmann in \cite{Nachtmann:1991ua}. 
The approach is based on the functional integral representation for 
the scattering matrix elements and the eikonal approximation 
separating the large energy from the small momentum transfer. 
The scattering is first considered in an external gluon 
potential, and in a second step one averages over the gluon field. 
In order to obtain a gauge invariant description one does not 
consider quark--quark scattering as the fundamental process. 
Instead the hadronic scattering amplitude is determined by the 
correlation function (or loosely speaking the scattering) of 
two Wegner--Wilson loops with light--like sides representing 
the trajectories of a quark--antiquark system. 
Along its trajectory $\Gamma$  the quark (or antiquark) collects 
a non--abelian phase factor 
\be
\label{eikonalphase}
\Vf = 
\mbox{P} \exp \left( 
-i g \int_\Gamma \Af_\mu (z) dz^\mu
\right)
\,,
\ee
where 
\be
\Af_\mu(z) = A^a_\mu(z) \frac{\lambda_a}{2}
\ee
is the Lie algebra valued gauge potential of the external gluon field. 
These nonabelian phase factors are nothing but the eikonal 
phases of the quarks. After the functional integration, or in other 
words the averaging, over the external gluon field is performed 
one can obtain the scattering amplitudes for hadrons (or photons) 
from the scattering amplitudes of clusters of Wegner--Wilson 
loops by averaging over the hadronic wave functions 
in transverse space. 
Hence the method allows one to study also effects of the spatial 
structure of the hadrons. 
We will give a few more technical details 
of this approach in section \ref{formalismssect} where we also 
describe how baryons can be modeled by a suitable choice of 
a cluster of Wegner--Wilson loops. 

The essential point in this approach is of course the averaging 
over the external gluon fields which requires 
information about the nonperturbative structure of QCD. 
This crucial step can be performed by making use a model of 
nonperturbative QCD, namely the so--called field correlator 
model or stochastic vacuum model 
\cite{Dosch:1987sk,Dosch:ha,Simonov:1987rn}, for a review of 
the model and its applications see \cite{Dosch:2000va}. In its 
original version that model was formulated in euclidean space. 
Its basic assumption is that the nonperturbative gluon vacuum field 
can be approximated by a Gaussian stochastic process in 
the field strengths $F_{\mu \nu}$. 
In a series of papers \cite{Kramer:tr}--\cite{Dosch:1994ym} 
the field correlator model has been extended and adapted for the 
use in the description of high energy scattering. 
This extension of the model includes the 
analytic continuation from euclidean to Minkowski space as 
well as other assumptions. Both the original version as well 
as the version that includes additional assumptions are 
often called stochastic vacuum model in the literature. 
In order to make a clear distinction 
we will in this review use the name 
stochastic vacuum model (SVM) only for the extended 
version, and refer to the original formulation of 
\cite{Dosch:1987sk,Dosch:ha,Simonov:1987rn} as the 
field correlator model. 

The extended version of the model in conjunction 
with the Nachtmann approach to high energy scattering 
has been applied to a variety of processes, see for 
example \cite{Dosch:pk}--\cite{Shoshi:2002fq}. 
It gives a good description of the data and succeeds in relating 
parameters of high energy scattering to those of hadron 
spectroscopy. In the model the mechanism leading to confinement 
induces a string--string interaction in high energy scattering. 
This in turn leads to an increase of the total cross section 
with the hadron size. Quark additivity on the other hand does not 
hold in this approach. In fact the different cross sections for 
pion--nucleon, kaon--nucleon and nucleon--nucleon scattering 
are well reproduced as an effect of their respective electromagnetic 
radii. 

In the Nachtmann approach one aims at applying the model 
to the averaging of the correlation function of two Wegner--Wilson 
loops. In order to do this one first transforms the line integrals 
over gluon potentials into surface integrals over the field strengths 
by means of the non--abelian Stokes theorem. In order to do so 
one has to introduce a common reference point $o$ for the two surfaces 
the boundaries of which are given by the two loops. The surface 
integrals can then be simplified by using the stochastic vacuum 
model which makes 
an assumption about the correlations of the field strengths tensors. 
Let us now briefly collect the main assumptions made in the model. 
For a more detailed description we refer the reader to 
\cite{Dosch:1994ym}. 
The basic object in the model is the correlator of two field strength 
tensors at points $x_1$ and $x_2$ which are parallel--transported 
to the common reference point $o$ along the curves $C_{x1}$ 
and $C_{x2}$, respectively, 
\be
\label{msvcorrelator}
\left\langle
\frac{g^2}{4 \pi^2}\, 
F_{\mu\nu}^a(o,x_1;C_{x1})  F_{\rho\sigma}^b(o,x_2;C_{x2}) 
\right\rangle
= 
\frac{1}{4} \,\delta^{ab} \, 
F_{\mu\nu\rho\sigma}(x_1,x_2,o;C_{x1},C_{x2})
\,.
\ee
The model now assumes that this quantity depends only weakly on the 
choice of the 
common reference point $o$ and of the two curves $C_{x1},C_{x2}$. 
Then Poincar\'e and parity invariance constrain $F_{\mu\nu\rho\sigma}$ 
to be of the general form 
\bea
\label{svmcorrpara}
F_{\mu\nu\rho\sigma}(z) &=& \frac{1}{24} \,G_2\,
\Bigg\{ (g_{\mu \rho} g_{\nu\sigma} - g_{\mu \sigma} g_{\nu\rho}) 
\left[\kappa D(z^2) + (1-\kappa)D_1(z^2)\right] +
\\
&& + (z_\sigma z_\nu g_{\mu \rho} - z_\rho z_\nu g_{\mu \sigma} 
+ z_\rho z_\mu g_{\nu\sigma}
- z_\sigma z_\mu g_{\nu\rho})(1-\kappa) \frac{dD_1(z^2)}{dz^2}
\Bigg\}
\,,
\nn 
\eea
where $z=x_1-x_2$. 
Here $G_2$ is proportional to the gluon condensate, 
$G_2 = \frac{1}{4\pi^2} \langle g^2 F F \rangle$. 
$D$ and $D_1$ are invariant functions with the 
normalization $D(0)=D_1(0)=1$. 
The quantity $\kappa$ is a parameter that measures in a sense 
the non--abelian character of the theory. In an abelian theory 
one would find $\kappa=0$. 
The functions $D$ and $D_1$ are required to decrease rapidly 
for large negative $z^2$ with a characteristic finite correlation 
length $a$. An ansatz with these properties that is also convenient 
to handle practically is 
\bea
\label{dpara}
D(z^2) &=& \int_{-\infty}^\infty 
\frac{d^4k}{(2\pi)^4} e^{-ikz} 
\frac{27 (2\pi)^4}{(8a)^2} 
\frac{i k^2}{(k^2-\lambda^{-2} + i \epsilon)^4}
\\
\label{d1para}
D_1(z^2)&=&\int_{-\infty}^\infty 
\frac{d^4k}{(2\pi)^4} e^{-ikz} 
\frac{2}{3} \frac{27 (2\pi)^4}{(8a)^2} 
\frac{i k^2}{(k^2-\lambda^{-2} + i \epsilon)^3}
\eea
with $\lambda=8a/(3\pi)$. The model thus contains only 
three parameters $G_2$, $\kappa$, and $a$ which need 
to be determined once. 
In euclidean space for example the functions $D$ and $D_1$ 
can be compared to lattice calculations of the correlator 
(\ref{msvcorrelator}) in order to determine these 
parameters \cite{DiGiacomo:1992df}--\cite{Meggiolaro:1998yn}. 
An appropriate set of values for the three parameters is for 
example 
\bea
G_2 &=& (496 \,\mbox{MeV})^4\\
\kappa &=& 0.74\\
a&=& 0.35 \,\mbox{fm}
\,,
\eea
but also slightly different values are in use. 

The central assumption is finally that nonperturbative 
vacuum fluctuations of the gluon field are determined by 
a Gaussian stochastic process. That means that any 
correlator of $n$ field strength tensors with $n>2$ 
factorizes into two--point functions of field strength tensors. 
In the example of $n=4$ this means in symbolic notation 
\be
\langle F_i F_j F_k F_l \rangle = \sum_{\mbox{\tiny pairings}} 
\langle F_i F_j \rangle \langle F_k F_l \rangle 
\,. 
\ee
The sum extends over all three possible pairings of the 
four field strength tensors. 
In the original field correlator model 
\cite{Dosch:1987sk,Dosch:ha,Simonov:1987rn} 
this assumption is made for the matrix--valued field strength 
tensors $\Ff_{\mu\nu}$. In the application to high energy scattering 
this assumption is extended to the color components $F^{a_i}_{\mu\nu}$ 
of the field strengths tensors. 
Due to color conservation the above assumption implies that the 
correlator of an odd number of field strength tensors vanishes. 

In order to study Odderon exchange processes in this model 
one expands the path--ordered exponential of the 
Wegner--Wilson loop into a power series which in 
particular is a series in color space. The color 
degrees of freedom can then be projected onto a 
symmetric color state appropriate for the Odderon. 
A number of Odderon induced processes has been computed 
using this model, and we will describe the corresponding 
results in more detail in later sections. 

The particular value of the SVM lies in the fact that 
it makes a wide class of soft high energy scattering 
processes accessible to a theoretical description 
on the basis of a non--abelian model of nonperturbative QCD. 
Presently there is no other method known that 
comes closer to a description of these soft processes in full QCD. 

An important property and probably a deficit of the model is 
that it predicts a trivial energy dependence of the cross 
sections, at least when the model is used in the form described above. 
This is true for processes with $C=+1$ exchange 
as well as for processes with $C=-1$ exchange. That means 
in particular that the Odderon intercept $\alpha_\soddi$ 
equals one in this approach. 
There have been attempts to include an energy dependence 
into the model, see for example \cite{Shoshi:2002in}. However, 
although successful in the description of the data the procedure 
used there appears not quite satisfactory. The energy dependence 
is not generated dynamically but constructed ad hoc, 
and two intercepts for the energy dependence appear as 
additional parameters. 
At the moment one has to say that the problem 
of an energy dependence in this model is not yet understood. 

It should be stressed again that the application of the SVM 
in the description of high energy scattering processes requires 
several additional assumptions. 
These assumptions are nontrivial and clearly go beyond the 
original formulation of the field correlator model in 
\cite{Dosch:1987sk,Dosch:ha,Simonov:1987rn}. 
A potential failure to describe certain high energy processes 
in this framework would therefore not imply that 
the original field correlator model does not give a valid picture of the 
nonperturbative QCD vacuum. 

\subsubsection{The Regge Trajectory of the Odderon}
\label{nptrajectorysect}

A fully nonperturbative approach aimed at a determination 
of the Odderon intercept was pursued in 
\cite{Kaidalov:1999yd,Kaidalov:1999de}. 
It relates the Odderon to glueball states via its Regge trajectory. 
Motivated by the observation of remarkably straight 
mesonic Regge trajectories (see section \ref{reggetheory})
is has long been speculated that also the Pomeron trajectory 
might be linear even at large $t$, 
\be
\label{ptrajnp}
 \alpha_\spommi (t) = \alpha_\spommi (0) + 
 \alpha'_\spommi t
\ee
with $\alpha'_\spommi = 0.25\,\mbox{GeV}^{-2}$. 
For comparatively small $t$ the linearity is well confirmed 
in the shrinkage of the forward peak in elastic $pp$ scattering. 
Since the Pomeron is described by two--gluon 
exchange one expects that the physical states on the Pomeron 
trajectory should be glueball states of largest spin. The lowest of 
these state should be a $2^{++}$ glueball consisting of two 
constituent gluons. 
Here it should be pointed out that the calculation of the 
meson\footnote{In this context we use `meson' only for 
$q\bar{q}$ bound states but not for glueballs.} 
spectrum is a very complicated bound state problem of QCD. 
Consequently, already the remarkable linearity of mesonic Regge 
trajectories is not fully understood theoretically. The situation 
of glueball trajectories is even less clear, theoretically as well 
as phenomenologically. A particular problem here is that 
the experimental identification of glueballs is very difficult due 
to their mixing with mesonic states. It is well possible that 
glueball trajectories are not linear in reality. 
However, the assumption of a linear Pomeron trajectory is certainly 
supported by the fact that a $2^{++}$ glueball candidate has 
been found \cite{Abatzis:1994ym} which exactly matches the 
linear Pomeron trajectory (\ref{ptrajnp}). 
Accepting the hypothesis of linear Regge trajectories also for 
gluonic bound states one can look for a glueball state with 
the quantum numbers of the Odderon and try to relate 
its mass to the Odderon intercept $\alpha_\soddi(0)$ via 
\be
\label{otrajnp}
 \alpha_\soddi (t) = \alpha_\soddi (0) + 
 \alpha'_\soddi t
\,.
\ee
A suitable candidate would 
be a $3^{--}$ state with three constituent gluons. 
In \cite{Kaidalov:1999yd,Kaidalov:1999de} the glueball 
spectrum is calculated using a method based on the area 
law for Wegner--Wilson loops at large distances. 
The predicted glueball masses 
vary with the string tension $\sigma=1/(2 \pi \alpha')$.  
The string tension depends on the representation of the particles 
connected by the string consisting of a color flux tube. For 
the fundamental representation, i.\,e.\ for a $q\bar{q}$ system, 
the string tension is $\sigma_f= 0.18\,\mbox{GeV}^2$. According 
to Casimir scaling the string tension in the adjoint representation, 
i.\,e.\ for gluons should be $\sigma_a= \frac{4}{9} \sigma_f$, 
resulting in $\alpha'\simeq 0.4 \,\mbox{GeV}^{-2}$. 
In \cite{Kaidalov:1999yd,Kaidalov:1999de} this value is used 
to predict a $3^{--}$ three--gluon glueball state 
with a mass of around $3.51 \,\mbox{GeV}$. For a string 
tension of $\sigma_f=0.238\,\mbox{GeV}^2$ that glueball 
is shifted to a mass of $4.03 \,\mbox{GeV}$. This is in good 
agreement with lattice simulations \cite{Morningstar:1999rf} 
using the same string tension which find a $3^{--}$ glueball 
at $4.1 \pm 0.29\,\mbox{GeV}$. 
Now the authors of \cite{Kaidalov:1999yd,Kaidalov:1999de} 
assume that the Regge slope $\alpha'_\soddi$ 
for a three--gluon glueball is the same as for the one for a 
two--gluon glueball, $\alpha'_\spommi$. That assumption is 
based on an analogy to the baryonic Regge slope which 
in a quark--diquark picture is expected to be the same 
as the mesonic Regge slope since the diquark acts like 
an antiquark. This assumption is rather speculative, 
firstly because of deviations from the 
exact quark--diquark configuration in a baryon, and secondly 
due to possible more intricate differences between a three--quark 
and a three--gluon bound state. 
For the parameters chosen in 
\cite{Kaidalov:1999yd,Kaidalov:1999de} 
the extrapolation of the Odderon trajectory from the $3^{--}$ state 
at $t=(3.51 \,\mbox{GeV})^2$ down to $t=0$ leads to an 
Odderon intercept as low as $\alpha_\soddi(0)=-1.5$. 
Even for lower values of the Odderon slope than chosen there 
one would still expect a negative Odderon intercept. 
That would imply that the Odderon exchange is very strongly 
suppressed at large energies, at least at low values of $t$. 
Such an object should strictly speaking not even be called 
an Odderon. 
The approach outlined here is to a large extent speculative. 
Still, the result is a very interesting possibility. 
It should be pointed out that the low Odderon intercept 
emerging from this picture applies only to the soft Odderon. 
It does by no means imply that a hard (perturbative) 
Odderon should also have a low intercept. 

In summary, one possible scenario 
of nonperturbative effects on the Odderon could be 
the following. In general we expect a similar situation as in the 
case of the Pomeron. There should be a hard Odderon which 
in first approximation can be identified with the BKP Odderon. 
It will still receive some nonperturbative 
contributions from low gluon momenta, but its intercept 
should remain in the vicinity of one. 
The hard Odderon should 
be visible in scattering processes dominated by only 
one hard scale. In addition there will be a soft Odderon 
dominated by nonperturbative physics. In the approach of 
\cite{Kaidalov:1999yd,Kaidalov:1999de} there are 
indications that its intercept might be very low, possibly 
making the soft Odderon invisible in low-$t$ processes. 
This might in fact explain the apparent absence of an 
Odderon contribution in forward scattering processes. 
We will turn to this and other phenomenological observations 
in the following section. 

\section{Phenomenology of the Odderon}
\label{phensect}
\setcounter{equation}{0} 

In the present section we will be concerned with the 
phenomenology of the Odderon. Here the main focus will 
be quite different from that of the preceding section. 
There the interest was concentrated mainly on 
the more theoretical questions related to the Odderon, 
namely its intercept and its r{\^o}le in the general 
picture of QCD in the Regge limit. Now we turn 
to experimental manifestations of the Odderon. 
The experimental evidence for the Odderon 
is at the moment rather scarce. It should therefore 
be the first and foremost goal to cleanly establish its 
existence experimentally. 
In this context the exact nature of the Odderon 
singularity in the complex angular momentum plane 
and the precise value of the 
Odderon intercept are not very important. 
At the present stage of Odderon phenomenology many 
studies are simply performed assuming an intercept 
around one, i.\,e.\ under the assumption that the 
Odderon is not suppressed at high energies. 
Small deviations of the intercept from one will in 
most cases have little effect on the typical 
observables\footnote{The asymmetries discussed in 
section \protect\ref{epasymmetrsect} are an exception 
to this rule. Here due to interference effects the results 
depend on the difference of the Pomeron and the Odderon 
intercepts in a nontrivial way.}.  
Nevertheless, the precise Odderon intercept remains 
of central interest from a theoretical point of view and 
will hopefully also become experimentally accessible 
when precise data on the Odderon eventually 
become available. 

For a long time the Odderon was discussed only in the 
context of $pp$ and $p\bar{p}$ scattering. This has two 
reasons. The first reason lies in the history of the Odderon 
which was first discussed in the context of asymptotic 
theorems in Regge theory. These theorems apply to 
the scattering of stable hadrons but not necessarily to 
processes involving photons for example. 
The second reason was simply the absence of suitable 
experimental data for other types of particle collisions,  
and there was obviously little incentive to consider 
the effects of the Odderon in other types of collisions. 
From a theoretical point of view, however, hadron--hadron 
scattering is the most difficult process to describe in QCD. 
Our knowledge of QCD is to a very large extent 
based on perturbation theory which is hardly applicable 
to soft hadronic processes. The proton and its antiparticle 
are complicated bound states, and their poorly understood 
internal structure unavoidably plays a crucial r{\^o}le in the 
description of $pp$ and $p\bar{p}$ scattering. 
Moreover, besides the Odderon there is a number of 
other exchanges that contribute in these scattering 
processes, all of which cannot be derived from first 
principles in QCD. The description of $pp$ and $p\bar{p}$ 
scattering therefore necessarily involves the fitting 
of a number of parameters to the data. Unfortunately, 
the most interesting observables for studying the Odderon 
are among those that are experimentally most difficult 
to measure, and despite the long history of $pp$ and $p\bar{p}$ 
scattering experiments there is in fact a lack of precise 
data in this field. Since the Odderon is mainly visible in 
the difference between $pp$ and $p\bar{p}$ cross sections 
it would be extremely helpful to have data for both reactions 
measured at the same energy and with the same experimental 
setup in order to reduce systematic uncertainties. 
Such a measurement of both reactions was performed 
only at the CERN ISR and also there only for a single energy. 
Evidence for the Odderon was found in precisely these 
data. But the difficulties inherent in the description of 
$pp$ and $p\bar{p}$ scattering and the lack of further data have 
made it impossible to establish the existence of the Odderon 
beyond reasonable doubt. Also for the future it is at least 
not obvious, that suitable observables in 
both $pp$ and $p\bar{p}$ scattering will be measured 
with sufficient accuracy at any present of future collider. 
However, there is a chance that $pp$ scattering data 
will become available from RHIC in an energy range 
that overlaps with that of the $p\pbar$ data from the 
CERN ISR. 
In such a situation it is natural to consider the possibility of finding 
the Odderon in other types of reactions. 
The cross sections for Odderon induced reactions in 
other scattering processes like electron--proton
or electron--positron scattering are however rather small. 
Only with the advent of HERA and with the high 
luminosity that this machine has accumulated 
one could realistically start to think about probing 
the Odderon in $ep$ collisions. Here one is particularly 
interested in exclusive processes to which the Pomeron 
does not contribute. This offers a much cleaner 
environment for identifying the effects of the Odderon. 
These processes 
are presently being studied at HERA and the first experimental 
data have just become available. The investigation 
of Odderon exchange in the even cleaner environment 
of real or virtual photon--photon collisions will only 
be possible at a future $e^+ e^-$ linear collider. 
The scattering of two photons occurs as a subprocess 
in $e^+ e^-$ collisions already at LEP, but the 
small cross sections for Odderon exchange processes 
require a higher energy and a much higher luminosity 
than LEP has reached. 

In this section we start with general considerations on the 
phenomenology of the Odderon in section 
\ref{genconsidsect}. We discuss the approximations 
and assumptions that are generally made 
and explain the origin of possible uncertainties in the 
predictions for Odderon--induced processes. 
In section \ref{formalismssect} 
we discuss how the scattering amplitude 
for processes involving Odderon exchange is 
obtained in different formalisms. 
In the remaining sections we discuss the different 
observables in which the Odderon can be studied, 
ordered by the initial state particles of the collisions: 
section \ref{ppsect} deals with $pp$ and $p\bar{p}$ 
scattering, section \ref{epsect} with electron--proton 
collisions, and section \ref{gammagammasect} 
finally with photon--photon collisions. 
So far the only evidence for the Odderon has been found 
in the difference of the differential cross sections 
for $pp$ and $p\bar{p}$ elastic scattering. 
Naturally, a large part of section \ref{ppsect} is 
devoted to this process. 
We also discuss the so--called $\rho$-parameter 
which is another interesting observable that can 
potentially indicate the existence of the Odderon. 
Since here the current experimental situation does 
not allow any firm conclusions we however keep that 
account short. Instead, 
in the remaining part of section \ref{ppsect} and in 
the following sections we turn to more exclusive 
processes, in which the Odderon is basically the 
only exchange that contributes to the production of the 
respective final states at high energy. 
These processes are very interesting because 
already their observation would be sufficient to 
cleanly establish the existence of the Odderon. 
Moreover, they offer particularly good chances 
to observe and to study the Odderon at current and 
future colliders. In section \ref{epsect} we 
discuss the first experimental 
results that have recently been obtained for some 
of these processes by the H1 collaboration in 
$ep$-collisions at HERA. 

Before we start we should mention that the Odderon 
has been discussed also in the context of other 
interesting processes which however will not be 
discussed in detail in the present review. Among them 
are pion--proton scattering 
\cite{Gauron:1983au}--\cite{Contogouris:di}, 
vector meson production in pion--proton scattering 
\cite{Struminsky:1991jg}, kaon--nucleon scattering 
\cite{Goloskokov:1986rw,Zakharovmeson}, 
and the structure function $F_3$ in deep inelastic 
neutrino--nucleon scattering \cite{Struminsky:ax}. 

\subsection{General Considerations}
\label{genconsidsect}

The phenomenology of the Odderon is in many respects 
still at a rather early stage of its development. 
On the one hand this is caused 
by the lack of guidance from experimental data on the 
Odderon. On the other hand the interest in more exclusive 
processes was strengthened only recently due to the 
possibility to observe them at HERA and at planned 
future colliders. The enormous progress in 
solving the BKP equation has also been made only recently, 
and the solutions discussed in the preceding section 
have so far been applied phenomenologically only in a few cases. 
Currently, the main aim in the phenomenology of the Odderon 
clearly is to establish its existence in an unambiguous way.
The exact nature of the Odderon singularity in the 
complex angular momentum plane is not known. 
Consequently, at the moment different perturbative 
and nonperturbative models are used. In most cases 
the choice of the model is motivated by the process 
under consideration and the perturbative or 
nonperturbative method used in the calculation. 
It is only in a few cases that different methods or 
models have been applied to the same process. 
Marked differences between different models have 
been found only when the two most extreme models 
for the Odderon are compared, namely the maximal 
Odderon and the simplest pole ansatz for the Odderon. 
But even here one has to have in mind that in all cases 
there are free parameters which leave some freedom 
as long as they cannot be fixed unambiguously for the 
different models due to the lack of data. 
As long as the intercept 
of the Odderon is in the vicinity of one even simple models 
for the Odderon should give at least a qualitatively correct 
picture. This applies especially in the case of exclusive 
reactions which can only be caused by Odderon exchange. 
Of course it would eventually be very desirable to 
determine also the precise nature of the Odderon singularity. 
But this appears to be very difficult in the light of what we 
know about the by now well--studied Pomeron. There 
a simple Regge picture with two poles describes the data 
well, but also a purely logarithmic fit gives a good (and 
even slightly better) fit to the available data. Moreover, 
perturbation theory indicates a much more complicated 
singularity structure. Hence we still do not know which 
kind of Regge singularity the Pomeron corresponds to --- 
despite the wealth of available data. 
We certainly have to expect that the situation for the Odderon 
will not be much simpler than for the Pomeron. 
Given the lack 
of data for the Odderon we probably have to accept that 
the chances for determining the type of its singularity 
are quite low at least at the moment. 

Let us now discuss which processes are suitable for 
observing and studying the Odderon. A very important 
property of the Odderon in this context is that it is a 
colorless object. In hadronic reactions the exchange of a 
colorless object in the $t$-channel generally leads to 
the formation of a rapidity gap in the final states, i.\,e.\ 
the outgoing particles are separated by a large region 
in rapidity in which there is no hadronic activity observed 
in the detector. Such processes are usually referred to as 
diffractive reactions\footnote{The notion of diffraction in 
hadronic reactions is often defined in different ways. 
In $ep$ collisions for example one sometimes demands that the 
proton stays intact for the event to be called diffractive. For the 
present review it will be sufficient to use a practical 
definition of diffraction based on the occurrence of a 
rapidity gap in the final state. In particular we include 
inelastic diffractive events in which for example the 
proton dissociates.}. In principle a rapidity gap can 
occur also if a colored object like a gluon is exchanged in the 
$t$-channel, but the probability for this to happen is exponentially 
suppressed with the size of the rapidity gap. 
If we concentrate on processes at large energy $\sqrt{s}$ 
(and neglect the hence small contribution of reggeon exchanges) 
we can say that a large rapidity gap in 
a hadronic reaction is a characteristic feature of Pomeron 
or Odderon exchange. 
Depending on the quantum numbers of the incoming and 
outgoing particles in diffractive reactions one can have 
cases in which only the Pomeron exchange contributes, 
others in which only the Odderon exchange contributes, 
or finally cases in which both contribute. 
The latter can occur if the scattering particles or the 
final state particles are 
not eigenstates of $C$ parity. This is the case for example 
for the proton or the antiproton. In these cases the Pomeron 
usually gives the dominant contribution and the Odderon 
is difficult to pin down in practice. The most important 
example for this is $pp$ and $p\bar{p}$ elastic scattering, 
which is also a special case of diffraction. In order to find 
the Odderon here one has to look at the difference of the 
differential cross sections for $pp$ and $p\bar{p}$ 
scattering, or in general for the respective particle and 
antiparticle cross sections. 
This difference is generally much smaller than the 
two cross sections themselves and hence difficult to measure, 
as we will see in more detail in section \ref{elasticppsect}.  
The situation is similar for total cross sections. Although 
the Odderon generally contributes to total cross sections 
they are by far dominated by the Pomeron. Also here 
the difference between the corresponding particle--particle 
and antiparticle--particle cross sections is much smaller 
than the total cross sections themselves, as we have 
already seen in section \ref{pomoddregge}. 
This problem is absent in processes that are induced only 
by Odderon exchange, and to which the Pomeron 
does not contribute. A typical example of such a reaction 
is the diffractive production of pseudoscalar mesons in 
$ep$ scattering at high energy. The electron emits a 
real or virtual photon which is transformed into the 
pseudoscalar meson by scattering on the proton. 
Both the photon and the pseudoscalar meson are eigenstates 
of $C$ parity, the photon carrying quantum number 
$C=-1$ and the pseudoscalar meson carrying $C=+1$. 
Consequently the scattering on the proton must be 
mediated by an object of negative $C$ parity. Hence it 
can be induced only by Odderon exchange and the Pomeron 
does indeed not contribute. It should be noted that in 
this reaction the proton can either stay intact or dissociate. 
As we will see the expected cross sections are larger in 
the latter case. In summary, we can say that exclusive 
diffractive reactions with suitable quantum numbers 
allowing only Odderon but not Pomeron exchange are 
the optimal place to look for the Odderon. 

It should be noted that besides the Odderon there are also 
other exchanges carrying negative $C$ parity, namely 
the photon and mesonic reggeons. Both exchanges are 
theoretically under good control but need to be taken 
into account in almost all calculations involving the 
Odderon. The intercept $\alpha_{\mbox{\scriptsize R}}$ 
of the mesonic Regge trajectory is 
at about $0.5$, and consequently the contribution of reggeon 
exchange vanishes rapidly with increasing energy. 
The characteristics of photon exchange on the other hand 
are in many cases very different from those of the Odderon, 
and the corresponding contributions are quite different 
in different regions of phase space. The two can therefore 
often be disentangled in sufficiently differential observables. 

Given the present state of the art in this field the theoretical 
predictions for Odderon induced processes are not very precise. 
The difficulties and problems in describing Odderon induced 
processes are typically very similar to those occurring in 
the case of the Pomeron. For nonperturbative approaches to 
the Odderon some of these difficulties and limitations have already 
been described in section \ref{nonpertoddsect}. In the following 
we will outline several points that are characteristic mainly for 
the perturbative Odderon. 

The conditions for the use of the perturbative Odderon are 
practically the same as for the perturbative Pomeron. 
A perturbative Odderon can be used with or without BKP 
resummation. In both cases the use of a perturbative Odderon 
requires a hard scale in the process, for example the large 
mass of a produced meson or a large momentum transfer $\sqrt{-t}$. 
In practice 
the involved momentum scales are often only moderately large, 
sometimes even just at the edge of the applicability of perturbative QCD. 
This induces an uncertainty which is usually very difficult to estimate. 
For the use of the resummed (BKP) Odderon the same conditions 
apply as for the BFKL Pomeron. These have been described in detail 
in section \ref{bfklapplicabsection}. Of special importance is again 
the problem of diffusion of transverse momenta. As in the BFKL 
Pomeron the transverse momenta in the three--gluon ladder of 
the BKP Odderon are not ordered. In the BFKL Pomeron this 
resulted in a diffusion or even a tunneling of momenta into the 
infrared region of small momenta as discussed in detail in 
section \ref{bfklapplicabsection}. A similar diffusion of momenta 
is clearly expected in the BKP Odderon as well. If the contribution 
of small momentum configurations in the Odderon becomes too 
large the perturbative approach is no longer justified. A detailed 
numerical study of the diffusion process in the BKP Odderon has 
not yet been performed, but would certainly be very valuable. 
In particular it would be interesting to see whether 
the transition to the tunneling 
process occurs at smaller or larger energies compared to the 
BFKL Pomeron when the external momenta are fixed at the 
same scales in both cases. Also for the Odderon the diffusion 
problem cannot be avoided completely. The best one can do is to 
suppress the contribution from the infrared region by choosing 
processes in which the momentum scales at both ends of the 
Odderon are sufficiently large. In this sense the best processes 
for the observation of the perturbative BKP Odderon would be 
scattering processes 
of two virtual photons of large and similar virtualities. 
Such processes are described in section \ref{gammagammasect} 
below. A typical example is the quasielastic scattering 
$\gamma^* \gamma^* \to \eta_c \eta_c$. The corresponding 
cross sections are rather small, but there is a good chance 
of observing these processes at a future linear collider. 
In most other processes the situation is less favorable for avoiding 
the diffusion problem, and the corresponding uncertainty 
is often difficult to control. 

Many calculations make use of Regge factorization or 
perturbative high energy factorization. Strictly speaking 
both are applicable only in the limit of high energy $\sqrt{s}$. 
Often it is difficult to determine to which accuracy 
factorization actually holds for a given process in a given 
kinematical situation. For perturbative factorization 
there is at least in principle a way to estimate the size 
of the corrections by comparing the resummed 
and factorized calculation with a fixed--order calculation. 
When one goes beyond the leading order approximation 
in the strong coupling constant a fixed--order calculation 
will naturally contain also factorization--breaking terms. 
For the Odderon, however, this is a rather difficult problem 
that has not yet been addressed so far. 

In any perturbative calculation there is an uncertainty 
associated with the choice of the scale of the strong coupling 
constant. In processes involving different momentum scales 
there is an ambiguity in the choice of the scale 
$\mu^2$ at which the running coupling constant $\alpha_s$ 
has to be evaluated. Since the coupling runs only logarithmically 
the difference between different scale choices is in most cases 
not too large. 
But for the Odderon the problem is considerably enhanced and 
in fact induces a rather large uncertainty in the theoretical 
predictions. 
This applies to simple three--gluon exchange as well as to the 
resummed BKP Odderon. 
The reason for this is very simple and lies in the 
fact that the coupling of three gluons to external particles 
involves a factor of $\alpha_s^3$ already on the amplitude level. 
That implies that the cross section is proportional to $\alpha_s^6$. 
Hence even small changes in the choice of the appropriate value 
of $\alpha_s$ lead to rather large changes in the cross section. 
In many practical applications the typical momentum scales 
are only moderately large such that the correct value for 
the coupling is difficult to find anyway. 
Changing the coupling $\alpha_s$ from $0.2$ to $0.3$ in such a 
situation implies that the cross section becomes larger by an order 
of magnitude. Unfortunately the problem can hardly be circumvented, 
and a theoretically clean scale setting would require a full 
next--to--leading order calculation. 
The cause of the problem is rather trivial and therefore the 
problem is very often considered not to be serious, hence it is often 
underestimated. But in many applications the uncertainty implied 
by the choice of $\alpha_s$ is actually the dominant one. 
We will therefore be confronted with this uncertainty throughout 
this section. 

So far most phenomenological studies of processes that can be 
calculated perturbatively use a simple model for the Odderon 
in which the three exchanged gluons do not interact with each 
other. If one uses a resummed (BKP) Odderon solution one has 
to study both known types of solutions, namely the JW solution 
and the BLV solution, see sections \ref{jwsolutionsect} and 
\ref{blvsolutionsect}, respectively. Since the two solutions have 
almost the same intercept their relative contribution to 
a scattering process strongly depends on the coupling to the 
external particles which can be very different for the two solutions. 
As we will see there are even scattering processes in which 
(at least in leading order) only one of these solutions can contribute. 
Another difference which is of potential phenomenological interest 
is that the two solutions have different twist, the JW solution 
having twist $4$ and the BLV solution having twist $3$ 
\cite{Grishaprivcomm}. 
See also section \ref{pertinterceptsect} for a discussion of the 
phenomenological properties of the two different solutions. 

The BKP Odderon has so far only been studied 
in LLA. But it should be expected that the NLL 
corrections to the Odderon will be of a similar size as 
for the Pomeron, and they will certainly be of phenomenological 
relevance. This should be kept in mind when interpreting 
effects that result from the use of an explicit Odderon 
solution of the BKP equation. At the moment not even 
the direction is known in which the NLL corrections 
will push the Odderon intercept. 

Another interesting question of phenomenological 
interest concerns the exchange of a three--gluon 
state of Pomeron type, carrying positive charge parity 
$C=+1$. It is relevant in processes in 
which both Odderon and the Pomeron exchange 
contribute. Usually, it is neglected in phenomenological studies. 
Depending on the process it is however not a priori 
clear that this is a consistent scheme. That issue is particularly 
pronounced in the perturbative framework to 
which we will now restrict the discussion of this point. 
The problem arises due to the fact that 
the leading exchanges in the $C=+1$ and $C=-1$ 
sectors (the BFKL Pomeron and the BKP Odderon, 
respectively) are not of the same order in the 
expansion in leading logarithms. The BFKL Pomeron 
resums terms of the order $(\alpha_s \log s)^n$, 
whereas the BKP Odderon includes terms of 
order $\alpha_s(\alpha_s \log s)^n$ only. 
A consistent approximation in the expansion in 
logarithms of the energy 
would therefore require to take into 
account also NLL corrections to the BFKL 
Pomeron as well as 
the three--gluon state in the $C=+1$ sector. 
The former is already nontrivial since it 
leads to new consistency problems 
concerning logarithms in transverse momentum 
which appear in the NLLA in the running of 
the coupling. The latter has not really been 
carried out so far. 
Whether in neglecting the three--gluon Pomeron state 
one obtains a consistent description 
depends very much on the observable. In the case of 
the charge asymmetries arising due to 
Pomeron--Odderon interference (see 
section \ref{epasymmetrsect}) the leading term 
is obtained by taking the leading terms in each sector, 
and it is of course a perfectly valid approximation scheme 
to neglect the $C=+1$ three--gluon state. 
In processes like $pp$ elastic scattering at large $t$, 
however, the amplitudes of Pomeron and Odderon exchange 
are added to obtain the total scattering amplitude and 
here neglecting the $C=+1$ three gluon state does not 
represent a consistent approximation scheme in the 
expansion of logarithms of the energy. 
The contribution of the $C=+1$ 
three--gluon state to scattering processes and in 
particular its coupling to external particles 
(its impact factors) have not been studied in much detail. 
But a very interesting effect has been 
observed in the extended GLLA in which additional 
terms are taken into account in order to 
ensure unitarity constraints also in subchannels of 
multi--reggeon exchange amplitudes, see section 
\ref{egllasect}. There it was found that the three--gluon 
Pomeron state reggeizes completely when it is coupled 
to certain impact factors. Here reggeization means 
that the $C=+1$ exchange with three gluons 
in the $t$-channel turns out to be a superposition of 
BFKL two--gluon exchanges, see eq.\ (\ref{d3reggeizes}). 
Due to the coupling to the impact factor the reggeizing part 
of the three--gluon state is singled out. 
This effect was shown for an impact factor which couples 
the three gluons to two virtual photons via a quark loop, 
but the same effect is expected also for similar impact 
factors like for example the $\gamma^* \to J/\psi$ transition. 
Since reggeization of gluons is a very deep phenomenon in 
nonabelian gauge theories it is well possible that a similar 
effects occurs also in the nonperturbative region of low momentum 
processes. 
If it would also apply to more complicated impact 
factors like the ones occurring in $pp$ and $p\bar{p}$ elastic 
scattering we would be led to a rather ironic conclusion. 
In that case it would be justified to neglect the $C=+1$ 
three--gluon exchange in fits to the cross section since 
its contribution would effectively be absorbed by 
the (two--gluon) Pomeron term in the fit. 
In summary we can say that the three--gluon state of 
Pomeron type can in many cases give contributions as 
large as the Odderon, and in some observables it even 
needs to be included in order to obtain a consistent approximation 
scheme in the sense of leading logarithms of the energy. 
It can be of  phenomenological importance in many 
cases and clearly deserves further study. 

\subsection{The Scattering Amplitude for Odderon Exchange}
\label{formalismssect}

Here we briefly sketch how the scattering amplitude $T^\soddi$ 
for an Odderon exchange process can be obtained in different 
formalisms: in Regge theory, in perturbative QCD, and 
in the functional approach in position space as developed in 
\cite{Nachtmann:1991ua}. Finally, we give an outline of how 
the latter approach can be used to implement a model for 
nonperturbative QCD like the stochastic vacuum model. 

The simplest case is the description of the Odderon in the 
formalism of Regge theory as described in section \ref{reggetheory}. 
Here the amplitude for Odderon exchange depends on the 
nature of the corresponding singularity in the complex 
angular momentum plane. In phenomenological investigations 
one often chooses a simple Odderon pole or the maximal 
Odderon, as we have described them in section \ref{pomoddregge}. 
In the case of a simple pole the amplitude is of the form 
(\ref{oddprop}) supplied with suitable factors for the coupling 
of the Odderon to the external particles. Examples of these 
couplings will be given in section \ref{Oddprotoncouplsect}, 
see for instance (\ref{reggeoddpcoupl}). 

In perturbation theory the amplitude for the 
exchange of an Odderon factorizes in the 
high energy limit. In analogy to the case of the 
perturbative Pomeron (see section \ref{bfklsect}) 
it can be written as a convolution of two impact 
factors and the Odderon Green function as 
has been visualized in figure \ref{figoddfact} 
in section \ref{bkpsect}. 
The dynamics reduces to the two--dimensional 
transverse plane, and the convolution is in the 
transverse momentum plane only. 
$\phi_1$ and $\phi_2$ denote the impact factors 
which couple the Odderon to the external particles. 
$\phi_\soddi$ represents the Green function of the 
Odderon in the sense of (\ref{oddigreen}). 
The scattering amplitude can be written symbolically as 
\be
A^\soddi(s,t) \sim \langle \phi_1 | \phi_\soddi | \phi_2\rangle 
\,,
\ee
measuring the overlap of the impact factors with 
the Odderon's Green function. 
In detail, it is given by 
\bea
\label{Aoddifact}
A^\soddi(s,t) = \frac{s}{32} \, \frac{5}{6} \, \frac{1}{3!} \,
\frac{1}{(2 \pi)^8}
\!\!\!\!\!\! \!\! &&\int 
d^2 \kf_1 \,d^2 \kf_2\, d^2 \kf_3\,
d^2 \kf'_1\,d^2 \kf'_2\,d^2 \kf'_3\,
\times \nn\\
&&\times
\,\phi_1(\kf_1,\kf_2,\kf_3) \,
\phi_\soddi(\kf_1,\kf_2,\kf_3;\kf'_1,\kf'_2,\kf'_3)\,
\phi_2(\kf'_1,\kf'_2,\kf'_3) \times\nn \\
&&\times
\,\delta(\qf - \kf_1-\kf_2-\kf_3) 
\delta(\qf - \kf'_1-\kf'_2-\kf'_3) 
\,,
\eea
where $\qf$ is the total transverse momentum transferred 
in the $t$-channel, and $t=-\qf^2$. 
The $1/3!$ is a symmetry factor reflecting the exchange 
of three identical bosons. 
The factor $5/6$ 
is a color factor originating from the contraction of the 
symmetric structure constants of $\mbox{SU(3)}$. 
One could have defined the impact factors also including 
a color tensor. Since we will be concerned only with 
Odderon impact factors the color tensor would always be 
a $d_{abc}$ tensor. For the present review we therefore 
prefer to pull out these tensors of $\phi_1$ and $\phi_2$ 
and to contract them already in this general formula. 
The impact factors $\phi_{1,2}$ 
depend on the scattering process and will be specified 
for different reactions in the following sections. 
The Green function for the Odderon is obtained from the 
solutions of the BKP equation as described in section 
\ref{oddpertsect}. 

In many phenomenological applications, however, 
a simpler picture for the Odderon is used, namely the 
perturbative exchange of three noninteracting gluons. 
In this case the Green function $\phi_\soddi$ of the Odderon 
becomes 
\be
\label{phioddisimple}
\phi_\soddi(\kf_1,\kf_2,\kf_3;\kf'_1,\kf'_2,\kf'_3)=
\frac{1}{\kf_1^2 \kf_2^2 \kf_3^2} \,
\delta(\kf_1-\kf'_1)
\delta(\kf_2-\kf'_2)
\delta(\kf_3-\kf'_3)
\,,
\ee
and the amplitude $A^\soddi(s,t)$ in (\ref{Aoddifact}) 
simplifies accordingly. 

Let us now turn to the functional approach of Nachtmann 
\cite{Nachtmann:1991ua} which we have outlined briefly 
already in section \ref{npheidelbergsect}. 
This approach is not as widely used as the well--known 
perturbative framework in general, but it plays a prominent 
r\^ole in applications to the phenomenology of the Odderon. 
We therefore find it useful to give a more detailed outline of 
that approach here. 
One starts from the scattering amplitude of two quarks 
in the high energy limit in a fixed external gluon field. 
The $S$-matrix element for two incoming quarks with 
momenta $p_1, p_2$ and two outgoing quarks with momenta 
$p_3,p_4$ can be expressed via the Lehmann--Symanzik--Zimmermann 
(LSZ) formalism as 
\bea
\label{ha1}
\langle\, p_3\,p_4|S|p_1\,p_2\,\rangle 
&=& \langle\, p_3\,p_4|p_1\,p_2\,\rangle + \nn \\
&&+
Z_\psi^{-2} \int d^4x_1\cdots d^4x_4
\exp \left[i(p_3x_3+p_4x_4-p_1x_1-p_2x_2)\right] \times
\nn \\
&& \times \langle\, {\rm T} \bar u(p_3) f(x_3)\bar u(p_4) f(x_4)
\bar f(x_1) u(p_1) \bar f(x_2) u(p_2)\,\rangle\,, 
\eea
where $f(x) = (i\gamma\del - m) \psi(x)$ and 
$Z_\psi$ is the wave function renormalization. The Green function 
in the last expression can then be expressed as a functional integral 
over the quark and gluon fields, $\psi$ and $A$ respectively, 
\be
\langle\,{\rm T} \psi(x_3)\psi(x_4)\bar \psi(x_1)\bar \psi(x_2)\,\rangle = 
\int \cD\psi\,
\cD \bar \psi \cD A \,\psi(x_3)\psi(x_4)\bar \psi(x_1)\bar \psi(x_2)
\exp[-i S_{\rm full QCD}]\,, 
\ee
where $S_{\rm full QCD}$ is the full QCD action.
The Gaussian fermion integration can be performed, giving 
\bea 
&&\hspace{-3mm}
\langle\,{\rm T} \psi(x_3)\psi(x_4)\bar \psi(x_1)\bar \psi(x_2)\,\rangle =
\int \cD A \det[-i(i\gamma \cdot D - m)]\times\\
&& \times \left[ S_F(x_3,x_1;B)\,S_F(x_4,x_2;B)+
S_F(x_3,x_2;B)\,S_F(x_4,x_1;B) \right]
\exp[-iS_{\rm pure QCD}] \,,
\nn 
\eea
which contains the functional determinant of the Dirac operator 
and the quark propagators $S_F(x_i,x_j;B)$ 
in the external color potential $A_\mu^F$. 
Here we are only left with the functional integration over the gluon 
fields with the pure QCD action, i.\,e.\ without quark contribution. 
For processes in which the momentum transfer is small compared 
to the total energy the second term (the $u$-channel term) in the 
sum in the integrand can be neglected. One thus obtains 
\be
\label{twoquarkinfield}
\langle\, p_3\,p_4|S|p_1\,p_2\,\rangle = 
\langle\, p_3\,p_4|p_1\,p_2\,\rangle + Z_\psi^{-2}\, \int \cD A\,
\cS(p_3,p_1;A) \cS(p_4,p_2;A) \exp[-i S_{\rm pure QCD}] \,,
\ee
where $\cS(p_i,p_j;A)$ is the scattering matrix element of a quark with 
momentum $p_j$ to one with momentum $p_i$ in an external color 
field $A$. The nonperturbative scattering amplitude for two quarks 
can thus be obtained from a functional integration of the product of the  
two scattering amplitudes of quarks in the gluon field over the latter. 
One can then show \cite{Nachtmann:1991ua} 
that these quark scattering matrix elements 
can be simplified in a generalized WKB approximation, 
\be
\cS(p_i,p_j;B) = \bar u(p_i) \gamma^\mu u(p_j) 
{\rm P}\exp\left[-ig\int_\Gamma \Af_\rho\,dx^\rho\right]
\left(1+O\left(\frac{1}{p^0_i}\right)\right)\,,
\ee
where $\Af$ is the Lie--algebra valued gauge potential. The
path--ordered integral is taken along the classical path $\Gamma$. 

We specifically consider two quarks moving at the speed of light 
in opposite directions with an impact vector $\vec{b}$ in the 
transverse $(x^1,x^2)$-plane. Denoting the paths of the quarks by 
$\Gamma_1, \Gamma_2$ we have 
\be
\Gamma_1=(x^0,\vec b/2, x^3=x^0) \quad \mbox{ and }\quad \Gamma_2=
(x^0, -\vec b/2, x^3= -x^0) 
\,.
\ee
The phases collected by the quarks along theses paths are 
\be
\label{eikonalphaseV}
\Vf_{i}(\pm\vec b/2)={\rm P}\exp\left[-ig \int_{\Gamma_{i}} \Af_\mu
(z)\ dz^\mu\right]~. 
\ee
Then the $S$-matrix element for two quarks with momenta $p_1$, $p_2$ 
and color indices $\alpha_1$, $\alpha_2$ leading to two quarks of 
momenta $p_3$, $p_4$ and colors  $\alpha_3$, $\alpha_4$ can 
be shown to be 
\be
S_{\alpha_3\alpha_4;\alpha_1\alpha_2}(s,t)
=\bar{u} (p_3) \gamma^\mu u(p_1) \bar{u} (p_4)
\gamma_\mu u(p_2)\,{\cal V}\,, 
\ee
with 
\be
{\cal V}=i Z^{-2}_\psi 
\left\langle\int d^2 b\ e^{-i\vec q \cdot \vec b}
\left[ {\bf V}_1\left(-{\vec b\over 2}\right)\right]_{\alpha_3\alpha_1}
\left[{\bf V}_2 \left(+{\vec b\over 2}\right)\right]_{\alpha_4\alpha_2}
\right\rangle\,. 
\ee
Here $\langle \,\cdot\,\rangle$ denotes functional integration over the
gluon field, and $\vec q$ is the momentum transfer $(p_1-p_3)$ projected
onto the transverse plane, and we assume $\vec q\,^2 \ll s$. 
The quark renormalization $Z_\psi$ is given by 
\be
\label{renormphase}
Z_\psi = \frac{1}{N_c} \,\tr  {\bf V}_1(0) 
= \frac{1}{N_c} \,\tr  {\bf V}_2(0) 
\,.
\ee
Finally, in the limit of high energies we have  helicity conservation, 
\begin{equation}
\bar u(p_3)\ \gamma^\mu u(p_1)\ \bar u(p_4)\ \gamma_\mu u(p_2)
\:\:\:\:
\mathop{\longrightarrow}_{s \to \infty}
\:\:\:\:
2s\delta_{\lambda_3\lambda_1}
\delta_{\lambda_4 \lambda_2} \,, 
\end{equation}
where $\lambda_i$ are the helicities of the quarks and $s=(p_1+p_2)^2$. 

The quark--quark scattering amplitude we have considered so far 
is of course explicitly gauge dependent. In order to apply the formalism 
to hadron--hadron scattering one has to describe the hadrons as 
color--neutral clusters of quarks and antiquarks. In high energy 
scattering these constituents of the hadrons move on parallel light--like 
lines. A meson can be described as a superposition of color dipoles 
(the quark--antiquark pairs) the size distribution of which is given 
by a transversal wave function. In high energy scattering the space--time 
behavior of these dipoles is then described by 
Wegner--Wilson loops ${\bf W}[C]$, 
\be
{\bf W}[C]= {\rm P} \exp 
\left [-ig \int_C \Af_\mu(z)\ dz^\mu\right]
\,.
\ee
The closed path $C$ consists of two 
lightlike sides formed by the quark and antiquark paths, and 
is closed by Schwinger strings extending to infinity which ensure gauge 
invariance. 

The loop--loop scattering amplitude depends not only on the 
impact factor but also on the transverse extension vectors of the loops. 
One obtains a reduced scattering amplitude $J$ (sometimes called 
the loop--loop profile function), related to the $S$-matrix element 
for the scattering of the two loops via $J=S-1$, 
\be
\label{looploopj}
J(\vec{b},\vec{R}_1,\vec{R_2}) = -\frac{1}{Z_1 Z_2} 
\langle W_1 W_2 \rangle
\ee
with 
\be
W_i = \frac{1}{N_c} \tr ( {\bf W}[C_i] - {\mathbf 1}) 
\,,
\ee
where the labels of the Wegner--Wilson loops refer to the two 
scattering dipoles. The $Z_i$ are renormalization factors for 
the loops defined in analogy to (\ref{renormphase})  
which replace the quark field renormalization constants $Z_\psi$. 
The scattering amplitude is finally obtained as 
\be
\label{amplitlooploop}
A(s,t) = 2 i s \int d^2b e^{-i \vec{q} \cdot \vec{b}}
\int d^2R_1 \,d^2R_2 \, J(\vec{b},\vec{R}_1,\vec{R_2}) 
\psi_i(\vec{R}_1) \psi^*_{i'}(\vec{R}_1) 
\psi_j(\vec{R}_2) \psi^*_{j'}(\vec{R}_2) 
\,,
\ee
where $i,j$ and $i',j'$ stand for the incoming and outgoing mesons. 
For elastic scattering we have $i=i'$ and $j=j'$, and the expression 
contains the transverse (anti)quark densities. If the incoming 
and outgoing mesons are not identical the integral (\ref{amplitlooploop}) 
measures the overlap of their wave functions. 

\begin{figure}[htb]
\begin{center}
\epsfysize 5cm
\epsfbox{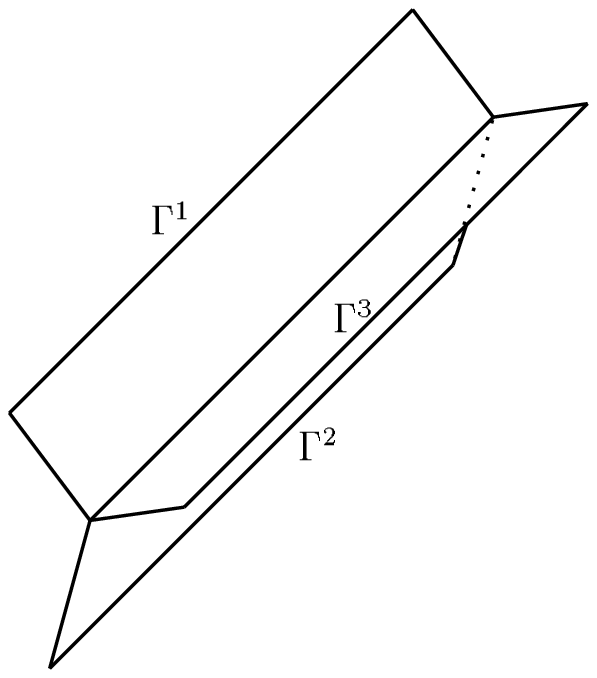}
\end{center}
\caption{The paths in a color neutral three--quark cluster
\label{baryon}}
\end{figure}

In order to describe a baryon one has to consider a cluster of three 
quarks moving on parallel light--like paths,
\be
\Gamma^a(x_0,\vec b/2+\vec x\,_1^a, x^3=x^0),
\:\:\: a=1,2,3 \,.
\ee
In order to ensure that these quark clusters asymptotically form 
color singlet states all colors are again 
parallel--transported in the remote past and future to a 
reference point of the cluster and there contracted antisymmetrically. 
This can be done for the baryon in meson--baryon scattering 
or baryon--baryon in scattering. Here we choose to give as an 
example the case of nucleon--nucleon scattering. 
One obtains \cite{Dosch:1994ym} for the $S$-matrix element 
for the scattering of color--neutral clusters, 
\bea
\label{smatrixbarybary}
\lefteqn{S\left(\vec{x}\,_1^1,\vec{x}\,_1^2,\vec{x}\,_1^3,
\vec{x}\,_2^1,\vec{x}\,_2^2,\vec{x}\,_2^3\right)= 
\frac{1}{36}\frac{1}{Z_1 Z_2} \times} 
\\
&&
\times \left\langle\epsilon_{\alpha\beta\gamma}
\left({\bf V}^1_1\right)_{\alpha\alpha'}\left({\bf V}^2_1\right)_{\beta\beta'}
\left({\bf V}^3_1\right)_{\gamma\gamma'}\epsilon_{\alpha'\beta'\gamma'}
\epsilon_{\rho\mu\nu}\left({\bf V}^1_2\right)_{\rho\rho'}
\left({\bf V}^2_2\right)_{\mu\mu'}
\left({\bf V}^3_2\right)_{\nu\nu'}\epsilon_{\rho'\mu'\nu'}\right\rangle \,. 
\nn
\eea 
The non-Abelian phase factors ${\bf V}_i^a$ are defined as 
in (\ref{eikonalphaseV}) with the U-shaped integration 
paths $\Gamma_i$ as indicated in figure \ref{baryon} for one cluster. 
The $Z_i$ denote again the wave function renormalization for the 
respective clusters. The reduced scattering amplitude $J$ is again 
obtained from the $S$-matrix element (\ref{smatrixbarybary}) 
via $J=S-1$. The latter can also be expressed in analogy 
to (\ref{looploopj}) as 
\be
\label{jbb} 
J(\vec{b},\vec{R}_1,\vec{R_2}) = - \frac{1}{Z_1 Z_2} 
\langle B_1 \cdot B_2 \rangle
\ee
with 
\be
B_i = \frac{1}{6}\, \epsilon_{abc} \epsilon_{a'b'c'} 
\left( W_{a'a}[C_{i1}]W_{b'b}[C_{i2}] W_{c'c}[C_{i3}] 
-\delta_{aa'} \delta_{bb'} \delta_{cc'} \right)
\,,
\ee
with $C_{ij}$ ($j=1,2,3$) being the paths bounding the three 
rectangular surfaces in figure \ref{baryon}. 
The amplitude for elastic nucleon--nucleon 
scattering is then 
\be
\label{Abaryfunct}
A(s,t)= 2i\, s \int d^2 b \,e^{-i \vec q \cdot \vec b} \int 
d^6{\mathbf{ R}}_1 
d^6{\mathbf{ R}}_2 
|\psi({\mathbf{ R}_1})|^2
|\psi({\mathbf{ R}_2})|^2 
J(\vec x\,_{1}^{1},\vec x\,_{1}^{2},\vec x\,_{1}^{3},\vec x\,_{2}^{1},
\vec x\,_{2}^{2},\vec x\,_{2}^{3})  \,,
\ee
where ${\mathbf{ R}}_i$ denotes the set of positions of the quarks relative 
to the center of nucleon~$i$, 
\be
\mathbf{ R}_i=\left(\vec R\,_i^1,\vec R\,_i^2,\vec R\,_i^3\right)\,, \:\:\:\:\:
\vec x\,_1^a = \frac{\vec b}2 + \vec R\,_1^a\,, \:\:\:\:\:
\vec x\,_2^a= -\frac{\vec b}2 +\vec R\,_2^a \,.
\ee
and $\vec b$ is the impact vector between the two nucleons. 
Note that so far the amplitudes 
(\ref{amplitlooploop}) and (\ref{Abaryfunct}) 
are still the full amplitudes and contain contributions from 
$C=+1$ as well as $C=-1$ exchanges. 

The functional integration over the gluon field in the above 
expressions is performed using the stochastic vacuum model (SVM) 
as described in section 
\ref{npheidelbergsect}. In order to do so one applies the 
nonabelian Stokes theorem to the Wegner--Wilson 
loops ${\bf W}[C]$ to get 
\be
W_{aa'}[S] = \left[ {\rm P_S} \exp 
\left (-\frac{1}{2} \,ig \int_S  {\bf F}_{\mu\nu}(z,w) 
d\sigma^{\mu\nu}(z)\right)
\right]_{aa'}
\ee
with the reference point $w$. The surface $S$ is bounded 
by the closed path $C$, $C=\del S$, and ${\rm P_S}$ 
indicates surface ordering. 
After this transformation the loops are then expanded 
in power series and inserted into the expressions 
(\ref{looploopj}) and (\ref{jbb}), respectively, 
and one contracts the color indices accordingly. 
One then makes use of the SVM assumption of Gaussian 
factorization for the field strengths and performs the 
surface integration using the SVM ansatz 
(\ref{svmcorrpara}) with (\ref{dpara}) and (\ref{d1para}) 
for the correlator of two field strengths. 
It can be shown that the leading contribution comes from 
terms in the expansion containing four field strengths, two 
from each hadron (dipole or baryon) in the scattering. This 
contribution has $C=+1$, is purely imaginary 
and contributes to Pomeron exchange \cite{Dosch:1994ym}. 
The Odderon contribution to the scattering amplitude 
is obtained from terms in the expansion containing three 
field strengths from each hadron. These terms contain 
$C=+1$ and $C=-1$ contributions, and the Odderon 
term is obtained by suitable projection in color space onto 
exchanges that are symmetric in the color labels. 
The corresponding expression becomes somewhat 
complicated due to different combinatorial factors as well 
as color factors. They are explicitly given for dipole--dipole, 
dipole--baryon, as well as for baryon--baryon scattering 
in \cite{Rueter:1996yb,Rueter:1998gj,Rueterthesis}. 
Taking into account in this way only the $C=-1$ contributions 
one obtains from (\ref{amplitlooploop}) and  (\ref{Abaryfunct}) 
the Odderon contribution $A^\soddi$ to the respective 
scattering amplitude. 

The functional approach can also be used to calculate the perturbative 
exchange of three gluons in a $C=-1$ state in position space, 
as it has been done for example in \cite{Dosch:2002ai}. 
To do so one expands the ${\bf V}^a_i$ in (\ref{smatrixbarybary}) 
up to order $g^3$ and in generators $\tau^a$ of $\mbox{SU}(3)$. 
One can then again project the result onto an exchange symmetric 
in the color labels which corresponds to the exchange of 
three gluons in a $C=-1$ state. 
The symmetry of the gluons in the $C=-1$ exchange makes 
the calculation much simpler as compared to the $C=+1$ 
exchange in this order in which the path ordering poses 
additional problems. (This also applies analogously to the 
general nonperturbative case discussed above.) 
One obtains 
\bea
\label{jpertk}
\lefteqn{J(\vec x\,_{1}^{1},\vec x\,_{1}^{2},\vec x\,_{1}^{3};
\vec x\,_{2}^{1},\vec x\,_{2}^{2},\vec x\,_{2}^{3})=}\nonumber \\
&& = g^6 
\sum_{a_i,b_i=1}^3 K(a_1,a_2,a_3;b_1,b_2,b_3)
\chi(\vec x\,_1^{a_1},\vec x\,_2^{b_1}) \chi(\vec x\,_1^{a_2},\vec x\,_2^{b_2})
\chi(\vec x\,_1^{a_3},\vec x\,_2^{b_3}) \,.
\eea
The factor $K$ contains combinatorial and color factors. 
For nucleon--nucleon scattering it can again be found 
in \cite{Rueter:1996yb,Rueterthesis}. 
Here $\chi$ is the gluon propagator in transverse space, 
\bea
\chi(\vec x,\vec y)&=&  \int \frac {d^2k}{(2 \pi)^2} \frac{1}{\vec
k\,^2+m^2} e^{-i \vec k \cdot (\vec x-\vec y)}\\
&=& \frac{1}{2 \pi} K_0\left( m \left|\vec x-\vec y \right|\right)
\,,
\eea
where $K_0$ is the modified Bessel function. 
The single diagrams occurring in the sum in (\ref{jpertk}) 
are infrared divergent. In order to 
regularize them one introduces a gluon mass $m$ which 
is possible in LO approximation. In all final gauge 
invariant expressions the divergences have to cancel 
and the gluon mass can safely be set to zero. 

\boldmath
\subsection{$pp$ and $p\bar{p}$ Scattering }
\label{ppsect}
\unboldmath

Scattering experiments with colliding beams of protons 
and antiprotons have a long tradition in particle physics. 
Today one usually speaks of high energies in 
$pp$ or $p\pbar$ scattering when referring to energies 
above $\sqrt{s} \simeq 20\,\mbox{GeV}$, which was 
the lowest energy of the CERN ISR, where $pp$ and 
$p\pbar$ scattering was studied. Higher energies were 
subsequently reached in $p\pbar$ scattering at the CERN SPS 
($\sqrt{s} \simeq 546\,\mbox{GeV}$), and at the 
Tevatron ($\sqrt{s} \simeq 1.8\,\mbox{TeV}$). 
At the Large Hadron Collider (LHC) 
we will see $pp$ collisions at an energy 
of $\sqrt{s}=14 \,\mbox{TeV}$. Data on $pp$ scattering, 
in particular with polarized beams, will soon also 
be taken at the Relativistic Heavy Ion Collider (RHIC) 
at energies up to $500\,\mbox{GeV}$. Possibly RHIC will 
offer the option of being operated in a $p\pbar$ mode as well. 
As we will see it would be most welcome for Odderon physics 
to have data on both processes at the same energy. 

In the present section we will deal with different observables 
and processes involving the Odderon in $pp$ and $p\pbar$ 
collisions. First we consider the Odderon--proton coupling 
which enters almost all of the observables which we will 
study. We then turn to elastic scattering and 
the $\rho$-parameter where we can compare our 
expectations with data. Finally, we turn to diffractive 
processes most of which have not yet been studied experimentally 
so far but will hopefully be observed in the not too distant future. 
Before we start we should point out another interesting observable 
involving the Odderon that has 
recently been suggested for the case of polarized $pp$ 
scattering but will not be covered in detail in the present review, 
namely the single--spin asymmetry of 
small--angle pion production in polarized $pp$ collisions 
\cite{Ahmedov:1999ne,Ahmedov:2002yh}. 

\subsubsection{Odderon-Proton Coupling and Proton Structure}
\label{Oddprotoncouplsect}

A good description of the coupling of the Odderon to the external 
particles in the scattering process is of central importance 
in the phenomenology of the Odderon. 
In soft processes this is obviously an extremely difficult task, 
but even in situations where perturbation theory is applied 
these couplings are sometimes difficult to describe. This is 
especially true when the Odderon couples 
to complicated hadronic bound states like a proton. 
The optimal situation is one in which the whole process can 
be calculated in perturbation theory, and there are in fact 
some processes of this kind, see section \ref{gammagammasect}. 
It turns out that the structure of the external particles 
can have very large effects 
especially in processes involving Odderon exchange. 
A diquark clustering in the proton for example can drastically 
suppress the Odderon coupling in processes in which the proton 
is scattered elastically, as we will see in detail below. 
If the external particles are hadrons the problem can usually 
only be approached by using some kind of model for the 
internal structure of the hadron. 

In Regge theory the couplings of the Odderon to the external 
particles are universal and need to be fixed from 
experiment. Once they are fixed they can be used for other processes. 
Unfortunately, the lack of experimental data on the Odderon 
is a serious obstacle for making real use of this approach. 
An additional unknown is the Dirac structure of the coupling. 
Often the Regge picture is used together with a parton picture 
of the colliding hadrons. In this case one needs to specify 
how the Odderon couples to the individual partons in a hadron. 
As we will discuss in more detail at the end of the present section 
the Odderon cannot couple to a single gluon because the gluon is 
an eigenstate of $C$ parity. One therefore has to specify only 
the coupling to the quarks. Usually this coupling is assumed to be 
vectorlike, i.\,e.\ to have the Dirac structure $\gamma^\mu$. 
The coupling to a quark thus reads 
\be
\label{oqcouplgamma}
-i\beta_\soddi \gamma^\mu
\,, 
\ee
with the Odderon--quark coupling constant $\beta_\soddi$. 
In addition one has to take into account the distribution of the 
quarks inside the proton which is conveniently done via 
the electromagnetic isoscalar form factor of the proton, 
$F_1(t)$. The contribution of the Pauli form factor $F_2$ 
can be neglected in the region of small $t$ that is relevant for 
most phenomenological applications. 
The Regge coupling of the Odderon to the proton 
then becomes 
\be
\label{reggeoddpcoupl}
  - 3 i \beta_\soddi F_1(t) \gamma^\mu 
\,. 
\ee
That expression should be used for the couplings $\beta_{ac}$ 
in eq.\ (\ref{factcouplregge}) where we did not yet 
specify the Dirac structure of the coupling. 
Note that the Pomeron--proton coupling can 
be obtained analogously and is given by exactly the same expression 
(\ref{reggeoddpcoupl}) 
with the obvious replacement $\beta_\soddi \to \beta_\spommi$. 

The situation is only slightly better in soft processes in which the 
Odderon is modeled by nonperturbative three--gluon exchange. 
Here the coupling depends very much on the form of the gluon 
propagators and on the model used for the proton. 
But if the Pomeron is modeled in a similar way by the exchange 
of two nonperturbative gluons, one can try to eliminate part 
of the model dependence by comparing the Odderon--proton 
coupling to the Pomeron--proton coupling. For some 
phenomenological applications it is in fact exactly the ratio 
of these two couplings that is relevant, like for example in 
the asymmetries arising from Pomeron--Odderon interference, 
see section \ref{epasymmetrsect}. 
One should certainly not expect any precise value in estimates 
of this ratio. But the emerging picture from studies in this direction 
is in general that the Odderon couples much more weakly to the proton 
than the Pomeron at low $t$, see for example \cite{Donnachie:wd}. 
That is also in agreement with the general observation that 
 at high energy the difference 
$\Delta \sigma = \sigma_T^{\bar{p}p} - \sigma_T^{pp} \sim \imag A_-$ 
is much smaller than the total cross sections $\sigma_T \sim \imag A_+$ 
themselves, where $A_-$ and $A_+$ are dominated by Odderon and 
Pomeron exchange, respectively. 
But it should be noted that the situation is very different at large $t$. 
In $pp$ elastic scattering for example the Odderon is 
actually the dominant exchange at large $t$ \cite{Donnachie:1979yu}, 
see section \ref{elasticppsect}. 
We will discuss the possible reasons for the apparently weak 
coupling of the Odderon to the proton in more detail 
at the end of this section. 

Let us now turn to processes that involve a hard scale and can 
hence be described in a perturbative approach. 
Here one typically makes a model for the proton impact 
factors in transverse momentum space. A perturbative 
calculation can of course also be done in configuration space, and 
then one needs to model the transverse wave function of the proton. 
An example of such a model will be discussed in the context 
of the functional approach to high energy scattering later in 
this section. 
In a perturbative framework also the appropriate choice of the 
strong coupling constant is extremely important as we have 
already discussed in section \ref{genconsidsect}. 

In the perturbative picture the Odderon is described 
by three gluons either with or without their pairwise 
interactions. Their coupling to the proton or antiproton 
is then described in high energy factorization 
by impact factors $\phi_p(\kf_1,\kf_2,\kf_3)$ 
in transverse momentum space, see eq.\ (\ref{Aoddifact}). 
Here we will consider only the elastic 
Odderon--proton impact factor which can be applied 
to reactions in which the proton stays intact. 
We will later see how this elastic impact factor 
is used in $pp$ and $p\bar{p}$ elastic scattering. 
Another important application is in diffractive 
$ep$ scattering when the proton does not break up. 
These are the main cases involving protons that 
have been considered in the perturbative framework. 
Of course one can also calculate impact factors in which the 
incoming particles break up into a more complicated 
system. As an example of this we will see the 
$\gamma^* \Od X$ impact factor 
in section \ref{gammagammasect}. 
Since the proton is a complicated bound state 
the proton impact factors cannot be calculated 
in perturbation theory. Hence some model assumptions 
need to be made even if the Odderon exchange is described 
in a perturbative framework. 
Fortunately general principles allow one to place 
considerable constraints on the form of the proton impact factors. 

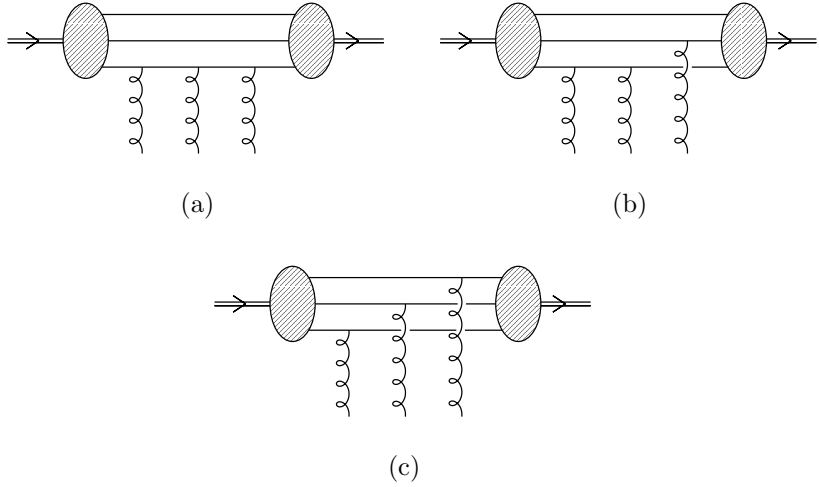
\begin{figure}[ht]
\vspace*{0.4cm}
\begin{center}
\input{impactdiag.pstex_t}
\end{center}
\caption{Diagrams contributing to the Odderon--proton impact factor
\label{fig:impactdiag}}
\end{figure}

At moderate momentum transfers $\sqrt{-t}$ at which these 
impact factors are actually used the proton can be described 
as a system of three constituent quarks to which the gluons 
are coupled. In order to preserve gauge invariance one has to 
take into account all possible combinations in which the three 
gluons can be coupled to the three quarks. There are three 
possible types of configurations in which the three gluons 
can be coupled to the proton. These are depicted in figure 
\ref{fig:impactdiag}. In (a) all three gluons couple 
to the same quark, in (b) only two couple to the 
same quark, and in (c) all three gluons couple to different 
quarks. 
All of them must be taken into account with their respective 
combinatorial factors. If this is not done correctly 
the natural cancellation 
of infrared divergences between the different diagrams ceases 
to work. Consequently, the integral over the 
transverse gluon momenta in (\ref{Aoddifact}) will 
in general become divergent. 

A gluon with vanishing transverse momentum cannot 
resolve the structure of the proton and sees only the 
total color charge. The color neutrality of the proton 
hence requires that the impact factor vanishes if one of the 
three transverse gluon momenta $\kf_i$ vanishes, 
\be
\left. 
\phi_p(\kf_1,\kf_2,\kf_3)
\right|_{ \kf_i=0 }= 0 \,,
\:\:\:\:\:\:\:\:\:\:
\:\:\; i \in \{1,2,3\}\,. 
\ee
If one further takes into account the requirement that the 
impact factor should be symmetric in the three gluon momenta 
one can show that the above conditions force the 
Odderon--proton form factor to be of the general form 
\be
\label{eq:impact}
\phi_p(\kf_1,\kf_2,\kf_3)  
=  8\,(2\pi)^2\, g^3 \,\left[F(\qf,0,0) - 
                \sum_{i=1}^3 F(\kf_i, \qf-\kf_i, 0)
                +2F(\kf_1,\kf_2,\kf_3)\right]
\,,
\ee
where $\qf= \sum_{i=1}^3 \kf_i$, and 
$F(\kf_1,\kf_2,\kf_3)$ is a form factor symmetric 
in its three arguments. The three terms in eq.\ (\ref{eq:impact}) 
can be easily identified with the three different types of 
diagrams in figure \ref{fig:impactdiag}. 
The first term corresponds to the diagram of type (a) in 
which all three gluons are coupled to the same quark, the 
second term corresponds to the diagram of type (b), and 
the last term to diagram (c). 

Different models for the form factor 
$F(\kf_1,\kf_2,\kf_3)$ have been proposed. Here we will 
explicitly give two of them which are the most popular ones. 
Besides the precise functional form of the form factor the overall 
normalization of the impact factor $\phi$ is a very 
important issue. It crucially depends on the choice of the 
strong coupling parameter $g$ in (\ref{eq:impact}), or equivalently 
on the choice of the strong coupling constant 
$\alpha_s = g^2/(4\pi)$. Therefore a model for the form factor 
necessarily needs to be supplemented with an appropriate 
choice of $\alpha_s$ which can also depend on the 
momentum transfer $\sqrt{-t}$ with $t=-\qf^2$.  
Here we only cite the values for $\alpha_s$ 
proposed together with the models for the form factor. 
For a critical discussion of these values we refer to 
sections \ref{elasticppsect} and \ref{epdiffsect} below. 

One model for the form factor $F$ was suggested by 
Fukugita and Kwieci\'nski in \cite{Fukugita:1979fe}, 
\be
\label{eq:formfact}
F(\kf_1,\kf_2,\kf_3) = \frac{A^2}{A^2 + \frac12
        [(\kf_1-\kf_2)^2 + (\kf_2-\kf_3)^2 + 
         (\kf_3-\kf_1)^2]}
\,.
\ee
The parameter $A$ is chosen to be half the $\rho$ meson mass, 
$A=384\,\mbox{MeV}$. 
In the original reference a rather large value of $\alpha_s = 1$ 
was proposed for the strong coupling constant. This value was 
motivated by the use of a similar value in an estimate of 
hadronic cross sections in the two--gluon model of 
\cite{Gunion:iy}. 

Another model for the form factor $F$ was proposed by Levin 
and Ryskin \cite{Levin:gg}. Their ansatz is motivated by 
a nonrelativistic quark model with oscillatory potential. 
Its explicit form is 
\be
\label{formfaclr}
F(\kf_1,\kf_2,\kf_3) = 
\exp \left( -R_p^2 \sum_{i=1}^3 \kf^2_i\right)
\,.
\ee
The parameter $R_p$ is supposed to be of the order of magnitude 
of the proton radius. In \cite{Levin:gg} a value of 
$R_p^2 = 2.75\,\mbox{GeV}^{-2}$ is proposed, and the 
authors suggest to choose $\alpha_s=1/3$. 

\begin{figure}[ht]
\vspace*{0.4cm}
\begin{center}
\epsfysize 3.5cm
\epsfbox{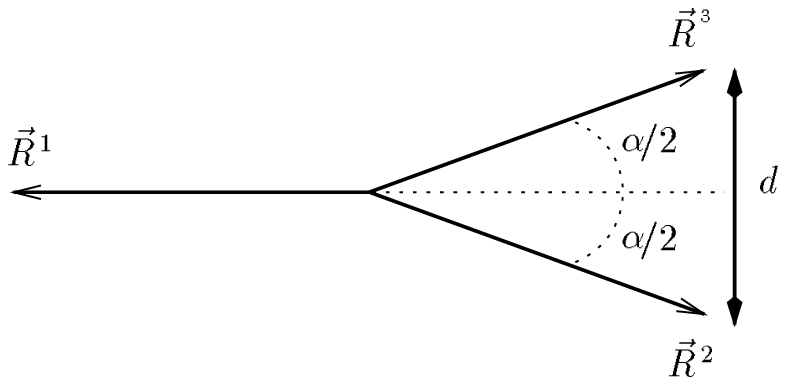}
\end{center}
\caption{Definition of the angle $\alpha$ characterizing the 
proton configuration
\label{star}}
\end{figure}

If a configuration space picture of high energy scattering like 
the one developed by Nachtmann (see section \ref{formalismssect}) 
is used one needs to describe the transverse structure of the 
proton by a suitable model for the transverse wave function. 
It should be emphasized that such a transverse wave function 
can also be used for a perturbative calculation in 
configuration space. 
An especially interesting aspect of the proton structure is 
the possibility that two of the three quarks in the proton form 
a relatively small cluster. As we will see such a quark--diquark 
structure of the proton strongly affects the coupling of the 
Odderon to the proton. This is especially true in elastic scattering 
processes. 
Let us here briefly describe a simple model that was used 
for example in \cite{Rueterthesis,Dosch:2002ai} in the context 
of the Odderon. 
In this model the quark density in the proton is in 
transverse position space given by the ansatz 
\begin{equation}
\label{wavefunction}
\left| \psi(\vec R_1,\vec R_2,\vec R_3)\right|^2=
\frac2\pi \frac1{S_p^2} \,\exp\left(-\frac{2R_1^2}{S_p^2}\right)\,
\delta^2(\vec R_2 - {\bf M}_\beta\,\vec R_1)\,
\delta^2(\vec R_3 - {\bf M}_{-\beta}\,\vec R_1)\,,
\end{equation}
where 
${\bf M}_\beta= \pmatrix{\cos\beta&-\sin\beta\cr\sin\beta&\cos\beta\cr}$
and $\beta=\pi-\alpha/2$. 
The quantity $S_p$ determines the electromagnetic radius of the
nucleon. The value $S_p= 0.8\,\mbox{fm}$ has proven to be an 
appropriate choice \cite{Dosch:2000jg,Rueterthesis}. 
The meaning of the angle $\alpha$ is illustrated in figure \ref{star}. 
The value $\alpha=2 \pi/3$ corresponds to a Mercedes
star configuration of the quarks in the nucleon, and 
$\alpha=0$ corresponds to a quark--diquark picture of 
the nucleon with an exactly pointlike diquark. 
For small angles $\alpha$ we can still speak of a diquark--cluster 
in the nucleon, and the distance $d$ between the two 
quarks in such a cluster can be called the diquark size, 
see figure \ref{star}. With the wave function (\ref{wavefunction}) 
one then obtains for the average diquark size $\langle d \rangle$ 
\be
 \langle d \rangle = \sqrt{\frac{\pi}{2}} \, 
\sin \frac{\alpha}{2} \, S_p \,.
\ee

Let us discuss what happens in the limit of a 
vanishing diquark size. Particularly interesting is 
the elastic case in which the proton stays intact. 
Remarkably it turns out that in this situation of 
a pointlike diquark cluster the 
Odderon does not couple to the proton at all for 
$t=0$, as was first observed in \cite{Zakharov:1989bh}. 
In that reference it was further shown that the Odderon--proton 
coupling vanishes also at $t\neq 0$ if the quark and 
diquark are assumed to have the same mass. 
The reason for this can be seen in a simple 
quantum mechanical argument. The coupling 
of the Odderon to the proton leads to a multiplication 
of the proton state with a phase factor which 
rotates is into a state almost orthogonal to the proton state, 
and the subsequent projection onto the outgoing state 
gives a very small value. If the Odderon is described by 
a three--gluon model this state is even exactly orthogonal 
to the original one. 
Specifically, it is the angular integration over the proton 
orientation in eq.\ (\ref{Abaryfunct}) that (almost) vanishes 
for the Odderon contribution. 
Another way of looking at this is the following. 
If one considers only the color degrees of freedom 
the point--like diquark acts like an antiquark. 
The proton thus looks like a quark--antiquark pair, 
or in other words like a meson. If it were 
only for the color degrees of freedom and if the 
quark and diquark had the same mass that meson 
would be an eigenstate of $C$ parity and as such 
would not couple elastically to the Odderon that carries 
negative $C$ parity. In reality of course the proton does not 
become an eigenstate of $C$ parity even in the limit 
of a point--like diquark since it is obviously not identical 
to its antiparticle the antiproton. Further, in reality also 
the masses of the quark and the diquark are not equal. 
But the Odderon only 
couples to the color degrees of freedom, and as far as those 
are concerned the proton in this limit acts like an eigenstate 
of $C$ parity, making the coupling to the Odderon impossible. 
In actual fact the system is only almost an eigenstate of $C$ parity 
with corrections originating from 
the difference in the quark and diquark masses. 
This argument again explains why the Odderon couples 
only very weakly to the proton in the limit of a point--like diquark 
if the proton stays intact in the scattering process. 

If the diquark cluster is not pointlike the elastic coupling 
of the Odderon to the proton does not vanish, but for 
small diquark sizes of less than about $0.3\,\mbox{fm}$ 
there is still a very strong suppression of the coupling 
\cite{Rueter:1996yb}. 
It should be emphasized that the suppression applies 
primarily to the elastic case. If the proton breaks up in the 
process there is a priori no reason to expect a similar suppression. 
If the proton structure in fact contains a diquark cluster 
of relatively small size one consequently expects that Odderon 
exchange leads to larger cross sections in processes 
in which the proton breaks up as compared to processes 
in which the proton scatters elastically. 

Another argument for the relative weakness of the $\Od p$ 
coupling as compared to the $\P p$ coupling was given in 
\cite{Ginzburg:qi}. It is based on the observation that due to 
the different quantum numbers under $C$ parity the 
Pomeron and the Odderon couple very differently to gluons. 
The Pomeron has positive $C$ parity and 
can couple to single quarks (or antiquarks) 
and to single gluons  in the proton. 
The Odderon can, like the Pomeron, couple to a single quark 
because the quark is not an eigenstate of $C$ parity. 
But the gluon is an eigenstate of $C$ parity and the Odderon 
can thus not couple to a single gluon. 
This leads in a sense to a decoupling of the Odderon from the 
gluon content of the proton which can induce a large effect 
since the parton densities of the proton are at high energy 
dominated by the gluon density. 
The coupling of the Odderon to different gluons (and quarks) 
is possible though, 
but requires a sufficiently large gluon density for correlations 
of several gluons to become relevant. At large gluon densities 
the decoupling of the Odderon from the gluons in the proton 
is hence not perfect, but there will still be 
a suppression relative to the Pomeron as long as multi--gluon 
correlations do not dominate the proton structure and the 
coupling to it. 
This argument for the weakness of the Odderon coupling 
is most convincing in a Regge picture of the Odderon in which 
it can only couple to single partons. It also gives a plausible picture 
for an Odderon consisting of three gluons. However, 
it should be noted that the coupling of a two-- or three--gluon 
system to the gluons in a proton is only poorly understood. 
In principle the three gluons in the 
Odderon couple to all gluons and quarks in the proton in all possible 
ways, and it is very difficult to estimate the contributions 
of different types of diagrams at given gluon and quark densities, 
and this is not even possible as a gauge invariant statement. 
It is therefore difficult to estimate how small the 
gluon density has to be in order to induce the decoupling of the 
Odderon described above. The physics of large 
gluon densities is often discussed in the context of saturation 
and recombination effects. A transition to a dense gluon system 
(often called a color glass condensate) is in fact expected to occur, 
but the exact energy needed for this transition is difficult to quantify. 
If the above argument really applies, however, it should also 
suppress the coupling of the Odderon in processes in which 
the proton breaks up --- in contrast to the diquark mechanism 
for Odderon suppression. 

\subsubsection{Elastic Scattering}
\label{elasticppsect}

Elastic $pp$ and $p\bar{p}$ scattering was for a long time the main 
study ground for Odderon physics after a marked difference 
in the differential cross sections for the two processes 
had been observed at the CERN ISR. 
At high energies the $C=-1$ reggeon exchanges can 
be neglected and the existence of such a difference 
is a typical sign of an Odderon. It is also in agreement 
with the Cornille--Martin theorem (\ref{cmtheoremreal}). 

\begin{figure}[ht]
\begin{center}
\vspace*{.4cm}
\input{allt.pstex_t}
\end{center}
\caption{Differential cross section for elastic $pp$ and $p\bar{p}$
for $\sqrt{s}=53\,\mbox{GeV}$ together with the Donnachie--Landshoff 
fit \protect\cite{Donnachie:1984hf}; data from 
\protect\cite{Amos:1985wx,Nagy:1979iw,Breakstone:1985pe}
\label{figallt}}
\end{figure}
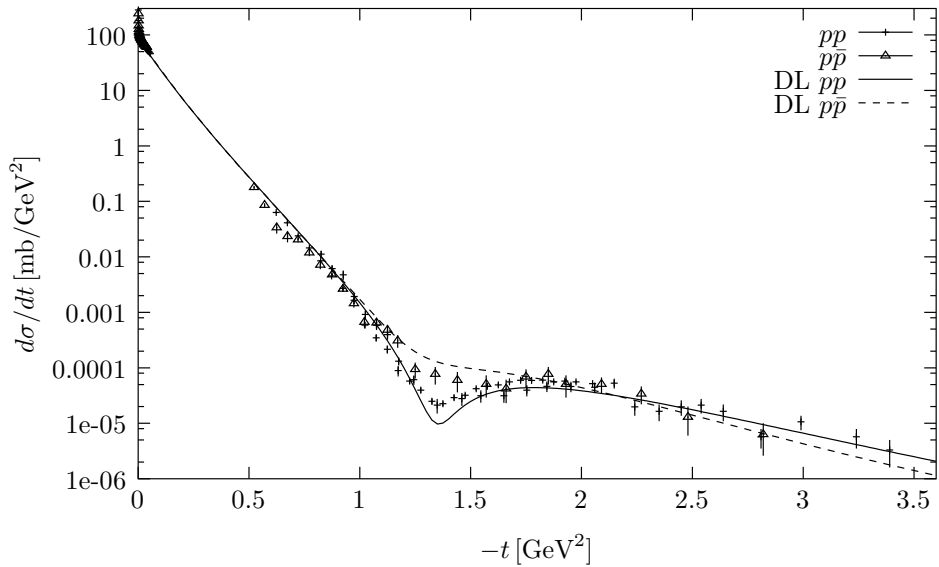

The $t$-dependence of elastic $pp$ scattering was measured 
at the ISR for five different center--of--mass energies 
$\sqrt{s}$ between $23.5$ and $62.5 \,\mbox{GeV}$ 
\cite{Amos:1985wx}--\cite{Amaldi:1980kd}. 
At all of these energies the differential cross section exhibits 
a characteristic dip at around $|t|\simeq 1.3\,\mbox{GeV}^2$. 
This $t$ region is hence often called the dip region or structure 
region of $pp$ elastic scattering. 
A successful description of the $pp$ differential cross section in 
terms of a Regge theory fit with only few parameters 
was given by Donnachie and Landshoff (DL) 
in \cite{Donnachie:1984hf}. Their fit included an Odderon modeled 
by the exchange of three noninteracting gluons with infrared--modified 
propagators. Remarkably, they predicted on the basis of their fit 
that the $p\bar{p}$ differential cross sections would not have the 
dip structure seen in $pp$ scattering but would instead only flatten 
off in the same region of $t$. 
The origin of this characteristic difference 
was in the DL fit almost exclusively attributed to the Odderon. 
Exactly the predicted behavior was soon afterwards 
observed \cite{Breakstone:1985pe,Erhan:1985mv} in elastic 
$p\bar{p}$ scattering at $\sqrt{s}=53\,\mbox{GeV}$. 
Figure \ref{figallt} shows both the $pp$ and $p\bar{p}$ data 
at that energy (from 
\cite{Amos:1985wx,Nagy:1979iw,Breakstone:1985pe}) 
together with the Donnachie--Landshoff fit to the $pp$ data 
and their prediction for $p\bar{p}$ scattering \cite{Donnachie:1984hf}. 

\begin{figure}[ht]
\begin{center}
\input{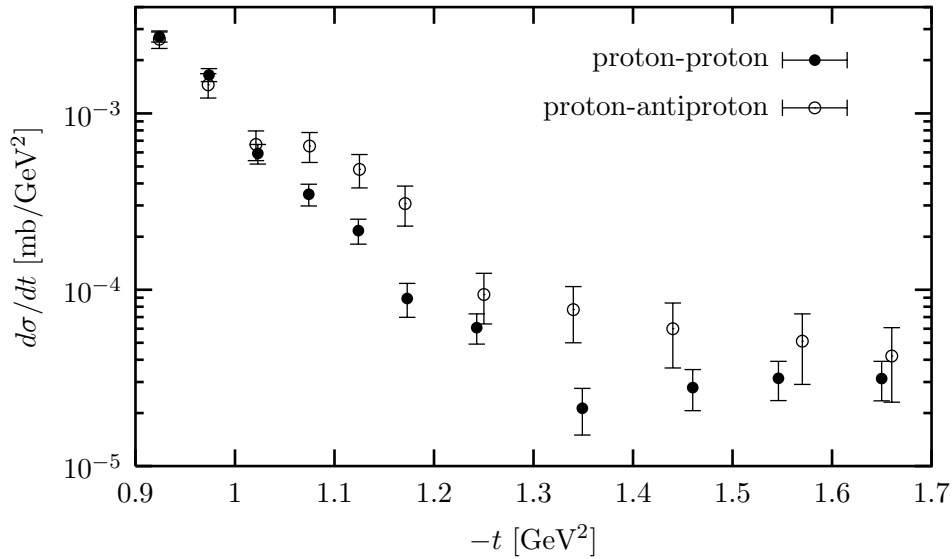}
\end{center}
\caption{Differential cross section for elastic $pp$ and $p\bar{p}$
scattering in the dip region for $\sqrt{s}=53\,\mbox{GeV}$;
data from \protect\cite{Breakstone:1985pe}
\label{fig:dipdiff}}
\end{figure}

To date the difference between $pp$ and $p\bar{p}$ elastic scattering 
in the dip region at $\sqrt{s}=53\,\mbox{GeV}$ remains 
the only real experimental evidence for the existence 
of the Odderon. We therefore show the data 
\cite{Breakstone:1985pe} once more in the relevant 
dip region in figure \ref{fig:dipdiff}. 
The figure shows that the difference between the 
$pp$ and $p\bar{p}$ data is actually not very large and depends 
on only a few data points. Moreover, the interpretation of that 
difference as evidence for the Odderon depends on the 
theoretical description of the data. The interpretation of the 
data has in fact been controversially discussed, and some 
scepticism remains. 

In order to get a better picture of Odderon effects in $pp$ 
and $p\bar{p}$ elastic scattering it would be tremendously 
helpful to have data for both reactions at the same energies, 
and if possible distributed over a wide energy range. 
But reality is different. 
Unfortunately, the $t$-dependence of $p\bar{p}$ scattering 
has been measured at only one energy in the ISR range 
over a sufficiently wide 
$t$ range to be useful for the study of Odderon effects. 
Even the data at $\sqrt{s}=53\,\mbox{GeV}$ have been taken 
only during the very last week of running of the ISR, and due to the 
comparatively low statistics they are not as precise as the 
corresponding $pp$ data\footnote{It should also be noted that 
there appears to be some discrepancy in the $pp$ data at that energy 
in the region around $|t|\simeq 1.7\,\mbox{GeV}^2$ between 
the two data sets of \protect\cite{Nagy:1979iw}
and \protect\cite{Breakstone:1985pe}. As is discussed in 
\cite{Amaldi:1980kd} the raw data sets for this observable are 
notoriously difficult to normalize.}. 
At higher energies, only $p\bar{p}$ elastic scattering data 
are available. Their differential cross section at 
$\sqrt{s}=546\,\mbox{GeV}$
has been measured at the CERN SPS 
collider \cite{Bozzo:1985th,Bernard:1986ye}, showing qualitatively 
the same behavior as the $p\bar{p}$ data at the ISR. 
The lack of data taken at the same energy for both reactions 
is one of the main limitations for Odderon physics in 
hadron--hadron scattering. 
The other limitation lies in the theoretical description of this 
process. The Odderon is just one of a number of different 
exchanges in the scattering process. All of them involve 
unknown parameters which need to be fixed phenomenologically. 
The different fit parameters are of course correlated and 
the Odderon contribution cannot be fixed uniquely. Again, 
this problem is to a large extent caused by the lack of 
sufficiently precise data over wide ranges in $s$ and $t$. 

The differential cross section for $pp$ and $p\bar{p}$ elastic 
scattering has in the meantime been studied in great detail 
by many authors in particular in the light of the Odderon 
hypothesis, see for example 
\cite{Gauron:1985gj,Gauron:1986nk,Gauron:1990cs,%
Dosch:2002ai,Levin:gg,Zakharov:1989bh}, 
\cite{Donnachie:iz}--\cite{Faissler:1980fk}. 
For the sake of the present review we will concentrate on two 
approaches only which can be viewed as the two extremes 
as far as the type of the Odderon singularity is concerned. 
The first one is the Regge fit approach by Donnachie and 
Landshoff initiated in \cite{Donnachie:1984hf}, and the second 
one is the approach by Gauron, Nicolescu and Leader 
related to the maximal Odderon \cite{Gauron:1990cs}. 
The Donnachie--Landshoff fit is the classic description of 
the elastic scattering data. It uses the framework of Regge 
theory and in particular a (Regge pole--like) 
three--gluon model for the Odderon. The maximal Odderon 
by construction saturates the asymptotic bounds and in this sense 
corresponds to a quite different type of Odderon. 
With these two approaches we hope 
to illustrate the potential but also the limitations of the 
differential $pp$ and $p\bar{p}$ elastic scattering data 
for a precise determination of the properties of the Odderon. 
We will in particular try to point out what the elastic scattering 
data teach us for the phenomenology of more exclusive processes 
mediated by Odderon exchange. 

We start with the Donnachie--Landshoff fit. Its original 
version \cite{Donnachie:1984hf} was obtained by fitting 
the then available $pp$ data from the ISR. As was already pointed out in 
\cite{Gauron:1985gj} this original version failed in predicting 
the correct magnitude of the $p\bar{p}$ data at 
$\sqrt{s}=546\,\mbox{GeV}$ 
measured at the SPS collider \cite{Bozzo:1985th}. 
In \cite{Donnachie:iz} a number of parameters of the DL fit 
were correspondingly adjusted, and the improved version 
of the fit gives a good description also of the $p\bar{p}$ data 
at the higher energy. In the ISR range the new fit deviates 
only slightly from the original one and gives an equally good 
description of the data. 
The fit is based on a number of exchanges in the $t$-channel: 
Pomeron, reggeon, Odderon, double Pomeron, triple Pomeron, 
Pomeron plus two gluons, and reggeon plus Pomeron. 
These contributions are added on the amplitude level, and 
the differential cross section is obtained via eq.\ (\ref{diffsigmaA}). 
The Odderon contribution turns out to be especially important 
at large $t$ and in the dip region. 

Before coming to the phenomenological aspects of the DL fit let 
us for a moment look at the Odderon term in the fit. 
Throughout reference \cite{Donnachie:1984hf} the Odderon 
exchange is called `three--gluon exchange' but is in fact in 
every sense an Odderon exchange with negative $C$ parity and 
with the three gluons in a symmetric color state. 
Although modeled as three--gluon exchange the Odderon 
in the DL fit is not a completely perturbative one. Instead, the 
authors use gluon propagators which are modified in the 
infrared region. The gluon propagators are cut off at 
$|t|\simeq 0.3\,\mbox{GeV}^2$ and below that value smoothly 
extrapolated with a parabola to zero at $t=0$. This procedure 
is not essential for the fit, but was mainly required due to 
the special choice made for the Odderon--proton impact factor 
in \cite{Donnachie:1984hf}. Motivated by the observation that 
the diagram (c) in figure \ref{fig:impactdiag} is the dominant 
one at large $t$ (see below), the authors have chosen to take into account 
only this diagram also at lower $t$. At $t\to 0$ the corresponding 
violation of gauge invariance leads to divergences which have 
then to be canceled by modifications of the gluon propagator. 
For a more detailed discussion of the use of infrared--modified 
gluon propagators see section \ref{nppropagatorsect}. 

\begin{figure}[ht]
\begin{center}
\epsfxsize=12cm
\epsfbox[88 460 475 760]{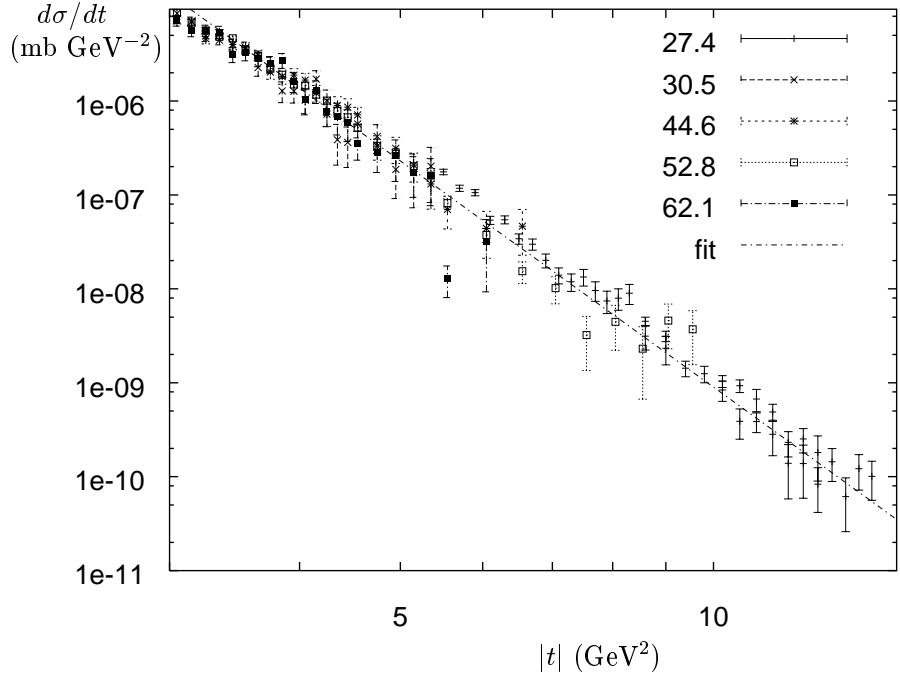}
\end{center}
\caption{Differential cross section for $pp$ elastic scattering at 
the largest available $t$ at different energies indicated in the 
figure as $\sqrt{s}$ in $\mbox{GeV}$, together 
with the fit (\protect\ref{largetppfit}); data from 
\protect\cite{Nagy:1979iw,Faissler:1980fk}, figure from 
\protect\cite{Donnachie:1996rq}
\label{largetfig}}
\end{figure}

Let us now first consider the large-$t$ region of $|t|> 3$ or 
$4 \,\mbox{GeV}^2$. Remarkably, in this region the 
data appear to be energy--independent over the whole ISR 
range and are extremely well described by the fit 
\be
\label{largetppfit}
\frac{d\sigma}{dt} = 0.09\, t^{-8}
\,,
\ee
see figure \ref{largetfig}. 
Exactly such a behavior, including the energy--independence, 
is expected from the perturbative exchange of three noninteracting 
gluons which gives \cite{Donnachie:1979yu} 
\be
\label{largetpppred}
\frac{d\sigma}{dt} \sim \alpha_s^6 t^{-8}
\,.
\ee
One power of $t^{-2}$ arises from external kinematic factors, 
and the remaining $t^{-6}$ from the three--gluon exchange. 
According to \cite{Donnachie:1979yu} the large $t$ region 
of elastic $pp$ and $p\pbar$ is dominated by Odderon exchange. 
There it is argued that this dominance of the Odderon comes 
about because the exchange of three gluons permits to distribute 
the momentum transfer evenly between the three quarks in the 
proton. Accordingly, the dominant contribution corresponds to a 
situation in which each of the three gluons is coupled to a different 
quark in the proton, as shown in diagram (c) in figure 
\ref{fig:impactdiag}. In the DL fit \cite{Donnachie:1984hf} the 
large-$t$ data were used to fix the normalization 
of the Odderon contribution to the scattering amplitude. 
For the prediction (\ref{largetpppred}) 
to match the observed behavior (\ref{largetppfit}) it appears necessary 
that the coupling $\alpha_s$ does not run with $t$. Due to 
the high power of $\alpha_s$ the effects of the running coupling\footnote{
Sometimes also the running of $\alpha_s$  with the energy $\sqrt{s}$ 
is considered in the literature. In our opinion this is a misconception. 
The scale of coupling constants is always given by a momentum scale, 
which only in some cases like $e^+e^-$ annihilation is equal to the 
energy $\sqrt{s}$. Therefore here the only possible scale on which the 
coupling could depend is in fact $\sqrt{-t}$.} 
would already lead to a substantial deviation from (\ref{largetppfit}). 
In \cite{Donnachie:1996rq} it was shown, however, that the effect of 
the running coupling could be almost compensated by nonperturbative 
corrections to the gluon propagator as they are suggested by 
Dyson--Schwinger equation analyses \cite{Cornwall:1981zr}. 

There are several open questions concerning the large $t$ region 
\cite{Donnachie:1996rq} of $pp$ elastic scattering. 
The first one is whether the differential cross section is really 
energy independent. Precise measurements at the LHC would 
be very helpful in this respect. 
It is likely that at high energies logarithmic corrections to 
the three--gluon exchange will become important, and the 
exchange of three noninteracting gluons should be replaced 
by the resummed (BKP) Odderon. 
At higher energies is is also feasible 
\cite{Donnachie:1996rq,Sotiropoulos:1993rd,Sotiropoulos:1994ub} 
that the triple--Pomeron exchange will eventually become the 
most important contribution at large $t$ 
so that the differential cross section would actually rise 
with increasing energy. 
Another puzzling question is why the behavior (\ref{largetpppred}) 
that is expected to hold asymptotically sets in already at rather 
low $t$, at least to a very good approximation. 
Although the large-$t$ elastic scattering cross sections are rather 
small they offer a very good opportunity for studying the Odderon, 
and future measurements in this kinematic region would be 
extremely valuable. 

The other region in which the Odderon contribution is important 
is the dip region around $|t|=1.3\,\mbox{GeV}^2$. The occurrence 
of the dip results from an interference of mainly the Pomeron, 
the double Pomeron, and the Odderon contributions to the 
amplitude. In the DL fit the imaginary part of the double Pomeron 
exchange cancels the imaginary part of the single Pomeron exchange 
at the position of the dip. Due to the different sign of the Odderon 
contribution in $pp$ and $p\pbar$ scattering the dip is filled by 
the Odderon in the latter case, but not in the former in which 
the Odderon term partially cancels the (small) real part of the 
single Pomeron exchange. The other 
contributions are not relevant for the existence of the dip, but 
are important for reproducing its exact shape. It should be emphasized 
that the other contributions with $C=-1$ in the DL fit (the reggeon 
pole and the reggeon--Pomeron cut) are far too small to account 
for the difference between the $pp$ and $p\pbar$ data in the 
dip region. 

Since the data in the dip region at ISR energies are the only 
data which show a more or less clear sign of the Odderon 
one can try to use these data to constrain the coupling of the 
Odderon to the proton. As we have seen in section 
\ref{Oddprotoncouplsect} that coupling can only be modeled, 
and a large uncertainty originating from the model assumptions 
is inherent in any calculation using it. An effort to extract 
as much information as possible 
on different models for the $\Od p$ coupling 
from the elastic $pp$ and $p\pbar$ data in the ISR range 
has been made in \cite{Dosch:2002ai}. We want to present the 
results of that study in some detail here since they give us 
valuable information on the basic parameters of the models. At the 
same time they can give us a feeling for the limitations of the elastic 
scattering data regarding the extraction of the Odderon contribution. 

In \cite{Dosch:2002ai} it is assumed that 
the Odderon can in the $t$-region around the dip be described 
by perturbative three--gluon exchange, and BKP resummation 
is not taken into account. In this way one can test the 
perturbative coupling of the Odderon via the models for the impact factors 
proposed by Fukugita and Kwieci\'nski (FK) and by Levin and Ryskin (LR), 
see section \ref{Oddprotoncouplsect}, as well as the 
geometric model for the transverse wave function of the proton 
given in that section. These models are widely used in other processes 
at similar (and in particular similarly low) momentum scales, for 
example in the diffractive production of pseudoscalar mesons in $ep$ 
collisions (see section \ref{excldiffppsect}). 
Clearly, $t$-values around $1.3\,\mbox{GeV}^2$ 
are at the lowest edge of applicability of perturbation theory, and 
one should keep this in mind when interpreting the results. 

\begin{figure}[pht]
\vspace*{0.4cm}
\def\size{\normalsize}
\begin{center}
\input{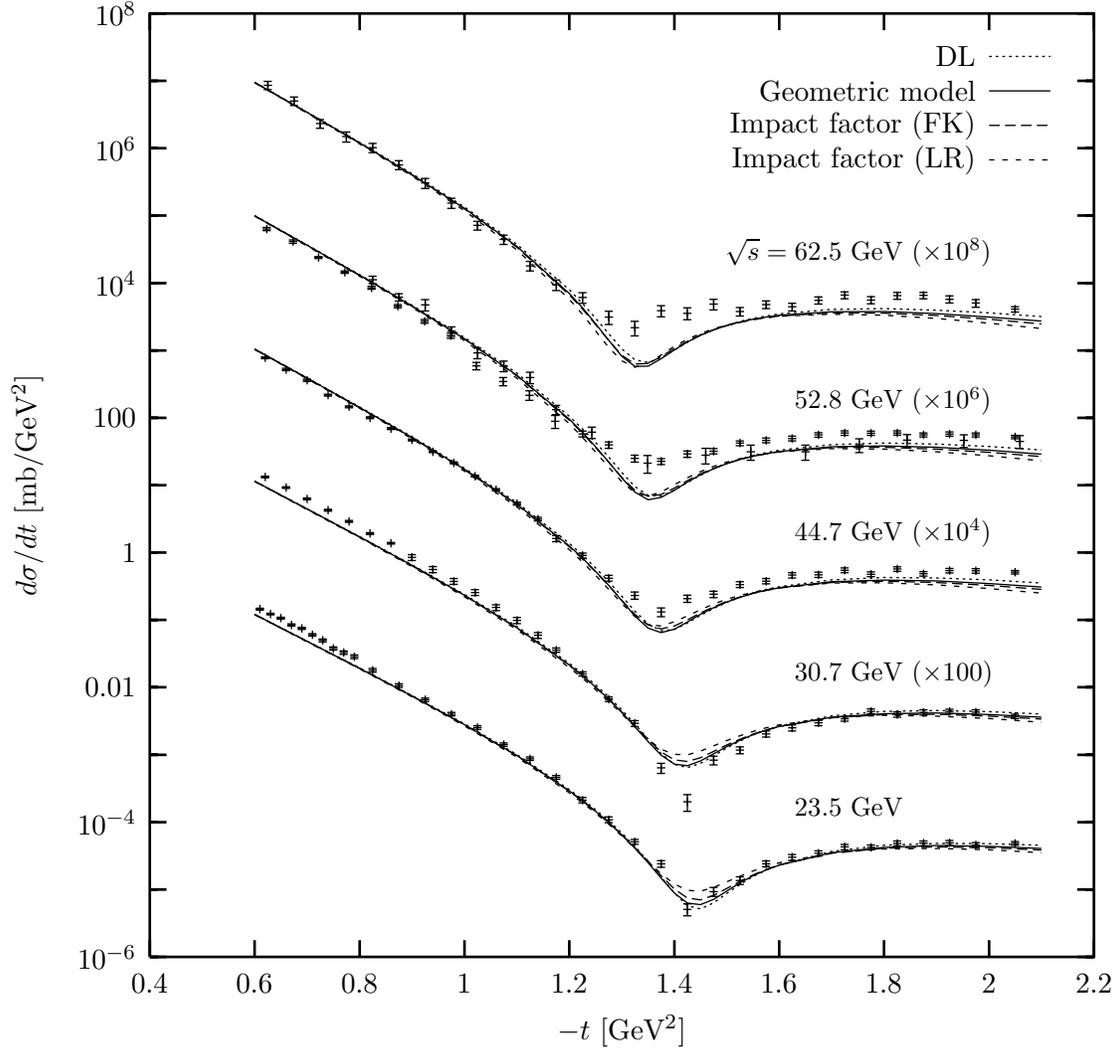}
\end{center}
\caption{Differential cross section for elastic $pp$ scattering
calculated using different couplings of the Odderon to the proton:
the original Donnachie--Landshoff fit (dotted), the geometrical
model for the proton eq.\ (\protect\ref{wavefunction}) (solid),
and the FK (long--dashed)and LR (short--dashed) impact factors; 
figure from \protect\cite{Dosch:2002ai}
\label{fig:allcurves}}
\end{figure}

\begin{figure}[htb]
\begin{center}
\input{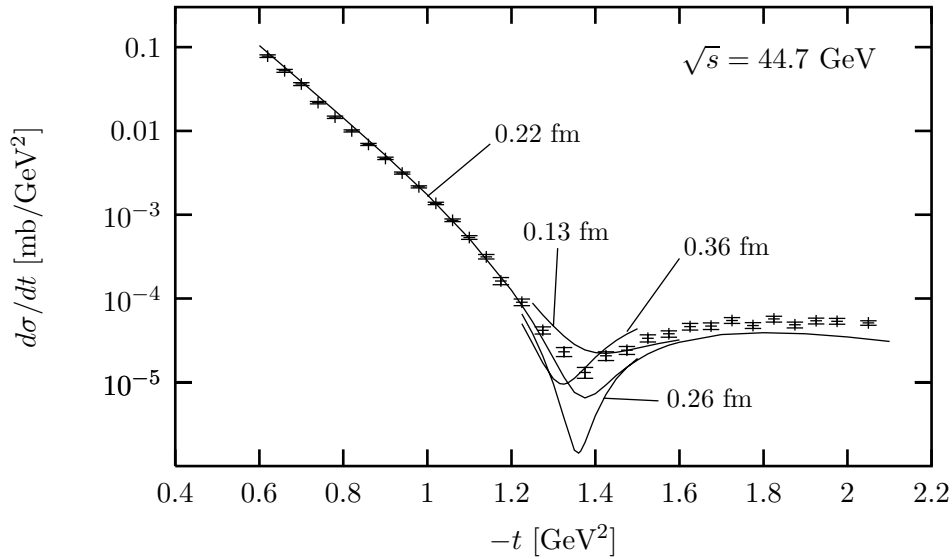}
\end{center}
\caption{Dependence of the differential cross section on the average
diquark size $\langle d \rangle$ chosen in the geometric
model of the proton for fixed coupling constant $\alpha_s=0.4$; 
figure from \protect\cite{Dosch:2002ai}
\label{fig:angles}}
\end{figure}

The DL fit is used in \cite{Dosch:2002ai} as a framework for testing 
different perturbative descriptions of the Odderon. One singles out 
the Odderon contribution $A^\soddi(s,t)$ to the DL fit and 
splits the amplitude accordingly, 
\be
\label{DLsplit}
A(s,t) = A^\soddi(s,t) + A^{\rm DL}(s,t) \,,
\ee
where $A^{\rm DL}$ denotes all other contributions 
to the scattering amplitude, including the $C$-odd 
reggeon contribution. All parameters of the DL fit are then kept fixed 
and only the Odderon term is replaced with other models for 
$A^\soddi(s,t)$ based on different $\Od p$ couplings and the 
perturbative exchange of three non--interacting gluons. 
The corresponding terms $A^\soddi(s,t)$ are calculated 
as described in sections \ref{formalismssect} and 
\ref{Oddprotoncouplsect}. The results for the differential 
cross section in the dip region are presented  together 
with the original DL fit and the relevant data 
in figure \ref{fig:allcurves}. 
The solid line in figure \ref{fig:allcurves} represents the result 
obtained with the geometric model (\ref{wavefunction}) 
for the proton. It almost coincides with 
the DL fit and gives a satisfactory description of all available data. 
The value of the strong coupling has been fixed at $\alpha_s=0.4$ 
and the angle $\alpha$ characterizing the proton configuration 
has been adjusted. 
The optimal description of the data is obtained for $\alpha =  0.14 \,\pi$, 
corresponding to an average diquark size (see figure \ref{star}) 
of $0.22\,\mbox{fm}$. 
For other choices of the average diquark size (or equivalently of 
the angle $\alpha$) and fixed $\alpha_s=0.4$ 
the description of the data becomes much 
worse as is illustrated in figure \ref{fig:angles} for one 
center--of--mass energy, $\sqrt{s} = 44.7\,\mbox{GeV}$. 
The parameters $\alpha_s$, $\alpha$, and $S_p$ in (\ref{wavefunction}) 
are of course strongly correlated in their effect on the differential 
cross section. Since $S_p$ is rather strictly constrained by the 
electromagnetic size of the nucleon it should not be varied. 
The constraints on the other two parameters in the model are only 
weak. 
The correct value of the strong coupling constant $\alpha_s$ 
is not known precisely in the dip region but has a strong effect 
on the cross section as it enters in the third power on the 
amplitude level already. The correct value of the angle $\alpha$ is 
even less constrained, and also the variation of $\alpha$ has a 
strong effect on the cross section. This is particularly true for small values 
of $\alpha$ (small diquark sizes) which are known to imply 
a strong suppression of the amplitude. In this framework 
it is not possible to determine 
$\alpha_s$ and the angle $\alpha$ independently. 
One can only determine the optimal value for $\alpha$ 
for different choices of $\alpha_s$ other than the $\alpha_s=0.4$ above. 
For the choice $\alpha_s=0.3$, for instance, one finds that 
the best description of the data results for $\alpha= 0.22\,\pi$, 
corresponding to an average diquark size of 
$\langle d \rangle = 0.34\,\mbox{fm}$. Choosing $\alpha_s=0.5$ 
instead, the optimal value is $\alpha= 0.095\,\pi$, 
corresponding to $\langle d \rangle = 0.15\,\mbox{fm}$. 
Is should be pointed out that the resulting sizes of the 
diquark cluster in the nucleon are rather small for all 
reasonable choices of $\alpha_s$ at the relevant momentum 
scale in the dip region.  
A Mercedes star configuration in the proton would in fact 
imply an unrealistically small value of $\alpha_s \simeq 0.17$. 
As we already said, this result of course assumes that LO 
perturbation theory can be applied in the dip region. 

\begin{figure}[htb]
\begin{center}
\input{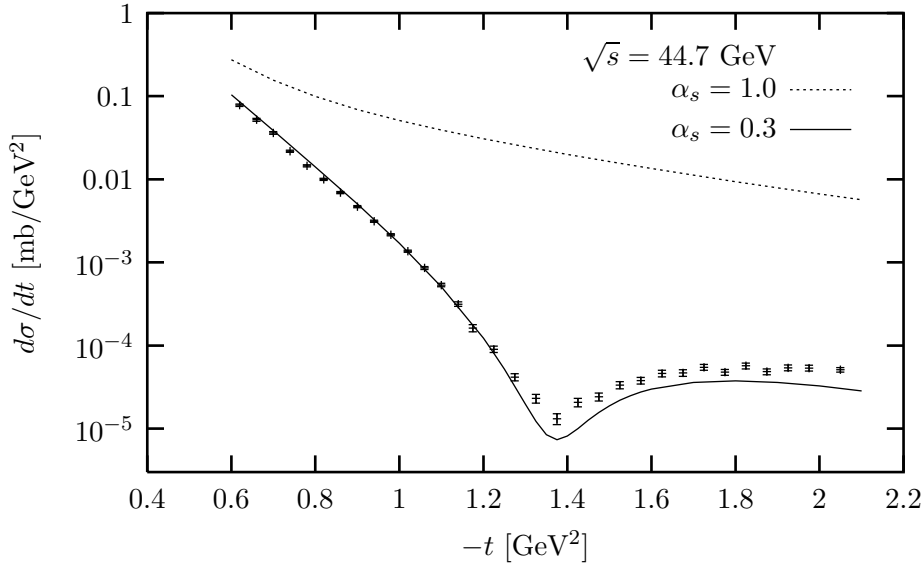}
\end{center}
\caption{Dependence of the differential cross section
obtained from the Fukugita--Kwieci\'nski impact factor
on the choice of $\alpha_s$; 
figure from \protect\cite{Dosch:2002ai}
\label{fig:impactdiffalpha}}
\end{figure}

Let us now turn to the models for the Odderon--proton 
impact factor (\ref{eq:impact}). 
Both models contain two parameters one of which is the strong 
coupling $\alpha_s$. The other one is 
in the case of the FK model the parameter $A=m_\rho /2$, 
in the case of the LR model it is the parameter $R_p$. 
The latter parameters are again related to the proton 
size and should thus be considered strongly constrained. 
Accordingly they should be kept at their original values. 
Hence only $\alpha_s$ is varied. 
The differential cross section obtained with the FK model 
(\ref{eq:formfact}) is 
shown as the long--dashed curve in figure \ref{fig:allcurves}. 
It gives an equally good description of the data as the DL fit 
and as the geometric model of the proton. 
In order to obtain this curve the authors of \cite{Dosch:2002ai} 
have chosen $\alpha_s=0.3$ 
instead of the value $\alpha_s=1.0$ originally proposed in 
\cite{Fukugita:1979fe}. If the latter value is chosen instead 
the differential cross section would dramatically overshoot 
the data and not even show a dip structure, 
as is illustrated for one center--of--mass energy 
($\sqrt{s}=44.7\,\mbox{GeV}$) in figure \ref{fig:impactdiffalpha}. 
This figure is a nice illustration of the strong dependence 
of the Odderon amplitude on the strong coupling 
constant $\alpha_s$. It is obvious here that the significance 
of that effect can hardly be overemphasized. 
Also the LR model (\ref{formfaclr}) 
for the impact factor leads to a good description of the 
data when the strong coupling constant is chosen as 
$\alpha_s=0.5$. The corresponding differential cross 
section is shown as the short--dashed curve in figure 
\ref{fig:allcurves}. Also here the dependence of the cross 
section on $\alpha_s$ is very strong, actually being the 
same as in the case of the FK model 
as can be easily seen from eq.\ (\ref{eq:impact}). 

\begin{figure}[htb]
\begin{center}
\input{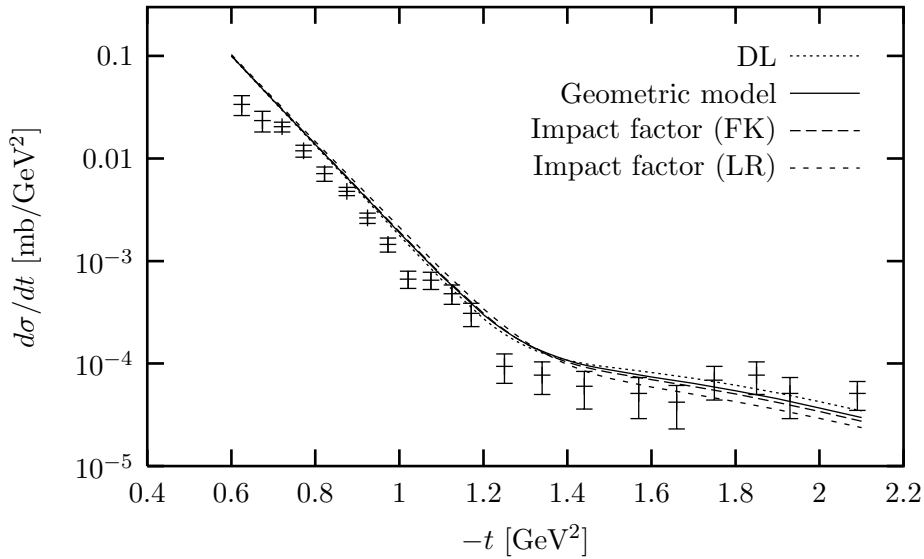}
\end{center}
\caption{Differential cross section for elastic $p\bar{p}$ scattering
at $\sqrt{s} = 53\,\mbox{GeV}$ as
calculated using different couplings of the Odderon to the proton:
the original DL fit (dotted),
the geometrical model for the proton (solid),
and the FK (long--dashed)
and LR (short--dashed) impact factors; 
figure from \protect\cite{Dosch:2002ai}, 
data from \protect\cite{Breakstone:1985pe}
\label{fig:ppbar}}
\end{figure}

Turning to the differential cross section for 
elastic $p\bar{p}$ scattering we show the corresponding 
results of \cite{Dosch:2002ai}  in figure \ref{fig:ppbar}. 
The same parameters are used as for the curves in 
figure \ref{fig:allcurves}. Again, the geometric model as well 
as the two models for the impact factors lead to a description 
of the data which is as good as the Donnachie--Landshoff fit, 
producing a shoulder rather than the dip observed in $pp$ scattering. 

In summary we can say that the experimental data available 
in the dip region are by far not precise enough to distinguish between 
different models for the coupling of the Odderon to the proton. 
All models for that coupling and the corresponding models 
for the proton structure lead to a satisfactory description of the 
data when the respective parameters are chosen appropriately. 
But for a given model these parameters are quite strongly 
constrained by the data. This applies in particular to the value 
of $\alpha_s$ in the two models using impact factors. 
This point is one of the most important conclusions we should 
draw from the analysis of the differential $pp$ and $p\pbar$ 
cross sections since all phenomenological predictions using the 
impact factors strongly depend on the value chosen for 
$\alpha_s$, see also the discussion of this point in 
section \ref{genconsidsect}. 

An excellent fit to all available $pp$ and $p\pbar$ in forward and 
nonforward direction was given by Gauron, Nicolescu and Leader 
(GNL) \cite{Gauron:1990cs} in 
the framework of the maximal Odderon which we have 
described in detail in section \ref{maxoddsect}. The author 
fit the almost 40 parameters of the model, in particular 
the constants $F_i$, $O_i$ and $b^\pm_i$ related to the 
maximal Odderon and Froissaron terms, 
(\ref{moterm}) and (\ref{froiterm}), respectively. The resulting 
description of the data is almost perfect and clearly better 
than the DL fit. This is shown in the differential $pp$ cross 
section in figure \ref{gnldlcompfig} below, where the 
dot--dashed line is the maximal Odderon fit \cite{Gauron:1990cs}, 
and the solid line is the DL fit \cite{Donnachie:1984hf}.
However, the maximal Odderon fit uses a larger number 
of parameters than the DL fit. Hence the better quality 
of the GNL fit does not imply that the maximal Odderon approach 
is favored over the DL approach by the data.  
Since the maximal Odderon has a strong energy dependence 
the GNL approach predicts deviations from the DL fit 
especially at very high energies. Therefore future experiments 
at TeV energies, for example at the LHC, will have a good chance 
of distinguishing the two fits. 

The central question is of course whether the Odderon can be 
identified unambiguously in the data on the differential 
cross section of $pp$ and $p\pbar$ elastic scattering. 
These data have been analyzed in a number of different ways. 
To the best of our knowledge no successful description of the 
data has been achieved without making use of an Odderon 
contribution of some kind.\footnote{A possible exception 
is the model of \protect
\cite{Goloskokov:1986rw,Goloskokov:1982mp,Goloskokov:1986sb,Goloskokov:ds}. 
There the data are well described by a fit in 
which the odd--under--crossing 
amplitude has a reggeon--type energy dependence. 
However, in that model the proton and the antiproton 
are surrounded by meson clouds inducing additional 
contributions to the scattering process even at high energies. 
Such a meson cloud is clearly a nonperturbative effect, 
the energy dependence of which difficult to estimate. 
This makes it somewhat difficult to compare the results of this 
model directly with other fits in which such a meson cloud 
is usually absent.} Therefore these data, in particular 
the small difference between $pp$ and $p\pbar$ 
scattering at $\sqrt{s}=53\,\mbox{GeV}$ in the dip region 
(see figure \ref{fig:dipdiff}), can really be regarded as strong 
evidence for the existence of the Odderon. But all successful 
description of the data also rely on a number of parameters 
for the Odderon and for the other contributions to the 
scattering amplitude which are not known a priori and need 
to be fitted. This clearly implies some limitations on the 
identification of the size and type of the Odderon contribution. 
The limitations have been demonstrated nicely in \cite{Volkerthesis}, 
as we will discuss now. 
The real Odderon, i.\,e.\ the actual singularity in the complex angular 
momentum plane, has to be universal in reality. The fits on the 
other hand describe the data successfully 
using quite different Odderon singularities, 
for example a single pole in the DL fit and a double pole in the GNL fit. 
This fact has two possible explanations: either the fit is not very 
sensitive to the Odderon contribution, or the fit is rather sensitive 
to the Odderon but the freedom in the other terms (including the 
$C=+1$ terms) leaves enough room for accommodating a given 
type of Odderon while still fitting the data. In \cite{Volkerthesis} it was 
shown that the latter is the case. This was done by simply taking the 
Odderon contribution from the DL fit and by implementing it 
into the GNL fit, and vice versa. In doing so all other parameters 
of the respective fits were left unchanged. The result is shown 
in figure \ref{gnldlcompfig} for the example of elastic $pp$ 
scattering at an energy of $\sqrt{s}=44.7\,\mbox{GeV}$ 
together with the relevant data. 
\begin{figure}[htb]
\vspace*{0.4cm}
\begin{center}
\input{gnldlcomp.pstex_t}
\end{center}
\caption{Non--universality of the Odderon contributions 
to the DL and GNL: the solid and dot--dashed lines represent 
the original fits, the short--dashed and long--dashed lines 
show the result of using the two fits with the other group's 
Odderon term, all curves and the data are for 
$\sqrt{s}=44.7\,\mbox{GeV}$; figure from \protect\cite{Volkerthesis}
\label{gnldlcompfig}}
\end{figure}
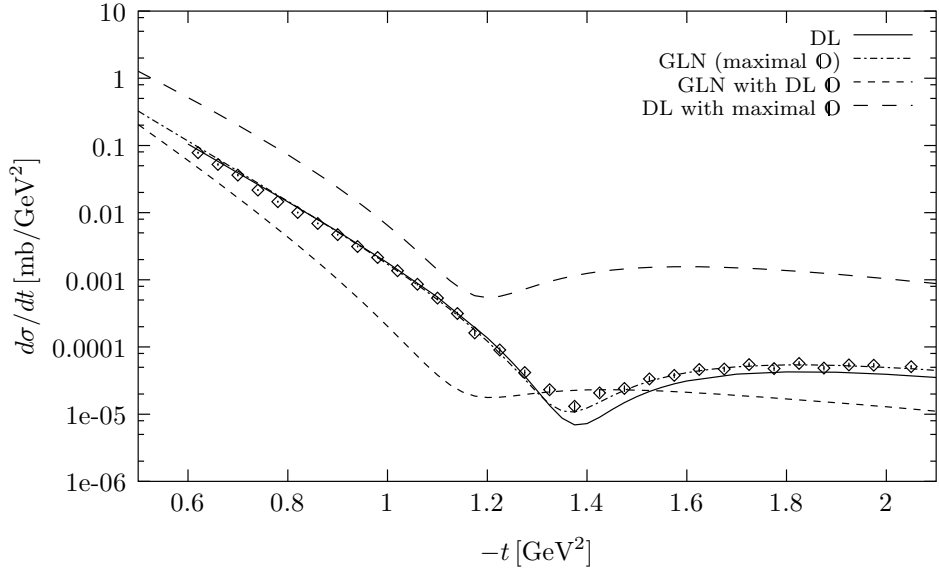
The solid and dot--dashed lines represent the original DL and GNL 
fits, respectively. The other two lines show the result of using 
the DL Odderon in the GNL fit and vice versa. Clearly, the 
Odderon contributions to the fits turn out to be non--universal, 
and the same immediately follows for the other contributions 
to the scattering amplitude, in particular for the $C=+1$ terms. 
Although the dip structure strongly constrains the fits, the Odderon 
cannot be unambiguously extracted from the fits, at least with 
the presently available data. 

In conclusion we can say that the differential cross sections of 
elastic $pp$ and $p\pbar$ scattering are an important 
study ground for Odderon physics, and in the framework of the 
different fits to the data it is possible to extract 
very important pieces of information. However, as we have tried 
to demonstrate, there are very severe limitations resulting from 
the simple fact that the Odderon is just one among many 
contributions to the scattering amplitude most of which are 
only poorly known. Therefore other processes are urgently 
needed which can offer complementary information about the 
Odderon. 

Another interesting possibility that has recently been proposed 
\cite{Leader:1999ua} is to look for Odderon effects in the 
spin dependence of elastic $pp$ scattering at small $t$. 
Let us recall that the Odderon gives a predominantly real 
contribution to the scattering amplitude whereas the contribution 
of the Pomeron is predominantly imaginary at small $t$, 
see the discussion before eq.\ (\ref{oddprop}). 
For the spin dependence of $pp$ scattering one has to 
consider not just one scattering amplitude but amplitudes which 
take into account spin flips. Accordingly, there are single--flip, 
double--flip and non--flip amplitudes. One can then construct 
a number of spin dependent asymmetries which can be expressed 
as real or imaginary parts of different products of those 
amplitudes. In this way one can find observables to which 
the Odderon gives the dominant contribution. 
The most promising among them seems to be the $t$-dependence 
of the double transverse spin asymmetry $A_{NN}$ at small 
$|t| <0.02\,\mbox{GeV}^2$, where the Odderon contribution 
is expected to lead to a characteristic change in the shape of the 
asymmetry. The exact size of this and related effects is very difficult to 
predict since they involve not only the coupling of the Odderon 
to the proton but also the spin dependence of this coupling. 
One can of course benefit from this sensitivity to get a handle 
on the spin dependence of the $\Od p$ coupling once measurements 
of these observables are available. There are in fact good prospects 
to measure these asymmetries in polarized $pp$ scattering at 
RHIC in the near future. 

\subsubsection{The $\rho$-Parameter and the Total Cross Section}
\label{rhosect}

A very interesting observable for the search for the Odderon 
is the so--called $\rho$ parameter. Let us recall its definition 
(\ref{rhodef}) as the ratio of the real part to the 
imaginary part of the forward scattering amplitude, 
\be
\rho (s) = \frac{\real A(s,t=0)}{\imag A(s,t=0)}
\,.
\ee
Here we will consider this ratio for $pp$ and $p\pbar$ scattering, 
but it can in principle be defined for other processes as well. 
Let us recall that the Pomeron contribution to the scattering 
amplitude is predominantly imaginary at high energy and small $t$, 
whereas the Odderon contribution is predominantly real there. 
The Odderon contribution changes sign when going from 
a particle--particle to an antiparticle--particle scattering process. 
Accordingly, effects of the Odderon can show up in 
the difference $\Delta \rho$ (see (\ref{deltarhodef})) 
of the $\rho$-parameters for $pp$ and $p\pbar$ scattering, 
\be
\Delta \rho (s)= \rho\,^{\bar{p}p}(s) - \rho\,^{pp}(s)
\,,
\ee
at high energies $\sqrt{s}$. 

\begin{figure}[ht]
\vspace*{0.4cm}
\begin{center}
\input{neurho.pstex_t}
\end{center}
\caption{Prediction of the maximal Odderon model 
\protect\cite{Gauron:1990cs} for the $\rho$-parameter 
for $pp$ and $p\pbar$ scattering 
in comparison with the high energy data \protect\cite{Hagiwara:fs}
\label{figrho}}
\end{figure}
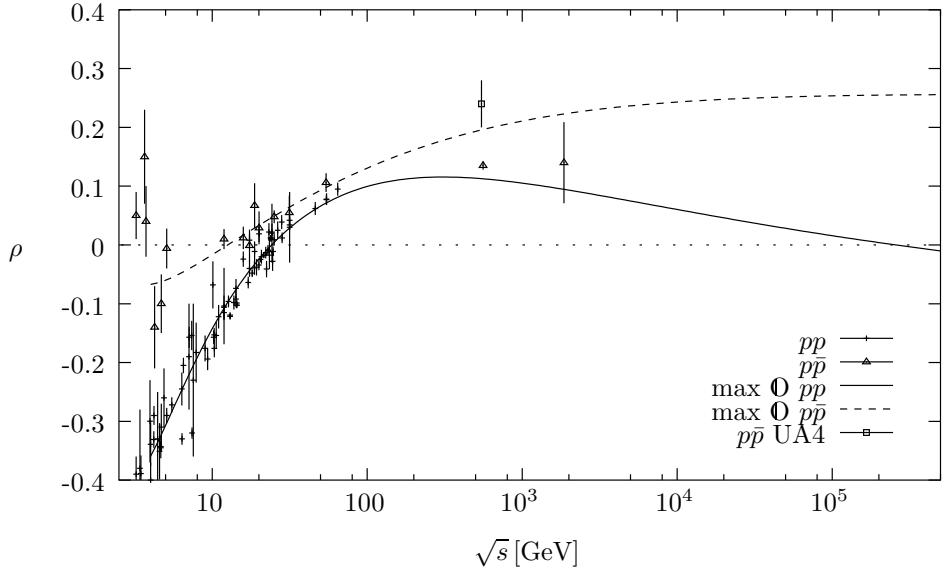

The $\rho$-parameter has been investigated by many authors, 
see for example 
\cite{Gauron:1990cs,Donnachie:iz,Gauron:1990cc}, 
\cite{Block:1982bv}--\cite{Cudell:2002xe}. 
The most dramatic effect is expected in the maximal Odderon model, 
see section \ref{maxoddsect}. Here the maximal possible 
energy dependence implies that $\Delta \rho$ slowly grows at 
large energies. Figure \ref{figrho} shows the prediction 
\cite{Gauron:1990cs} of the maximal Odderon approach 
for the $\rho$-parameters for $pp$ and $p\pbar$ scattering 
together with the relevant data taken at high energies \cite{Hagiwara:fs}. 
These predictions are obtained with the same parameters in the 
maximal Odderon fit which have led to a successful description 
of the differential $pp$ and $p\pbar$ elastic cross sections, 
see previous section. In the Donnachie--Landshoff fit, 
on the other hand, the Odderon contribution stays constant 
with energy whereas the Pomeron contribution grows 
with energy. Consequently, in the DL fit $\Delta \rho(s)$ 
becomes negligible above energies of about 
$\sqrt{s} \simeq 300\,\mbox{GeV}$ \cite{Donnachie:iz}. 
Although $\Delta \rho \to 0$ at high energies would probably 
rule out the maximal Odderon, a small value of $\Delta \rho$ 
at large energies does in general not imply the absence of 
the Odderon. 

Measurements of the $\rho$ parameter at high energies have been 
performed at all $pp$ and $p\pbar$ colliders, see for example 
\cite{Amos:1985wx}, \cite{Foley:1967nk}--\cite{Avila:2002bp}. 
The usual method to determine this parameter experimentally 
starts from the observed elastic differential distribution 
$dN/dt$ at the smallest possible $|t|$. In order to extract $\rho$ 
one first needs to separate the strong interaction contribution $f_S$ 
from the Coulomb term $f_C$ (which is well--known)  
\be
\frac{1}{L} \frac{dN}{dt} = 
\frac{1}{\pi} |f_C + f_S|^2
\,,
\ee
where $L$ is the integrated accelerator luminosity. 
Then a theoretical model for the extrapolation of the 
distribution down to $t=0$ is needed. The usual choice is 
\be
\label{rhoassume}
f_S = \frac{\sigma_T}{4 \pi} (i + \rho) \exp(-b |t|/2)
\,,
\ee
where $b$ is the slope parameter of the diffraction peak, 
and $\rho$ and $b$ are assumed to be constant. The 
determination of $\rho$ obviously depends on this model, 
and therefore $\rho$ is sometimes called a semitheoretical 
parameter. Note that it is typically the combination 
$ (1+\rho^2) \sigma_T$ which occurs in the measurement, 
and the determination of $\rho$ is always closely related to 
the measurement of the total cross section $\sigma_T$. 
At most energies measurements have only been performed for 
either $pp$ or $p\pbar$ scattering. The determination of $\Delta \rho$ 
then requires to compute $\rho^{pp}$ from $\rho^{\pbar p}$ 
(or vice versa) via dispersion relations. This involves quite a 
large uncertainty, and the resulting values for $\Delta \rho$ 
should be interpreted only very carefully. 

Considerable excitement was caused by the observation 
\cite{Bernard:1987vq} of a large value of  $\rho^{\pbar p}=0.24\pm0.04$ 
at $\sqrt{s}=546\,\mbox{GeV}$ 
by the UA4 collaboration at the CERN SPS. In our figure 
\ref{figrho} this datum is indicated as a small box. 
For reasons that become clear from the figure it 
was interpreted \cite{Bernard:1987kf} 
as strong indication for the maximal Odderon.\footnote{Also 
other possible explanations for that datum were discussed, 
in particular models with new thresholds, see for example 
\protect \cite{Kang:1991zt}.} 
It should be pointed out that this measurement was made 
in a single run of the SPS of only about two days, and systematic 
effects could not be thoroughly studied. A later measurement 
\cite{Augier:1993sz} in fact found a much lower value 
$\rho^{\pbar p}=0.135\pm0.015$ 
(see also figure  \ref{figrho}) which does not favor the maximal 
Odderon hypothesis. This latter measurement was made 
with an eleven times higher statistics and with much better 
control of systematic effects. According to \cite{Augier:1993sz} 
the earlier measurement \cite{Bernard:1987vq} `should be 
considered superseded'. The usual interpretation of the 
data shown in figure \ref{figrho} is usually that $\Delta \rho$ 
is very small, and the usual though not too precise estimate 
is 
\be
 |\Delta \rho| \le 0.05
\ee
at energies in the TeV range. 

An alternative interpretation in particular of the measurement 
\cite{Augier:1993sz} was given in \cite{Gauron:1996sm}. 
There it was pointed out that the theoretical assumption 
(\ref{rhoassume}) usually made in the experimental 
determination of $\rho$ is suggestive but possibly wrong. 
According to the Auberson--Kinoshita--Martin 
theorem (\ref{akmformula}) the 
$t$-dependence of the differential distribution $dN/dt$ 
can in fact exhibit damped oscillations at small $|t|$. 
This would clearly affect the extraction of the $\rho$ 
parameter that relies on the extrapolation to $t=0$. 
The authors of \cite{Gauron:1996sm} observe that the 
$dN/dt$  data in \cite{Augier:1993sz} seem to show 
a small bump which could well be interpreted as a sign 
of such a damped oscillation. It is then argued that 
by extrapolating the oscillation backwards to $t=0$ one 
could find a value up to $\rho^{\pbar p} = 0.23$. 
This scenario is certainly interesting, but appears 
rather speculative. However, it shows that the data do 
not actually rule out the maximal Odderon approach. 

\begin{figure}[ht] 
\vspace*{0.4cm}
\begin{center} 
\input{drhodata.pstex_t}
\end{center}
\caption{$\Delta \rho$ at UA4/2 energy for proton-(anti)proton 
scattering as a function of the diquark size $d$ according to 
\protect\cite{Rueter:1996yb}
\label{figdrhodata}}
\end{figure}
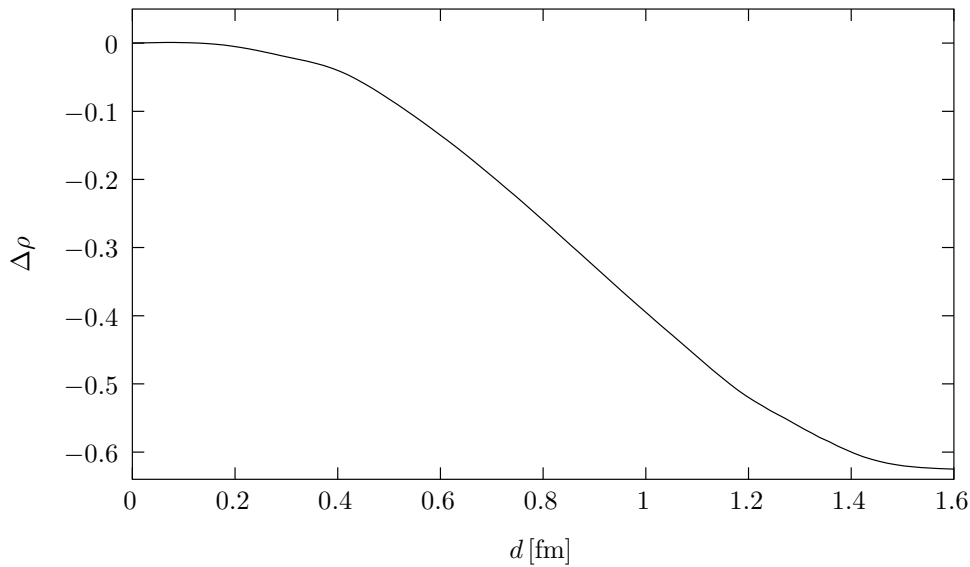

In \cite{Rueter:1996yb} it was shown that a low value of 
$\Delta \rho$ can also be explained by the suppression 
of the Odderon coupling to the proton due 
to a possible diquark clustering in the proton. 
Using the stochastic vacuum model for high energy scattering 
(see sections \ref{npheidelbergsect} and \ref{formalismssect}) 
and the geometric model (\ref{wavefunction}) for the 
transverse wave function of the proton one finds a strong 
suppression of $\Delta \rho$ with the average diquark 
size $\langle d \rangle$. This is shown in figure 
\ref{figdrhodata} for an energy of $\sqrt{s}=541\,\mbox{GeV}$. 
Already diquark sizes of about $0.3\,\mbox{fm}$ drastically 
reduce the resulting value for $\Delta \rho$. 

As we have seen in section \ref{pomoddregge} the Odderon can 
cause a difference $\Delta \sigma$ between the total 
cross sections for $pp$ and $p\pbar$ scattering which 
can potentially even grow logarithmically with energy, see 
(\ref{deltasigmalog}). The experimental data for the 
total cross section in $p\pbar$ scattering at the Tevatron 
are currently not conclusive, with two mutually contradicting 
values \cite{Amos:1989at,Abe:1993xy}. It seems difficult 
to find a clear signal of the Odderon in this situation. 
But a measurement of the total cross section at the LHC 
will be very interesting in this respect. The maximal 
Odderon approach predicts $\Delta \sigma = - 6 \,\mbox{mb}$ 
at LHC energies \cite{Gauron:1990cs}. 
Such an effect should probably show up as a disagreement 
of the corresponding $pp$ cross section with conventional fits. 
In addition we recall that also the correlation of the signs 
of $\Delta \rho$ and $\Delta \sigma$ could give 
a hint to the Odderon, see the discussion at the end of 
section \ref{basicoddsect}. 

The $\rho$-parameter and the total cross sections have 
also been investigated using cosmic ray data, see for 
example \cite{Avila:2002tk}. The cosmic ray data are 
included in different simultaneous fits of the 
$\rho$-parameter and of the total cross section based 
on different models like the DL model etc. It turns out 
that also with the inclusion of theses data the fits do not 
show any clear sign of the Odderon. 
However, it should be pointed out that 
cosmic ray data have in general much larger errors 
than $pp$ data from collider experiments, and the 
corresponding results cannot be expected to be very 
precise, especially when one is looking for only one 
of many interfering contributions to the scattering amplitude. 

In summary we can say that the presently available data 
on the $\rho$-parameter and on total cross sections 
do not show any clear sign of the existence of the Odderon. 
In the case of the $\rho$-parameter the main problem is 
the absence of simultaneous measurements of 
$\rho\,^{\bar{p}p}(s)$ and $\rho\,^{pp}(s)$ at the same energy 
$\sqrt{s}$ in the TeV range. The extraction of $\Delta \rho$ from 
$\rho\,^{\bar{p}p}(s)$ only is simply not sufficiently precise. 
It should also be emphasized again that even $\Delta \rho=0$ 
at high energies does not imply that there is no Odderon. 
In the case of total cross sections the Tevatron data themselves 
do not give a clear enough picture from which any conclusions 
about the Odderon could be drawn. 
In order to really find out whether there are Odderon effects 
in the $\rho$-parameter and in the total cross section one 
would need to measure both quantities in both $pp$ and $p\pbar$ 
scattering at the same energy preferably in the TeV range. 
Given the present colliders and the current plans for future 
colliders the chances for such a measurement in the TeV range 
are low. At RHIC, however, $pp$ scattering will be studied at energies 
up to $\sqrt{s}=500\,\mbox{GeV}$, which overlaps with the 
energy range of $p\pbar$ scattering at the CERN SPS, 
and the comparison of the corresponding data offers a very good 
chance of improving our picture of Odderon effects in the 
$\rho$ parameter and in the total cross sections. 

\subsubsection{Double-Diffractive Vector Meson Production}
\label{excldiffppsect}

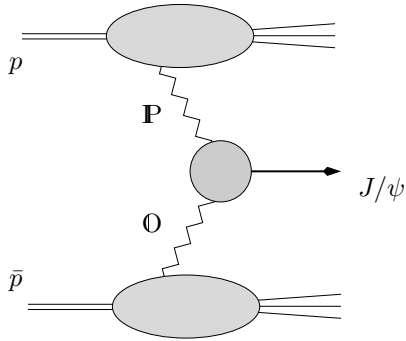
\begin{figure}[ht]
\vspace*{0.4cm}
\begin{center}
\input{ppjpsi.pstex_t}
\end{center}
\caption{Pomeron--Odderon fusion mechanism for double--diffractive 
$J/\psi$ production in $p\pbar$ scattering
\label{pomoddfusion}}
\end{figure}

An interesting process that would permit a rather 
clean identification of the Odderon is the double--diffractive 
production of unflavored vector mesons like $\phi$ and $J/\psi$ 
in $pp$ or $p\pbar$ scattering, for example 
\be
p + p \longrightarrow p + J/\psi + p
\,,
\ee
with the $+$ signs indicating rapidity gaps in the final state. 
This process has been investigated 
in the framework of Regge theory in \cite{Schafer:na}. 
At high energies reggeon exchange is suppressed with energy and 
thus negligible. If the coupling of the Odderon to the proton 
were very small, however, reggeon exchange of for instance 
an $\omega$ could still be 
of comparable size. But in the case of $J/\psi$ production 
reggeon exchange is also forbidden by Zweig's rule. Therefore 
$J/\psi$ production is in principle a cleaner test for the Odderon 
than $\phi$ production.  
Apart from this point $\phi$ and $J/\psi$ production can 
be treated analogously, and here we will only discuss the case of 
double--diffractive $J/\psi$ production. 
Due to the quantum numbers of the $J/\psi$ 
the only strong interaction mechanism for this process is 
Pomeron--Odderon fusion as shown in figure \ref{pomoddfusion}. 
In addition to the diagram in the figure there is another diagram 
in which the Pomeron and the Odderon are interchanged. In the 
diagram we show the process for $p\pbar$ scattering, but 
the same process can of course also occur in $pp$ scattering. 
Assuming simple Regge poles for the Pomeron and the Odderon 
with the corresponding propagators as given in section \ref{pomoddregge} 
one finds, however, that in the latter case it would be required that 
the Pomeron and Odderon intercepts are different, 
$\alpha_\spommi \neq \alpha_\soddi$. Otherwise the two 
diagrams mentioned above add destructively and the cross section 
vanishes for $pp$ scattering. In general the cross section is 
expected to be larger for $p\pbar$ scattering than for $pp$ 
scattering by about an order of magnitude. 

The calculation \cite{Schafer:na} uses a vector--like coupling 
of both the Pomeron and the Odderon to the quarks in the vector 
meson, see eq.\ (\ref{oqcouplgamma}). 
The total cross section for double--diffractive 
 $J/\psi$ production in $p\pbar$ scattering is then estimated to be 
$\sigma=3 \,\mbox{mb}\,\cdot c_0^2 \cdot N$ at $\sqrt{s}=2\,\mbox{TeV}$. 
Here $c_0$ is the ratio of $C=-1$ and $C=+1$ exchanges for which 
an upper bound can be estimated using the data on the $\rho$-parameter 
described in the previous section, $c_0 \le 0.05$. The constant $N$ is 
supposed to take into account some effects that have been neglected 
in the calculation of \cite{Schafer:na}, and the authors give the 
estimate $N=0.01$. The most important of these effects is 
probably that the couplings of the Pomeron and Odderon are not 
point--like as assumed in the Regge framework. In reality some 
form factor for the meson should be taken into account. 

The process should have a rather clean signature experimentally 
with the $J/\psi$ decaying for example into a lepton pair and 
clearly separated from the $p$ and $\pbar$ directions by 
rapidity gaps originating from the colorless exchanges of the 
Pomeron and the Odderon. The chances of observing 
this process at the Tevatron and also at the LHC should be good. 

It should be noted here that a 
considerable uncertainty in the prediction of this process is 
related to the occurrence of the two rapidity gaps. It was found 
that at the Tevatron the survival probability of rapidity gaps 
is considerably lower than in comparable processes at HERA. 
This problem is currently discussed in much detail in the 
context of double--diffractive Higgs production at the 
Tevatron which is considered to be an especially suitable 
process for a clean discovery of the Higgs boson. The problem 
is caused due to additional soft emissions filling the rapidity gaps 
in the process of hadronization. A really profound understanding 
of the reduced gap survival probability in hadron--hadron 
collisions has not yet been achieved, for a recent review see 
for example \cite{Khoze:2002py} and references therein. 
So far this question has only been addressed for the gap survival 
probability in the case of Pomeron exchange. A priori it is not 
clear that Pomeron and Odderon exchange are equivalent in this 
respect, but it appears likely that the gaps resulting from Odderon 
exchange will be suppressed in the same way as those originating 
from Pomeron exchange. If one were very optimistic one could 
even think of the Odderon as a test for possible mechanism 
for the gap suppression that are currently discussed for the Pomeron. 
But that would first require that the Odderon is actually 
discovered and tested in much detail. So this is certainly a 
project for the more distant future. 
The same problem of the gap survival probability 
will also occur at the LHC. But there, 
due to the high design luminosity one also has to 
deal with the additional problem of disentangling the signal 
from the underlying event, i.\,e.\ from additional activity 
in the detector due to multiple interactions in the bunch 
crossings. 

In \cite{Schafer:na} the double--diffractive production 
of $J/\psi$ was considered for the case that the proton and 
antiproton both stay intact. It should be noted that if 
the suppression of the Odderon--nucleon 
coupling is in fact due to the possible diquark structure 
of the nucleon as discussed in section \ref{Oddprotoncouplsect} 
then the cross section should be larger if one or both (anti)protons 
break up, of course without filling the rapidity gap. 

\begin{figure}[ht]
\vspace*{.4cm}
\begin{center}
\input{pofupert2.pstex_t}
\end{center}
\caption{Perturbative description of $J/\psi$ production via 
Pomeron--Odderon fusion
\label{pertpofusion}}
\end{figure}
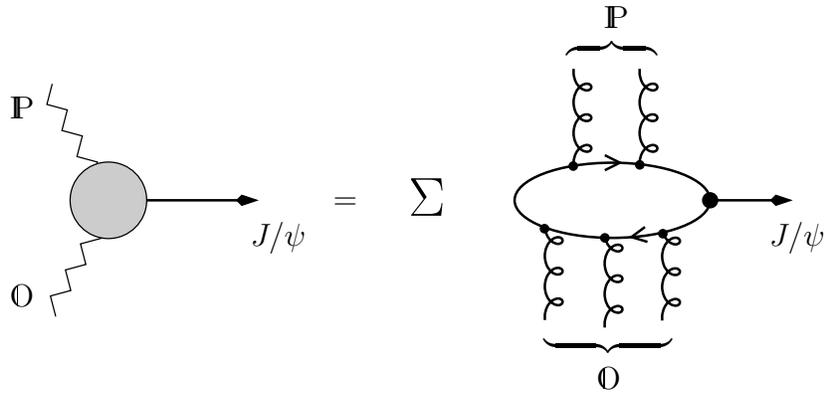

One of the uncertainties in the calculation \cite{Schafer:na} 
is due to the pointlike coupling of the Odderon 
to the quarks in the vector meson. For a heavy meson like 
the $J/\psi$ (but not necessarily for the $\phi$) the large mass 
of the quark provides a hard scale, and the 
calculation can be improved considerably by using perturbation 
theory instead of the Regge framework used in \cite{Schafer:na}. 
One would then model the Pomeron and Odderon by 
perturbative two-- and three--gluon exchange, respectively. 
The Pomeron exchange can be associated with the unintegrated 
gluon structure function of the proton, and for the Odderon coupling 
to the proton one can use the impact factors discussed in section 
\ref{Oddprotoncouplsect}. 
The Pomeron--Odderon fusion process, however, is a rather 
challenging object in perturbation theory. In the corresponding 
diagrams the two gluons of the Pomeron and the 
three gluons of the Odderon are coupled to the vector 
meson through a loop formed by a charm quark as is 
shown in figure \ref{pertpofusion}. 
The summation symbol in the figure indicates that 
the gluons have to be coupled to the quark loop in 
all possible ways in order to obtain a gauge invariant 
result. Unfortunately, this phenomenologically very 
important calculation has not yet been performed. 
The exact perturbative result for the 
Pomeron--Odderon fusion process might well differ considerably 
from the simplified Regge picture used in \cite{Schafer:na}. 

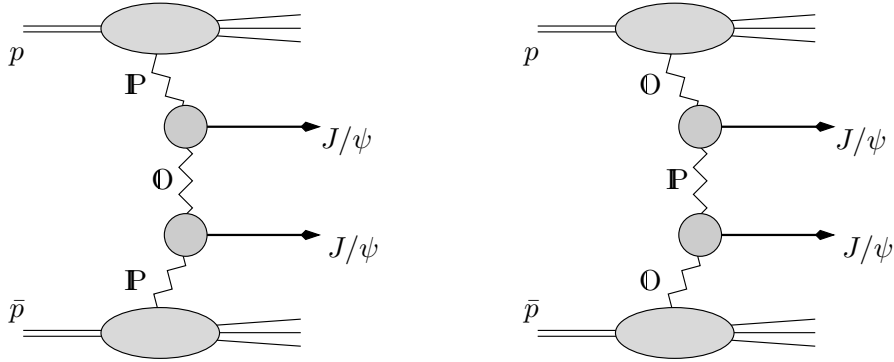
\begin{figure}[ht]
\vspace*{.4cm}
\begin{center}
\input{ppjj.pstex_t}
\end{center}
\caption{Diagrams contributing to the triple--diffractive 
production of two $J/\psi$ mesons in $p\pbar$ scattering
\label{ppjjfig}}
\end{figure}

The experimental setup of the Tevatron and the LHC should 
in principle also make it possible to observe multi--diffractive 
events. A very interesting process for studying the Odderon 
would then be the triple--diffractive production of two 
$J/\psi$ mesons, 
\be
p \pbar\to p+ J/\psi+ J/\psi +\pbar
\,,
\ee 
with the $+$ signs again indicating rapidity gaps between the 
particles in the final state. Again, this process can also 
be studied in $pp$ scattering, and also the production of 
two $\phi$ mesons or of one $J/\psi$ and one $\phi$ 
will be an equally good possibility. 
Also the breakup of the (anti)protons 
can be considered as long as the rapidity gaps are not filled 
by the breakup. At high energy this process is described by 
the two diagrams in figure \ref{ppjjfig}. 
The advantage of this process compared to the 
double--diffractive production of vector mesons can be seen 
in the first diagram in the figure: in this diagram the Odderon does not 
couple to the (anti)proton at all. This process can therefore 
be used to find the Odderon even if its coupling to the 
(anti)proton were extremely small for one of the reasons discussed 
in section \ref{Oddprotoncouplsect}. In that case 
one could even neglect the second diagram in figure 
\ref{ppjjfig} in the calculation of this process. For this process 
it would again be very valuable to know precisely the perturbative 
Pomeron--Odderon fusion process of figure \ref{pertpofusion} 
such that a fully perturbative calculation would become possible. 

In principle the triple--diffractive production of two  $J/\psi$ 
mesons would offer a very clean experimental signature, 
for example with the two $J/\psi$ mesons decaying into two 
lepton pairs of well--defined mass. 
So far this process has not yet been calculated nor even estimated. 
Even if this were done, there would still be the same uncertainties 
as in the case of the double--diffractive production, in particular 
the problem of the gap survival probability. The latter could even be 
enhanced for the triple--gap events. 
But with the high luminosity of the LHC there should be a very 
good chance of seeing this process, although the observation of 
triple--gap events still poses a serious experimental challenge. 

Finally, we should again point out that the potential uncertainties 
in predictions of the processes discussed in this section do not 
pose a problem for the detection of the Odderon. It is the mere 
observation of these processes that would be sufficient to 
establish the existence of the Odderon. 

\subsection{Electron-Proton Scattering}
\label{epsect}

The HERA machine at DESY has been very successful over the 
past decade in studying electron--proton scattering at high energies. 
One of the main early discoveries 
at HERA was the observation that a considerable fraction 
of about $15 \!-\! 20 \%$ of the events are diffractive, 
i.\,e.\ exhibit a large rapidity gap without hadronic 
activity in the detector. These events have been observed 
in photoproduction as well as in electroproduction 
processes. The main cause for the emergence of a rapidity 
gap is the exchange of a colorless object. 
Accordingly, the diffractive cross sections measured at HERA 
are well described in the framework of Pomeron exchange. 
Also Odderon exchange leads to a rapidity gap in the final 
state, and some (presumably small) fraction of the diffractive 
events should actually be caused by the Odderon. 
One class of such events should be characterized by a diffractively 
produced system which carries positive charge parity. In another 
class of events the diffractively produced final state does not 
have a definite charge parity. The latter events can be induced 
by both Pomeron and Odderon exchange. In the present 
section we will consider both possibilities and 
discuss several diffractive processes and asymmetries 
which offer a good chance of finding the Odderon at HERA 
in the near future. 

\subsubsection{Diffractive Processes}
\label{epdiffsect}

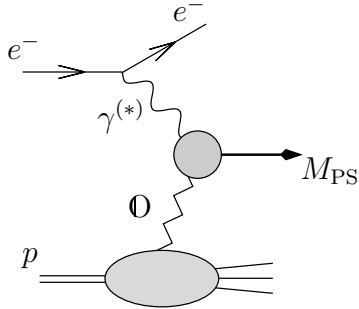
\begin{figure}[ht]
\vspace*{.4cm}
\begin{center}
\input{diffmes.pstex_t}
\end{center}
\caption{Diffractive production of a pseudoscalar meson in $ep$ scattering
\label{diffmesfig}}
\end{figure}

Here we will be concerned with exclusive diffractive 
$ep$ scattering processes in which the diffractively 
produced system carries 
positive charge parity $C=+1$. The real or virtual photon 
emitted by the electron carries negative $C$ parity and 
its transformation into a diffractive final state system of 
positive $C$ parity hence 
requires the $t$-channel exchange of an object of negative 
$C$ parity. Pomeron exchange thus cannot contribute to this 
process. It can only be mediated by the exchange of an 
Odderon, of a reggeon, or of a photon. 
The cleanest diffractive process involving Odderon exchange 
is the exclusive diffractive production of a single meson 
with positive charge parity. The mesons with the suitable 
quantum numbers are pseudoscalar and tensor mesons. 
For the case of Odderon exchange this process 
is illustrated in figure \ref{diffmesfig}. 
In addition to the diagram in figure \ref{diffmesfig} there is 
a similar diagram for photon exchange in which the Odderon 
is replaced by a photon. The reggeon contribution is very 
small at large energies and can often be neglected in this process. 
It turns out that photon exchange is important 
depending on the kinematical situation and needs 
to be considered when cross sections are calculated. In the 
present review our main focus is the Odderon and we will 
discuss in most cases only the Odderon contribution to the 
respective processes, having in mind that the photon contribution 
is under comparatively good theoretical control and is usually 
taken into account when necessary. 

The cross sections for these processes obviously depend 
on the masses of the mesons and on the virtuality of the 
photon. An early proposal for the study of these processes 
was made in \cite{Barakhovsky:ra}. But it was only 
more recently that the expected cross sections have been 
estimated in more detailed calculations. 
The largest cross sections are obviously expected for the 
production of light pseudoscalar and tensor 
mesons in photoproduction, 
i.\,e.\ in the case that the photon emitted from the electron is 
real. We will begin with this process. 

\begin{figure}[ht]
\begin{center}
\vspace*{.4cm}
\input{kina.pstex_t}
\end{center}
\caption{$p_T$ distribution for diffractive pion production 
in the photoproduction region according to 
\protect\cite{Kilian:1998ew}
\label{kina}}
\end{figure}
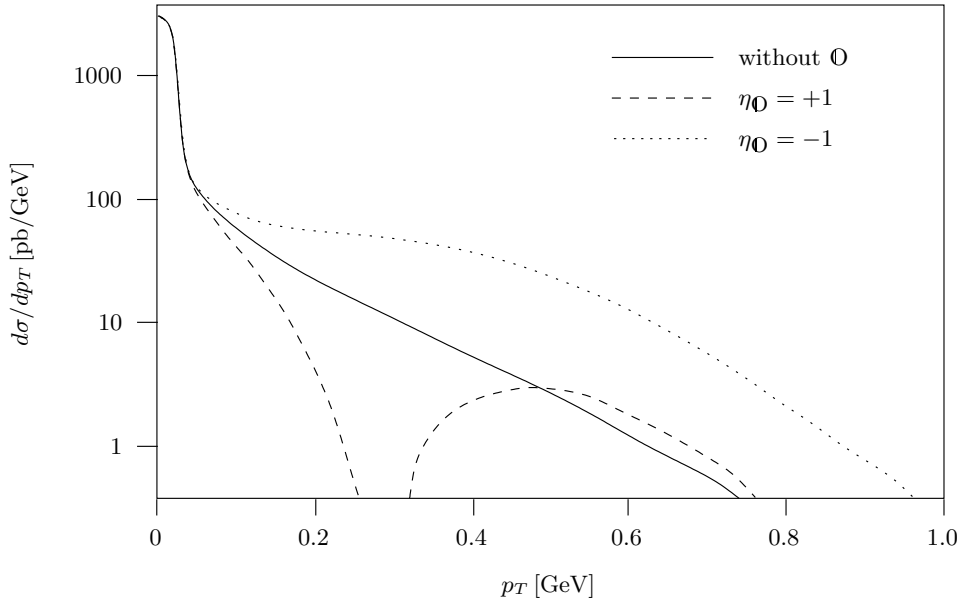

The diffractive photoproduction of light mesons is clearly 
a soft process which cannot be treated in perturbative QCD 
and requires the use of nonperturbative methods. 
The first investigation was done in the framework of 
Regge theory in \cite{Kilian:1998ew} where the 
production of $\pi^0$, $\eta$, $\eta'$, and also of $\eta_c$ 
mesons was considered. A $\gamma^\mu$-type coupling 
of the Odderon to the quarks in the meson and to the proton 
is used, see eqs.\ (\ref{oqcouplgamma}) and (\ref{reggeoddpcoupl}). 
The parameters in a calculation in the framework of Regge 
theory are rather difficult to determine without suitable 
data that could be used to determine them in other processes. 
The authors estimate the unknown Odderon couplings 
to the quarks in terms of those of the Pomeron, 
\be
\beta_\soddi^2 = 0.05 \, \beta_\spommi^2
\,,
\ee
and further assume an Odderon intercept of exactly one. 
Both possible values for the unknown Odderon phase 
$\eta_\soddi = \pm 1$ are considered. 
The authors then concentrate on the photoproduction 
region for which the cuts are usually chosen as 
\bea
\label{photoprodregion}
0.3 < &y& < 0.7 \\
\label{smallq2}
0<&Q^2& <0.01\,\mbox{GeV}^2 
\eea
for the fractional energy loss of the electron $y$ and 
for the photon virtuality $Q^2$. 
An interesting result is that in the case of pion production 
there is a characteristic 
difference between the transverse momentum ($p_T$) 
spectra of the pions for the two choices $\eta_\soddi = \pm 1$ 
for the Odderon phase, see figure \ref{kina}. 
For $\eta_\soddi = +1$ there is 
a prominent dip at around $p_T=0.3 \,\mbox{GeV}$ 
which is absent in the case $\eta_\soddi = -1$. 
The estimated cross sections are necessarily rather 
uncertain in this approach. In the photoproduction 
region the cross section for $\pi^0$ production is 
for instance found to be around $75\,\mbox{pb}$, and for 
$\eta_c$ it is $4\,\mbox{pb}$. Also the case of deep 
inelastic scattering has been studied in \cite{Kilian:1998ew}. 
If one keeps the constraint (\ref{photoprodregion}) 
and chooses the photon virtuality $Q^2> 1\,\mbox{GeV}^2$ 
one finds a reduction of the cross sections by about two 
orders of magnitude as compared to the photoproduction region. 

\begin{figure}[ht]
\vspace*{0.4cm}
\begin{center}
\input{rdn.pstex_t}
\end{center}
\caption{Total cross section for diffractive $\pi^0$ production 
as a function of the photon virtuality, shown for two different sizes 
for the diquark structure in the proton and for the case of 
proton dissociation; according to \protect\cite{Rueter:1998gj} 
\label{rdnfig}}
\end{figure}
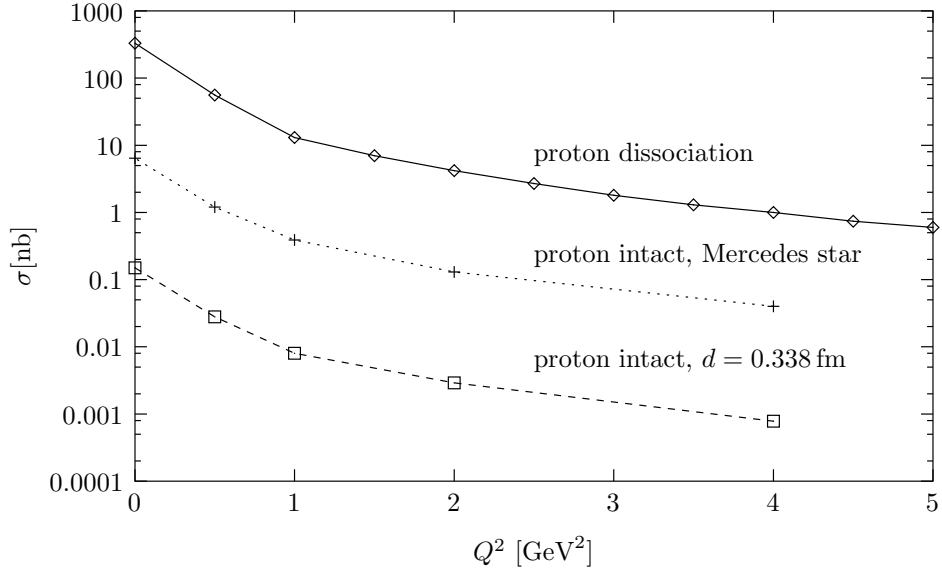

A more ambitious study was performed in 
\cite{Rueter:1998gj,Berger:1999ca,Berger:2000wt} 
where the cross sections for light $C=+1$ mesons 
were considered in the framework of the stochastic vacuum model. 
For the details of the formalism we refer to sections 
\ref{npheidelbergsect} and \ref{formalismssect} where 
the stochastic vacuum model has been described in detail. 
The main ingredients specific to the diffractive production 
of for example $\pi^0$ mesons are wave functions 
for the photon and pion which enter in (\ref{amplitlooploop}). 
The photon wave function is computed using light cone perturbation 
theory, and the pion wave function is modeled in transverse and 
longitudinal momentum space. It is assumed to be 
exponentially decreasing in the longitudinal momentum, 
and the transverse momentum dependence is assumed to 
behave as proposed by Wirbel, Stech and Bauer \cite{Wirbel:1985ji}. 
The proton is modeled as composed of a quark and 
a scalar diquark. Here one has in mind the apparent 
absence of an effect of the Odderon in the $\rho$ parameter, 
see section \ref{rhosect}, which can be explained \cite{Rueter:1996yb} 
by a diquark cluster of a size of about $0.3\,\mbox{fm}$ 
in the proton. Of course one has to test whether the 
assumption of a quark--diquark structure is essential 
for the results of the calculation. 
The  dependence of the diffractive production 
of $\pi^0$ mesons on the diquark size was studied in 
\cite{Rueter:1998gj}. 
As expected (see section \ref{Oddprotoncouplsect}) 
it was found that the cross sections are much lower for smaller 
diquark sizes. In addition, it was found that the cross sections 
are not suppressed due to the diquark clustering if the proton 
dissociates or is excited in the scattering process. 
This is illustrated in figure \ref{rdnfig} which shows the 
total cross section for exclusive diffractive $\pi^0$ production 
as a function of the photon virtuality $Q^2$. 
The elastic cross section is in fact at least a factor of 50 smaller 
than the inelastic cross section. This indicates that the assumption 
of a diquark cluster in the proton is in fact not very essential 
for processes in which the proton dissociates or is excited. 

\begin{figure}[ht]
\centering
\includegraphics[width=0.6\textwidth,clip]{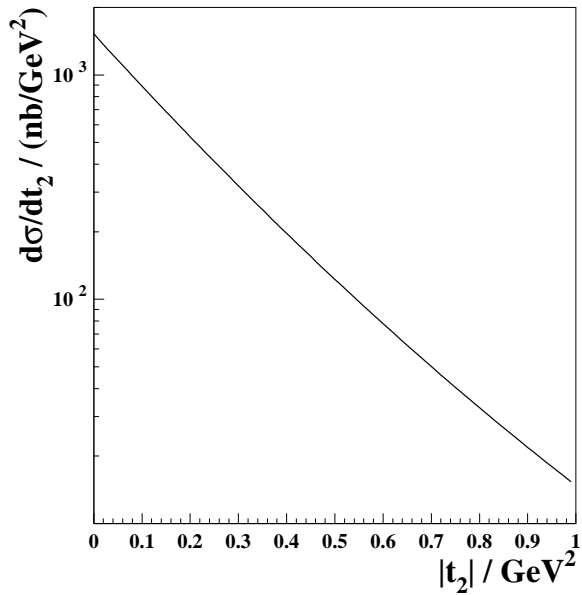}
\caption{Differential cross section for the process 
$\gamma p \to \pi^0 N^*$; figure from \protect\cite{Berger:1999ca} 
\label{tdepsvm}}
\end{figure}

\begin{figure}[ht]
\centering
\includegraphics[width=0.6\textwidth,clip]{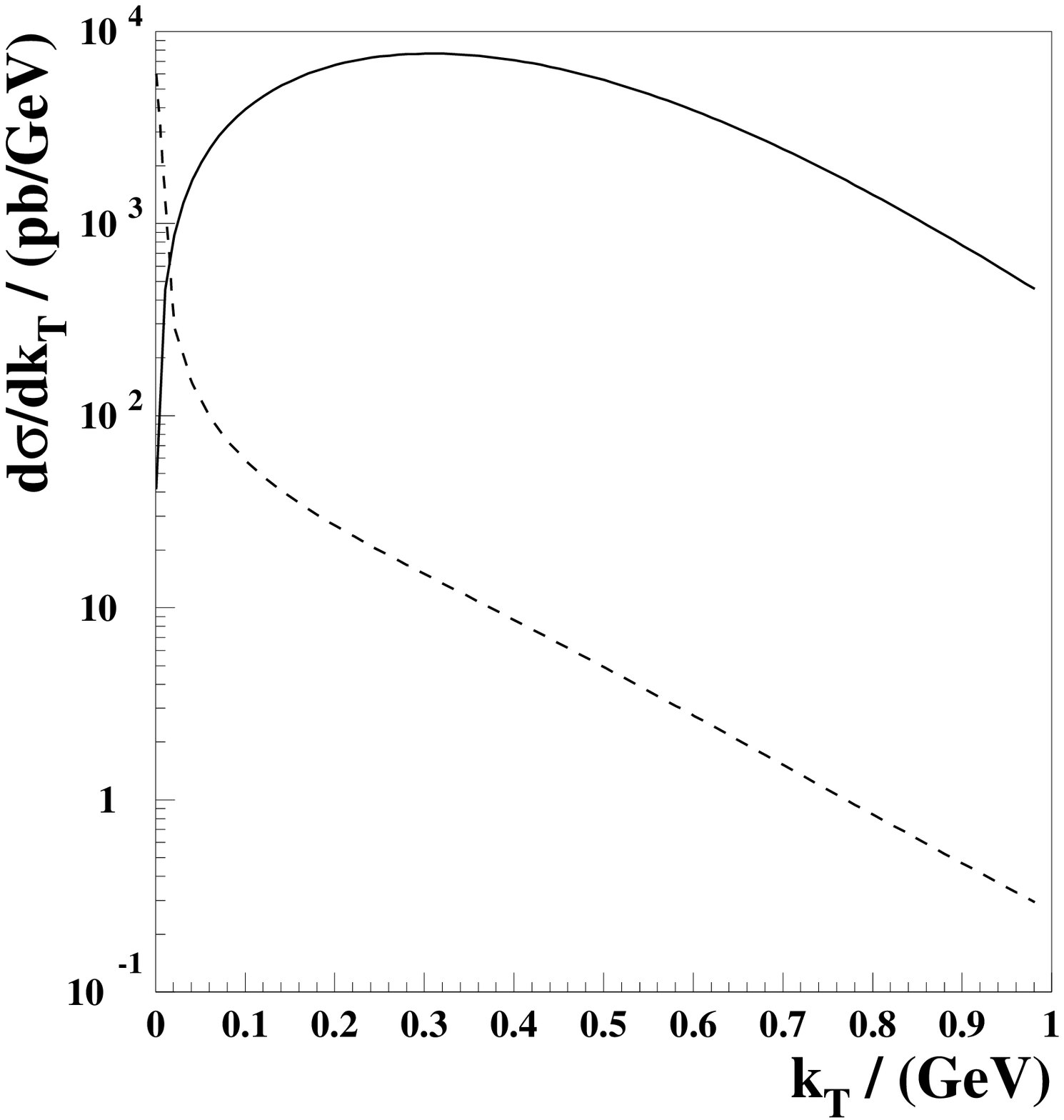}
\caption{Transverse momentum ($k_T$) distribution of the 
pion for Odderon exchange (solid line) and photon exchange 
(dashed line); figure from \protect\cite{Berger:1999ca} 
\label{ktdepsvm}}
\end{figure}

The production of pseudoscalar and tensor mesons in the 
photoproduction region (\ref{photoprodregion}), (\ref{smallq2}) 
was studied using the stochastic vacuum model 
in \cite{Berger:1999ca} and \cite{Berger:2000wt}, 
respectively. In particular, the process was considered for the 
excitation of the proton into the negative parity resonances 
$N(1520)$ and $N(1535)$, which are both 
compatible with the diquark picture. 
The differential cross section for 
$\gamma p \to \pi^0 N^*$ including these two resonances 
for $N^*$ is shown in figure \ref{tdepsvm}, 
where $t_2$ is the squared momentum transfer carried 
by the Odderon. 
It is interesting to note that the Odderon and photon contributions 
to the scattering process behave quite differently here. This can 
be seen in the differential distribution in the transverse momentum 
$k_T$ of the diffractively produced pion in figure 
\ref{ktdepsvm}. 
Interference contributions are not taken into account in this 
figure in order to make the comparison more transparent. 

\begin{figure}[ht]
\centering
\includegraphics[width=0.75\textwidth,clip]{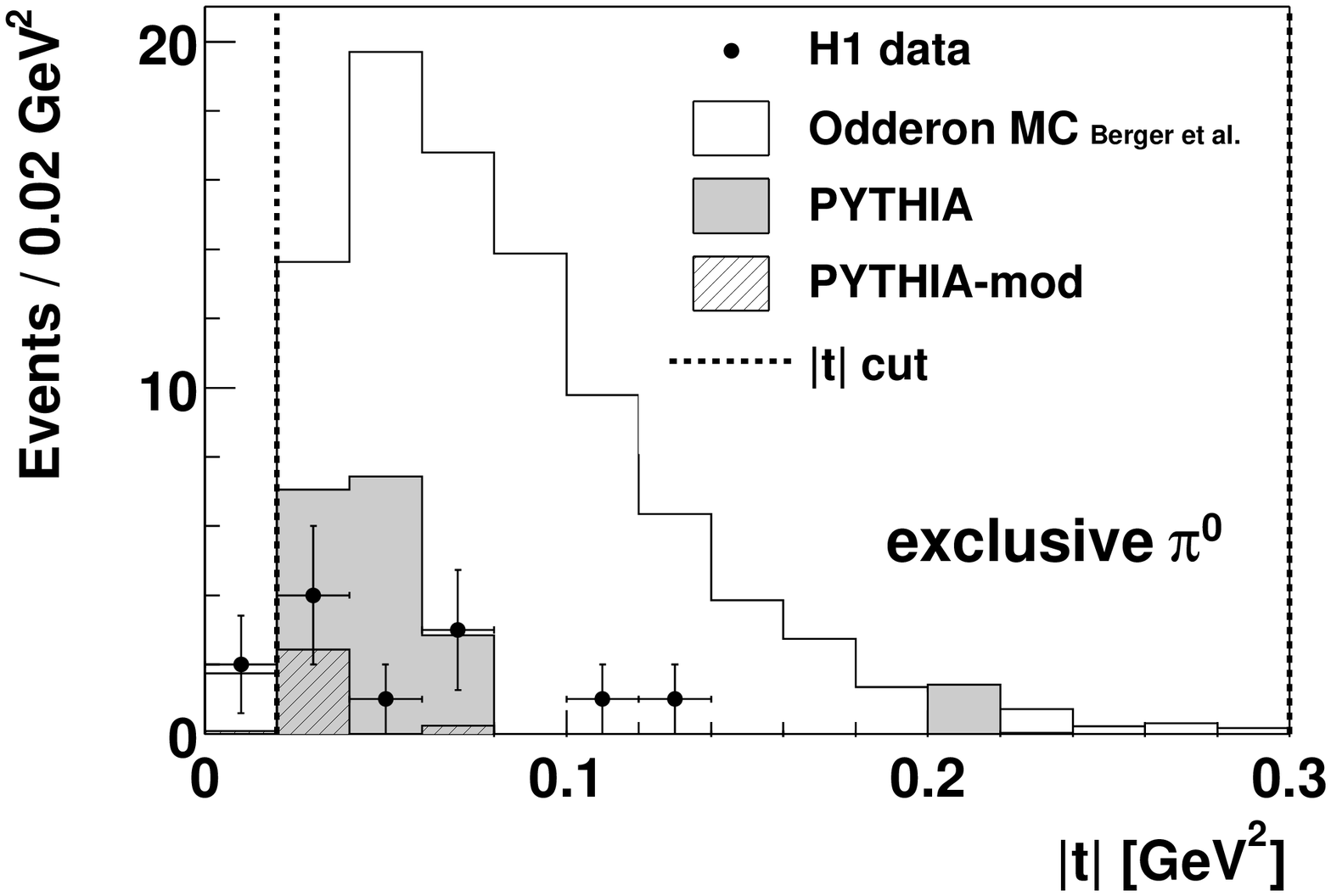}
\caption{Measured $t$-distribution for Odderon candidate 
events with $M_{\gamma \gamma}< 335\,\mbox{MeV}$; 
figure from \protect\cite{Adloff:2002dw}
\label{h1fig}}
\end{figure}

Some of the processes studied in 
\cite{Berger:1999ca} and \cite{Berger:2000wt} have 
recently been investigated by the H1 collaboration. 
The cross sections expected in the photoproduction 
region in the framework of the stochastic vacuum model 
for these processes are 
\bea
\label{pionprediction}
\sigma(\gamma p \to \pi^0 N^*) &=& 200 \,\mbox{nb} \\
\label{a2prediction}
\sigma(\gamma p \to a_2 N^*) &=& 190 \,\mbox{nb} \\
\label{f2prediction}
\sigma(\gamma p \to f_2 N^*) &=& 21 \,\mbox{nb} 
\,.
\eea
It should be pointed out that the authors of 
\cite{Berger:1999ca,Berger:2000wt} estimate the 
error in these predictions to be of the order of $50 \%$. 
The experimental search for this process uses the fact 
that the positive $C$ parity of the diffractively produced 
system forces the number of photons to be even if the 
system decays completely into photons, and it is checked for 
the case of Pomeron exchange processes that the detection of 
photons is as efficient as expected. 
The search for Odderon induced processes has so far 
not been successful. In \cite{Adloff:2002dw} 
an upper bound at $95 \%$ confidence level 
was determined for the case of pion production 
\be
\sigma(\gamma p \to \pi^0 N^*) < 49 \,\mbox{nb} 
\,.
\ee
This is far smaller than the expectation (\ref{pionprediction}). 
Figure \ref{h1fig} shows the measured $t$-distribution 
of $2 \gamma$ events which should originate from the 
decay of the pion. This is compared to a Monte Carlo simulation 
based on the expectations of \cite{Berger:1999ca}. 
The data are clearly in agreement with background 
expectations, and there is no sign of 
an Odderon induced production of pions. 
Preliminary data are also available for the production 
of $a_2$ and $f_2$ tensor mesons 
\cite{Olsson:2001nm,Golling:2001ju}. 
Also here no signal 
has been observed, and the upper bounds for the cross sections 
are 
\bea
\sigma(\gamma p \to a_2 N^*) &=& 96 \,\mbox{nb} \\
\sigma(\gamma p \to f_2 N^*) &=& 16 \,\mbox{nb} 
\,.
\eea
These bounds are again below the expectations (\ref{a2prediction}) 
and (\ref{f2prediction}), but not as significantly as in the case of 
$\pi^0$ mesons. In all three cases the reason for the failure 
of the prediction made in the framework of the stochastic vacuum 
model is unclear. In principle, there are uncertainties related 
to the couplings of the Odderon and to its energy dependence. 
It was pointed out in \cite{Adloff:2002dw} that the experimental 
bounds would be reconciled with the theoretical expectations 
if the latter are scaled down using an energy dependence of 
the Odderon given by an intercept of $\alpha_\soddi(0) =0.7$. 
As we have discussed in section \ref{nonpertoddsect} such 
an intercept is certainly possible for the soft Odderon. 
It would clearly seem premature to draw any firm 
conclusions concerning the existence of the Odderon 
from these experimental findings. One should keep in 
mind that the diffractive photoproduction of pions 
is a very soft process, and probably the most difficult 
Odderon mediated process to describe in QCD. 
It will be very interesting to see whether the experimental 
bounds can be improved in the case of the tensor mesons, 
where they are still in the range of the theoretical 
expectations. 

In view of the large uncertainties in the predictions for the 
photoproduction of light mesons it is natural to consider also 
cases which can be treated perturbatively. One possibility is 
the electroproduction of mesons of large or 
intermediate mass at sufficiently large photon virtuality. 
The other possibility is to concentrate on the diffractive 
production of heavy mesons like the $\eta_c$. In the latter 
case the meson mass provides a hard scale that makes a 
perturbative calculation possible even in the photoproduction 
limit. 

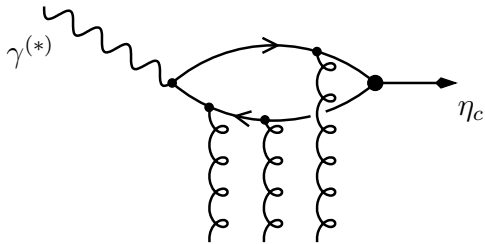
\begin{figure}[ht]
\vspace*{0.4cm}
\begin{center}
\input{etaimpact.pstex_t}
\caption{One of the diagrams contributing 
to the $\gamma^{(*)} \eta_c \Od$ impact factor\label{getaO}
}
\end{center}
\end{figure}

Such calculations have been performed for the photo- or 
electroproduction of $\eta_c$ mesons \cite{Bartels:2001hw}, 
\cite{Schafer:1992pq}--\cite{Engel:1997cg}, 
or the leptoproduction of $f_2(1270)$ or $\eta(548)$ mesons 
\cite{Ryskin:1998kt}. Here we will discuss as an example the 
case of $\eta_c$ mesons. Again we consider the exclusive 
diffractive production of these mesons with the constraint 
that the proton stays intact. 
The perturbative calculation involves as a new 
element the $\gamma \eta_c \Od$ impact factor. 
It was calculated for the 
first time in \cite{Czyzewski:1996bv}. It carries a vector index 
$\mu$ due to the photon entering it. It turns out that only 
its transverse components $\mu=i$ ($i=1,2$) are different from zero. 
One of the diagrams contributing to the impact factor is 
shown in figure \ref{getaO}. 
Again, the gluons have to be attached to the quark lines in all 
possible ways in order to ensure gauge invariance. 
Using again $\qf=\kf_1+\kf_2+\kf_3$ the impact factor 
becomes 
\be
\label{etaoimpact}
\phi_\gamma^i = b \,\epsilon_{ij} 
\frac{q_j}{\qf^2} 
\left(
\frac{\qf^2}{\qf^2 + Q^2 + 4 m_c^2} 
+ \sum_{k=1}^3 
\frac{(2 \kf_k -\qf) \cdot \qf}{(2\kf_k-\qf)^2 + Q^2 + 4 m_c^2} 
\right)
\,,
\ee
where
\be
b=\frac{16}{\pi} e_c g_s^3 \,\frac{1}{2}\,m_{\eta_c} b_0
\,,
\ee
with $m_{\eta_c}= 2.98\,\mbox{GeV}$, $m_c=1.4\,\mbox{GeV}$, 
and $b_0$ is related to the radiative width 
$\Gamma_{\eta_c \to \gamma \gamma} = 7\,\mbox{keV}$ 
of the $\eta_c$ meson, 
\be
b_0=\frac{16 \pi^3}{3 e_c^2} 
\sqrt{ \frac{\pi \Gamma_{\eta_c \to \gamma \gamma}}{m_{\eta_c}}} 
\,.
\ee
With this impact factor at hand one can use the usual perturbative 
framework as described in section \ref{formalismssect}. 
In \cite{Czyzewski:1996bv,Engel:1997cg} the Odderon is 
described by the perturbative exchange of three non--interacting 
gluons. The $\Od p$ coupling is described by the Fukugita--Kwieci\'nski form 
factor (\ref{eq:formfact}) with the choice $\alpha_s = 1$ for the 
strong coupling constant. The two papers use slightly different 
choices for $\alpha_s$ in the $\gamma \eta_c \Od$ impact factor. 
The calculations give total cross sections of around $11 \,\mbox{pb}$ 
for photoproduction and only $0.1\,\mbox{pb}$ at a photon 
virtuality of $Q^2= 25\,\mbox{GeV}^2$. These cross sections 
are energy independent due to the simple model of three 
non--interacting gluons for the Odderon. 

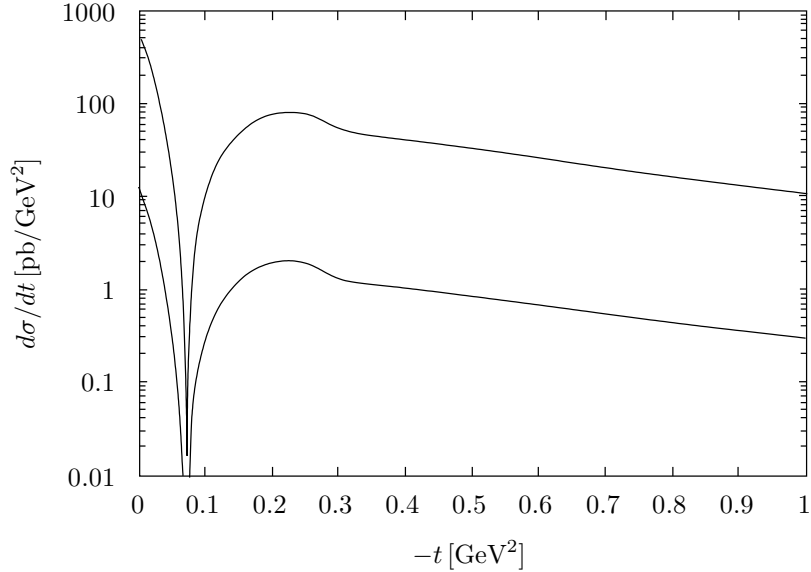
\begin{figure}[ht]
\vspace*{0.4cm}
\begin{center}
\input{etacfinal.pstex_t}
\caption{The differential cross sections 
for exclusive diffractive $\eta_c$ production 
according to \protect\cite{Bartels:2001hw}. 
The upper curve refers to $Q^2=0$, the lower curve to 
$Q^2= 25\,\mbox{GeV}^2$. 
\label{vaccafig}
}
\end{center}
\end{figure}

In \cite{Bartels:2001hw} the same process was calculated using 
a solution of the BKP equation for the Odderon. This was the 
first phenomenological application of a resummed Odderon solution. 
The authors first noticed that one cannot use the 
Janik--Wosiek solution in this process because its coupling to the 
$\gamma \eta_c \Od$ impact factor (\ref{etaoimpact})
vanishes. This is due to the 
special dependence of the impact factor on the transverse momenta 
of the gluons. One can easily see that in each term of the impact 
factor two gluon momenta enter only as a sum. The Fourier 
transformation of such a term to impact parameter space 
yields a delta function of the difference of the 
two corresponding gluon coordinates. But this delta function 
immediately implies that any Odderon solution of the form 
(\ref{odderonwavefct}) vanishes when convoluted with 
this impact factor, and this applies in particular to the 
Janik--Wosiek solution. We should add here that this 
decoupling of the JW solution from the $\gamma \eta_c \Od$ impact 
factor according to the argument just given holds only in leading 
order, i.\,e.\ in the approximation that the impact factor is represented 
by a quark loop diagram, see figure \ref{getaO}. In higher order 
diagrams there will be gluon corrections to the loop diagram, and 
these can give a nonvanishing contribution to the $\gamma \eta_c \Od$ 
impact factor. The situation turns out to be different for the 
Bartels--Lipatov--Vacca solution of the BKP equation. 
This solution does couple to the $\gamma \eta_c \Od$ impact 
factor (\ref{etaoimpact}) already in leading order. 
However, it has an intercept of exactly one and one would hence 
expect only a small effect in comparison to the exchange 
of three non--interacting gluons. 
Surprisingly though, it was found that 
using exactly the same parameters as in \cite{Czyzewski:1996bv} 
the cross section for diffractive $\eta_c$ production 
is enhanced by about an order of magnitude due to the use of the 
resummed Odderon solution. The total cross section now 
becomes $\sim 50\,\mbox{pb}$ for photoproduction, and 
$1.3\,\mbox{pb}$ at a photon virtuality of $Q^2= 25\,\mbox{GeV}$. 
The corresponding differential cross section is shown in 
figure \ref{vaccafig}. 
The dip in the differential cross section occurs only when the 
BLV solution for the Odderon is used, but is absent in the 
calculations \cite{Czyzewski:1996bv,Engel:1997cg} 
where the Odderon is modeled by three non-interacting gluons. 
In all three calculations the differential cross section vanishes 
in the exact forward direction $t=0$, but starts to decrease 
only at very small $|t|$ such that this property is not visible 
in figure \ref{vaccafig}. Due to that fact the cross section is 
at small $|t|$ dominated by photon exchange. The very forward 
direction is hence not well suited for finding the Odderon in this 
process.  

It should be pointed out that all three perturbative calculations of this 
process have been performed assuming a coupling constant 
$\alpha_s=1$ in the $\Od p$ impact factor. According to the 
findings of \cite{Dosch:2002ai} this is a very optimistic choice. 
Calculating the differential cross section 
for elastic $pp$ scattering in the dip region, 
which corresponds to an at least similar kinematical situation, 
one finds that this value if far too large, and $\alpha_s \simeq 0.3$ 
would be more appropriate, 
see figure \ref{fig:impactdiffalpha} and the corresponding 
discussion in section \ref{elasticppsect}. 
We recall here that the difference between these two values for $\alpha_s$ 
causes a large difference for the cross section since the squared 
Odderon--proton impact factor entering it 
is proportional to $\alpha_s^3$, and the more realistic choice 
of $\alpha_s \simeq 0.3$ would reduce the cross section by 
a factor of about $30$. With this more realistic estimate the 
perturbatively calculated cross sections are 
in a range similar to the estimate obtained using the 
framework of Regge theory in \cite{Kilian:1998ew}, see above. 

In an alternative approach one can make use of the large mass of 
the charm quark to describe the diffractive production of $\eta_c$ 
mesons using techniques typical for treating heavy quarks. 
This approach has been pursued in \cite{Ma:2003py} where the 
photoproduction region of this process was considered. The $\eta_c$ 
meson is treated in the framework of nonrelativistic QCD, and 
heavy quark effective theory is used to describe the exchange of 
soft gluons. Then a systematic expansion in inverse powers of the 
charm mass $m_c$ can be applied. The cross section can be 
expressed in terms of four functions parametrizing the 
matrix element of a twist-3 gluon operator taken between 
the ingoing and outgoing proton states. These four functions 
are of nonperturbative origin and are presently not 
known numerically. A rough estimate of the diffractive $\eta_c$ 
photoproduction cross section is possible though when the process 
is compared to the diffractive photoproduction of $J/\psi$ mesons, 
which has been studied in the same framework in \cite{Ma:2003py}. 
Such a rough estimate gives a differential cross section of 
$2\,\mbox{nb}/\mbox{GeV}^2$ in the forward direction $t=0$ 
at HERA. This is at least an order of magnitude larger than the 
values obtained in the perturbative approach 
\cite{Czyzewski:1996bv,Engel:1997cg,Bartels:2001hw}, 
and may be even larger when a smaller coupling in the 
Odderon--proton impact factor is chosen in those calculations. 
Interestingly, it is found in the approach of \cite{Ma:2003py} that in 
contrast to the perturbative results the cross section for this process 
due to Odderon exchange does not vanish in the limit $t=0$. 

Even with an optimistic choice of $\alpha_s$ the cross section 
for diffractive $\eta_c$ production is unfortunately 
too small to be observed at HERA. 
Nevertheless, the investigation of this process is very 
instructive for the following reasons. It has been found here that 
the $\gamma \Od \eta_c$ impact factor has a very special 
dependence on the three gluon momenta. As a consequence, 
it does not couple to the Janik--Wosiek solution of the BKP 
equation for the Odderon. But it does couple to the 
Bartels--Lipatov--Vacca solution. This observation is 
quite relevant for the interpretation of different types of solutions 
of the BKP equation, and in fact it probably implies that the 
Hilbert space for admissible solutions of the BKP equations 
has to be extended as compared to previous assumptions, 
see also section \ref{blvsolutionsect}. 
In addition, it is a very interesting result that 
the inclusion of the resummed Odderon solution 
increases the cross section as 
compared to the exchange of three non--interacting 
gluons. This is very unexpected since the BLV solution used 
here has an intercept exactly equal to one, i.\,e.\ has the same 
intercept as the exchange of three non--interacting gluons. 
It is conceivable that the use of resummed 
Odderon solutions might in general lead to higher cross 
sections also in other processes. It would be very important 
to study this question in more detail. 

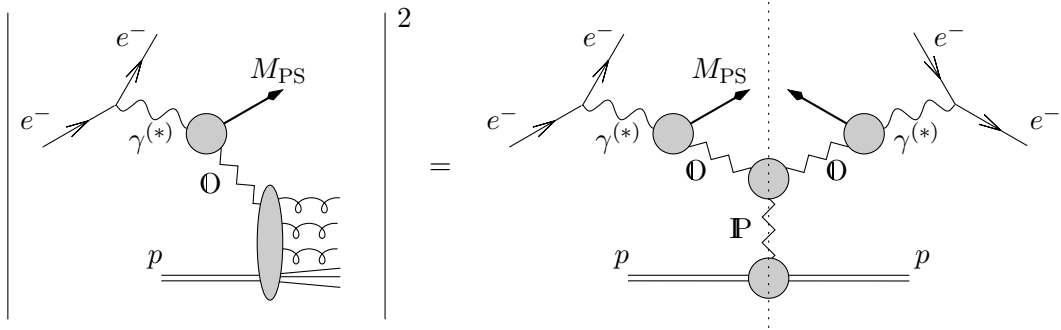
\begin{figure}[ht]
\vspace*{.4cm}
\begin{center}
\input{poocombine.pstex_t}
\end{center}
\caption{The Pomeron--Odderon--Odderon vertex in diffractive 
production of pseudoscalar mesons with a breakup of the proton into 
a large--mass system
\label{poomesonfig}}
\end{figure}

So far we have considered the diffractive production of 
pseudoscalar and tensor mesons under the assumption that 
the proton stays intact or is excited into 
a small--mass state. A very interesting possibility is the other 
extreme, namely the case in which the proton breaks up into 
a large--mass system $X$. The breakup of the proton into 
a large--mass system certainly reduces the probability of 
having a rapidity gap, but there should still be a considerable fraction 
of events left which have a rapidity gap in the final state 
between the meson and the large--mass proton remnant, 
ensuring that the reaction is mediated by Odderon exchange. 
The corresponding amplitude is shown on the left hand 
side of figure \ref{poomesonfig}. 
The large--mass system is generated mainly by gluon 
emissions. If one now squares this amplitude to obtain 
the cross section for this process (right hand side in figure 
\ref{poomesonfig}) one finds that the cross section 
involves the Pomeron--Odderon--Odderon ($\P \Od \Od$) 
vertex. The possibility of observing the $\P \Od \Od$ vertex 
in this process has first been mentioned in \cite{Barakhovsky:ra}. 
An advantage of this process is that it does not involve the 
poorly known coupling of the Odderon to the proton. Instead 
it involves the coupling of the two Odderons to the Pomeron 
and the comparatively well--known coupling of the Pomeron 
to the proton. If the diffractively produced meson is heavy, 
one can use perturbation theory 
to calculate the cross section. In particular, one can use 
the perturbative $\P \Od \Od$ vertex \cite{Bartels:1999aw}, 
see section \ref{poosect}. 
Note also that the cross section is in this case expected to be enhanced 
at large energies due to the occurrence of a hard Pomeron. 
The case of inclusive diffractive $\eta_c$ production, 
$\gamma^{(*)} + p \to \eta_c + X$, has been studied in \cite{Bartels:2003zu}. 
Here the BLV solution for the perturbative Odderon has been used, 
and the perturbative $\P \Od \Od$ vertex for this solution is 
calculated in the large-$N_c$ limit. 
The coupling of the Pomeron to the proton can be determined using 
a Pomeron--based fit of the gluon density of the proton at small $x$. 
A suitable scale for the strong coupling in the $\P \Od \Od$ vertex is 
estimated based on the diffusion mechanism of the momenta 
in the perturbative Pomeron, see section \ref{bfklapplicabsection}. 
In addition to the diagram in figure \ref{poomesonfig} (representing 
the triple-Regge contribution) there 
is another contribution to this process in which the three--gluon 
states from the $\gamma \Od \eta_c$ 
impact factors are coupled directly to the Pomeron. This second 
contribution comes from the part of the six--gluon amplitude $D_6$ 
which reggeizes, see section \ref{egllasect}. Also this contribution 
is computed in the large-$N_c$ limit based on the results of 
\cite{Bartels:1999aw}. Both contributions are of the same order 
of magnitude but behave differently with the transferred momentum. 
It should be pointed out that the cross section is large in the 
region of phase space in which the Pomeron fills the whole 
rapidity range between the proton and the $\eta_c$ meson. Therefore 
it would be necessary to apply suitable cuts on the rapidity in order to 
focus only on processes where there is an Odderon, 
i.\,e.\ a gap between the $\eta_c$ meson 
and the proton remnant $X$. In \cite{Bartels:2003zu} the total 
cross section integrated over the mass of the system $X$ is 
calculated for the kinematics of HERA ($\sqrt{s}=300\,\mbox{GeV}$). 
In the case of photoproduction the total cross section is 
found to be $65\,\mbox{pb}$, whereas at $Q^2=25\,\mbox{GeV}^2$ 
it is only $1.5\,\mbox{pb}$. This however includes contributions 
with only a small or no gap between the $\eta_c$ and the system $X$, 
and it should hence only be interpreted as an estimate of the order 
of magnitude. As expected these cross sections are 
larger than those for the quasi--elastic process 
$\gamma^{(*)} + p \to \eta_c + p$ 
in which the proton stays intact. The difference is even more 
pronounced when a realistic value for $\alpha_s$ is chosen in that 
process, as has been discussed above. 

\subsubsection{Asymmetries}
\label{epasymmetrsect}

The processes discussed in the preceding section typically 
involve the exchange of an Odderon on the amplitude 
level. The corresponding cross sections are thus 
quadratic in the factor describing the Odderon exchange 
and in the couplings of the Odderon to the external 
particles. Both of these dependencies imply some 
uncertainties, in particular when we are dealing with 
soft processes. 
There might be 
unexpected effects hidden in the Odderon couplings 
as well as in the Odderon propagation, 
as we have seen in the case of diffractive 
photoproduction of $\pi^0$ mesons where 
the theoretical estimates turned out to be rather poor. 
It is therefore desirable to consider also observables 
in which these uncertainties are reduced. 
A class of such observables has been proposed 
recently. These observables make use of the fact that 
the Odderon can interfere with the Pomeron in 
suitably defined asymmetries of final state particles 
if the final state does not have a well--defined charge parity, 
for example if the final state consists of a pair of charged pions 
which can be in a $C=+1$ state as well as in a $C=-1$ 
state. The asymmetry is then proportional to the product 
of the amplitudes for Pomeron and for Odderon exchange, 
and thus linear in the Odderon amplitude. Since the Pomeron 
is a comparatively well--understood object such asymmetries 
are expected to have a much smaller uncertainty caused 
by the unknowns of the Odderon. 
Moreover, some of the asymmetries discussed so far 
are estimated to be large enough to offer a substantial 
discovery potential for the Odderon in these 
observables at HERA. 

\begin{figure}[ht]
\vspace*{.4cm}
\begin{center}
\input{diffcc.pstex_t}
\end{center}
\caption{Mechanism for the diffractive production of a 
$c\bar{c}$ pair
\label{diffccfig}}
\end{figure}
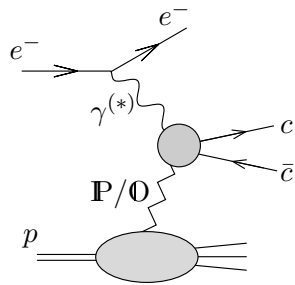

The first example of such an asymmetry was proposed 
in \cite{Brodsky:1999mz}. We will use it to explain the 
basic idea behind the Pomeron--Odderon interference. 
Consider the diffractive production of a charm 
quark--antiquark pair as sketched in figure \ref{diffccfig}. 
Here the photon can be either real or virtual. To be specific 
we consider the case of real photons in which also the 
expected cross sections are larger. 
A $c\bar{c}$ pair as such is not necessarily an eigenstate 
of charge conjugation $C$. 
The pair can be produced in a $C=-1$ state 
via Pomeron exchange as well as in a $C=+1$ state 
via Odderon exchange. 
We denote the amplitudes for the two production 
mechanism as $A_\spommi$ and $A_\soddi$, respectively. 
In the calculation of total cross sections the two amplitudes 
do obviously not interfere. However, on the level of the 
differential cross section there is still an interference 
term, 
\be
d\sigma = 
|A|^2 = |A_\spommi + A_\soddi|^2 = 
|A_\spommi|^2 + 2 \real (A_\spommi A_\soddi^*) + |A_\soddi|^2 
\,.
\ee
The idea is to isolate this interference term in suitable asymmetries 
defined through differential cross sections. 
A first look at the interference term tells us 
that it is expected to be small. The 
reason lies in the fact that Pomeron exchange gives a predominantly 
imaginary contribution to the amplitude while Odderon exchange 
gives a predominantly real contribution. Hence the two contributions 
are almost orthogonal to each other and the interference term is small. 
As we have seen in section \ref{crossoddsect} this orthogonality of 
the two contributions is even exact in 
the case of two Regge poles for the Pomeron and Odderon with 
exactly the same intercept. We therefore expect that the 
asymmetries constructed from the interference term vanish in 
the limit $\alpha_\soddi \to \alpha_\spommi$. 
However, it turns out that this cancellation is effective only on the 
parton level, as was observed in \cite{Ivanov:2001zc}. 
If one considers asymmetries on the hadron level (see the example 
of di--pion production below) there are additional Breit--Wigner 
phases associated with the formation of the hadrons which can 
eliminate the cancellation. This should be especially important 
for light hadrons, and probably would have little effect in 
charm production. In addition, we do not expect the 
Pomeron and Odderon intercepts to be equal and hence one can 
hope to see an effect also in the process of diffractive open charm 
production, to which we now return. 

\begin{figure}[ht]
\centering
\includegraphics[width=0.6\textwidth,clip]{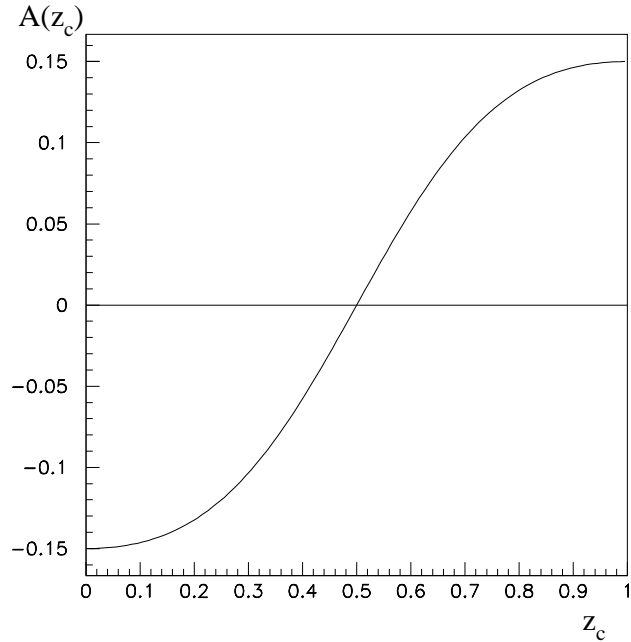}
\caption{The asymmetry in fractional energy $z_c$ of the 
charm versus the anticharm jets for $s_{\gamma p}/M_X^2 =100$; 
figure from \protect\cite{Brodsky:1999mz}
\label{brodskyfig}}
\end{figure}

In \cite{Brodsky:1999mz} it was shown that the charge asymmetry 
\be
\label{brodskyasymm}
\cA (t,M_x^2,z_c) = 
\frac{{\displaystyle
\frac{d\sigma}{dt \, dM_X^2 \, dz_c}
-\frac{d\sigma}{dt \, dM_X^2 \, dz_{\bar{c}}}}
}{{\displaystyle 
\frac{d\sigma}{dt \, dM_X^2 \, dz_c}
+\frac{d\sigma}{dt \, dM_X^2 \, dz_{\bar{c}}}}
}
\ee
is in fact proportional to the interference term of the Pomeron and 
Odderon amplitudes. Here $M_X$ is the invariant mass of the 
charm quark--antiquark pair, and $z_c$ ($z_{\bar{c}}$) denotes 
the momentum sharing of the $c\bar{c}$ pair. 
Denoting the photon momentum by $q$ and the proton momentum 
by $p$, and the charm quark (antiquark) momentum by 
$p_c$ ($p_{\bar{c}}$) we have 
\be
z_c = \frac{p_c \cdot p}{q\cdot p}
\,,
\ee
and an analogous relation for the antiquark. $z_c + z_{\bar{c}} = 1$ 
holds in Born approximation on the parton level. We are interested 
in the kinematic region of a large energy 
of the $\gamma p$ system, $s_{\gamma p} \gg M_X^2$. 
The authors of \cite{Brodsky:1999mz} use a Regge picture of the 
Pomeron and Odderon to estimate the asymmetry 
(\ref{brodskyasymm}) which can then be expressed as 
\be
\label{brodskylong}
\cA (t,M_x^2,z_c) = 
\frac{{\displaystyle
g^\spommi_{pp'} g^\soddi_{pp'} 
\left(\frac{s_{\gamma p}}{M_X^2}\right)^{\alpha_\tpommi+\alpha_\toddi}
\frac{
2 \sin\left[\frac{\pi}{2} (\alpha_\toddi - \alpha_\tpommi) \right]
}{
\sin \frac{\pi \alpha_\tpommi}{2} \cos \frac{\pi \alpha_\toddi}{2}} 
\,\,g_\spommi^{\gamma c \bar{c}} g_\soddi^{\gamma c \bar{c}} 
}}{{\displaystyle
\left[
g^\spommi_{pp'}\left(\frac{s_{\gamma p}}{M_X^2}\right)^{\alpha_\tpommi}
g_\spommi^{\gamma c \bar{c}}/ \sin \frac{\pi \alpha_\tpommi}{2} 
\right]^2 
+
\left[
g^\soddi_{pp'}\left(\frac{s_{\gamma p}}{M_X^2}\right)^{\alpha_\toddi}
g_\soddi^{\gamma c \bar{c}}/ \cos \frac{\pi \alpha_\toddi}{2} 
\right]^2 
}}
\,,
\ee
where the meaning of the different couplings 
$g_{\spommi/ \soddi}$ should be evident. We see that as expected 
the asymmetry is proportional to 
$\sin [ (\alpha_\soddi - \alpha_\spommi) \pi/2]$ and 
thus vanishes if the two intercepts are equal. 
One now assumes that the Pomeron and the Odderon 
are coupled to the charm quark and antiquark individually 
and to the single quarks in the proton according to 
(\ref{oqcouplgamma}) and (\ref{reggeoddpcoupl}), 
respectively. The dependence of the asymmetry on 
$z_c$ and $z_{\bar{c}}$ enters while 
coupling the Pomeron and Odderon 
separately to the charm quark and antiquark. 
The authors then further consider $t \simeq 0$ and 
use for the (hard) Pomeron and Odderon intercepts the values 
$\alpha_\spommi =1.2$ and $\alpha_\soddi =0.95$, respectively. . 
They estimate the relative couplings of the Pomeron and 
Odderon to the proton as $g^\soddi_{pp'}/g^\spommi_{pp'}=0.1$, 
and assume for their relative coupling to the charm quark 
$\kappa_\soddi^{\gamma c\bar{c}}/\kappa_\spommi^{\gamma c\bar{c}}=0.6$. 
Then the asymmetry becomes 
\be
\cA (t\simeq 0,M_x^2,z_c) \simeq 
0.45 \, \left(\frac{s_{\gamma p}}{M_X^2}\right)^{-0.25} 
\frac{2 z_c -1}{z_c^2 + (1-z_c)^2}
\,.
\ee
For a typical value of $s_{\gamma p}/M_X^2 =100$ this results 
in a $\sim 15 \%$ asymmetry for large $z_c$ as illustrated in 
figure \ref{brodskyfig}. 
It should be emphasized that the estimates for the different couplings 
made here are quite uncertain. Experimentally, the identification 
of charm quarks is nontrivial, and the measurement of the 
fractional energy of charm jets in diffraction is very challenging. 
Therefore the charge asymmetry for diffractive open charm 
production is probably not the best possible asymmetry for actually 
finding an Odderon effect. But it is conceptually the simplest, and 
therefore we have chosen to present it in some detail here. 

Better prospects for an actual discovery of an Odderon--Pomeron 
interference effect at HERA can be expected from the exclusive 
diffractive production of charged pion pairs. 
The process is in principle similar to the production of open 
charm in figure \ref{diffccfig}, but now one considers 
the production of a $\pi^+\pi^-$ pair instead of a $c\bar{c}$ 
pair. Again, the pion pair can be produced in a $C=+1$ state 
via Odderon exchange as well as in a $C=-1$ state via 
Pomeron exchange. One can construct different asymmetries 
involving the interference term of the two corresponding 
amplitudes. 
Asymmetries in this process have been studied both in 
electroproduction \cite{Hagler:2002nh,Hagler:2002nf} and 
in photoproduction \cite{Ivanov:2001zc,Ginzburg:2002zd}. 
The advantage of photoproduction is clearly the much larger 
cross section for this process. In electroproduction, on the other 
hand, the large photon virtuality allows one to calculate the 
process in a perturbative approach which is under better 
theoretical control. 

\begin{figure}[ht]
\vspace*{0.4cm}
\centering
\includegraphics[width=0.7\textwidth,clip]{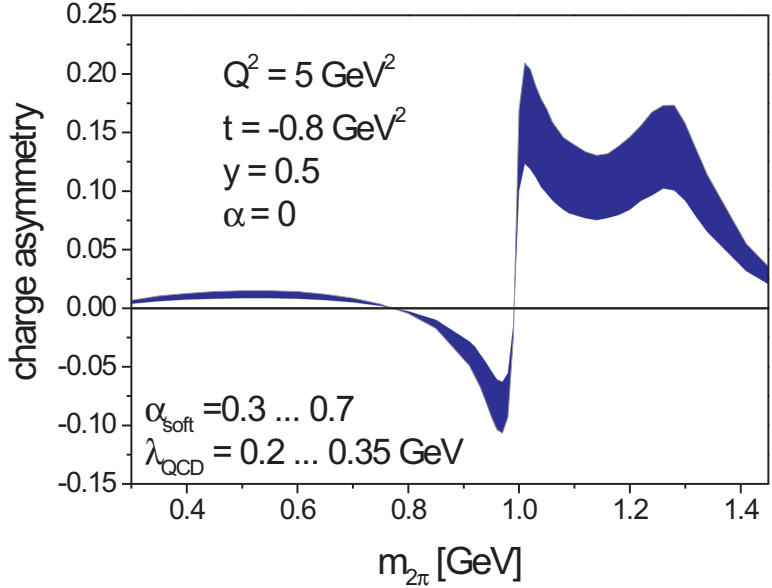}
\caption{
Charge asymmetry in the diffractive 
electroproduction of charged pion pairs at $Q^2=5\,\mbox{GeV}^2$ 
as a function of the dipion mass $m_{2 \pi}$; 
figure from \protect\cite{Hagler:2002nf}
\label{haeglercharge}}
\end{figure}

Let us first consider the case of electroproduction. 
In \cite{Hagler:2002nh} only the contribution of longitudinally 
polarized photons was considered which dominate at very 
large photon virtualities $Q^2$. In this approximation the 
charge asymmetry of the pion pair was calculated. 
In \cite{Hagler:2002nf} this calculation was improved by also 
taking into account the contribution of transversely polarized photons. 
With the latter contribution it was also possible to calculate a
spin asymmetry which only occurs as an interference 
effect between longitudinally and transversely polarized photons. 
The calculation of the Pomeron and Odderon amplitudes 
is done using the perturbative exchange of two or three 
non--interacting gluons, respectively. The coupling of the 
Odderon to the proton is modeled using the 
Fukugita--Kwieci\'nski form factor (\ref{eq:formfact}), 
and a similar model is used for the Pomeron--proton impact 
factor. The $\gamma^* \Od \pi^+ \pi^-$ and 
$\gamma^* \P \pi^+ \pi^-$ impact factors are calculated 
by coupling the three or two perturbative gluons to a quark--antiquark 
pair in all possible ways, and the resulting expression is then 
convoluted with appropriate generalized two pion distribution 
amplitudes (GDAs) \cite{Diehl:1998dk}--\cite{Diehl:2000uv}. 
They describe the transition of the $q\bar{q}$ pair into the 
$\pi^+ \pi^-$ final state. The one used in the Odderon amplitude 
involves the $s$ and $d$ wave contributions 
corresponding to $f_0$ and $f_2$ meson poles in the $\pi\pi$ 
elastic amplitude. The GDA for the Pomeron amplitude involves 
the $p$ wave described by the phase shift and 
Breit--Wigner distribution for the $\rho$ meson. 
The GDAs are nonperturbative objects which contain the full 
information about the production of the pion pair in the final 
state. Accordingly, there is a large theoretical uncertainty 
related to these amplitudes which is rather difficult to estimate. 
The charge asymmetry is defined as 
\be
{\cal A_C} (Q^2,t,m_{\pi^+\pi^-}^2,y,\alpha) =
\frac{ {\displaystyle
\sum_{\lambda =\pm}
\int \cos \theta \, 
d\sigma (s,Q^2,t,m_{\pi^+\pi^-}^2,y,\alpha,\theta,\lambda)
}
}{ {\displaystyle 
\sum_{\lambda =\pm}
\int d\sigma (s,Q^2,t,m_{\pi^+\pi^-}^2,y,\alpha,\theta,\lambda) 
}
}
\,,
\ee
where $s$ is the squared energy of the photon--proton system, 
$y$ is the fractional energy loss of the electron, $\alpha$ is the 
angle between the electron scattering plane and the hadronic 
scattering plane, $\theta$ is 
the polar angle of the $\pi^+$ in the dipion rest frame, 
and $\lambda$ is the initial electron helicity. 
Note that in the approximation used in 
\cite{Hagler:2002nh,Hagler:2002nf} the charge asymmetry 
${\cal A_C}$ is independent of $\sqrt{s}$ at sufficiently large energies. 
In figure \ref{haeglercharge} we show the numerical results 
\cite{Hagler:2002nf} for the charge asymmetry as a 
function of the two--pion invariant mass and for 
values for the other parameters which are typical for 
the HERA kinematic region. 
The error band corresponds to a variation of 
$\Lambda_{\mbox{\scriptsize QCD}}$ and of the strong 
coupling constant in the $\Od p$ impact factor. 
Here the values of the strong coupling were chosen in a 
range compatible with the results obtained for the 
Odderon--proton coupling obtained in the description of 
$pp$ elastic scattering in \cite{Dosch:2002ai}, see section 
\ref{elasticppsect}. It was found that the size of the 
charge asymmetry decreases with increasing 
photon virtuality $Q^2$. When going from $Q^2=3\,\mbox{GeV}^2$ 
to $Q^2=10\,\mbox{GeV}^2$, for example, the charge asymmetry 
varies roughly in the range of the band in figure \ref{haeglercharge}. 
The shape of the asymmetry as a function of the dipion mass, 
however, is almost invariant under changes in $Q^2$. 
Depending on the dipion mass the charge asymmetry is sizable and 
in a range that could well be accessible experimentally. 
The shape of the charge distribution can largely be understood 
in terms of the $\pi \pi$ phase shifts. 
It is largest around the $f_0$ and $f_2$ masses. 
Especially the zeros of the charge asymmetry should be 
robust against changes in most of the parameters. 

\begin{figure}[ht]
\vspace*{0.4cm}
\centering
\includegraphics[width=0.7\textwidth,clip]{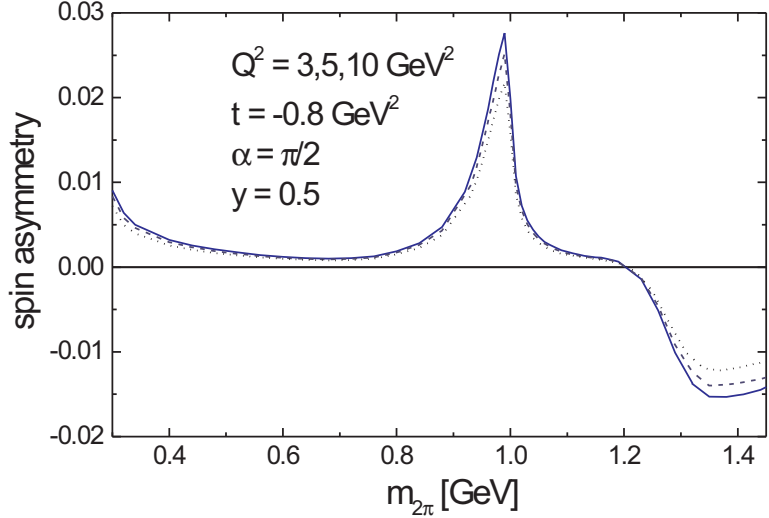}
\caption{Single spin asymmetry in the diffractive 
electroproduction of charged pion pairs 
as a function of the dipion mass for different photon 
virtualities $Q^2$; 
figure from \protect\cite{Hagler:2002nf}
\label{haglerspin}}
\end{figure}

The single spin asymmetry 
\be
{\cal A_S} (Q^2,t,m_{\pi^+\pi^-}^2,y,\alpha) =
\frac{ {\displaystyle
\sum_{\lambda =\pm} \lambda
\int \cos \theta \, 
d\sigma (s,Q^2,t,m_{\pi^+\pi^-}^2,y,\alpha,\theta,\lambda)
}}{{\displaystyle 
\sum_{\lambda =\pm}
\int d\sigma(s,Q^2,t,m_{\pi^+\pi^-}^2,y,\alpha,\theta,\lambda)
}}
\,.
\ee
has been calculated in the same way, and its dependence on the 
dipion mass is shown in figure \ref{haglerspin}. 
The figure clearly shows that the spin asymmetry depends 
only very weakly on the photon virtuality. 
The spin asymmetry is considerably smaller than the charge 
asymmetry. An experimental verification of the Odderon 
in the spin asymmetry will hence be rather difficult. 

A similar analysis for the diffractive photoproduction 
of a charged pion pair was performed in 
\cite{Ivanov:2001zc,Ginzburg:2002zd} in the framework 
of Regge theory. Again it is crucial to observe that the 
Pomeron amplitude for the production of a pion pair is 
dominated by the intermediate formation of a $\rho$ 
meson, whereas the Odderon amplitude is dominated 
by the $f_2(1270)$ resonance in the region above 
$\sim 1\,\mbox{GeV}$. Accordingly, the result can 
be expressed in terms of an overlap function involving 
the Breit--Wigner distributions of the respective resonances. 
An important ingredient is the production cross section 
for $f_2$ mesons, $\sigma(\gamma p \to f_2 p)$, which is 
assumed to be larger than $1\,\mbox{nb}$. 
In addition to the charge asymmetry the authors of 
\cite{Ginzburg:2002zd} study also 
the transverse asymmetry and a forward--backward 
asymmetry. They conclude that especially the charge 
asymmetry should lead to a very significant effect 
in the dipion mass region of $1.1 \! - \! 1.5\,\mbox{GeV}$. 
It should again be emphasized that in the case of diffractive 
dipion production the orthogonality of the Odderon and 
Pomeron contributions at the parton level is lifted when 
going to the hadron level due to the additional Breit--Wigner 
phases occurring in the hadronization. 
In a sense the large expected 
effects in the charge asymmetry can thus be viewed as a 
maximal violation of parton--hadron duality. 
One should here be able to observe an effect on the hadron 
level which is almost absent on the parton level. 

\subsection{Photon-Photon Processes}
\label{gammagammasect}

In all processes that we have considered so far one of the 
largest uncertainties comes from the coupling of the 
Odderon to the proton which is of nonperturbative 
nature and cannot be calculated from first principles. 
Moreover, it is well possible that this coupling is very 
small, and different possible mechanism for this have been 
discussed in section \ref{Oddprotoncouplsect}. 
The processes which we want to discuss now are not 
affected by this problem. Moreover, they can in many 
cases be calculated perturbatively. Therefore 
photon--photon processes can be considered the cleanest 
possible test of the Odderon, at least from a theoretical 
perspective. 

Photon--photon scattering occurs as a 
subprocess in electron--positron collisions when both 
the electron and the positron radiate a photon, 
\be
e^+ e^- \longrightarrow e^+ e^- \gamma^{(*)} \gamma^{(*)} 
\,,
\ee
with the subsequent reaction of the two photons. Here the 
photons can be both real or virtual, or one can be real 
and the other one can be virtual. In the following we will 
only consider the subsequent scattering of the two photons. 
If the electron or positron 
are tagged in the detector one is in a position to reconstruct 
the four--momenta of the photons and their virtuality. 
By implementing suitable cuts on the photon virtualities in 
tagged events one can select only events in which both 
photons are virtual. 

The process of interest for the Odderon is the quasi--diffractive 
production of pseudoscalar or tensor mesons. Here it is possible 
that two pseudoscalar or tensor mesons are produced, 
\bea
 \gamma^{(*)} \gamma^{(*)} &\longrightarrow &
M_{\mbox{\scriptsize PS }} + M_{\mbox{\scriptsize PS }}
\\
\gamma^{(*)} \gamma^{(*)} &\longrightarrow &
M_{\mbox{\scriptsize T}} + M_{\mbox{\scriptsize T }}
\,,
\eea
where for the pseudoscalar mesons we can have for example 
$M_{\mbox{\scriptsize PS }}= \pi$, $\eta$, $\eta'$, or $\eta_c$. 
For the tensor mesons one can consider for instance 
$M_{\mbox{\scriptsize T}}= a_2$ or $f_2$. 
It is also possible that only one pseudoscalar or tensor 
meson is produced quasidiffractively while the other 
photon dissociates into a hadronic system $X$ of small mass, 
\be
\gamma^{(*)} \gamma^{(*)} \longrightarrow 
M_{\mbox{\scriptsize PS/T}} + X
\,.
\ee
In all of these processes the momentum transfer should be 
small compared to the energy of the two--photon system. 
In this case all of the above processes can only be caused by 
Odderon or photon exchange, and the exchange of photons 
is theoretically well--understood. Due to the quantum numbers 
of the produced mesons Pomeron exchange is excluded. 

In the case of heavy meson production, like for example 
for $\eta_c$ mesons, the meson mass provides a hard scale 
in the process and one can apply perturbation theory even 
in the case of real photons. 
For the production of light mesons one can in double--tagged 
events choose to consider only the scattering of two highly 
virtual photons such that again perturbation theory can 
be applied. In theses cases the Odderon can safely be 
modeled by the exchange of three gluons. So far this has been 
done only using three non--interacting gluons, but in an 
improved calculation one could also make use of the 
known solutions of the BKP equation. 

Early studies of the above processes have been performed in 
\cite{Barakhovsky:ra,Ginzburg:gy,Ginzburg:1991hd}. 
The processes of interest have a rather small cross section 
in particular in situations where we can apply perturbation 
theory. This is either due to the restricted phase space for 
the production of heavy mesons, or due to the strong suppression 
of photon fluxes when one goes from real to virtual photons. 
Unfortunately, the energy of the LEP collider was not 
sufficient to study the Odderon in photon--photon interactions 
there. Accordingly, the interest in these processes was low 
for quite some time. Renewed interest was raised in the light 
of plans for a future linear $e^+e^-$-collider with a center--of-mass 
energy of $\sqrt{s}= 500\,\mbox{GeV}$ with a very high luminosity. 
Such a machine would in fact offer excellent opportunities for studying 
high energy QCD. Some processes of interest for Pomeron physics 
have already been mentioned in section \ref{bfklapplicabsection}. 
For the Odderon it is likely that all of the processes mentioned above 
can be observed at such a linear collider. 

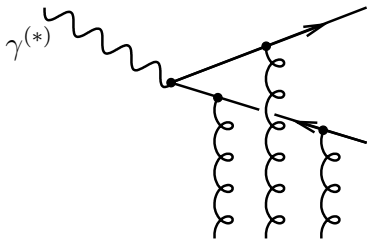
\begin{figure}[ht]
\vspace*{0.4cm}
\begin{center}
\input{gximpact.pstex_t}
\caption{One of the diagrams contributing 
to the $\gamma^{(*)} X \Od$ impact factor
\label{goximpact}}
\end{center}
\end{figure}

Let us consider as an example the production of $\eta_c$ mesons 
as it has been studied in \cite{Motyka:1998kb}. 
The amplitude for $\gamma \gamma \to \eta_c \eta_c$ is calculated 
perturbatively according to (\ref{Aoddifact}) and the Odderon 
is modeled by the exchange of three non--interacting gluons according 
to (\ref{phioddisimple}). 
The $\gamma^{(*)} \eta_c \Od$ 
impact factors are calculated as described in section \ref{epdiffsect}, 
see figure \ref{getaO}. In the case of the reaction 
$\gamma \gamma \to \eta_c X$ one has to use in addition the 
$\gamma X \Od$ impact factor which can also be computed 
perturbatively. One of the diagrams for this impact factor is shown 
in figure \ref{goximpact}, 
and here again the gluons have to be attached to the quark lines 
in all possible ways in order to arrive at a gauge invariant expression. 

\begin{figure}[ht]
\centering
\hbox{
\epsfxsize = 7cm
\epsfysize = 7cm
\epsfbox[30 230 520 730]{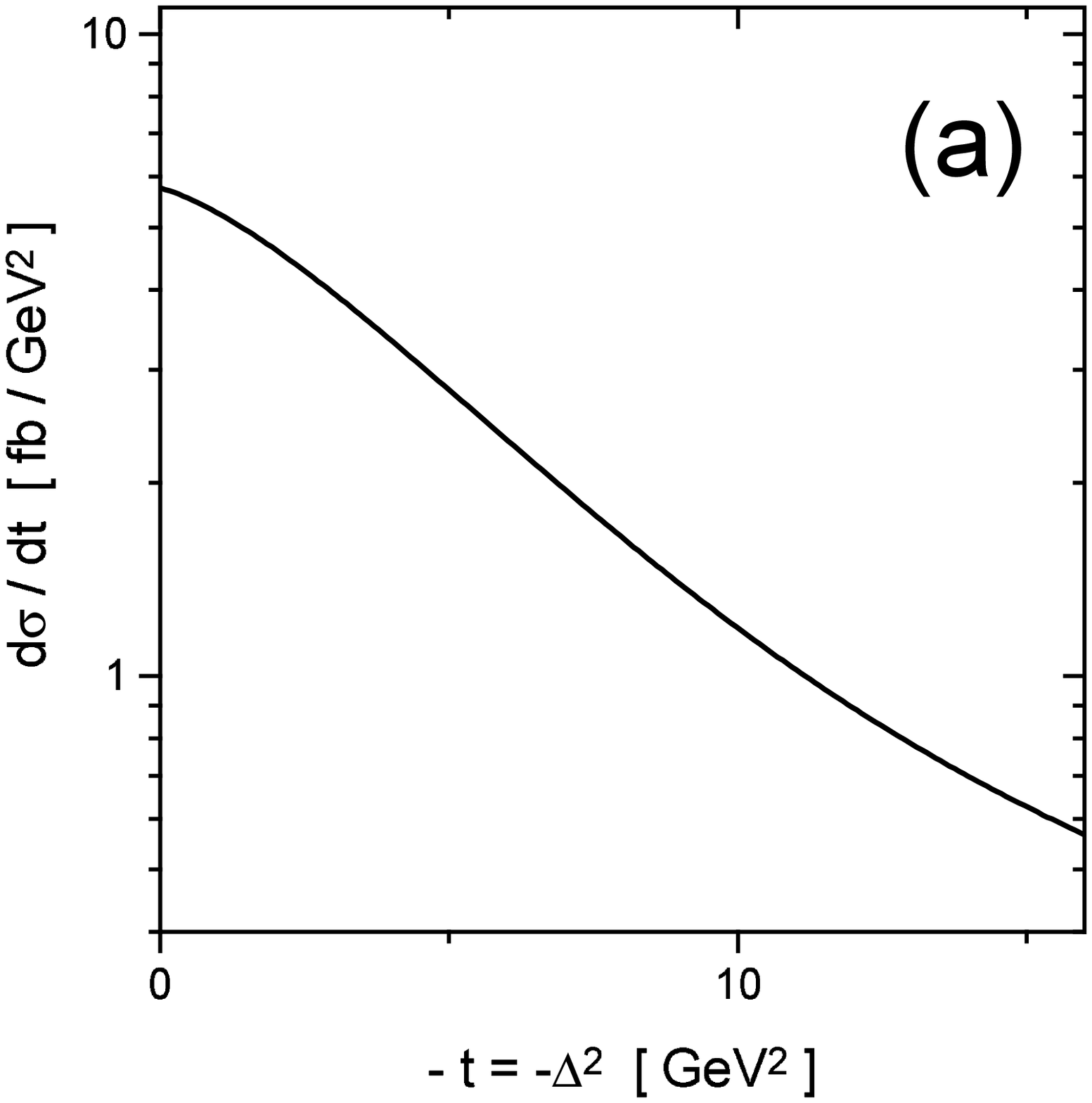}
\epsfxsize = 7cm
\epsfysize = 7cm
\epsfbox[30 230 520 730]{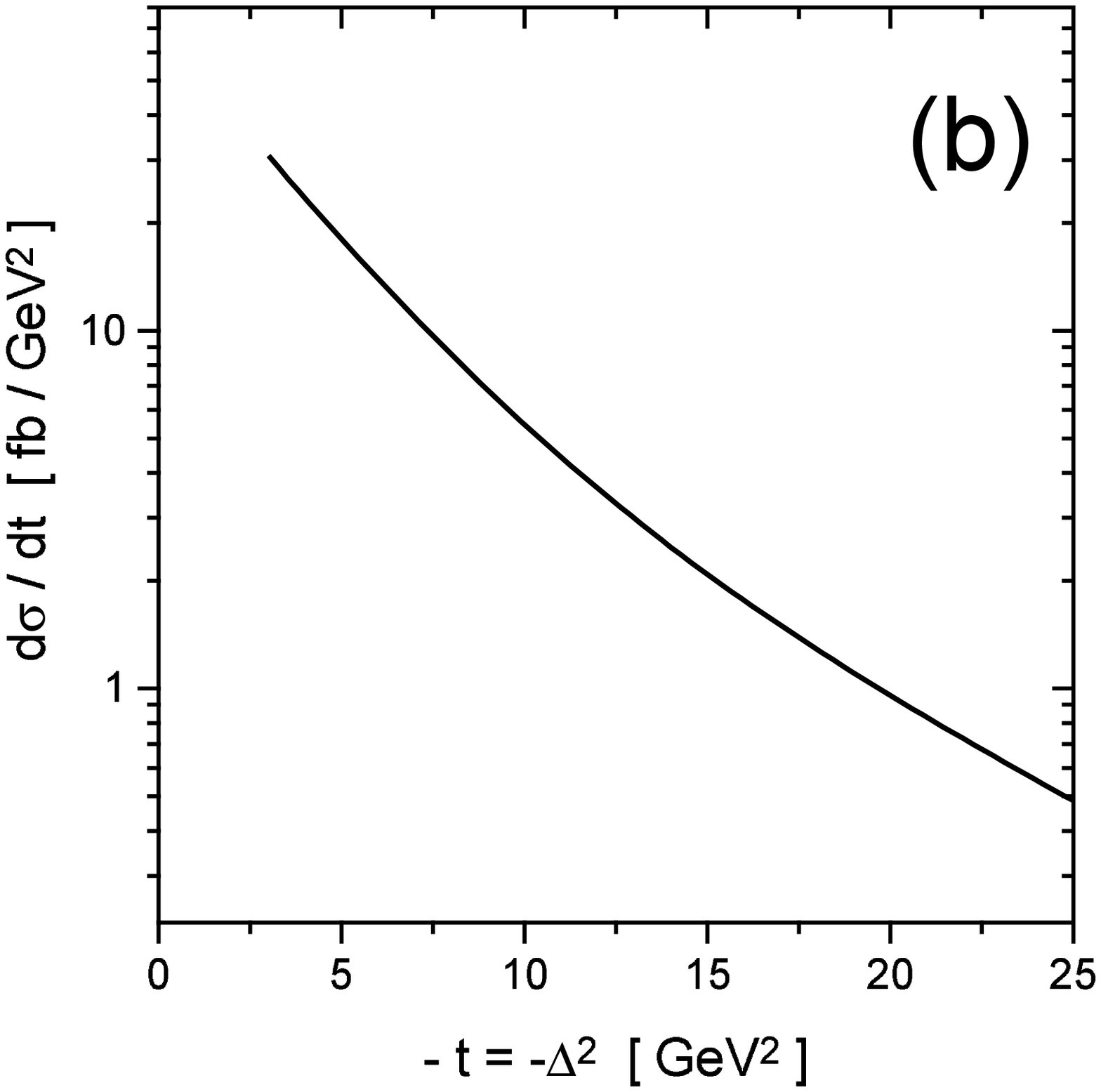}
}                                                                               
\caption{Differential cross section for the processes 
$\gamma \gamma \to \eta_c \eta_c$ (a) and 
$\gamma \gamma \to \eta_c X$ (b); 
figure from \protect\cite{Motyka:1998kb} 
\label{ggetaetafig}}
\end{figure}

In \cite{Motyka:1998kb} the case of real photons is considered. 
The differential cross sections for the two processes due to Odderon 
exchange are shown in figure \ref{ggetaetafig}. 
The total cross sections for the two processes are 
$\sigma_T^{\eta_c \eta_c}= 43\,\mbox{fb}$, 
and 
$\sigma_T^{\eta_c X} (|t|>3\,\mbox{GeV}^2) = 120\,\mbox{fb}$, 
respectively. 
These values are obtained assuming $\alpha_s(m_c^2)=0.38$, 
and the authors of \cite{Motyka:1998kb} 
assume that the strong coupling is given by 
$\tilde{\alpha}_s=0.3$ in the $\gamma^{(*)} X \Od$ impact factor. 
We recall that the cross sections obtained using a simple three--gluon 
exchange model of the Odderon do not depend on the energy. 
For untagged events in which the electron and positron are 
scattered under a very small angle $\theta < 30 \,\mbox{mrad}$ 
in the detector one finds for a $\sqrt{s}=500 \,\mbox{GeV}$ 
Linear Collider cross sections of 
$\sigma(e^+ e^-\to e^+e^- \eta_c \eta_c) = 3.5\,\mbox{fb}$, 
and 
$\sigma(e^+ e^-\to e^+ e^- \eta_c X) = 10\,\mbox{fb}$. 

Finally, let us point out that photon--photon processes offer 
also the possibility of studying the Pomeron--Odderon--Odderon 
vertex. This is in complete analogy to the occurrence of this 
vertex in the diffractive production of pseudoscalar mesons 
in $ep$ collisions when the proton dissociates into a large--mass 
system, see figure \ref{poomesonfig} and the corresponding 
discussion in section \ref{epdiffsect}. 
In photon--photon collisions the analogous process is the 
quasidiffractive production of pseudoscalar or tensor mesons
\be
 \gamma^{(*)}  \gamma^{(*)} 
\longrightarrow M_{\mbox{\scriptsize PS/T}} + X
\,,
\ee
with $X$ now being a large--mass system separated from the 
meson by a large rapidity gap. The photons can again be 
either real or virtual. For the case of highly virtual photons 
or the production of a heavy meson one can apply perturbation 
theory. Hence it would be possible to measure in these processes 
the perturbative $\P \Od \Od$ vertex \cite{Bartels:1999aw} 
(see section \ref{poosect}). Again it is likely that the cross section 
for the production of a large--mass system $X$ is enhanced as 
compared to a small--mass system $X$. 

\section{Conclusions}
\label{summary}

The Odderon is an interesting object. 
While initially widely regarded only as a rather theoretical 
possibility in the framework of Regge theory 
its existence is now considered almost unavoidable 
if our QCD--based picture of high energy scattering is correct. 
This dramatic change in the general opinion on the Odderon 
was mainly caused by the impressive progress that has been 
made recently in perturbative studies of the Odderon. 
These results clearly indicated the existence of an Odderon in 
high energy scattering. 

The perturbative Odderon can be understood as the $t$-channel exchange 
of three interacting reggeized gluons. In the leading logarithmic 
approximation it is described by the BKP equation which is 
the analogue of the BFKL equation for the Pomeron. The BKP 
equation has been solved in the leading logarithmic approximation. 
The intercept of the Odderon has been found to be very close to 
one or exactly one, depending on the scattering process. 
The study of the BKP Odderon has led to amazing 
and unexpected results which go far beyond the Odderon problem 
itself. The BKP Odderon Hamiltonian has been found to be 
equivalent to an integrable model, namely the XXX Heisenberg 
model of $\mbox{SL}(2,\C)$ spin zero. 
It was possible to extend these results also to the large-$N_c$ 
limit of  $t$-channel exchanges with arbitrary numbers of reggeized 
gluons. The problem, also in this wider context, has been 
found to be closely related to two--dimensional conformal field 
theory, to the theory of elliptic functions, and to string theory. 
To understand the Regge limit of hadronic scattering processes 
in the framework of QCD is one of the longstanding and 
most difficult problems in high energy physics. These new results 
have opened new and promising ways to approach that problem, 
and we are, at least in perturbation theory, beginning to understand 
how Pomerons and Odderons behave and interact. However, 
a full understanding of the Regge limit based only on perturbative 
QCD will not be possible, and eventually nonperturbative effects 
will play an important r\^ole. Here the theoretical progress is 
still very slow. Our knowledge of the Pomeron and the Odderon 
in nonperturbative QCD is rather limited, and we have to retract 
to models for describing soft processes. 

Experimentally, the Odderon has proven to be an elusive object. 
For a long time the search for the Odderon concentrated on differences 
between particle and antiparticle cross sections, in particular in $pp$ and 
$p\pbar$ scattering, as well as on the $\rho$ parameter. This was 
motivated by the discovery of a sign of the Odderon in the differential 
cross sections for $pp$ and $p\pbar$ elastic scattering at the CERN ISR, 
which to date remains the only experimental evidence for the Odderon. 
But exactly in these processes the Odderon is particularly difficult 
to find because it is just one of many contributions to the amplitude 
most of which are not a priori known and need to be fitted. 
The identification of the Odderon in these processes almost 
unavoidably remains ambiguous, especially in view of the 
rather limited amount and precision of the available data. 
Much better for finding the Odderon are processes in which it is (besides 
the photon) the 
only possible exchange, and a variety of them has been suggested recently. 
These processes or observables fall into two classes. One is the 
diffractive production of mesons with suitable quantum numbers 
which cannot be produced via Pomeron exchange. The other one 
contains asymmetries which originate from Pomeron--Odderon 
interference effects. Remarkably, such processes exist for 
all types of particle collisions that can be studied at present and 
future high--energy colliders. 

The phenomenological estimates for Odderon--induced 
processes still involve considerable uncertainties. They mainly originate 
from the unknown intercept of the Odderon in soft processes and from 
the couplings of the Odderon to external particles which can only 
be modeled in most cases. The situation is complicated by the 
fact that, in contrast to the Pomeron, there are no or only very 
few experimental data that could guide us in developing better 
phenomenological models of the Odderon and its couplings. 
Different mechanism have been proposed 
which can potentially cause a suppression of Odderon effects 
in certain kinematical situations, for example in soft processes 
or in processes at small momentum transfer. In forward 
scattering, for example, a low Odderon intercept could 
easily explain a strong suppression of Odderon--mediated 
processes. It is also feasible 
that the Odderon couples only very weakly to protons, and also 
here different mechanism appear possible. 
Finally, it would not be surprising if the Odderon looked different 
in different reactions and in different kinematical situations. 
The actual or apparent absence of the Odderon in some reactions 
or observables does therefore not imply that it does not exist. 
If the Odderon cannot be found in hard processes, in particular at 
large momentum transfer, however, we would be faced with a 
real puzzle. It is therefore vital for a good understanding of the 
Odderon to look for it in all possible reactions. 

There are many open questions related to the Odderon. 
The investigation of the perturbative Odderon is currently 
a very active field. It is likely that further relations 
of the Odderon to integrable models, conformal field theory 
and string theory will be found. Another interesting aspect 
is the study of the BKP Odderon in next--to--leading 
logarithmic approximation, as it has been done for the Pomeron. 
There the corrections turned out to be large, and the same 
might hold for the Odderon. Clearly, this would also be 
of phenomenological relevance. A very important 
but also very difficult problem is to improve our understanding 
of nonperturbative QCD which plays in important r\^ole 
in almost all phenomenological applications of the Odderon. 
The phenomenology of the Odderon in general 
is in many respects still 
in the early stage. We have tried to indicate open problems 
and possible directions throughout this review. 
A very big step forward would certainly be possible as soon 
as the Odderon is observed experimentally. Precise data 
on one of the processes involving only Odderon exchange 
would immediately reduce many uncertainties of the 
present phenomenological predictions. 

The first and foremost question remains, of course, 
whether the Odderon really exists. 
A positive answer to this question was eagerly awaited 
from the first attempt to observe an exclusive 
processes which can only be caused by Odderon exchange. 
But a search for diffractive pion production in $ep$ 
collisions by the H1 collaboration at HERA did not show 
any sign of the Odderon. This result is rather surprising, 
and its cause is difficult to identify at the moment. 
Despite this disappointing result the chances of finding 
the Odderon are very good. The search can be continued in a 
variety of processes at all current and future collider experiments. 
The Odderon is an important piece in our theoretical 
understanding of high energy QCD. Although it seems to 
be hiding I am optimistic that it can and will be found. 

\section*{Acknowledgements}
I would like to thank my collaborators J.\ Bartels, H.\,G.\ Dosch, 
and V.\ Schatz for numerous most helpful discussions on subjects 
related to this report and for the pleasant working atmosphere. 
For instructive discussions I am very grateful to 
G.\ Kor\-chemsky, J.\ Kwieci{\'n}ski, 
L.\,N.\ Lipatov, O.\ Nachtmann, R.\ Peschanski, 
G.\ Salam, and G.\,P.\ Vacca. 
I would like to express my gratitude to T.\ Berndt, S.\ Munier, and 
V.\ Schatz for valuable comments on the 
manuscript. 
Finally, I am especially grateful to  H.\,G.\ Dosch and to O.\ Nachtmann 
for their encouragement to write this report and for their comments 
on the manuscript.

\end{document}

%% file: reggeabcd.pstex_t
\begin{picture}(0,0)%
\includegraphics{reggeabcd.pstex}%
\end{picture}%
\setlength{\unitlength}{4144sp}%
\begingroup\makeatletter\ifx\SetFigFont\undefined%
\gdef\SetFigFont#1#2#3#4#5{%
  \reset@font\fontsize{#1}{#2pt}%
  \fontfamily{#3}\fontseries{#4}\fontshape{#5}%
  \selectfont}%
\fi\endgroup%
\begin{picture}(2082,1194)(889,-1063)
\put(1306,-1051){\makebox(0,0)[lb]{\smash{\SetFigFont{12}{14.4}{\rmdefault}{\mddefault}{\updefault}$b$}}}
\put(1306,-376){\makebox(0,0)[lb]{\smash{\SetFigFont{12}{14.4}{\rmdefault}{\mddefault}{\updefault}$a$}}}
\put(2971,-1051){\makebox(0,0)[lb]{\smash{\SetFigFont{12}{14.4}{\rmdefault}{\mddefault}{\updefault}$d$}}}
\put(2971,-376){\makebox(0,0)[lb]{\smash{\SetFigFont{12}{14.4}{\rmdefault}{\mddefault}{\updefault}$c$}}}
\put(901,-601){\makebox(0,0)[lb]{\smash{\SetFigFont{12}{14.4}{\rmdefault}{\mddefault}{\updefault}$s$}}}
\put(2251,-61){\makebox(0,0)[lb]{\smash{\SetFigFont{12}{14.4}{\rmdefault}{\mddefault}{\updefault}$t$}}}
\end{picture}

%% file: reggefact.pstex_t
\begin{picture}(0,0)%
\includegraphics{reggefact.pstex}%
\end{picture}%
\setlength{\unitlength}{4144sp}%
\begingroup\makeatletter\ifx\SetFigFont\undefined%
\gdef\SetFigFont#1#2#3#4#5{%
  \reset@font\fontsize{#1}{#2pt}%
  \fontfamily{#3}\fontseries{#4}\fontshape{#5}%
  \selectfont}%
\fi\endgroup%
\begin{picture}(1665,1275)(631,-736)
\put(631,-511){\makebox(0,0)[lb]{\smash{\SetFigFont{12}{14.4}{\rmdefault}{\mddefault}{\updefault}$b$}}}
\put(631,164){\makebox(0,0)[lb]{\smash{\SetFigFont{12}{14.4}{\rmdefault}{\mddefault}{\updefault}$a$}}}
\put(2296,-511){\makebox(0,0)[lb]{\smash{\SetFigFont{12}{14.4}{\rmdefault}{\mddefault}{\updefault}$d$}}}
\put(2296,164){\makebox(0,0)[lb]{\smash{\SetFigFont{12}{14.4}{\rmdefault}{\mddefault}{\updefault}$c$}}}
\put(1396,-736){\makebox(0,0)[lb]{\smash{\SetFigFont{12}{14.4}{\rmdefault}{\mddefault}{\updefault}$\beta_{bd}$}}}
\put(1396,344){\makebox(0,0)[lb]{\smash{\SetFigFont{12}{14.4}{\rmdefault}{\mddefault}{\updefault}$\beta_{ac}$}}}
\end{picture}

%% file: halbleiter.pstex_t
\begin{picture}(0,0)%
\includegraphics{halbleiter.pstex}%
\end{picture}%
\setlength{\unitlength}{4144sp}%
\begingroup\makeatletter\ifx\SetFigFont\undefined%
\gdef\SetFigFont#1#2#3#4#5{%
  \reset@font\fontsize{#1}{#2pt}%
  \fontfamily{#3}\fontseries{#4}\fontshape{#5}%
  \selectfont}%
\fi\endgroup%
\begin{picture}(1339,2278)(1801,-1380)
\put(2836,-601){\makebox(0,0)[lb]{\smash{\SetFigFont{12}{14.4}{\rmdefault}{\mddefault}{\updefault}$\vdots$}}}
\put(1801,344){\makebox(0,0)[lb]{\smash{\SetFigFont{12}{14.4}{\rmdefault}{\mddefault}{\updefault}$\kf_1, x_1$}}}
\put(1801,-61){\makebox(0,0)[lb]{\smash{\SetFigFont{12}{14.4}{\rmdefault}{\mddefault}{\updefault}$\kf_2, x_2$}}}
\put(1801,-1006){\makebox(0,0)[lb]{\smash{\SetFigFont{12}{14.4}{\rmdefault}{\mddefault}{\updefault}$\kf_n, x_n$}}}
\end{picture}

%% file: regge2gluon.pstex_t
\begin{picture}(0,0)%
\includegraphics{regge2gluon.pstex}%
\end{picture}%
\setlength{\unitlength}{4144sp}%
\begingroup\makeatletter\ifx\SetFigFont\undefined%
\gdef\SetFigFont#1#2#3#4#5{%
  \reset@font\fontsize{#1}{#2pt}%
  \fontfamily{#3}\fontseries{#4}\fontshape{#5}%
  \selectfont}%
\fi\endgroup%
\begin{picture}(1559,1724)(1699,-1778)
\put(2476,-961){\makebox(0,0)[b]{\smash{\SetFigFont{10}{12.0}{\rmdefault}{\mddefault}{\updefault}$\phi_\tpommi$}}}
\put(2476,-1636){\makebox(0,0)[b]{\smash{\SetFigFont{10}{12.0}{\rmdefault}{\mddefault}{\updefault}$\phi_2$}}}
\put(2476,-286){\makebox(0,0)[b]{\smash{\SetFigFont{10}{12.0}{\rmdefault}{\mddefault}{\updefault}$\phi_1$}}}
\end{picture}

%% file: limit2to2.pstex_t
\begin{picture}(0,0)%
\includegraphics{limit2to2.pstex}%
\end{picture}%
\setlength{\unitlength}{3947sp}%
\begingroup\makeatletter\ifx\SetFigFont\undefined
% extract first six characters in \fmtname
\def\x#1#2#3#4#5#6#7\relax{\def\x{#1#2#3#4#5#6}}%
\expandafter\x\fmtname xxxxxx\relax \def\y{splain}%
\ifx\x\y   % LaTeX or SliTeX?
\gdef\SetFigFont#1#2#3{%
  \ifnum #1<17\tiny\else \ifnum #1<20\small\else
  \ifnum #1<24\normalsize\else \ifnum #1<29\large\else
  \ifnum #1<34\Large\else \ifnum #1<41\LARGE\else
     \huge\fi\fi\fi\fi\fi\fi
  \csname #3\endcsname}%
\else
\gdef\SetFigFont#1#2#3{\begingroup
  \count@#1\relax \ifnum 25<\count@\count@25\fi
  \def\x{\endgroup\@setsize\SetFigFont{#2pt}}%
  \expandafter\x
    \csname \romannumeral\the\count@ pt\expandafter\endcsname
    \csname @\romannumeral\the\count@ pt\endcsname
  \csname #3\endcsname}%
\fi
\fi\endgroup
\begin{picture}(624,624)(1489,-1498)
\end{picture}

%% file: ladder.pstex_t
\begin{picture}(0,0)%
\includegraphics{ladder.pstex}%
\end{picture}%
\setlength{\unitlength}{3947sp}%
\begingroup\makeatletter\ifx\SetFigFont\undefined
% extract first six characters in \fmtname
\def\x#1#2#3#4#5#6#7\relax{\def\x{#1#2#3#4#5#6}}%
\expandafter\x\fmtname xxxxxx\relax \def\y{splain}%
\ifx\x\y   % LaTeX or SliTeX?
\gdef\SetFigFont#1#2#3{%
  \ifnum #1<17\tiny\else \ifnum #1<20\small\else
  \ifnum #1<24\normalsize\else \ifnum #1<29\large\else
  \ifnum #1<34\Large\else \ifnum #1<41\LARGE\else
     \huge\fi\fi\fi\fi\fi\fi
  \csname #3\endcsname}%
\else
\gdef\SetFigFont#1#2#3{\begingroup
  \count@#1\relax \ifnum 25<\count@\count@25\fi
  \def\x{\endgroup\@setsize\SetFigFont{#2pt}}%
  \expandafter\x
    \csname \romannumeral\the\count@ pt\expandafter\endcsname
    \csname @\romannumeral\the\count@ pt\endcsname
  \csname #3\endcsname}%
\fi
\fi\endgroup
\begin{picture}(502,980)(3486,-1715)
\end{picture}

%% file: poperator.pstex_t
\begin{picture}(0,0)%
\includegraphics{poperator.pstex}%
\end{picture}%
\setlength{\unitlength}{4144sp}%
\begingroup\makeatletter\ifx\SetFigFont\undefined%
\gdef\SetFigFont#1#2#3#4#5{%
  \reset@font\fontsize{#1}{#2pt}%
  \fontfamily{#3}\fontseries{#4}\fontshape{#5}%
  \selectfont}%
\fi\endgroup%
\begin{picture}(417,1033)(1261,-704)
\put(1261,-646){\makebox(0,0)[lb]{\smash{\SetFigFont{12}{14.4}{\rmdefault}{\mddefault}{\updefault}$\phi_1$}}}
\put(1666,-646){\makebox(0,0)[lb]{\smash{\SetFigFont{12}{14.4}{\rmdefault}{\mddefault}{\updefault}$\phi_2$}}}
\put(1576,164){\makebox(0,0)[lb]{\smash{\SetFigFont{12}{14.4}{\rmdefault}{\mddefault}{\updefault}$\P$}}}
\end{picture}

%% file: pgreen.pstex_t
\begin{picture}(0,0)%
\includegraphics{pgreen.pstex}%
\end{picture}%
\setlength{\unitlength}{4144sp}%
\begingroup\makeatletter\ifx\SetFigFont\undefined%
\gdef\SetFigFont#1#2#3#4#5{%
  \reset@font\fontsize{#1}{#2pt}%
  \fontfamily{#3}\fontseries{#4}\fontshape{#5}%
  \selectfont}%
\fi\endgroup%
\begin{picture}(417,1339)(1261,-704)
\put(1261,-646){\makebox(0,0)[lb]{\smash{\SetFigFont{12}{14.4}{\rmdefault}{\mddefault}{\updefault}$\phi_{1'}$}}}
\put(1666,-646){\makebox(0,0)[lb]{\smash{\SetFigFont{12}{14.4}{\rmdefault}{\mddefault}{\updefault}$\phi_{2'}$}}}
\put(1261,479){\makebox(0,0)[lb]{\smash{\SetFigFont{12}{14.4}{\rmdefault}{\mddefault}{\updefault}$\phi_1$}}}
\put(1666,479){\makebox(0,0)[lb]{\smash{\SetFigFont{12}{14.4}{\rmdefault}{\mddefault}{\updefault}$\phi_2$}}}
\end{picture}

%% file: diffusion.pstex_t
\begin{picture}(0,0)%
\includegraphics{diffusion.pstex}%
\end{picture}%
\setlength{\unitlength}{3729sp}%
\begingroup\makeatletter\ifx\SetFigFont\undefined%
\gdef\SetFigFont#1#2#3#4#5{%
  \reset@font\fontsize{#1}{#2pt}%
  \fontfamily{#3}\fontseries{#4}\fontshape{#5}%
  \selectfont}%
\fi\endgroup%
\begin{picture}(3510,3612)(316,-4021)
\put(631,-2311){\makebox(0,0)[lb]{\smash{\SetFigFont{11}{13.2}{\rmdefault}{\mddefault}{\updefault}$t$}}}
\put(316,-1726){\makebox(0,0)[lb]{\smash{\SetFigFont{11}{13.2}{\rmdefault}{\mddefault}{\updefault}$\ln \kf^2$}}}
\put(3826,-2311){\makebox(0,0)[lb]{\smash{\SetFigFont{11}{13.2}{\rmdefault}{\mddefault}{\updefault}$t'$}}}
\put(2251,-4021){\makebox(0,0)[b]{\smash{\SetFigFont{11}{13.2}{\rmdefault}{\mddefault}{\updefault}$\Delta Y$}}}
\put(2251,-3526){\makebox(0,0)[b]{\smash{\SetFigFont{11}{13.2}{\rmdefault}{\mddefault}{\updefault}NP region}}}
\end{picture}

%% file: regge3gluon.pstex_t
\begin{picture}(0,0)%
\includegraphics{regge3gluon.pstex}%
\end{picture}%
\setlength{\unitlength}{3947sp}%
\begingroup\makeatletter\ifx\SetFigFont\undefined%
\gdef\SetFigFont#1#2#3#4#5{%
  \reset@font\fontsize{#1}{#2pt}%
  \fontfamily{#3}\fontseries{#4}\fontshape{#5}%
  \selectfont}%
\fi\endgroup%
\begin{picture}(2274,2264)(289,-1943)
\put(1426,-1756){\makebox(0,0)[b]{\smash{\SetFigFont{10}{12.0}{\rmdefault}{\mddefault}{\updefault}$\phi_2$}}}
\put(1426,-848){\makebox(0,0)[b]{\smash{\SetFigFont{10}{12.0}{\rmdefault}{\mddefault}{\updefault}$\phi_\soddi$}}}
\put(1426, 44){\makebox(0,0)[b]{\smash{\SetFigFont{10}{12.0}{\rmdefault}{\mddefault}{\updefault}$\phi_1$}}}
\end{picture}

%% file: oddleiter.pstex_t
\begin{picture}(0,0)%
\includegraphics{oddleiter.pstex}%
\end{picture}%
\setlength{\unitlength}{4144sp}%
\begingroup\makeatletter\ifx\SetFigFont\undefined%
\gdef\SetFigFont#1#2#3#4#5{%
  \reset@font\fontsize{#1}{#2pt}%
  \fontfamily{#3}\fontseries{#4}\fontshape{#5}%
  \selectfont}%
\fi\endgroup%
\begin{picture}(924,1284)(799,-793)
\end{picture}

%% file: oddioperator.pstex_t
\begin{picture}(0,0)%
\includegraphics{oddioperator.pstex}%
\end{picture}%
\setlength{\unitlength}{4144sp}%
\begingroup\makeatletter\ifx\SetFigFont\undefined%
\gdef\SetFigFont#1#2#3#4#5{%
  \reset@font\fontsize{#1}{#2pt}%
  \fontfamily{#3}\fontseries{#4}\fontshape{#5}%
  \selectfont}%
\fi\endgroup%
\begin{picture}(540,1033)(1216,-704)
\put(1576,164){\makebox(0,0)[lb]{\smash{\SetFigFont{12}{14.4}{\rmdefault}{\mddefault}{\updefault}$\Od$}}}
\put(1756,-646){\makebox(0,0)[lb]{\smash{\SetFigFont{12}{14.4}{\rmdefault}{\mddefault}{\updefault}$\phi_3$}}}
\put(1216,-646){\makebox(0,0)[lb]{\smash{\SetFigFont{12}{14.4}{\rmdefault}{\mddefault}{\updefault}$\phi_1$}}}
\put(1486,-646){\makebox(0,0)[lb]{\smash{\SetFigFont{12}{14.4}{\rmdefault}{\mddefault}{\updefault}$\phi_2$}}}
\end{picture}

%% file: oddigreen.pstex_t
\begin{picture}(0,0)%
\includegraphics{oddigreen.pstex}%
\end{picture}%
\setlength{\unitlength}{4144sp}%
\begingroup\makeatletter\ifx\SetFigFont\undefined%
\gdef\SetFigFont#1#2#3#4#5{%
  \reset@font\fontsize{#1}{#2pt}%
  \fontfamily{#3}\fontseries{#4}\fontshape{#5}%
  \selectfont}%
\fi\endgroup%
\begin{picture}(540,1339)(1216,-704)
\put(1486,479){\makebox(0,0)[lb]{\smash{\SetFigFont{12}{14.4}{\rmdefault}{\mddefault}{\updefault}$\phi_2$}}}
\put(1756,-646){\makebox(0,0)[lb]{\smash{\SetFigFont{12}{14.4}{\rmdefault}{\mddefault}{\updefault}$\phi_{3'}$}}}
\put(1216,-646){\makebox(0,0)[lb]{\smash{\SetFigFont{12}{14.4}{\rmdefault}{\mddefault}{\updefault}$\phi_{1'}$}}}
\put(1216,479){\makebox(0,0)[lb]{\smash{\SetFigFont{12}{14.4}{\rmdefault}{\mddefault}{\updefault}$\phi_1$}}}
\put(1756,479){\makebox(0,0)[lb]{\smash{\SetFigFont{12}{14.4}{\rmdefault}{\mddefault}{\updefault}$\phi_3$}}}
\put(1486,-646){\makebox(0,0)[lb]{\smash{\SetFigFont{12}{14.4}{\rmdefault}{\mddefault}{\updefault}$\phi_{2'}$}}}
\end{picture}

%% file: solutiond41.pstex_t
\begin{picture}(0,0)%
\includegraphics{solutiond41.pstex}%
\end{picture}%
\setlength{\unitlength}{3947sp}%
\begingroup\makeatletter\ifx\SetFigFont\undefined
% extract first six characters in \fmtname
\def\x#1#2#3#4#5#6#7\relax{\def\x{#1#2#3#4#5#6}}%
\expandafter\x\fmtname xxxxxx\relax \def\y{splain}%
\ifx\x\y   % LaTeX or SliTeX?
\gdef\SetFigFont#1#2#3{%
  \ifnum #1<17\tiny\else \ifnum #1<20\small\else
  \ifnum #1<24\normalsize\else \ifnum #1<29\large\else
  \ifnum #1<34\Large\else \ifnum #1<41\LARGE\else
     \huge\fi\fi\fi\fi\fi\fi
  \csname #3\endcsname}%
\else
\gdef\SetFigFont#1#2#3{\begingroup
  \count@#1\relax \ifnum 25<\count@\count@25\fi
  \def\x{\endgroup\@setsize\SetFigFont{#2pt}}%
  \expandafter\x
    \csname \romannumeral\the\count@ pt\expandafter\endcsname
    \csname @\romannumeral\the\count@ pt\endcsname
  \csname #3\endcsname}%
\fi
\fi\endgroup
\begin{picture}(865,966)(2373,-2016)
\end{picture}

%% file: solu42.pstex_t
\begin{picture}(0,0)%
\includegraphics{solu42.pstex}%
\end{picture}%
\setlength{\unitlength}{4144sp}%
\begingroup\makeatletter\ifx\SetFigFont\undefined%
\gdef\SetFigFont#1#2#3#4#5{%
  \reset@font\fontsize{#1}{#2pt}%
  \fontfamily{#3}\fontseries{#4}\fontshape{#5}%
  \selectfont}%
\fi\endgroup%
\begin{picture}(907,1398)(3875,-2389)
\put(4681,-1951){\makebox(0,0)[lb]{\smash{\SetFigFont{10}{12.0}{\rmdefault}{\mddefault}{\updefault}$V_{2 \to 4}$}}}
\end{picture}

%% file: impactdiag.pstex_t
\begin{picture}(0,0)%
\epsfig{file=impactdiag.pstex}%
\end{picture}%
\setlength{\unitlength}{2072sp}%
\begingroup\makeatletter\ifx\SetFigFont\undefined%
\gdef\SetFigFont#1#2#3#4#5{%
  \reset@font\fontsize{#1}{#2pt}%
  \fontfamily{#3}\fontseries{#4}\fontshape{#5}%
  \selectfont}%
\fi\endgroup%
\begin{picture}(9719,5639)(395,-5236)
\put(7876,-2086){\makebox(0,0)[b]{\smash{\SetFigFont{10}{12.0}{\rmdefault}{\mddefault}{\updefault}(b)}}}
\put(2701,-2086){\makebox(0,0)[b]{\smash{\SetFigFont{10}{12.0}{\rmdefault}{\mddefault}{\updefault}(a)}}}
\put(5176,-5236){\makebox(0,0)[b]{\smash{\SetFigFont{10}{12.0}{\rmdefault}{\mddefault}{\updefault}(c)}}}
\end{picture}

%% file: allt.pstex_t
\begin{picture}(0,0)%
\includegraphics{allt.pstex}%
\end{picture}%
\setlength{\unitlength}{3947sp}%
\begingroup\makeatletter\ifx\SetFigFont\undefined%
\gdef\SetFigFont#1#2#3#4#5{%
  \reset@font\fontsize{#1}{#2pt}%
  \fontfamily{#3}\fontseries{#4}\fontshape{#5}%
  \selectfont}%
\fi\endgroup%
\begin{picture}(5798,3477)(305,-3061)
\put(1005,-2627){\makebox(0,0)[rb]{\smash{\SetFigFont{10}{12.0}{\familydefault}{\mddefault}{\updefault}1e-06}}}
\put(1005,-2278){\makebox(0,0)[rb]{\smash{\SetFigFont{10}{12.0}{\familydefault}{\mddefault}{\updefault}1e-05}}}
\put(1005,-1930){\makebox(0,0)[rb]{\smash{\SetFigFont{10}{12.0}{\familydefault}{\mddefault}{\updefault}0.0001}}}
\put(1005,-1581){\makebox(0,0)[rb]{\smash{\SetFigFont{10}{12.0}{\familydefault}{\mddefault}{\updefault}0.001}}}
\put(1005,-1232){\makebox(0,0)[rb]{\smash{\SetFigFont{10}{12.0}{\familydefault}{\mddefault}{\updefault}0.01}}}
\put(1005,-883){\makebox(0,0)[rb]{\smash{\SetFigFont{10}{12.0}{\familydefault}{\mddefault}{\updefault}0.1}}}
\put(1005,-535){\makebox(0,0)[rb]{\smash{\SetFigFont{10}{12.0}{\familydefault}{\mddefault}{\updefault}1}}}
\put(1005,-186){\makebox(0,0)[rb]{\smash{\SetFigFont{10}{12.0}{\familydefault}{\mddefault}{\updefault}10}}}
\put(1005,163){\makebox(0,0)[rb]{\smash{\SetFigFont{10}{12.0}{\familydefault}{\mddefault}{\updefault}100}}}
\put(1079,-2751){\makebox(0,0)[b]{\smash{\SetFigFont{10}{12.0}{\familydefault}{\mddefault}{\updefault}0}}}
\put(1775,-2751){\makebox(0,0)[b]{\smash{\SetFigFont{10}{12.0}{\familydefault}{\mddefault}{\updefault}0.5}}}
\put(2471,-2751){\makebox(0,0)[b]{\smash{\SetFigFont{10}{12.0}{\familydefault}{\mddefault}{\updefault}1}}}
\put(3167,-2751){\makebox(0,0)[b]{\smash{\SetFigFont{10}{12.0}{\familydefault}{\mddefault}{\updefault}1.5}}}
\put(3863,-2751){\makebox(0,0)[b]{\smash{\SetFigFont{10}{12.0}{\familydefault}{\mddefault}{\updefault}2}}}
\put(4560,-2751){\makebox(0,0)[b]{\smash{\SetFigFont{10}{12.0}{\familydefault}{\mddefault}{\updefault}2.5}}}
\put(5256,-2751){\makebox(0,0)[b]{\smash{\SetFigFont{10}{12.0}{\familydefault}{\mddefault}{\updefault}3}}}
\put(5952,-2751){\makebox(0,0)[b]{\smash{\SetFigFont{10}{12.0}{\familydefault}{\mddefault}{\updefault}3.5}}}
\put(425,-1149){\rotatebox{90.0}{\makebox(0,0)[b]{\smash{\SetFigFont{10}{12.0}{\familydefault}{\mddefault}{\updefault}$d\sigma/dt \,[\mbox{mb}/\mbox{GeV}^2]$}}}}
\put(3585,-3061){\makebox(0,0)[b]{\smash{\SetFigFont{10}{12.0}{\familydefault}{\mddefault}{\updefault}$-t\, [\mbox{GeV}^2]$}}}
\put(5529,180){\makebox(0,0)[rb]{\smash{\SetFigFont{10}{12.0}{\familydefault}{\mddefault}{\updefault}$pp$}}}
\put(5529, 32){\makebox(0,0)[rb]{\smash{\SetFigFont{10}{12.0}{\familydefault}{\mddefault}{\updefault}$p\bar{p}$}}}
\put(5529,-116){\makebox(0,0)[rb]{\smash{\SetFigFont{10}{12.0}{\familydefault}{\mddefault}{\updefault}DL $pp$}}}
\put(5529,-264){\makebox(0,0)[rb]{\smash{\SetFigFont{10}{12.0}{\familydefault}{\mddefault}{\updefault}DL $p\bar{p}$}}}
\end{picture}

%% file: gnldlcomp.pstex_t
\begin{picture}(0,0)%
\includegraphics{gnldlcomp.pstex}%
\end{picture}%
\setlength{\unitlength}{3947sp}%
\begingroup\makeatletter\ifx\SetFigFont\undefined%
\gdef\SetFigFont#1#2#3#4#5{%
  \reset@font\fontsize{#1}{#2pt}%
  \fontfamily{#3}\fontseries{#4}\fontshape{#5}%
  \selectfont}%
\fi\endgroup%
\begin{picture}(5798,3495)(305,-3061)
\put(1005,-2627){\makebox(0,0)[rb]{\smash{\SetFigFont{10}{12.0}{\familydefault}{\mddefault}{\updefault}1e-06}}}
\put(1005,-2205){\makebox(0,0)[rb]{\smash{\SetFigFont{10}{12.0}{\familydefault}{\mddefault}{\updefault}1e-05}}}
\put(1005,-1782){\makebox(0,0)[rb]{\smash{\SetFigFont{10}{12.0}{\familydefault}{\mddefault}{\updefault}0.0001}}}
\put(1005,-1360){\makebox(0,0)[rb]{\smash{\SetFigFont{10}{12.0}{\familydefault}{\mddefault}{\updefault}0.001}}}
\put(1005,-938){\makebox(0,0)[rb]{\smash{\SetFigFont{10}{12.0}{\familydefault}{\mddefault}{\updefault}0.01}}}
\put(1005,-516){\makebox(0,0)[rb]{\smash{\SetFigFont{10}{12.0}{\familydefault}{\mddefault}{\updefault}0.1}}}
\put(1005,-93){\makebox(0,0)[rb]{\smash{\SetFigFont{10}{12.0}{\familydefault}{\mddefault}{\updefault}1}}}
\put(1005,329){\makebox(0,0)[rb]{\smash{\SetFigFont{10}{12.0}{\familydefault}{\mddefault}{\updefault}10}}}
\put(1392,-2751){\makebox(0,0)[b]{\smash{\SetFigFont{10}{12.0}{\familydefault}{\mddefault}{\updefault}0.6}}}
\put(2019,-2751){\makebox(0,0)[b]{\smash{\SetFigFont{10}{12.0}{\familydefault}{\mddefault}{\updefault}0.8}}}
\put(2645,-2751){\makebox(0,0)[b]{\smash{\SetFigFont{10}{12.0}{\familydefault}{\mddefault}{\updefault}1}}}
\put(3272,-2751){\makebox(0,0)[b]{\smash{\SetFigFont{10}{12.0}{\familydefault}{\mddefault}{\updefault}1.2}}}
\put(3898,-2751){\makebox(0,0)[b]{\smash{\SetFigFont{10}{12.0}{\familydefault}{\mddefault}{\updefault}1.4}}}
\put(4525,-2751){\makebox(0,0)[b]{\smash{\SetFigFont{10}{12.0}{\familydefault}{\mddefault}{\updefault}1.6}}}
\put(5151,-2751){\makebox(0,0)[b]{\smash{\SetFigFont{10}{12.0}{\familydefault}{\mddefault}{\updefault}1.8}}}
\put(5778,-2751){\makebox(0,0)[b]{\smash{\SetFigFont{10}{12.0}{\familydefault}{\mddefault}{\updefault}2}}}
\put(425,-1149){\rotatebox{90.0}{\makebox(0,0)[b]{\smash{\SetFigFont{10}{12.0}{\familydefault}{\mddefault}{\updefault}$d\sigma/dt\, [\mbox{mb}/\mbox{GeV}^2]$}}}}
\put(3585,-3061){\makebox(0,0)[b]{\smash{\SetFigFont{10}{12.0}{\familydefault}{\mddefault}{\updefault}$-t\, [\mbox{GeV}^2]$}}}
\put(5498,180){\makebox(0,0)[rb]{\smash{\SetFigFont{8}{9.6}{\familydefault}{\mddefault}{\updefault}DL}}}
\put(5498, 32){\makebox(0,0)[rb]{\smash{\SetFigFont{8}{9.6}{\familydefault}{\mddefault}{\updefault}GLN (maximal $\Od$)}}}
\put(5498,-116){\makebox(0,0)[rb]{\smash{\SetFigFont{8}{9.6}{\familydefault}{\mddefault}{\updefault}GLN with DL $\Od$}}}
\put(5498,-264){\makebox(0,0)[rb]{\smash{\SetFigFont{8}{9.6}{\familydefault}{\mddefault}{\updefault}DL with maximal $\Od$}}}
\end{picture}

%% file: neurho.pstex_t
\begin{picture}(0,0)%
\includegraphics{neurho.pstex}%
\end{picture}%
\setlength{\unitlength}{3947sp}%
\begingroup\makeatletter\ifx\SetFigFont\undefined%
\gdef\SetFigFont#1#2#3#4#5{%
  \reset@font\fontsize{#1}{#2pt}%
  \fontfamily{#3}\fontseries{#4}\fontshape{#5}%
  \selectfont}%
\fi\endgroup%
\begin{picture}(5802,3540)(301,-3106)
\put(857,-2627){\makebox(0,0)[rb]{\smash{\SetFigFont{10}{12.0}{\familydefault}{\mddefault}{\updefault}-0.4}}}
\put(857,-2258){\makebox(0,0)[rb]{\smash{\SetFigFont{10}{12.0}{\familydefault}{\mddefault}{\updefault}-0.3}}}
\put(857,-1888){\makebox(0,0)[rb]{\smash{\SetFigFont{10}{12.0}{\familydefault}{\mddefault}{\updefault}-0.2}}}
\put(857,-1519){\makebox(0,0)[rb]{\smash{\SetFigFont{10}{12.0}{\familydefault}{\mddefault}{\updefault}-0.1}}}
\put(857,-1149){\makebox(0,0)[rb]{\smash{\SetFigFont{10}{12.0}{\familydefault}{\mddefault}{\updefault}0}}}
\put(857,-779){\makebox(0,0)[rb]{\smash{\SetFigFont{10}{12.0}{\familydefault}{\mddefault}{\updefault}0.1}}}
\put(857,-410){\makebox(0,0)[rb]{\smash{\SetFigFont{10}{12.0}{\familydefault}{\mddefault}{\updefault}0.2}}}
\put(857,-40){\makebox(0,0)[rb]{\smash{\SetFigFont{10}{12.0}{\familydefault}{\mddefault}{\updefault}0.3}}}
\put(857,329){\makebox(0,0)[rb]{\smash{\SetFigFont{10}{12.0}{\familydefault}{\mddefault}{\updefault}0.4}}}
\put(1517,-2751){\makebox(0,0)[b]{\smash{\SetFigFont{10}{12.0}{\familydefault}{\mddefault}{\updefault}10}}}
\put(2490,-2751){\makebox(0,0)[b]{\smash{\SetFigFont{10}{12.0}{\familydefault}{\mddefault}{\updefault}100}}}
\put(3464,-2751){\makebox(0,0)[b]{\smash{\SetFigFont{10}{12.0}{\familydefault}{\mddefault}{\updefault}$10^3$}}}
\put(4437,-2751){\makebox(0,0)[b]{\smash{\SetFigFont{10}{12.0}{\familydefault}{\mddefault}{\updefault}$10^4$}}}
\put(5411,-2751){\makebox(0,0)[b]{\smash{\SetFigFont{10}{12.0}{\familydefault}{\mddefault}{\updefault}$10^5$}}}
\put(3511,-3061){\makebox(0,0)[b]{\smash{\SetFigFont{10}{12.0}{\familydefault}{\mddefault}{\updefault}$\sqrt{s}\, [\mbox{GeV}]$}}}
\put(5386,-1736){\makebox(0,0)[rb]{\smash{\SetFigFont{10}{12.0}{\familydefault}{\mddefault}{\updefault}$pp$}}}
\put(5386,-1884){\makebox(0,0)[rb]{\smash{\SetFigFont{10}{12.0}{\familydefault}{\mddefault}{\updefault}$p\bar{p}$}}}
\put(5386,-2032){\makebox(0,0)[rb]{\smash{\SetFigFont{10}{12.0}{\familydefault}{\mddefault}{\updefault}max $\Od$ $pp$}}}
\put(5386,-2180){\makebox(0,0)[rb]{\smash{\SetFigFont{10}{12.0}{\familydefault}{\mddefault}{\updefault}max $\Od$ $p\bar{p}$}}}
\put(5386,-2328){\makebox(0,0)[rb]{\smash{\SetFigFont{10}{12.0}{\familydefault}{\mddefault}{\updefault}$p\bar{p}$ UA4}}}
\put(301,-1149){\makebox(0,0)[b]{\smash{\SetFigFont{10}{12.0}{\familydefault}{\mddefault}{\updefault}$\rho$}}}
\end{picture}

%% file: drhodata.pstex_t
\begin{picture}(0,0)%
\includegraphics{drhodata.pstex}%
\end{picture}%
\setlength{\unitlength}{3947sp}%
\begingroup\makeatletter\ifx\SetFigFont\undefined%
\gdef\SetFigFont#1#2#3#4#5{%
  \reset@font\fontsize{#1}{#2pt}%
  \fontfamily{#3}\fontseries{#4}\fontshape{#5}%
  \selectfont}%
\fi\endgroup%
\begin{picture}(5937,3464)(166,-3061)
\put(931,-2751){\makebox(0,0)[b]{\smash{\SetFigFont{10}{12.0}{\rmdefault}{\mddefault}{\updefault}0}}}
\put(2221,-2751){\makebox(0,0)[b]{\smash{\SetFigFont{10}{12.0}{\rmdefault}{\mddefault}{\updefault}0.4}}}
\put(2866,-2751){\makebox(0,0)[b]{\smash{\SetFigFont{10}{12.0}{\rmdefault}{\mddefault}{\updefault}0.6}}}
\put(3511,-2751){\makebox(0,0)[b]{\smash{\SetFigFont{10}{12.0}{\rmdefault}{\mddefault}{\updefault}0.8}}}
\put(4156,-2751){\makebox(0,0)[b]{\smash{\SetFigFont{10}{12.0}{\rmdefault}{\mddefault}{\updefault}1}}}
\put(4801,-2751){\makebox(0,0)[b]{\smash{\SetFigFont{10}{12.0}{\rmdefault}{\mddefault}{\updefault}1.2}}}
\put(5446,-2751){\makebox(0,0)[b]{\smash{\SetFigFont{10}{12.0}{\rmdefault}{\mddefault}{\updefault}1.4}}}
\put(6091,-2751){\makebox(0,0)[b]{\smash{\SetFigFont{10}{12.0}{\rmdefault}{\mddefault}{\updefault}1.6}}}
\put(1576,-2751){\makebox(0,0)[b]{\smash{\SetFigFont{10}{12.0}{\rmdefault}{\mddefault}{\updefault}0.2}}}
\put(857,-1170){\makebox(0,0)[rb]{\smash{\SetFigFont{10}{12.0}{\familydefault}{\mddefault}{\updefault}$-0.3$}}}
\put(857,-742){\makebox(0,0)[rb]{\smash{\SetFigFont{10}{12.0}{\familydefault}{\mddefault}{\updefault}$-0.2$}}}
\put(857,115){\makebox(0,0)[rb]{\smash{\SetFigFont{10}{12.0}{\familydefault}{\mddefault}{\updefault}0}}}
\put(3511,-3061){\makebox(0,0)[b]{\smash{\SetFigFont{10}{12.0}{\familydefault}{\mddefault}{\updefault}$d$\,[\mbox{fm}]}}}
\put(857,-1599){\makebox(0,0)[rb]{\smash{\SetFigFont{10}{12.0}{\familydefault}{\mddefault}{\updefault}$-0.4$}}}
\put(857,-2027){\makebox(0,0)[rb]{\smash{\SetFigFont{10}{12.0}{\familydefault}{\mddefault}{\updefault}$-0.5$}}}
\put(857,-314){\makebox(0,0)[rb]{\smash{\SetFigFont{10}{12.0}{\familydefault}{\mddefault}{\updefault}$-0.1$}}}
\put(857,-2456){\makebox(0,0)[rb]{\smash{\SetFigFont{10}{12.0}{\familydefault}{\mddefault}{\updefault}$-0.6$}}}
\put(301,-1149){\rotatebox{90.0}{\makebox(0,0)[b]{\smash{\SetFigFont{10}{12.0}{\familydefault}{\mddefault}{\updefault}$\Delta \rho$}}}}
\end{picture}

%% file: ppjpsi.pstex_t
\begin{picture}(0,0)%
\includegraphics{ppjpsi.pstex}%
\end{picture}%
\setlength{\unitlength}{3315sp}%
\begingroup\makeatletter\ifx\SetFigFont\undefined%
\gdef\SetFigFont#1#2#3#4#5{%
  \reset@font\fontsize{#1}{#2pt}%
  \fontfamily{#3}\fontseries{#4}\fontshape{#5}%
  \selectfont}%
\fi\endgroup%
\begin{picture}(2610,2511)(226,-2216)
\put(226,-196){\makebox(0,0)[lb]{\smash{\SetFigFont{10}{12.0}{\rmdefault}{\mddefault}{\updefault}$p$}}}
\put(226,-1816){\makebox(0,0)[lb]{\smash{\SetFigFont{10}{12.0}{\rmdefault}{\mddefault}{\updefault}$\pbar$}}}
\put(2836,-1141){\makebox(0,0)[lb]{\smash{\SetFigFont{10}{12.0}{\rmdefault}{\mddefault}{\updefault}$J/\psi$}}}
\put(1216,-601){\makebox(0,0)[lb]{\smash{\SetFigFont{10}{12.0}{\rmdefault}{\mddefault}{\updefault}$\P$}}}
\put(1216,-1411){\makebox(0,0)[lb]{\smash{\SetFigFont{10}{12.0}{\rmdefault}{\mddefault}{\updefault}$\Od$}}}
\end{picture}

%% file: pofupert2.pstex_t
\begin{picture}(0,0)%
\includegraphics{pofupert2.pstex}%
\end{picture}%
\setlength{\unitlength}{4144sp}%
\begingroup\makeatletter\ifx\SetFigFont\undefined%
\gdef\SetFigFont#1#2#3#4#5{%
  \reset@font\fontsize{#1}{#2pt}%
  \fontfamily{#3}\fontseries{#4}\fontshape{#5}%
  \selectfont}%
\fi\endgroup%
\begin{picture}(4702,2325)(-359,-2761)
\put(-359,-2266){\makebox(0,0)[lb]{\smash{\SetFigFont{12}{14.4}{\rmdefault}{\mddefault}{\updefault}$\Od$}}}
\put(-359,-1141){\makebox(0,0)[lb]{\smash{\SetFigFont{12}{14.4}{\rmdefault}{\mddefault}{\updefault}$\P$}}}
\put(1081,-1906){\makebox(0,0)[lb]{\smash{\SetFigFont{12}{14.4}{\rmdefault}{\mddefault}{\updefault}$J/\psi$}}}
\put(4186,-1906){\makebox(0,0)[lb]{\smash{\SetFigFont{12}{14.4}{\rmdefault}{\mddefault}{\updefault}$J/\psi$}}}
\put(2836,-2446){\makebox(0,0)[lb]{\smash{\SetFigFont{12}{14.4}{\rmdefault}{\mddefault}{\updefault}$\underbrace{\hspace*{1.7cm}}$}}}
\put(2971,-781){\makebox(0,0)[lb]{\smash{\SetFigFont{12}{14.4}{\rmdefault}{\mddefault}{\updefault}$\overbrace{\hspace*{1.2cm}}$}}}
\put(3196,-601){\makebox(0,0)[lb]{\smash{\SetFigFont{12}{14.4}{\rmdefault}{\mddefault}{\updefault}$\P$}}}
\put(2026,-1726){\makebox(0,0)[lb]{\smash{\SetFigFont{12}{14.4}{\rmdefault}{\mddefault}{\updefault}{\huge $\Sigma$}}}}
\put(1576,-1681){\makebox(0,0)[lb]{\smash{\SetFigFont{12}{14.4}{\rmdefault}{\mddefault}{\updefault}$=$}}}
\put(3151,-2761){\makebox(0,0)[lb]{\smash{\SetFigFont{12}{14.4}{\rmdefault}{\mddefault}{\updefault}$\Od$}}}
\end{picture}

%% file: ppjj.pstex_t
\begin{picture}(0,0)%
\includegraphics{ppjj.pstex}%
\end{picture}%
\setlength{\unitlength}{3729sp}%
\begingroup\makeatletter\ifx\SetFigFont\undefined%
\gdef\SetFigFont#1#2#3#4#5{%
  \reset@font\fontsize{#1}{#2pt}%
  \fontfamily{#3}\fontseries{#4}\fontshape{#5}%
  \selectfont}%
\fi\endgroup%
\begin{picture}(5535,2379)(496,-2150)
\put(2566,-736){\makebox(0,0)[lb]{\smash{\SetFigFont{11}{13.2}{\rmdefault}{\mddefault}{\updefault}$J/\psi$}}}
\put(496,-151){\makebox(0,0)[lb]{\smash{\SetFigFont{11}{13.2}{\rmdefault}{\mddefault}{\updefault}$p$}}}
\put(1261,-376){\makebox(0,0)[lb]{\smash{\SetFigFont{11}{13.2}{\rmdefault}{\mddefault}{\updefault}$\P$}}}
\put(1441,-1006){\makebox(0,0)[lb]{\smash{\SetFigFont{11}{13.2}{\rmdefault}{\mddefault}{\updefault}$\Od$}}}
\put(496,-1861){\makebox(0,0)[lb]{\smash{\SetFigFont{11}{13.2}{\rmdefault}{\mddefault}{\updefault}$\pbar$}}}
\put(2611,-1456){\makebox(0,0)[lb]{\smash{\SetFigFont{11}{13.2}{\rmdefault}{\mddefault}{\updefault}$J/\psi$}}}
\put(1261,-1681){\makebox(0,0)[lb]{\smash{\SetFigFont{11}{13.2}{\rmdefault}{\mddefault}{\updefault}$\P$}}}
\put(5986,-736){\makebox(0,0)[lb]{\smash{\SetFigFont{11}{13.2}{\rmdefault}{\mddefault}{\updefault}$J/\psi$}}}
\put(3916,-151){\makebox(0,0)[lb]{\smash{\SetFigFont{11}{13.2}{\rmdefault}{\mddefault}{\updefault}$p$}}}
\put(4681,-376){\makebox(0,0)[lb]{\smash{\SetFigFont{11}{13.2}{\rmdefault}{\mddefault}{\updefault}$\Od$}}}
\put(4861,-1006){\makebox(0,0)[lb]{\smash{\SetFigFont{11}{13.2}{\rmdefault}{\mddefault}{\updefault}$\P$}}}
\put(3916,-1861){\makebox(0,0)[lb]{\smash{\SetFigFont{11}{13.2}{\rmdefault}{\mddefault}{\updefault}$\pbar$}}}
\put(6031,-1456){\makebox(0,0)[lb]{\smash{\SetFigFont{11}{13.2}{\rmdefault}{\mddefault}{\updefault}$J/\psi$}}}
\put(4681,-1681){\makebox(0,0)[lb]{\smash{\SetFigFont{11}{13.2}{\rmdefault}{\mddefault}{\updefault}$\Od$}}}
\end{picture}

%% file: diffmes.pstex_t
\begin{picture}(0,0)%
\includegraphics{diffmes.pstex}%
\end{picture}%
\setlength{\unitlength}{4144sp}%
\begingroup\makeatletter\ifx\SetFigFont\undefined%
\gdef\SetFigFont#1#2#3#4#5{%
  \reset@font\fontsize{#1}{#2pt}%
  \fontfamily{#3}\fontseries{#4}\fontshape{#5}%
  \selectfont}%
\fi\endgroup%
\begin{picture}(1777,1894)(586,-2150)
\put(676,-1861){\makebox(0,0)[lb]{\smash{\SetFigFont{12}{14.4}{\rmdefault}{\mddefault}{\updefault}$p$}}}
\put(2341,-1366){\makebox(0,0)[lb]{\smash{\SetFigFont{12}{14.4}{\rmdefault}{\mddefault}{\updefault}$M_{\mbox{\scriptsize PS}}$}}}
\put(1576,-421){\makebox(0,0)[lb]{\smash{\SetFigFont{12}{14.4}{\rmdefault}{\mddefault}{\updefault}$e^-$}}}
\put(1126,-1051){\makebox(0,0)[lb]{\smash{\SetFigFont{12}{14.4}{\rmdefault}{\mddefault}{\updefault}$\gamma^{(*)}$}}}
\put(1306,-1591){\makebox(0,0)[lb]{\smash{\SetFigFont{12}{14.4}{\rmdefault}{\mddefault}{\updefault}$\Od$}}}
\put(586,-646){\makebox(0,0)[lb]{\smash{\SetFigFont{12}{14.4}{\rmdefault}{\mddefault}{\updefault}$e^-$}}}
\end{picture}

%% file: kina.pstex_t
\begin{picture}(0,0)%
\includegraphics{kina.pstex}%
\end{picture}%
\setlength{\unitlength}{3729sp}%
\begingroup\makeatletter\ifx\SetFigFont\undefined%
\gdef\SetFigFont#1#2#3#4#5{%
  \reset@font\fontsize{#1}{#2pt}%
  \fontfamily{#3}\fontseries{#4}\fontshape{#5}%
  \selectfont}%
\fi\endgroup%
\begin{picture}(6222,3962)(3559,-5597)
\put(8416,-2041){\makebox(0,0)[lb]{\smash{\SetFigFont{9}{10.8}{\rmdefault}{\mddefault}{\updefault}without $\Od$}}}
\put(6643,-5236){\makebox(0,0)[b]{\smash{\SetFigFont{9}{10.8}{\rmdefault}{\mddefault}{\updefault}0.4}}}
\put(7672,-5236){\makebox(0,0)[b]{\smash{\SetFigFont{9}{10.8}{\rmdefault}{\mddefault}{\updefault}0.6}}}
\put(8719,-5236){\makebox(0,0)[b]{\smash{\SetFigFont{9}{10.8}{\rmdefault}{\mddefault}{\updefault}0.8}}}
\put(9772,-5236){\makebox(0,0)[b]{\smash{\SetFigFont{9}{10.8}{\rmdefault}{\mddefault}{\updefault}1.0}}}
\put(5593,-5236){\makebox(0,0)[b]{\smash{\SetFigFont{9}{10.8}{\rmdefault}{\mddefault}{\updefault}0.2}}}
\put(4543,-5236){\makebox(0,0)[b]{\smash{\SetFigFont{9}{10.8}{\rmdefault}{\mddefault}{\updefault}0}}}
\put(7153,-5548){\makebox(0,0)[b]{\smash{\SetFigFont{9}{10.8}{\rmdefault}{\mddefault}{\updefault}$p_T\,[\mbox{GeV}]$}}}
\put(8416,-2311){\makebox(0,0)[lb]{\smash{\SetFigFont{9}{10.8}{\rmdefault}{\mddefault}{\updefault}$\eta_\soddi=+1$}}}
\put(8416,-2581){\makebox(0,0)[lb]{\smash{\SetFigFont{9}{10.8}{\rmdefault}{\mddefault}{\updefault}$\eta_\soddi=-1$}}}
\put(4321,-2161){\makebox(0,0)[rb]{\smash{\SetFigFont{9}{10.8}{\rmdefault}{\mddefault}{\updefault}1000}}}
\put(4321,-2986){\makebox(0,0)[rb]{\smash{\SetFigFont{9}{10.8}{\rmdefault}{\mddefault}{\updefault}100}}}
\put(4321,-3803){\makebox(0,0)[rb]{\smash{\SetFigFont{9}{10.8}{\rmdefault}{\mddefault}{\updefault}10}}}
\put(4321,-4628){\makebox(0,0)[rb]{\smash{\SetFigFont{9}{10.8}{\rmdefault}{\mddefault}{\updefault}1}}}
\put(3691,-3298){\rotatebox{90.0}{\makebox(0,0)[b]{\smash{\SetFigFont{9}{10.8}{\rmdefault}{\mddefault}{\updefault}$d\sigma/dp_T\,[\mbox{pb}/\mbox{GeV}]$}}}}
\end{picture}

%% file: rdn.pstex_t
\begin{picture}(0,0)%
\includegraphics{rdn.pstex}%
\end{picture}%
\setlength{\unitlength}{3947sp}%
\begingroup\makeatletter\ifx\SetFigFont\undefined%
\gdef\SetFigFont#1#2#3#4#5{%
  \reset@font\fontsize{#1}{#2pt}%
  \fontfamily{#3}\fontseries{#4}\fontshape{#5}%
  \selectfont}%
\fi\endgroup%
\begin{picture}(5835,3495)(305,-3061)
\put(1005,-2627){\makebox(0,0)[rb]{\smash{\SetFigFont{10}{12.0}{\familydefault}{\mddefault}{\updefault}0.0001}}}
\put(1005,-2205){\makebox(0,0)[rb]{\smash{\SetFigFont{10}{12.0}{\familydefault}{\mddefault}{\updefault}0.001}}}
\put(1005,-1782){\makebox(0,0)[rb]{\smash{\SetFigFont{10}{12.0}{\familydefault}{\mddefault}{\updefault}0.01}}}
\put(1005,-1360){\makebox(0,0)[rb]{\smash{\SetFigFont{10}{12.0}{\familydefault}{\mddefault}{\updefault}0.1}}}
\put(1005,-938){\makebox(0,0)[rb]{\smash{\SetFigFont{10}{12.0}{\familydefault}{\mddefault}{\updefault}1}}}
\put(1005,-516){\makebox(0,0)[rb]{\smash{\SetFigFont{10}{12.0}{\familydefault}{\mddefault}{\updefault}10}}}
\put(1005,-93){\makebox(0,0)[rb]{\smash{\SetFigFont{10}{12.0}{\familydefault}{\mddefault}{\updefault}100}}}
\put(1005,329){\makebox(0,0)[rb]{\smash{\SetFigFont{10}{12.0}{\familydefault}{\mddefault}{\updefault}1000}}}
\put(1079,-2751){\makebox(0,0)[b]{\smash{\SetFigFont{10}{12.0}{\familydefault}{\mddefault}{\updefault}0}}}
\put(2081,-2751){\makebox(0,0)[b]{\smash{\SetFigFont{10}{12.0}{\familydefault}{\mddefault}{\updefault}1}}}
\put(3084,-2751){\makebox(0,0)[b]{\smash{\SetFigFont{10}{12.0}{\familydefault}{\mddefault}{\updefault}2}}}
\put(4086,-2751){\makebox(0,0)[b]{\smash{\SetFigFont{10}{12.0}{\familydefault}{\mddefault}{\updefault}3}}}
\put(5089,-2751){\makebox(0,0)[b]{\smash{\SetFigFont{10}{12.0}{\familydefault}{\mddefault}{\updefault}4}}}
\put(6091,-2751){\makebox(0,0)[b]{\smash{\SetFigFont{10}{12.0}{\familydefault}{\mddefault}{\updefault}5}}}
\put(3585,-3061){\makebox(0,0)[b]{\smash{\SetFigFont{10}{12.0}{\familydefault}{\mddefault}{\updefault}$Q^2$ [$\mbox{GeV}^2$]}}}
\put(3585,-556){\makebox(0,0)[lb]{\smash{\SetFigFont{10}{12.0}{\familydefault}{\mddefault}{\updefault}proton dissociation}}}
\put(425,-1149){\rotatebox{90.0}{\makebox(0,0)[b]{\smash{\SetFigFont{10}{12.0}{\familydefault}{\mddefault}{\updefault}$\sigma [\mbox{nb}]$}}}}
\put(3585,-1186){\makebox(0,0)[lb]{\smash{\SetFigFont{10}{12.0}{\familydefault}{\mddefault}{\updefault}proton intact, Mercedes star}}}
\put(3585,-1848){\makebox(0,0)[lb]{\smash{\SetFigFont{10}{12.0}{\familydefault}{\mddefault}{\updefault}proton intact, $d=0.338\,\mbox{fm}$}}}
\end{picture}

%% file: etaimpact.pstex_t
\begin{picture}(0,0)%
\includegraphics{etaimpact.pstex}%
\end{picture}%
\setlength{\unitlength}{4144sp}%
\begingroup\makeatletter\ifx\SetFigFont\undefined%
\gdef\SetFigFont#1#2#3#4#5{%
  \reset@font\fontsize{#1}{#2pt}%
  \fontfamily{#3}\fontseries{#4}\fontshape{#5}%
  \selectfont}%
\fi\endgroup%
\begin{picture}(2722,1455)(1621,-2611)
\put(1621,-1501){\makebox(0,0)[lb]{\smash{\SetFigFont{12}{14.4}{\rmdefault}{\mddefault}{\updefault}$\gamma^{(*)}$}}}
\put(4321,-1816){\makebox(0,0)[lb]{\smash{\SetFigFont{12}{14.4}{\rmdefault}{\mddefault}{\updefault}$\eta_c$}}}
\end{picture}

%% file: etacfinal.pstex_t
\begin{picture}(0,0)%
\includegraphics{etacfinal.pstex}%
\end{picture}%
\setlength{\unitlength}{4144sp}%
\begingroup\makeatletter\ifx\SetFigFont\undefined%
\gdef\SetFigFont#1#2#3#4#5{%
  \reset@font\fontsize{#1}{#2pt}%
  \fontfamily{#3}\fontseries{#4}\fontshape{#5}%
  \selectfont}%
\fi\endgroup%
\begin{picture}(4742,3350)(1141,-3181)
\put(1876,-2881){\makebox(0,0)[b]{\smash{\SetFigFont{10}{12.0}{\rmdefault}{\mddefault}{\updefault}0}}}
\put(2266,-2881){\makebox(0,0)[b]{\smash{\SetFigFont{10}{12.0}{\rmdefault}{\mddefault}{\updefault}0.1}}}
\put(2666,-2881){\makebox(0,0)[b]{\smash{\SetFigFont{10}{12.0}{\rmdefault}{\mddefault}{\updefault}0.2}}}
\put(3061,-2881){\makebox(0,0)[b]{\smash{\SetFigFont{10}{12.0}{\rmdefault}{\mddefault}{\updefault}0.3}}}
\put(3461,-2881){\makebox(0,0)[b]{\smash{\SetFigFont{10}{12.0}{\rmdefault}{\mddefault}{\updefault}0.4}}}
\put(3861,-2881){\makebox(0,0)[b]{\smash{\SetFigFont{10}{12.0}{\rmdefault}{\mddefault}{\updefault}0.5}}}
\put(5056,-2881){\makebox(0,0)[b]{\smash{\SetFigFont{10}{12.0}{\rmdefault}{\mddefault}{\updefault}0.8}}}
\put(4256,-2881){\makebox(0,0)[b]{\smash{\SetFigFont{10}{12.0}{\rmdefault}{\mddefault}{\updefault}0.6}}}
\put(5461,-2881){\makebox(0,0)[b]{\smash{\SetFigFont{10}{12.0}{\rmdefault}{\mddefault}{\updefault}0.9}}}
\put(5866,-2881){\makebox(0,0)[b]{\smash{\SetFigFont{10}{12.0}{\rmdefault}{\mddefault}{\updefault}1}}}
\put(1741,-2151){\makebox(0,0)[rb]{\smash{\SetFigFont{10}{12.0}{\rmdefault}{\mddefault}{\updefault}0.1}}}
\put(1741,-2716){\makebox(0,0)[rb]{\smash{\SetFigFont{10}{12.0}{\rmdefault}{\mddefault}{\updefault}0.01}}}
\put(1741,-1051){\makebox(0,0)[rb]{\smash{\SetFigFont{10}{12.0}{\rmdefault}{\mddefault}{\updefault}10}}}
\put(1741,-486){\makebox(0,0)[rb]{\smash{\SetFigFont{10}{12.0}{\rmdefault}{\mddefault}{\updefault}100}}}
\put(1741, 64){\makebox(0,0)[rb]{\smash{\SetFigFont{10}{12.0}{\rmdefault}{\mddefault}{\updefault}1000}}}
\put(1741,-1601){\makebox(0,0)[rb]{\smash{\SetFigFont{10}{12.0}{\rmdefault}{\mddefault}{\updefault}1}}}
\put(4661,-2881){\makebox(0,0)[b]{\smash{\SetFigFont{10}{12.0}{\rmdefault}{\mddefault}{\updefault}0.7}}}
\put(3861,-3181){\makebox(0,0)[b]{\smash{\SetFigFont{10}{12.0}{\rmdefault}{\mddefault}{\updefault}$-t\,[\mbox{GeV}^2]$}}}
\put(1261,-1326){\rotatebox{90.0}{\makebox(0,0)[b]{\smash{\SetFigFont{10}{12.0}{\rmdefault}{\mddefault}{\updefault}$d\sigma/dt \,[\mbox{pb}/\mbox{GeV}^2]$}}}}
\end{picture}

%% file: poocombine.pstex_t
\begin{picture}(0,0)%
\includegraphics{poocombine.pstex}%
\end{picture}%
\setlength{\unitlength}{3522sp}%
\begingroup\makeatletter\ifx\SetFigFont\undefined%
\gdef\SetFigFont#1#2#3#4#5{%
  \reset@font\fontsize{#1}{#2pt}%
  \fontfamily{#3}\fontseries{#4}\fontshape{#5}%
  \selectfont}%
\fi\endgroup%
\begin{picture}(7437,2364)(529,-2233)
\put(631,-781){\makebox(0,0)[lb]{\smash{\SetFigFont{11}{13.2}{\rmdefault}{\mddefault}{\updefault}$e^-$}}}
\put(1306,-196){\makebox(0,0)[lb]{\smash{\SetFigFont{11}{13.2}{\rmdefault}{\mddefault}{\updefault}$e^-$}}}
\put(1531,-1726){\makebox(0,0)[lb]{\smash{\SetFigFont{11}{13.2}{\rmdefault}{\mddefault}{\updefault}$p$}}}
\put(2251,-421){\makebox(0,0)[lb]{\smash{\SetFigFont{11}{13.2}{\rmdefault}{\mddefault}{\updefault}$M_{\mbox{\scriptsize PS}}$}}}
\put(1891,-1231){\makebox(0,0)[lb]{\smash{\SetFigFont{11}{13.2}{\rmdefault}{\mddefault}{\updefault}$\Od$}}}
\put(1396,-916){\makebox(0,0)[lb]{\smash{\SetFigFont{11}{13.2}{\rmdefault}{\mddefault}{\updefault}$\gamma^{(*)}$}}}
\put(3286,-61){\makebox(0,0)[lb]{\smash{\SetFigFont{11}{13.2}{\rmdefault}{\mddefault}{\updefault}$2$}}}
\put(3511,-1096){\makebox(0,0)[lb]{\smash{\SetFigFont{11}{13.2}{\rmdefault}{\mddefault}{\updefault}$=$}}}
\put(4816,-1726){\makebox(0,0)[lb]{\smash{\SetFigFont{11}{13.2}{\rmdefault}{\mddefault}{\updefault}$p$}}}
\put(5626,-1546){\makebox(0,0)[lb]{\smash{\SetFigFont{11}{13.2}{\rmdefault}{\mddefault}{\updefault}$\P$}}}
\put(5311,-1141){\makebox(0,0)[lb]{\smash{\SetFigFont{11}{13.2}{\rmdefault}{\mddefault}{\updefault}$\Od$}}}
\put(6301,-1141){\makebox(0,0)[lb]{\smash{\SetFigFont{11}{13.2}{\rmdefault}{\mddefault}{\updefault}$\Od$}}}
\put(7111,-916){\makebox(0,0)[rb]{\smash{\SetFigFont{11}{13.2}{\rmdefault}{\mddefault}{\updefault}$\gamma^{(*)}$}}}
\put(3916,-781){\makebox(0,0)[lb]{\smash{\SetFigFont{11}{13.2}{\rmdefault}{\mddefault}{\updefault}$e^-$}}}
\put(4591,-196){\makebox(0,0)[lb]{\smash{\SetFigFont{11}{13.2}{\rmdefault}{\mddefault}{\updefault}$e^-$}}}
\put(4681,-916){\makebox(0,0)[lb]{\smash{\SetFigFont{11}{13.2}{\rmdefault}{\mddefault}{\updefault}$\gamma^{(*)}$}}}
\put(7291,-196){\makebox(0,0)[rb]{\smash{\SetFigFont{11}{13.2}{\rmdefault}{\mddefault}{\updefault}$e^-$}}}
\put(7966,-781){\makebox(0,0)[rb]{\smash{\SetFigFont{11}{13.2}{\rmdefault}{\mddefault}{\updefault}$e^-$}}}
\put(6931,-1726){\makebox(0,0)[lb]{\smash{\SetFigFont{11}{13.2}{\rmdefault}{\mddefault}{\updefault}$p$}}}
\put(5356,-421){\makebox(0,0)[lb]{\smash{\SetFigFont{11}{13.2}{\rmdefault}{\mddefault}{\updefault}$M_{\mbox{\scriptsize PS}}$}}}
\end{picture}

%% file: diffcc.pstex_t
\begin{picture}(0,0)%
\includegraphics{diffcc.pstex}%
\end{picture}%
\setlength{\unitlength}{3729sp}%
\begingroup\makeatletter\ifx\SetFigFont\undefined%
\gdef\SetFigFont#1#2#3#4#5{%
  \reset@font\fontsize{#1}{#2pt}%
  \fontfamily{#3}\fontseries{#4}\fontshape{#5}%
  \selectfont}%
\fi\endgroup%
\begin{picture}(1800,1894)(586,-2150)
\put(676,-1861){\makebox(0,0)[lb]{\smash{\SetFigFont{11}{13.2}{\rmdefault}{\mddefault}{\updefault}$p$}}}
\put(1576,-421){\makebox(0,0)[lb]{\smash{\SetFigFont{11}{13.2}{\rmdefault}{\mddefault}{\updefault}$e^-$}}}
\put(586,-646){\makebox(0,0)[lb]{\smash{\SetFigFont{11}{13.2}{\rmdefault}{\mddefault}{\updefault}$e^-$}}}
\put(1126,-1051){\makebox(0,0)[lb]{\smash{\SetFigFont{11}{13.2}{\rmdefault}{\mddefault}{\updefault}$\gamma^{(*)}$}}}
\put(2386,-1141){\makebox(0,0)[lb]{\smash{\SetFigFont{11}{13.2}{\rmdefault}{\mddefault}{\updefault}$c$}}}
\put(2386,-1456){\makebox(0,0)[lb]{\smash{\SetFigFont{11}{13.2}{\rmdefault}{\mddefault}{\updefault}$\bar{c}$}}}
\put(1126,-1591){\makebox(0,0)[lb]{\smash{\SetFigFont{11}{13.2}{\rmdefault}{\mddefault}{\updefault}$\P/\Od$}}}
\end{picture}

%% file: gximpact.pstex_t
\begin{picture}(0,0)%
\includegraphics{gximpact.pstex}%
\end{picture}%
\setlength{\unitlength}{4144sp}%
\begingroup\makeatletter\ifx\SetFigFont\undefined%
\gdef\SetFigFont#1#2#3#4#5{%
  \reset@font\fontsize{#1}{#2pt}%
  \fontfamily{#3}\fontseries{#4}\fontshape{#5}%
  \selectfont}%
\fi\endgroup%
\begin{picture}(2182,1427)(1621,-2726)
\put(1621,-1591){\makebox(0,0)[lb]{\smash{\SetFigFont{12}{14.4}{\rmdefault}{\mddefault}{\updefault}$\gamma^{(*)}$}}}
\end{picture}